\documentclass[12pt]{article}
\begin{document}
\newcommand{\beq}{\begin{equation}}
\newcommand{\eeq}{\end{equation}}
\newcommand{\beqa}{\begin{eqnarray}}
\newcommand{\eeqa}{\end{eqnarray}}
\newcommand{\beqar}{\begin{eqnarray*}}
\newcommand{\eeqar}{\end{eqnarray*}}
\newcommand{\al}{\alpha}
\newcommand{\be}{\beta}
\newcommand{\del}{\delta}
\newcommand{\D}{\Delta}
\newcommand{\eps}{\epsilon}
\newcommand{\ga}{\gamma}
\newcommand{\Ga}{\Gamma}
\newcommand{\ka}{\kappa}
\newcommand{\nn}{\nonumber}
\newcommand{\inn}{\!\cdot\!}
\newcommand{\h}{\eta}
\newcommand{\ii}{\iota}
\newcommand{\kk}{\varphi}
\newcommand\F{{}_3F_2}
\newcommand{\la}{\lambda}
\newcommand{\La}{\Lambda}
\newcommand{\na}{\prt}
\newcommand{\Om}{\Omega}
\newcommand{\om}{\omega}
\newcommand{\p}{\phi}
\newcommand{\sig}{\sigma}
\renewcommand{\t}{\theta}
\newcommand{\z}{\zeta}
\newcommand{\ssc}{\scriptscriptstyle}
\newcommand{\eg}{{\it e.g.,}\ }
\newcommand{\ie}{{\it i.e.,}\ }
\newcommand{\labell}[1]{\label{#1}} %{\label{#1}} %
\newcommand{\reef}[1]{(\ref{#1})}
\newcommand\prt{\partial}
\newcommand\veps{\varepsilon}
\newcommand{\pol}{\varepsilon}
\newcommand\vp{\varphi}
\newcommand\ls{\ell_s}
\newcommand\cF{{\cal F}}
\newcommand\cA{{\cal A}}
\newcommand\cS{{\cal S}}
\newcommand\cT{{\cal T}}
\newcommand\cV{{\cal V}}
\newcommand\cL{{\cal L}}
\newcommand\cM{{\cal M}}
\newcommand\cN{{\cal N}}
\newcommand\cG{{\cal G}}
\newcommand\cH{{\cal H}}
\newcommand\cI{{\cal I}}
\newcommand\cJ{{\cal J}}
\newcommand\cl{{\iota}}
\newcommand\cP{{\cal P}}
\newcommand\cQ{{\cal Q}}
\newcommand\cg{{\it g}}
\newcommand\cR{{\cal R}}
\newcommand\cB{{\cal B}}
\newcommand\cO{{\cal O}}
\newcommand\tcO{{\tilde {{\cal O}}}}
\newcommand\bg{\bar{g}}
\newcommand\bb{\bar{b}}
\newcommand\bH{\bar{H}}
\newcommand\bX{\bar{X}}
\newcommand\bK{\bar{K}}
\newcommand\bA{\bar{A}}
\newcommand\bZ{\bar{Z}}
\newcommand\bxi{\bar{\xi}}
\newcommand\bphi{\bar{\phi}}
\newcommand\bpsi{\bar{\psi}}
\newcommand\bprt{\bar{\prt}}
\newcommand\bet{\bar{\eta}}
\newcommand\btau{\bar{\tau}}
\newcommand\hF{\hat{F}}
\newcommand\hA{\hat{A}}
\newcommand\hT{\hat{T}}
\newcommand\htau{\hat{\tau}}
\newcommand\hD{\hat{D}}
\newcommand\hf{\hat{f}}
\newcommand\hg{\hat{g}}
\newcommand\hp{\hat{\phi}}
\newcommand\hi{\hat{i}}
\newcommand\ha{\hat{a}}
\newcommand\hb{\hat{b}}
\newcommand\hQ{\hat{Q}}
\newcommand\hP{\hat{\Phi}}
\newcommand\hS{\hat{S}}
\newcommand\hX{\hat{X}}
\newcommand\tL{\tilde{\cal L}}
\newcommand\hL{\hat{\cal L}}
\newcommand\tG{{\widetilde G}}
\newcommand\tg{{\widetilde g}}
\newcommand\tphi{{\widetilde \phi}}
\newcommand\tPhi{{\widetilde \Phi}}
\newcommand\te{{\tilde e}}
\newcommand\tk{{\tilde k}}
\newcommand\tf{{\tilde f}}
\newcommand\ta{{\tilde a}}
\newcommand\tb{{\tilde b}}
\newcommand\tR{{\tilde R}}
\newcommand\teta{{\tilde \eta}}
\newcommand\tF{{\widetilde F}}
\newcommand\tK{{\widetilde K}}
\newcommand\tE{{\widetilde E}}
\newcommand\tpsi{{\tilde \psi}}
\newcommand\tX{{\widetilde X}}
\newcommand\tD{{\widetilde D}}
\newcommand\tO{{\widetilde O}}
\newcommand\tS{{\tilde S}}
\newcommand\tB{{\widetilde B}}
\newcommand\tA{{\widetilde A}}
\newcommand\tT{{\widetilde T}}
\newcommand\tC{{\widetilde C}}
\newcommand\tV{{\widetilde V}}
\newcommand\thF{{\widetilde {\hat {F}}}}
\newcommand\Tr{{\rm Tr}}
\newcommand\tr{{\rm tr}}
\newcommand\STr{{\rm STr}}
\newcommand\hR{\hat{R}}
\newcommand\M[2]{M^{#1}{}_{#2}}

\newcommand\bS{\textbf{ S}}
\newcommand\bI{\textbf{ I}}
\newcommand\bJ{\textbf{ J}}

%\begin{document}
\begin{titlepage}
\begin{center}

\vskip 2 cm
{\LARGE \bf   Minimal  gauge invariant couplings\\  \vskip 0.25 cm at  order $\ell_p^6$ in M-theory
 }\\
\vskip 1.25 cm
 Mohammad R. Garousi\footnote{garousi@um.ac.ir}
 
\vskip 1 cm
{{\it Department of Physics, Faculty of Science, Ferdowsi University of Mashhad\\}{\it P.O. Box 1436, Mashhad, Iran}\\}
\vskip .1 cm
 \end{center}
\begin{abstract}
 Removing   the field redefinitions, the  Bianchi identities and the total derivative  freedoms from the general form of  the gauge invariant couplings  at order $\ell_p^6$ for the bosonic fields of M-theory,   we have found that the minimum number of  independent couplings in the structures with even number of the three-form, is  1062. We find that there are  schemes in which there is no coupling involving   $R,\,R_{\mu\nu},\,\nabla_\mu F^{\mu\alpha\beta\gamma}$. In these schemes,  there are sub-schemes in which, except one coupling which has the second derivative of $F^{(4)}$, the couplings can have no term with more than two derivatives. We  find some of the parameters  by  dimensionally reducing the couplings  on a circle and comparing them    with the known couplings of the one-loop effective action  of type IIA superstring theory. In particular, we find the coupling which has term with more than two derivatives is  zero.

\end{abstract}
\end{titlepage}
%\newpage
\tableofcontents

%\newpage
\section{Introduction} \label{intro}

M-theory is a consistent  quantum theory of  gravity  which includes all types of superstring theories at different limits \cite{Schwarz:1996bh}. In particular, the compactification of M-theory on a circle produces  the type IIA superstring theory.
A convinent  way to study different phenomena in this theory is to use an effective action  which is a derivative-expansionin of the theory  in terms of    its  massless fields \cite{Howe:1983sra,Witten:1995ex}. The leading order terms in this  expansion  is the 11-dimensional supergravity and the next to the leading order terms are at eight-derivative order or $\ell_p^6$-order in which we are interested. 
There are different techniques   to find such couplings \cite{Green:1997di,Green:1997as,Anguelova:2004pg,Cederwall:2000ye,Cederwall:2004cg,deHaro:2002vk,Howe:2003cy,Rajaraman:2005ag,Hyakutake:2007sm,Peeters:2005tb,Bakhtiarizadeh:2017ojz}. One of them is  the S-matrix method in which various S-matrix elements are calculated in the M-theory and then they are compared  with the corresponding S-matrix elements in the M-theory effective action. Another method is to use the dimensional reduction of the $\ell_p^6$-order  couplings on the circle and compare them with the one-loop effective action of the  type IIA superstring  theory at order $\alpha'^3$. To use  these methods, one needs to know the independent gauge invariant couplings at  order $\ell_p^6$.

 To find such independent couplings, one has to impose various Bianchi identities, use  field redefinitions freedom \cite{Gross:1986iv,Tseytlin:1986ti,Deser:1986xr} and remove total derivative terms from the most general gauge invariant couplings. Hence,  one should first write  all  gauge invariant couplings at  order $\ell_p^6$ and then imposes the above freedoms to reduce them  to the minimal couplings. The parameters in the gauge invariant  action are  either ambiguous or unambiguous depending on whether or not they are changed under these freedoms.   Some combinations of the ambiguous  parameters, however, remain invariant.  This allows one to separate the ambiguous parameters to essential parameters, and some arbitrary parameters. Depending on which set of parameters are choosing as essential parameters and how to choose the arbitrary parameters, one has different schemes. The minimum number of independent couplings are found in the schemes that all the arbitrary parameters are set to zero. This method has been used in \cite{Garousi:2019cdn} to find  60  independent gauge invariant couplings   at order $\alpha'^2$ in the bosonic string theory, and  in \cite{Garousi:2020mqn} to find 872 independent NS-NS couplings   at order $\alpha'^3$ in type II superstring theories. These parameters are then fixed in the tree-level effective action by the T-duality \cite{Garousi:2020gio,Garousi:2020lof}. We  are interested in  finding such independent couplings at order $\ell_p^6$ in M-theory for  bosonic fields. 
 The effective action of M-theory at each order of $\ell_p$ has two  sectors: The Chern-Simons sector which has odd number of the three-form and another  sector which has even number of the three-form. Each sector should be invariant under the parity transformation  which changes the sign of the three-form \cite{Duff:1986hr}.  In this paper we are interested in finding the independent couplings in the second sector at order $\ell_p^6$ . 

The outline of the paper is as follows: In section 2, using the package   "xAct" \cite{Nutma:2013zea}, we write the most general gauge invariant couplings involving the 11-dimensional metric $g_{\mu\nu}$ and the three-form  $A^{(3)}$ at order $\ell_p^6$. There are 17746 such  couplings. Then we add to them the most general total derivative terms and field redefinitions with arbitrary parameters. To impose various Bianchi identities, we rewrite  them in the local inertial frame in which the partial derivative of the metric is zero, and rewrite the terms which have derivatives of three-field strength $F^{(4)}$,  in terms of potential, \ie $F^{(4)}=dA^{(3)}$.   We then use the arbitrary parameters in the total derivative terms and in the field redefinitions  to show that there are only 1062 unambiguous and essential parameters and all other parameters are arbitrary which can be set to zero.  We show that there are minimal schemes in which there are 1061 couplings  which have no term with  more than two derivatives and  no term  involving $R,R_{\mu\nu},\nabla_\mu F^{\mu\alpha\beta\gamma}$. There is  one essential coupling which has two derivatives on $F$.  We write the explicit form of   this coupling, and  the couplings with structures  $F^6R$, $F^4 R^2$, $F^2R^3$, $R^4$, $F^2(\nabla F)^2$, and $(\nabla F)^4$ where $R$ stands for the Riemann curvature,  in this section, and the remaining couplings with the structures $F^8$, $F^4(\nabla F)^2$, and $RF^2(\nabla F)^2$,  in the Appendix. In section 3, we briefly discuss the dimensional reduction of the couplings on a circle to find  some of the  parameters involving four fields by comparing them with the known couplings in the one-loop effective action of type IIA superstring theory. In particular, this comparison dictates that the coupling which has  term with more than two derivatives is  zero.

\section{Minimal couplings at order  $\ell_p^6$}\label{sec.2}

The bosonic part of the effective action of M-theory has the following derivative-expansion or $\ell_p$-expansion where $\ell_p$ is the 11-dimensional Plank-length:
\beqa
S_{\rm eff}&=& \frac{1}{2\kappa_{11}^2}\Big[\int d^{11}x\sqrt{-g}( \cL_0+\ell_p^6\cL_6+\cdots)+\int( \cL_0^{CS}+\ell_p^6\cL_6^{CS}+\cdots)\Big]\labell{seff}
\eeqa
where we have used the fact that the  M-theory has no effective action at four and six derivative orders. In fact, as we will argue  in the next section, the orders of the derivative terms in the M-theory effective action are at $\ell_p^0,\ell_p^6,\ell_p^{12},\ell_P^{18},\ell_p^{24},\cdots$. 

The effective action must be invariant under the coordinate transformations and under the $A$-field  gauge transformations. The metric and $A$-field      must  appear in the Lagrangian $ \mathcal{L}_n$  trough their field strengths and their covariant derivatives, \eg
the Lagrangian $\cL$ at the leading order of $\ell_p$  is
\beqa
\mathcal{L}_0&= R-\frac{1}{2.4!}F_{\mu\nu\alpha\beta}F^{\mu\nu\alpha\beta}
\eeqa
 The  Chern-Simons form at the leading order is not invariant under the gauge transformation, \ie 
 \beqa
\cL^{CS}_0&=&-\frac{1}{6}A^{(3)}\wedge F^{(4)}\wedge F^{(4)}
 \eeqa
However, its corresponding action is invariant. The  11-dimensional supergravity is invariant under the parity transformation   which changes the sign of the $A$-field \cite{Duff:1986hr}. This  should be the symmetry of all higher-derivative terms in \reef{seff}. The parity symmetry then constrains $\cL$ to have even number of $A$-field and $\cL^{CS}$ to have odd number of $A$-field.  In this paper, we are interested only on the couplings in $\cL_6$. 
 A systematic method has been used in \cite{Garousi:2019cdn,Garousi:2020mqn} to find the minimum number of independent couplings at order $\alpha'^2$ and $\alpha'^3$ in the effective action of the bosonic string theory. It has been found that  there are 60 couplings at  order $\alpha'^2$ and 872 couplings at order $\alpha'^3$.  In this section,  we are going to use this method to find similar couplings in  $\cL$  at order $\ell_p^6$ in the M-theory.

Following \cite{Garousi:2019cdn}, one first should write all gauge invariant  couplings at eight-derivative order which has  even number of the three-form. Using the package   "xAct" \cite{Nutma:2013zea}, one finds there are 17746 such couplings in 40 different structures, \ie 
\beqa
 L'_6=& m'_1 F_{\alpha\nu}{}^{\delta \epsilon} F^{\alpha \beta \gamma\nu} \
F_{\beta\phi}{}^{\varepsilon \mu} F_{\gamma}{}^{\zeta \eta\phi} F_{\
\delta \varepsilon\sigma}{}^{\theta} F_{\epsilon \zeta}{}^{\iota\sigma} \
F_{\mu \iota}{}^{\kappa\lambda} F_{\eta \theta \kappa\lambda}+\cdots\labell{L3}
\eeqa
where $m'_1,\cdots, m'_{17746}$ are some parameters\footnote{Using a  computer with 32 GB RAM,  the package can generate all couplings excepts the couplings with structure $F^8$. The couplings in this structure which include $F_{\mu\nu\alpha\beta}F^{\mu\nu\alpha\beta}$, can be also generated by finding all couplings with structure $F^6$ and then multiplying them with $F_{\mu\nu\alpha\beta}F^{\mu\nu\alpha\beta}$. We found the couplings  which have no factor  $F_{\mu\nu\alpha\beta}F^{\mu\nu\alpha\beta}$,  as follows: We first find all couplings with structure $H^8$ where $H$ is a three-form. The package can generate such couplings. We then replace each three-form $H$  with the four-form  $F$ with one free index. The resulting couplings then  each has 8 free indices. We then contract all possible contractions of the free indices, and remove the couplings which have the factor  $F_{\mu\nu\alpha\beta}F^{\mu\nu\alpha\beta}$. Note that the number of latter   couplings  is two less than the number of  the couplings constructed by multiplying all contractions of $F^6$ with  $F_{\mu\nu\alpha\beta}F^{\mu\nu\alpha\beta}$. }. The above couplings however are  not all independent. Some of them are related by total derivative terms, some of them are related by field redefinitions, and some others are related by various Bianchi identities.

To remove the total derivative terms from the above couplings, we consider the most general total derivative terms at order $\ell_p^6$ which have the following structure:
\beqa
\frac{\ell_p^6}{2\kappa_{11}^2}\int d^{11}x \sqrt{-g} \mathcal{J}_6=\frac{\ell_p^6}{2\kappa_{11}^2}\int d^{11}x\sqrt{-g} \nabla_\alpha ({\cal I}_6^\alpha) \labell{J3}
\eeqa
where the vector ${\cal I}_6^\alpha$ is   all possible  covariant and gauge invariant  terms at seven-derivative level which has  even number of the three-form, \ie, 
\beqa
{\cal I}_6^\alpha= &J_1 F^{\gamma\delta\epsilon\mu}R^{\alpha\beta}R_{\beta\varepsilon\epsilon\theta}\nabla_{\delta}F_{\gamma\mu}{}^{\varepsilon\theta}+\cdots
\eeqa
where the coefficients  $J_1,\cdots, J_{7760}$ are 7760 arbitrary parameters. Adding the total derivative terms with arbitrary coefficients  to $L'_6$, one finds the same Lagrangian     but with different parameters $m''_1, m''_2, \cdots$. We call the new Lagrangian  ${ L}''_6$. Hence 
\beqa
\Delta''_6-{\cal J}_6&=&0\labell{DL}
\eeqa
where $\Delta''_6={ L}''_6-L'_6$ is the same as $L'_6$ but with coefficients $\delta m''_1,\delta m''_2,\cdots$ where $\delta m''_i= m''_i-m'_i$. Solving the above equation, one finds some linear  relations between  only $\delta m''_1,\delta m''_2,\cdots$ which indicate how the couplings are related among themselves by the total derivative terms. The above equation also gives some relations between the coefficients of the total derivative terms and $\delta m''_1,\delta m''_2,\cdots$ in which we are not interested.

The couplings in \reef{DL}, however, are in a  fixed field variables. One is free to change the field variables as 
\begin{eqnarray}
g_{\mu\nu}&\rightarrow &g_{\mu\nu}+\ell_p^6 \delta g^{(6)}_{\mu\nu}\nn\\
A_{\mu\nu\alpha}&\rightarrow &A_{\mu\nu\alpha}+ \ell_p^6\delta A^{(6)}_{\mu\nu\alpha}\labell{gbp}
\end{eqnarray}
where the tensors $\delta g^{(6)}_{\mu\nu}$ and $\delta A^{(6)}_{\mu\nu\alpha}$  are all possible covariant and gauge invariant terms at 6-derivative level.  The parity symmetry constrains $\delta g^{(6)}_{\mu\nu}$ to have even number of the three-form  and $\delta A^{(6)}_{\mu\nu\alpha}$ to have  odd number of the three-form, \ie,
\beqa
 \delta g^{(6)}_{\alpha\beta}&=& g_1 F_{\{\alpha}{}^{\gamma\delta\mu}F_{\beta\}\gamma}{}^{\epsilon\nu}F_{\delta\mu}{}^{\varepsilon\theta}F_{\epsilon\varepsilon\nu}{}^{\eta}F_{\theta}{}^{\lambda\kappa\sigma}F_{\eta\lambda\kappa\sigma}+\cdots \nn\\
 \delta A^{(6)}_{\alpha\beta\mu}&=& e_1 R^{\gamma\delta}R_{\delta\epsilon\varepsilon [\alpha}\nabla_{\beta }F_{\mu]\gamma}{}^{\epsilon\varepsilon}+\cdots \labell{eq.12}
\eeqa
The coefficients $g_1,\cdots, g_{1987}$ and  $e_1,\cdots, e_{2679}$  are arbitrary parameters. When the field variables in $\cL_6$  are changed according to the above  field redefinitions, they produce some couplings at orders $\ell_p^8$ and higher in which we are not interested in this paper. However, when the field variables in $S_0$  are changed,  up to some total derivative terms, the following   couplings  at order $\ell_p^6$ are produced:
\beqa
\delta S_0&\!\!\!\!\!=\!\!\!\!\!\!&\frac{\delta S_0}{\delta g_{\alpha\beta}}\delta g^{(3)}_{\alpha\beta}+\frac{\delta S_0}{\delta A_{\alpha\beta\mu}}\delta A^{(3)}_{\alpha\beta\mu}\equiv \frac{\ell_p^6}{2\kappa_{11}^2}\int d^{11}x\sqrt{-g}\mathcal{K}_6\nn\\
&\!\!\!\!\!=\!\!\!\!\!\!& \frac{\ell_p^6}{2\kappa_{11}^2}\int d^{11} x\sqrt{-g}\Big[\left(\frac{1}{6} \nabla_{\gamma}F^{\gamma\alpha \beta\mu }-\frac{1}{48.144}\epsilon^{\alpha\beta\mu\nu\gamma\lambda\theta\tau\kappa\sigma\zeta}F_{\nu\gamma\lambda\theta}F_{\tau\kappa\sigma\zeta}\right)\delta A^{(6)}_{\alpha\beta\mu} 
\labell{eq.13}\\
&&\qquad\qquad\qquad\qquad-(  R^{\alpha \beta}-\frac{1}{12}F^{\alpha \gamma \delta\mu} F^{\beta}{}_{\gamma \delta\mu})\delta g^{(6)}_{\alpha\beta}+( \frac{1}{2}R -\frac{1}{48} F_{\alpha \beta \gamma\nu} F^{\alpha \beta \gamma\nu})\delta g^{(6)\mu}{}_\mu) \Big]\nn
\eeqa
 Note that if $\delta A^{(6)}_{\mu\nu\alpha}$ included the even number of the $A$-field, then the couplings in the second line  would not be invariant under the parity. The second term in the second line above produces couplings in the Chern-Simons sector in which we are not interested in this paper, hence, we do not consider the effect of field redefinition $\delta A^{(6)}_{\alpha\beta\mu}$  on this term. Adding the total derivative terms and field redefinition terms  to $L'_6$, one finds the same Lagrangian     but with different parameters $m_1, m_2, \cdots$. We call the new Lagrangian  ${\cal L}_6$. Hence 
\beqa
\Delta_6-{\cal J}_6-{\cal K}_6&=&0\labell{DLK}
\eeqa
where $\Delta_6={ \cal L}_6-L'_6$ is the same as $L'_6$ but with coefficients $\delta m_1,\delta m_2,\cdots$ where $\delta m_i= m_i-m'_i$. Solving the above equation, one finds some linear  relations between  only $\delta m_1,\delta m_2,\cdots$ which indicate how the couplings are related among themselves by the total derivative and field redefinition terms. There are also many relations between $\delta m_1,\delta m_2,\cdots$ and the coefficients of total derivative terms and field redefinitions in which we are not interested,   

However, to solve the equation \reef{DLK} one should write it in terms of independent couplings, \ie    one has to impose the following Bianchi identities as well:
\beqa
 R_{\alpha[\beta\gamma\delta]}&=&0\nn\\
 \nabla_{[\mu}R_{\alpha\beta]\gamma\delta}&=&0\labell{bian}\\
\nabla_{[\mu}F_{\nu\alpha\beta\gamma]}&=&0\nn\\
{[}\nabla,\nabla{]}\mathcal{O}-R\mathcal{O}&=&0\nn
\eeqa
To impose the  Bianchi identities in non-gauge invariant form, one may rewrite the terms in \reef{DLK} in  the local frame in which the first partial derivative of metric is zero, and  rewrite the terms in \reef{DLK} which have derivatives of $F$ in terms of A-field, \ie $F=dA$.  In this way,  the Bianchi identities satisfy automatically \cite{Garousi:2019cdn}. In fact, writing the couplings in terms of potential rather than field strength, there would be no Bianchi identity at all. This way of imposing  the Bianchi identities is very easy to perform by the computer.    

Using the above  steps, one can rewrite the different terms on the left-hand side of \reef{DLK} in terms of independent but non-gauge invariant couplings. The solution to the equation \reef{DLK} then has two parts. One part is 1062 relations between only $\delta m_i$'s, and the other part is some relations between the coefficients of the total derivative terms, field redefinitions  and $\delta m_i$'s in which we are not interested. The number of relations in the first part gives the number of independent couplings in ${\cal L}_6$. In a particular scheme, one may set some of the coefficients in  $L_6'$ to zero, however, after replacing the non-zero terms in \reef{DLK}, the number of   relations between only $\delta m_i$'s should not be changed, \ie there must be always 1062 relations.  We set the coefficients of the couplings in $L_6'$ in which each term that has $R$, $R_{\mu\nu}$ or $\nabla_\mu F^{\mu\nu\alpha\beta}$ to be  zero.   After setting these coefficients to zero, there are still 1062 relations between  $\delta m_i$'s.  This means we are  allowed  to  remove these terms. 

%We find that there are two couplings which involve the scalar curvature that their parameters are unambiguous, \ie 
%\beqa
%{\cal L}_6\supset
%m_{133}^{} F_{\alpha \beta }{}^{\epsilon \varepsilon } 
%F^{\alpha \beta \gamma \delta } F_{\gamma \delta \epsilon 
%}{}^{\kappa } F_{\varepsilon }{}^{\lambda \mu \nu } F_{\kappa 
%\lambda }{}^{\sigma \tau } F_{\mu \nu \sigma \tau } R 
%+ m_{134}^{} F_{\alpha \beta }{}^{\epsilon \varepsilon } 
%F^{\alpha \beta \gamma \delta } F_{\gamma \delta \epsilon 
%\varepsilon } F_{\kappa \lambda }{}^{\sigma \tau } F^{\kappa 
%\lambda \mu \nu } F_{\mu \nu \sigma \tau } R\labell{mm}
%\eeqa
%Obviously there is no total derivative term with the above structure because it does not involve derivative of the scalar curvature  and $F$. Moreover,  the Bianchi identities does not affect these terms. The field redefinition  can not produce the above couplings either. Apart from the above two terms, one can set to zero the coefficients of all other terms that have the scalar curvature. 

We then try to set zero the couplings in  $L_6'$ which have term with more then two derivatives. Imposing this condition and  then solving \reef{DLK} again, one would find 1061 relations between only  $\delta m_i$'s. It means that at least  one of the independent couplings has terms with more than two derivatives. We have found this independent coupling to be 
\beqa
{\cal L}_6&\supset& m_{315}^{} F_{\epsilon }{}^{\mu \nu \sigma } R_{\alpha \gamma 
}{}^{\epsilon \varepsilon } R^{\alpha \beta \gamma \delta } 
\nabla_{\sigma }\nabla_{\delta }F_{\beta \varepsilon \mu \nu 
}\labell{T53}
\eeqa
The way we have found the above coupling is that we divided the couplings involving more than two derivatives to two parts. We then set the coefficients of one part to zero. If the corresponding equations in \reef{DLK} gives 1062 relations between the remaining  $\delta m_i$'s then that choice is allowed, otherwise the other part is allowed to be zero. Again we divided the  non-zero part to two parts and set half of them to zero. If the  corresponding equations in \reef{DLK} gives 1062 relations between the remaining  $\delta m_i$'s then that choice is allowed, otherwise the other part is allowed to be zero. Repeating this strategy one finds the above couplings is one of the independent couplings. Apart from the above coupling, all other couplings which have terms with more than two derivatives are allowed to be zero. There are still 3304 couplings which have no term with more than two derivatives and have no terms with structures $R,\,R_{\mu\nu},\,\nabla_\mu F^{\mu\nu\alpha\beta}$. Hence, there are still  many choices for choosing the  non-zero  coefficients such that they satisfy the 1062 relations $\delta m_i=0$.    In the particular scheme that we have chosen, there is  one coupling  appears in \reef{T53}, and the other 1061 couplings appear in the  9 structures  $F^6R$, $F^4 R^2$, $F^2R^3$, $R^4$, $F^2(\nabla F)^2$, $(\nabla F)^4$, $F^8$, $F^4(\nabla F)^2$,  $R F^2(\nabla F)^2$.

We have found there are 47 couplings with structure of one Riemann curvature and six $F$, \ie
\beqa
{\cal L}_6^{F^6R}&=&  m_{135}^{} F_{\alpha }{}^{\epsilon \varepsilon \mu } F_{\beta 
}{}^{\nu \sigma \lambda } F_{\gamma \epsilon }{}^{\kappa \tau 
} F_{\delta \nu }{}^{\omega \varphi } F_{\varepsilon \sigma 
\kappa \omega } F_{\mu \lambda \tau \varphi } R^{\alpha \beta 
\gamma \delta } + \nn\\&& m_{136}^{} F_{\alpha }{}^{\epsilon \varepsilon 
\mu } F_{\beta }{}^{\nu \sigma \lambda } F_{\gamma \epsilon 
\nu }{}^{\kappa } F_{\delta }{}^{\tau \omega \varphi } 
F_{\varepsilon \sigma \kappa \tau } F_{\mu \lambda \omega 
\varphi } R^{\alpha \beta \gamma \delta } + \nn\\&& m_{144}^{} 
F_{\alpha }{}^{\epsilon \varepsilon \mu } F_{\beta \epsilon 
\varepsilon }{}^{\nu } F_{\gamma }{}^{\sigma \lambda \kappa } 
F_{\delta }{}^{\tau \omega \varphi } F_{\mu \sigma \lambda 
\tau } F_{\nu \kappa \omega \varphi } R^{\alpha \beta \gamma 
\delta } + \nn\\&& m_{145}^{} F_{\alpha \beta }{}^{\epsilon \varepsilon 
} F_{\gamma \epsilon }{}^{\mu \nu } F_{\delta }{}^{\sigma 
\lambda \kappa } F_{\varepsilon }{}^{\tau \omega \varphi } 
F_{\mu \sigma \lambda \tau } F_{\nu \kappa \omega \varphi } 
R^{\alpha \beta \gamma \delta } + \nn\\&& m_{146}^{} F_{\alpha \beta 
}{}^{\epsilon \varepsilon } F_{\gamma \delta }{}^{\mu \nu } F_{
\epsilon }{}^{\sigma \lambda \kappa } F_{\varepsilon }{}^{\tau 
\omega \varphi } F_{\mu \sigma \lambda \tau } F_{\nu \kappa 
\omega \varphi } R^{\alpha \beta \gamma \delta } + \nn\\&& m_{139}^{} 
F_{\alpha \beta }{}^{\epsilon \varepsilon } F_{\gamma }{}^{\mu 
\nu \sigma } F_{\delta }{}^{\lambda \kappa \tau } F_{\epsilon 
\mu }{}^{\omega \varphi } F_{\varepsilon \lambda \omega 
\varphi } F_{\nu \sigma \kappa \tau } R^{\alpha \beta \gamma 
\delta } + \nn\\&& m_{140}^{} F_{\alpha }{}^{\epsilon \varepsilon \mu } 
F_{\beta \epsilon }{}^{\nu \sigma } F_{\gamma \varepsilon }{}^{
\lambda \kappa } F_{\delta }{}^{\tau \omega \varphi } F_{\mu 
\lambda \tau \omega } F_{\nu \sigma \kappa \varphi } 
R^{\alpha \beta \gamma \delta } + \nn\\&& m_{141}^{} F_{\alpha 
}{}^{\epsilon \varepsilon \mu } F_{\beta \epsilon }{}^{\nu 
\sigma } F_{\gamma \varepsilon }{}^{\lambda \kappa } F_{\delta 
\lambda }{}^{\tau \omega } F_{\mu \tau \omega }{}^{\varphi } 
F_{\nu \sigma \kappa \varphi } R^{\alpha \beta \gamma \delta 
} + \nn\\&& m_{137}^{} F_{\alpha }{}^{\epsilon \varepsilon \mu } 
F_{\beta }{}^{\nu \sigma \lambda } F_{\gamma \epsilon 
\varepsilon }{}^{\kappa } F_{\delta }{}^{\tau \omega \varphi } 
F_{\mu \tau \omega \varphi } F_{\nu \sigma \lambda \kappa } 
R^{\alpha \beta \gamma \delta } + \nn\\&& m_{138}^{} F_{\alpha 
}{}^{\epsilon \varepsilon \mu } F_{\beta \epsilon }{}^{\nu 
\sigma } F_{\gamma \varepsilon }{}^{\lambda \kappa } F_{\delta 
}{}^{\tau \omega \varphi } F_{\mu \tau \omega \varphi } 
F_{\nu \sigma \lambda \kappa } R^{\alpha \beta \gamma \delta 
} + \nn\\&& m_{142}^{} F_{\alpha }{}^{\epsilon \varepsilon \mu } 
F_{\beta \epsilon }{}^{\nu \sigma } F_{\gamma }{}^{\lambda 
\kappa \tau } F_{\delta \lambda }{}^{\omega \varphi } 
F_{\varepsilon \mu \kappa \tau } F_{\nu \sigma \omega \varphi 
} R^{\alpha \beta \gamma \delta } + \nn\\&& m_{143}^{} F_{\alpha 
}{}^{\epsilon \varepsilon \mu } F_{\beta \epsilon }{}^{\nu 
\sigma } F_{\gamma \varepsilon }{}^{\lambda \kappa } F_{\delta 
}{}^{\tau \omega \varphi } F_{\mu \lambda \kappa \tau } 
F_{\nu \sigma \omega \varphi } R^{\alpha \beta \gamma \delta 
} + \nn\\&& m_{153}^{} F_{\alpha }{}^{\epsilon \varepsilon \mu } 
F_{\beta \epsilon \varepsilon }{}^{\nu } F_{\gamma }{}^{\sigma 
\lambda \kappa } F_{\delta \sigma \lambda }{}^{\tau } F_{\mu 
\kappa }{}^{\omega \varphi } F_{\nu \tau \omega \varphi } 
R^{\alpha \beta \gamma \delta } + \nn\\&& m_{154}^{} F_{\alpha \beta 
}{}^{\epsilon \varepsilon } F_{\gamma \epsilon }{}^{\mu \nu } 
F_{\delta }{}^{\sigma \lambda \kappa } F_{\varepsilon \sigma 
\lambda }{}^{\tau } F_{\mu \kappa }{}^{\omega \varphi } F_{\nu 
\tau \omega \varphi } R^{\alpha \beta \gamma \delta } + 
\nn\\&& m_{155}^{} F_{\alpha \beta }{}^{\epsilon \varepsilon } F_{\gamma 
\delta }{}^{\mu \nu } F_{\epsilon }{}^{\sigma \lambda \kappa } 
F_{\varepsilon \sigma \lambda }{}^{\tau } F_{\mu \kappa 
}{}^{\omega \varphi } F_{\nu \tau \omega \varphi } R^{\alpha 
\beta \gamma \delta } + \nn\\&& m_{150}^{} F_{\alpha }{}^{\epsilon 
\varepsilon \mu } F_{\beta \epsilon \varepsilon }{}^{\nu } 
F_{\gamma }{}^{\sigma \lambda \kappa } F_{\delta \sigma 
}{}^{\tau \omega } F_{\mu \lambda \kappa }{}^{\varphi } F_{\nu 
\tau \omega \varphi } R^{\alpha \beta \gamma \delta } + 
\nn\\&& m_{151}^{} F_{\alpha \beta }{}^{\epsilon \varepsilon } F_{\gamma 
\epsilon }{}^{\mu \nu } F_{\delta }{}^{\sigma \lambda \kappa } 
F_{\varepsilon \sigma }{}^{\tau \omega } F_{\mu \lambda \kappa 
}{}^{\varphi } F_{\nu \tau \omega \varphi } R^{\alpha \beta 
\gamma \delta } + \nn\\&& m_{152}^{} F_{\alpha \beta }{}^{\epsilon 
\varepsilon } F_{\gamma \delta }{}^{\mu \nu } F_{\epsilon }{}^{
\sigma \lambda \kappa } F_{\varepsilon \sigma }{}^{\tau \omega 
} F_{\mu \lambda \kappa }{}^{\varphi } F_{\nu \tau \omega 
\varphi } R^{\alpha \beta \gamma \delta } + \nn\\&& m_{147}^{} 
F_{\alpha }{}^{\epsilon \varepsilon \mu } F_{\beta \epsilon 
\varepsilon }{}^{\nu } F_{\gamma }{}^{\sigma \lambda \kappa } 
F_{\delta }{}^{\tau \omega \varphi } F_{\mu \sigma \lambda 
\kappa } F_{\nu \tau \omega \varphi } R^{\alpha \beta \gamma 
\delta } + \nn\\&& m_{148}^{} F_{\alpha \beta }{}^{\epsilon \varepsilon 
} F_{\gamma \epsilon }{}^{\mu \nu } F_{\delta }{}^{\sigma 
\lambda \kappa } F_{\varepsilon }{}^{\tau \omega \varphi } 
F_{\mu \sigma \lambda \kappa } F_{\nu \tau \omega \varphi } 
R^{\alpha \beta \gamma \delta } + \nn\\&& m_{149}^{} F_{\alpha \beta 
}{}^{\epsilon \varepsilon } F_{\gamma \delta }{}^{\mu \nu } F_{
\epsilon }{}^{\sigma \lambda \kappa } F_{\varepsilon }{}^{\tau 
\omega \varphi } F_{\mu \sigma \lambda \kappa } F_{\nu \tau 
\omega \varphi } R^{\alpha \beta \gamma \delta } + \nn\\&& m_{177}^{} 
F_{\alpha \beta }{}^{\epsilon \varepsilon } F_{\gamma }{}^{\mu 
\nu \sigma } F_{\delta }{}^{\lambda \kappa \tau } F_{\epsilon 
\mu \nu }{}^{\omega } F_{\varepsilon \lambda \omega 
}{}^{\varphi } F_{\sigma \kappa \tau \varphi } R^{\alpha \beta 
\gamma \delta } + \nn\\&& m_{176}^{} F_{\alpha \beta }{}^{\epsilon 
\varepsilon } F_{\gamma }{}^{\mu \nu \sigma } F_{\delta 
}{}^{\lambda \kappa \tau } F_{\epsilon \mu \lambda }{}^{\omega 
} F_{\varepsilon \nu \omega }{}^{\varphi } F_{\sigma \kappa 
\tau \varphi } R^{\alpha \beta \gamma \delta } + \nn\\&& m_{178}^{} F_{
\alpha \beta }{}^{\epsilon \varepsilon } F_{\gamma }{}^{\mu 
\nu \sigma } F_{\delta }{}^{\lambda \kappa \tau } F_{\epsilon 
\varepsilon }{}^{\omega \varphi } F_{\mu \nu \lambda \omega } 
F_{\sigma \kappa \tau \varphi } R^{\alpha \beta \gamma \delta 
} + \nn\\&& m_{179}^{} F_{\alpha \beta }{}^{\epsilon \varepsilon } 
F_{\gamma }{}^{\mu \nu \sigma } F_{\delta \mu }{}^{\lambda 
\kappa } F_{\epsilon \nu }{}^{\tau \omega } F_{\varepsilon 
\lambda \tau }{}^{\varphi } F_{\sigma \kappa \omega \varphi } 
R^{\alpha \beta \gamma \delta } + \nn\\&& m_{156}^{} F_{\alpha 
}{}^{\epsilon \varepsilon \mu } F_{\beta }{}^{\nu \sigma 
\lambda } F_{\gamma \epsilon }{}^{\kappa \tau } F_{\delta \nu 
}{}^{\omega \varphi } F_{\varepsilon \mu \omega \varphi } 
F_{\sigma \lambda \kappa \tau } R^{\alpha \beta \gamma \delta 
} + \nn\\&& m_{157}^{} F_{\alpha \beta }{}^{\epsilon \varepsilon } 
F_{\gamma }{}^{\mu \nu \sigma } F_{\delta }{}^{\lambda \kappa 
\tau } F_{\epsilon \mu }{}^{\omega \varphi } F_{\varepsilon 
\nu \omega \varphi } F_{\sigma \lambda \kappa \tau } 
R^{\alpha \beta \gamma \delta } + \nn\\&& m_{158}^{} F_{\alpha \beta 
}{}^{\epsilon \varepsilon } F_{\gamma }{}^{\mu \nu \sigma } F_{
\delta \mu }{}^{\lambda \kappa } F_{\epsilon \nu }{}^{\tau 
\omega } F_{\varepsilon \tau \omega }{}^{\varphi } F_{\sigma 
\lambda \kappa \varphi } R^{\alpha \beta \gamma \delta } + 
\nn\\&& m_{159}^{} F_{\alpha }{}^{\epsilon \varepsilon \mu } F_{\beta 
}{}^{\nu \sigma \lambda } F_{\gamma \epsilon \varepsilon 
}{}^{\kappa } F_{\delta }{}^{\tau \omega \varphi } F_{\mu \nu 
\tau \omega } F_{\sigma \lambda \kappa \varphi } R^{\alpha 
\beta \gamma \delta } + \nn\\&& m_{160}^{} F_{\alpha }{}^{\epsilon 
\varepsilon \mu } F_{\beta \epsilon }{}^{\nu \sigma } 
F_{\gamma \varepsilon }{}^{\lambda \kappa } F_{\delta }{}^{\tau 
\omega \varphi } F_{\mu \nu \tau \omega } F_{\sigma \lambda 
\kappa \varphi } R^{\alpha \beta \gamma \delta } + \nn\\&& m_{161}^{} 
F_{\alpha \beta }{}^{\epsilon \varepsilon } F_{\gamma \epsilon 
}{}^{\mu \nu } F_{\delta }{}^{\sigma \lambda \kappa } 
F_{\varepsilon }{}^{\tau \omega \varphi } F_{\mu \nu \tau 
\omega } F_{\sigma \lambda \kappa \varphi } R^{\alpha \beta 
\gamma \delta } + \nn\\&& m_{162}^{} F_{\alpha }{}^{\epsilon \varepsilon 
\mu } F_{\beta }{}^{\nu \sigma \lambda } F_{\gamma \epsilon 
\nu }{}^{\kappa } F_{\delta \varepsilon }{}^{\tau \omega } 
F_{\mu \tau \omega }{}^{\varphi } F_{\sigma \lambda \kappa 
\varphi } R^{\alpha \beta \gamma \delta } + \nn\\&& m_{163}^{} 
F_{\alpha }{}^{\epsilon \varepsilon \mu } F_{\beta }{}^{\nu 
\sigma \lambda } F_{\gamma \epsilon \varepsilon }{}^{\kappa } 
F_{\delta \nu }{}^{\tau \omega } F_{\mu \tau \omega 
}{}^{\varphi } F_{\sigma \lambda \kappa \varphi } R^{\alpha 
\beta \gamma \delta } + \nn\\&& m_{164}^{} F_{\alpha }{}^{\epsilon 
\varepsilon \mu } F_{\beta \epsilon }{}^{\nu \sigma } 
F_{\gamma \varepsilon }{}^{\lambda \kappa } F_{\delta \nu }{}^{
\tau \omega } F_{\mu \tau \omega }{}^{\varphi } F_{\sigma 
\lambda \kappa \varphi } R^{\alpha \beta \gamma \delta } + 
\nn\\&& m_{165}^{} F_{\alpha \beta }{}^{\epsilon \varepsilon } F_{\gamma 
\epsilon }{}^{\mu \nu } F_{\delta }{}^{\sigma \lambda \kappa } 
F_{\varepsilon \mu }{}^{\tau \omega } F_{\nu \tau \omega }{}^{
\varphi } F_{\sigma \lambda \kappa \varphi } R^{\alpha \beta 
\gamma \delta } + \nn\\&& m_{166}^{} F_{\alpha }{}^{\epsilon \varepsilon 
\mu } F_{\beta }{}^{\nu \sigma \lambda } F_{\gamma \epsilon 
}{}^{\kappa \tau } F_{\delta \nu }{}^{\omega \varphi } 
F_{\varepsilon \mu \kappa \omega } F_{\sigma \lambda \tau 
\varphi } R^{\alpha \beta \gamma \delta } + \nn\\&& m_{169}^{} 
F_{\alpha \beta }{}^{\epsilon \varepsilon } F_{\gamma }{}^{\mu 
\nu \sigma } F_{\delta \mu \nu }{}^{\lambda } F_{\epsilon 
}{}^{\kappa \tau \omega } F_{\varepsilon \kappa \tau 
}{}^{\varphi } F_{\sigma \lambda \omega \varphi } R^{\alpha 
\beta \gamma \delta } + \nn\\&& m_{167}^{} F_{\alpha }{}^{\epsilon 
\varepsilon \mu } F_{\beta }{}^{\nu \sigma \lambda } F_{\gamma 
\epsilon }{}^{\kappa \tau } F_{\delta \nu }{}^{\omega \varphi 
} F_{\varepsilon \mu \kappa \tau } F_{\sigma \lambda \omega 
\varphi } R^{\alpha \beta \gamma \delta } + \nn\\&& m_{168}^{} 
F_{\alpha }{}^{\epsilon \varepsilon \mu } F_{\beta }{}^{\nu 
\sigma \lambda } F_{\gamma \epsilon \nu }{}^{\kappa } 
F_{\delta }{}^{\tau \omega \varphi } F_{\varepsilon \mu \kappa 
\tau } F_{\sigma \lambda \omega \varphi } R^{\alpha \beta 
\gamma \delta } + \nn\\&& m_{171}^{} F_{\alpha }{}^{\epsilon \varepsilon 
\mu } F_{\beta }{}^{\nu \sigma \lambda } F_{\gamma \epsilon 
\varepsilon \nu } F_{\delta }{}^{\kappa \tau \omega } F_{\mu 
\kappa \tau }{}^{\varphi } F_{\sigma \lambda \omega \varphi } 
R^{\alpha \beta \gamma \delta } + \nn\\&& m_{170}^{} F_{\alpha 
}{}^{\epsilon \varepsilon \mu } F_{\beta }{}^{\nu \sigma 
\lambda } F_{\gamma \epsilon \varepsilon }{}^{\kappa } 
F_{\delta }{}^{\tau \omega \varphi } F_{\mu \nu \kappa \tau } 
F_{\sigma \lambda \omega \varphi } R^{\alpha \beta \gamma 
\delta } + \nn\\&& m_{172}^{} F_{\alpha }{}^{\epsilon \varepsilon \mu } 
F_{\beta \epsilon \varepsilon }{}^{\nu } F_{\gamma \mu 
}{}^{\sigma \lambda } F_{\delta }{}^{\kappa \tau \omega } 
F_{\nu \kappa \tau }{}^{\varphi } F_{\sigma \lambda \omega 
\varphi } R^{\alpha \beta \gamma \delta } + \nn\\&& m_{173}^{} 
F_{\alpha \beta }{}^{\epsilon \varepsilon } F_{\gamma \epsilon 
}{}^{\mu \nu } F_{\delta \mu }{}^{\sigma \lambda } 
F_{\varepsilon }{}^{\kappa \tau \omega } F_{\nu \kappa \tau 
}{}^{\varphi } F_{\sigma \lambda \omega \varphi } R^{\alpha 
\beta \gamma \delta } + \nn\\&& m_{174}^{} F_{\alpha \beta 
}{}^{\epsilon \varepsilon } F_{\gamma \delta }{}^{\mu \nu } F_{
\epsilon \mu }{}^{\sigma \lambda } F_{\varepsilon }{}^{\kappa 
\tau \omega } F_{\nu \kappa \tau }{}^{\varphi } F_{\sigma 
\lambda \omega \varphi } R^{\alpha \beta \gamma \delta } + 
\nn\\&& m_{175}^{} F_{\alpha \beta }{}^{\epsilon \varepsilon } F_{\gamma 
\delta }{}^{\mu \nu } F_{\epsilon \varepsilon }{}^{\sigma 
\lambda } F_{\mu }{}^{\kappa \tau \omega } F_{\nu \kappa \tau 
}{}^{\varphi } F_{\sigma \lambda \omega \varphi } R^{\alpha 
\beta \gamma \delta } + \nn\\&& m_{180}^{} F_{\alpha \beta 
}{}^{\epsilon \varepsilon } F_{\gamma \delta }{}^{\mu \nu } F_{
\epsilon \varepsilon \mu }{}^{\sigma } F_{\kappa \tau \omega 
\varphi } F_{\nu }{}^{\lambda \kappa \tau } F_{\sigma \lambda 
}{}^{\omega \varphi } R^{\alpha \beta \gamma \delta } + 
\nn\\&& m_{181}^{} F_{\alpha \beta }{}^{\epsilon \varepsilon } F_{\gamma 
\delta }{}^{\mu \nu } F_{\epsilon \varepsilon \mu \nu } 
F_{\kappa \tau \omega \varphi } F_{\sigma \lambda }{}^{\omega 
\varphi } F^{\sigma \lambda \kappa \tau } R^{\alpha \beta 
\gamma \delta }\labell{T2}
\eeqa 

There are 63 couplings with structure of two Riemann curvatures and four $F$, \ie
\beqa
{\cal L}_6^{F^4R^2}&\!\!\!\!\!=\!\!\!\!\!& m_{182}^{} F_{\epsilon \varepsilon }{}^{\sigma \lambda } 
F^{\epsilon \varepsilon \mu \nu } F_{\mu \sigma }{}^{\kappa 
\tau } F_{\nu \lambda \kappa \tau } R_{\alpha \gamma \beta 
\delta } R^{\alpha \beta \gamma \delta } +\nn\\&& m_{183}^{} 
F_{\epsilon \varepsilon }{}^{\sigma \lambda } F^{\epsilon 
\varepsilon \mu \nu } F_{\mu \nu }{}^{\kappa \tau } F_{\sigma 
\lambda \kappa \tau } R_{\alpha \gamma \beta \delta } 
R^{\alpha \beta \gamma \delta } + \nn\\&& m_{184}^{} F_{\epsilon 
\varepsilon \mu }{}^{\sigma } F^{\epsilon \varepsilon \mu \nu 
} F_{\nu }{}^{\lambda \kappa \tau } F_{\sigma \lambda \kappa 
\tau } R_{\alpha \gamma \beta \delta } R^{\alpha \beta \gamma 
\delta } + \nn\\&& m_{188}^{} F_{\beta }{}^{\mu 
\nu \sigma } F_{\delta \mu \nu \sigma } F_{\epsilon 
}{}^{\lambda \kappa \tau } F_{\varepsilon \lambda \kappa \tau 
} R_{\alpha }{}^{\epsilon }{}_{\gamma }{}^{\varepsilon } 
R^{\alpha \beta \gamma \delta } + \nn\\&& m_{187}^{} F_{\beta }{}^{\mu 
\nu \sigma } F_{\delta \mu \nu }{}^{\lambda } F_{\epsilon 
\sigma }{}^{\kappa \tau } F_{\varepsilon \lambda \kappa \tau } 
R_{\alpha }{}^{\epsilon }{}_{\gamma }{}^{\varepsilon } R^{\alpha 
\beta \gamma \delta } + \nn\\&& m_{186}^{} F_{\beta }{}^{\mu \nu 
\sigma } F_{\delta \mu }{}^{\lambda \kappa } F_{\epsilon \nu 
\lambda }{}^{\tau } F_{\varepsilon \sigma \kappa \tau } 
R_{\alpha }{}^{\epsilon }{}_{\gamma }{}^{\varepsilon } R^{\alpha 
\beta \gamma \delta } + \nn\\&& m_{189}^{} F_{\beta \varepsilon 
}{}^{\mu \nu } F_{\delta \epsilon }{}^{\sigma \lambda } F_{\mu 
\sigma }{}^{\kappa \tau } F_{\nu \lambda \kappa \tau } 
R_{\alpha }{}^{\epsilon }{}_{\gamma }{}^{\varepsilon } R^{\alpha 
\beta \gamma \delta } + \nn\\&& m_{190}^{} F_{\beta \epsilon }{}^{\mu 
\nu } F_{\delta \varepsilon }{}^{\sigma \lambda } F_{\mu 
\sigma }{}^{\kappa \tau } F_{\nu \lambda \kappa \tau } 
R_{\alpha }{}^{\epsilon }{}_{\gamma }{}^{\varepsilon } R^{\alpha 
\beta \gamma \delta } + \nn\\&& m_{191}^{} F_{\beta \delta }{}^{\mu 
\nu } F_{\epsilon \varepsilon }{}^{\sigma \lambda } F_{\mu 
\sigma }{}^{\kappa \tau } F_{\nu \lambda \kappa \tau } 
R_{\alpha }{}^{\epsilon }{}_{\gamma }{}^{\varepsilon } R^{\alpha 
\beta \gamma \delta } + \nn\\&& m_{192}^{} F_{\beta \varepsilon 
}{}^{\mu \nu } F_{\delta \epsilon }{}^{\sigma \lambda } F_{\mu 
\nu }{}^{\kappa \tau } F_{\sigma \lambda \kappa \tau } 
R_{\alpha }{}^{\epsilon }{}_{\gamma }{}^{\varepsilon } R^{\alpha 
\beta \gamma \delta } + \nn\\&& m_{193}^{} F_{\beta \epsilon }{}^{\mu 
\nu } F_{\delta \varepsilon }{}^{\sigma \lambda } F_{\mu \nu 
}{}^{\kappa \tau } F_{\sigma \lambda \kappa \tau } R_{\alpha 
}{}^{\epsilon }{}_{\gamma }{}^{\varepsilon } R^{\alpha \beta 
\gamma \delta } + \nn\\&& m_{194}^{} F_{\beta \delta }{}^{\mu \nu } F_{
\epsilon \varepsilon }{}^{\sigma \lambda } F_{\mu \nu 
}{}^{\kappa \tau } F_{\sigma \lambda \kappa \tau } R_{\alpha 
}{}^{\epsilon }{}_{\gamma }{}^{\varepsilon } R^{\alpha \beta 
\gamma \delta } + \nn\\&& m_{195}^{} F_{\beta \delta \epsilon 
\varepsilon } F_{\mu \nu }{}^{\kappa \tau } F^{\mu \nu \sigma 
\lambda } F_{\sigma \lambda \kappa \tau } R_{\alpha 
}{}^{\epsilon }{}_{\gamma }{}^{\varepsilon } R^{\alpha \beta 
\gamma \delta } + \nn\\&& m_{196}^{} F_{\beta \varepsilon }{}^{\mu \nu 
} F_{\delta \epsilon \mu }{}^{\sigma } F_{\nu }{}^{\lambda 
\kappa \tau } F_{\sigma \lambda \kappa \tau } R_{\alpha 
}{}^{\epsilon }{}_{\gamma }{}^{\varepsilon } R^{\alpha \beta 
\gamma \delta } + \nn\\&& m_{197}^{} F_{\beta \delta }{}^{\mu \nu } F_{
\epsilon \varepsilon \mu }{}^{\sigma } F_{\nu }{}^{\lambda 
\kappa \tau } F_{\sigma \lambda \kappa \tau } R_{\alpha 
}{}^{\epsilon }{}_{\gamma }{}^{\varepsilon } R^{\alpha \beta 
\gamma \delta } + \nn\\&& m_{200}^{} 
F_{\beta \gamma }{}^{\nu \sigma } F_{\delta \nu }{}^{\lambda 
\kappa } F_{\epsilon \varepsilon \sigma }{}^{\tau } F_{\mu 
\lambda \kappa \tau } R_{\alpha }{}^{\epsilon \varepsilon \mu 
} R^{\alpha \beta \gamma \delta } + \nn\\&& m_{202}^{} F_{\beta \gamma 
}{}^{\nu \sigma } F_{\delta \varepsilon \nu \sigma } 
F_{\epsilon }{}^{\lambda \kappa \tau } F_{\mu \lambda \kappa 
\tau } R_{\alpha }{}^{\epsilon \varepsilon \mu } R^{\alpha 
\beta \gamma \delta } + \nn\\&& m_{201}^{} F_{\beta \gamma }{}^{\nu 
\sigma } F_{\delta \varepsilon \nu }{}^{\lambda } F_{\epsilon 
\sigma }{}^{\kappa \tau } F_{\mu \lambda \kappa \tau } 
R_{\alpha }{}^{\epsilon \varepsilon \mu } R^{\alpha \beta 
\gamma \delta } + \nn\\&& m_{203}^{} F_{\beta \varepsilon }{}^{\nu 
\sigma } F_{\gamma \epsilon \nu }{}^{\lambda } F_{\delta \mu 
}{}^{\kappa \tau } F_{\sigma \lambda \kappa \tau } R_{\alpha 
}{}^{\epsilon \varepsilon \mu } R^{\alpha \beta \gamma \delta 
} + \nn\\&& m_{207}^{} F_{\beta \gamma \epsilon }{}^{\nu } F_{\delta 
\varepsilon \nu }{}^{\sigma } F_{\mu }{}^{\lambda \kappa \tau 
} F_{\sigma \lambda \kappa \tau } R_{\alpha }{}^{\epsilon 
\varepsilon \mu } R^{\alpha \beta \gamma \delta } + \nn\\&& m_{204}^{} 
F_{\beta \gamma \varepsilon }{}^{\nu } F_{\delta \epsilon }{}^{
\sigma \lambda } F_{\mu \nu }{}^{\kappa \tau } F_{\sigma 
\lambda \kappa \tau } R_{\alpha }{}^{\epsilon \varepsilon \mu 
} R^{\alpha \beta \gamma \delta } + \nn\\&& m_{205}^{} F_{\beta \gamma 
\epsilon }{}^{\nu } F_{\delta \varepsilon }{}^{\sigma \lambda } 
F_{\mu \nu }{}^{\kappa \tau } F_{\sigma \lambda \kappa \tau } 
R_{\alpha }{}^{\epsilon \varepsilon \mu } R^{\alpha \beta 
\gamma \delta } + \nn\\&& m_{206}^{} F_{\beta \gamma \epsilon 
\varepsilon } F_{\delta }{}^{\nu \sigma \lambda } F_{\mu \nu 
}{}^{\kappa \tau } F_{\sigma \lambda \kappa \tau } R_{\alpha 
}{}^{\epsilon \varepsilon \mu } R^{\alpha \beta \gamma \delta 
} + \nn\\&& m_{208}^{} F_{\beta \gamma \epsilon }{}^{\nu } F_{\delta 
\varepsilon \mu }{}^{\sigma } F_{\nu }{}^{\lambda \kappa \tau 
} F_{\sigma \lambda \kappa \tau } R_{\alpha }{}^{\epsilon 
\varepsilon \mu } R^{\alpha \beta \gamma \delta } + \nn\\&& m_{235}^{} F_{\alpha \epsilon 
}{}^{\sigma \lambda } F_{\beta \mu \sigma }{}^{\kappa } 
F_{\gamma \varepsilon \lambda }{}^{\tau } F_{\delta \nu \kappa 
\tau } R^{\alpha \beta \gamma \delta } R^{\epsilon \varepsilon 
\mu \nu } + \nn\\&& m_{238}^{} F_{\alpha \epsilon }{}^{\sigma \lambda } 
F_{\beta \varepsilon \sigma \lambda } F_{\gamma \mu 
}{}^{\kappa \tau } F_{\delta \nu \kappa \tau } R^{\alpha 
\beta \gamma \delta } R^{\epsilon \varepsilon \mu \nu } + 
\nn\\&& m_{237}^{} F_{\alpha \epsilon }{}^{\sigma \lambda } F_{\beta 
\varepsilon \sigma }{}^{\kappa } F_{\gamma \mu \lambda 
}{}^{\tau } F_{\delta \nu \kappa \tau } R^{\alpha \beta 
\gamma \delta } R^{\epsilon \varepsilon \mu \nu } + \nn\\&& m_{236}^{} 
F_{\alpha \epsilon }{}^{\sigma \lambda } F_{\beta \varepsilon 
}{}^{\kappa \tau } F_{\gamma \mu \sigma \lambda } F_{\delta 
\nu \kappa \tau } R^{\alpha \beta \gamma \delta } R^{\epsilon 
\varepsilon \mu \nu } + \nn\\&& m_{243}^{} F_{\alpha \beta }{}^{\sigma 
\lambda } F_{\gamma \epsilon \sigma \lambda } F_{\delta 
\varepsilon }{}^{\kappa \tau } F_{\mu \nu \kappa \tau } 
R^{\alpha \beta \gamma \delta } R^{\epsilon \varepsilon \mu 
\nu } + \nn\\&& m_{242}^{} F_{\alpha \beta }{}^{\sigma \lambda } 
F_{\gamma \epsilon \sigma }{}^{\kappa } F_{\delta \varepsilon 
\lambda }{}^{\tau } F_{\mu \nu \kappa \tau } R^{\alpha \beta 
\gamma \delta } R^{\epsilon \varepsilon \mu \nu } + \nn\\&& m_{246}^{} 
F_{\alpha \beta }{}^{\sigma \lambda } F_{\gamma \delta \sigma 
\lambda } F_{\epsilon \varepsilon }{}^{\kappa \tau } F_{\mu 
\nu \kappa \tau } R^{\alpha \beta \gamma \delta } R^{\epsilon 
\varepsilon \mu \nu } + \nn\\&& m_{245}^{} F_{\alpha \beta }{}^{\sigma 
\lambda } F_{\gamma \delta \sigma }{}^{\kappa } F_{\epsilon 
\varepsilon \lambda }{}^{\tau } F_{\mu \nu \kappa \tau } 
R^{\alpha \beta \gamma \delta } R^{\epsilon \varepsilon \mu 
\nu } + \nn\\&& m_{244}^{} F_{\alpha \beta }{}^{\sigma \lambda } 
F_{\gamma \delta }{}^{\kappa \tau } F_{\epsilon \varepsilon 
\sigma \lambda } F_{\mu \nu \kappa \tau } R^{\alpha \beta 
\gamma \delta } R^{\epsilon \varepsilon \mu \nu } + \nn\\&& m_{240}^{} 
F_{\alpha \beta }{}^{\sigma \lambda } F_{\gamma \epsilon 
\sigma }{}^{\kappa } F_{\delta \varepsilon \kappa }{}^{\tau } 
F_{\mu \nu \lambda \tau } R^{\alpha \beta \gamma \delta } R^{
\epsilon \varepsilon \mu \nu } + \nn\\&& m_{241}^{} F_{\alpha \beta 
}{}^{\sigma \lambda } F_{\gamma \delta }{}^{\kappa \tau } 
F_{\epsilon \varepsilon \sigma \kappa } F_{\mu \nu \lambda 
\tau } R^{\alpha \beta \gamma \delta } R^{\epsilon \varepsilon 
\mu \nu } + \nn\\&& m_{239}^{} F_{\alpha \beta }{}^{\sigma \lambda } 
F_{\gamma \epsilon }{}^{\kappa \tau } F_{\delta \varepsilon 
\kappa \tau } F_{\mu \nu \sigma \lambda } R^{\alpha \beta 
\gamma \delta } R^{\epsilon \varepsilon \mu \nu } + \nn\\&& m_{252}^{} 
F_{\alpha \beta }{}^{\sigma \lambda } F_{\gamma \epsilon \mu 
\sigma } F_{\delta \varepsilon }{}^{\kappa \tau } F_{\nu 
\lambda \kappa \tau } R^{\alpha \beta \gamma \delta } 
R^{\epsilon \varepsilon \mu \nu } + \nn\\&& m_{251}^{} F_{\alpha \beta 
}{}^{\sigma \lambda } F_{\gamma \epsilon \mu }{}^{\kappa } 
F_{\delta \varepsilon \sigma }{}^{\tau } F_{\nu \lambda \kappa 
\tau } R^{\alpha \beta \gamma \delta } R^{\epsilon \varepsilon 
\mu \nu } + \nn\\&& m_{257}^{} F_{\alpha \beta \epsilon }{}^{\sigma } 
F_{\gamma \varepsilon \mu \sigma } F_{\delta }{}^{\lambda 
\kappa \tau } F_{\nu \lambda \kappa \tau } R^{\alpha \beta 
\gamma \delta } R^{\epsilon \varepsilon \mu \nu } + \nn\\&& m_{254}^{} 
F_{\alpha \beta \epsilon }{}^{\sigma } F_{\gamma \varepsilon 
\sigma }{}^{\lambda } F_{\delta \mu }{}^{\kappa \tau } F_{\nu 
\lambda \kappa \tau } R^{\alpha \beta \gamma \delta } 
R^{\epsilon \varepsilon \mu \nu } + \nn\\&& m_{253}^{} F_{\alpha \beta 
\epsilon }{}^{\sigma } F_{\gamma \varepsilon }{}^{\lambda 
\kappa } F_{\delta \mu \sigma }{}^{\tau } F_{\nu \lambda 
\kappa \tau } R^{\alpha \beta \gamma \delta } R^{\epsilon 
\varepsilon \mu \nu } + \nn\\&& m_{255}^{} F_{\alpha \beta \epsilon 
}{}^{\sigma } F_{\gamma \varepsilon \mu }{}^{\lambda } 
F_{\delta \sigma }{}^{\kappa \tau } F_{\nu \lambda \kappa 
\tau } R^{\alpha \beta \gamma \delta } R^{\epsilon \varepsilon 
\mu \nu } + \nn\\&& m_{256}^{} F_{\alpha \beta \epsilon \varepsilon } 
F_{\gamma \mu }{}^{\sigma \lambda } F_{\delta \sigma 
}{}^{\kappa \tau } F_{\nu \lambda \kappa \tau } R^{\alpha 
\beta \gamma \delta } R^{\epsilon \varepsilon \mu \nu } + 
\nn\\&& m_{261}^{} F_{\alpha \beta \epsilon }{}^{\sigma } F_{\gamma 
\delta \mu \sigma } F_{\varepsilon }{}^{\lambda \kappa \tau } 
F_{\nu \lambda \kappa \tau } R^{\alpha \beta \gamma \delta } 
R^{\epsilon \varepsilon \mu \nu } + \nn\\&& m_{259}^{} F_{\alpha \beta 
\epsilon }{}^{\sigma } F_{\gamma \delta \sigma }{}^{\lambda } 
F_{\varepsilon \mu }{}^{\kappa \tau } F_{\nu \lambda \kappa 
\tau } R^{\alpha \beta \gamma \delta } R^{\epsilon \varepsilon 
\mu \nu } + \nn\\&& m_{258}^{} F_{\alpha \beta \epsilon }{}^{\sigma } 
F_{\gamma \delta }{}^{\lambda \kappa } F_{\varepsilon \mu 
\sigma }{}^{\tau } F_{\nu \lambda \kappa \tau } R^{\alpha 
\beta \gamma \delta } R^{\epsilon \varepsilon \mu \nu } + 
\nn\\&& m_{260}^{} F_{\alpha \beta \epsilon }{}^{\sigma } F_{\gamma 
\delta \mu }{}^{\lambda } F_{\varepsilon \sigma }{}^{\kappa 
\tau } F_{\nu \lambda \kappa \tau } R^{\alpha \beta \gamma 
\delta } R^{\epsilon \varepsilon \mu \nu } + \nn\\&& m_{262}^{} 
F_{\alpha \beta \epsilon }{}^{\sigma } F_{\gamma \delta 
\varepsilon }{}^{\lambda } F_{\mu \sigma }{}^{\kappa \tau } F_{
\nu \lambda \kappa \tau } R^{\alpha \beta \gamma \delta } R^{
\epsilon \varepsilon \mu \nu } + \nn\\&& m_{263}^{} F_{\alpha \beta 
\epsilon \varepsilon } F_{\gamma \delta }{}^{\sigma \lambda } 
F_{\mu \sigma }{}^{\kappa \tau } F_{\nu \lambda \kappa \tau } 
R^{\alpha \beta \gamma \delta } R^{\epsilon \varepsilon \mu 
\nu } + \nn\\&& m_{249}^{} F_{\alpha \beta \epsilon }{}^{\sigma } 
F_{\gamma \varepsilon \mu }{}^{\lambda } F_{\delta \lambda 
}{}^{\kappa \tau } F_{\nu \sigma \kappa \tau } R^{\alpha 
\beta \gamma \delta } R^{\epsilon \varepsilon \mu \nu } + 
\nn\\&& m_{248}^{} F_{\alpha \beta \epsilon }{}^{\sigma } F_{\gamma 
\varepsilon }{}^{\lambda \kappa } F_{\delta \mu \lambda 
}{}^{\tau } F_{\nu \sigma \kappa \tau } R^{\alpha \beta 
\gamma \delta } R^{\epsilon \varepsilon \mu \nu } + \nn\\&& m_{250}^{} 
F_{\alpha \beta \epsilon }{}^{\sigma } F_{\gamma \delta 
}{}^{\lambda \kappa } F_{\varepsilon \mu \lambda }{}^{\tau } 
F_{\nu \sigma \kappa \tau } R^{\alpha \beta \gamma \delta } 
R^{\epsilon \varepsilon \mu \nu } + \nn\\&& m_{247}^{} F_{\alpha \beta 
}{}^{\sigma \lambda } F_{\gamma \epsilon \mu }{}^{\kappa } 
F_{\delta \varepsilon \kappa }{}^{\tau } F_{\nu \sigma \lambda 
\tau } R^{\alpha \beta \gamma \delta } R^{\epsilon \varepsilon 
\mu \nu } + \nn\\&& m_{264}^{} F_{\alpha \beta \epsilon }{}^{\sigma } 
F_{\gamma \varepsilon \mu }{}^{\lambda } F_{\delta \nu 
}{}^{\kappa \tau } F_{\sigma \lambda \kappa \tau } R^{\alpha 
\beta \gamma \delta } R^{\epsilon \varepsilon \mu \nu } + 
\nn\\&& m_{265}^{} F_{\alpha \beta \epsilon \varepsilon } F_{\gamma \mu 
}{}^{\sigma \lambda } F_{\delta \nu }{}^{\kappa \tau } 
F_{\sigma \lambda \kappa \tau } R^{\alpha \beta \gamma \delta 
} R^{\epsilon \varepsilon \mu \nu } + \nn\\&& m_{266}^{} F_{\alpha 
\beta \epsilon }{}^{\sigma } F_{\gamma \delta \varepsilon }{}^{
\lambda } F_{\mu \nu }{}^{\kappa \tau } F_{\sigma \lambda 
\kappa \tau } R^{\alpha \beta \gamma \delta } R^{\epsilon 
\varepsilon \mu \nu } + \nn\\&& m_{267}^{} F_{\alpha \beta \epsilon 
\varepsilon } F_{\gamma \delta }{}^{\sigma \lambda } F_{\mu 
\nu }{}^{\kappa \tau } F_{\sigma \lambda \kappa \tau } 
R^{\alpha \beta \gamma \delta } R^{\epsilon \varepsilon \mu 
\nu } + \nn\\&& m_{268}^{} F_{\alpha \beta \epsilon \varepsilon } 
F_{\gamma \delta \mu }{}^{\sigma } F_{\nu }{}^{\lambda \kappa 
\tau } F_{\sigma \lambda \kappa \tau } R^{\alpha \beta \gamma 
\delta } R^{\epsilon \varepsilon \mu \nu } + \nn\\&& m_{269}^{} 
F_{\alpha \beta \epsilon \varepsilon } F_{\gamma \delta \mu 
\nu } FF R^{\alpha \beta \gamma \delta } R^{\epsilon 
\varepsilon \mu \nu }+  m_{198}^{} F_{\beta \varepsilon }{}^{\mu \nu 
} F_{\delta \epsilon \mu \nu } FF R_{\alpha }{}^{\epsilon }{}_{
\gamma }{}^{\varepsilon } R^{\alpha \beta \gamma \delta } + 
\nn\\&& m_{199}^{} F_{\beta \delta }{}^{\mu \nu } F_{\epsilon 
\varepsilon \mu \nu } FF R_{\alpha }{}^{\epsilon }{}_{\gamma 
}{}^{\varepsilon } R^{\alpha \beta \gamma \delta }  +m_{209}^{} 
F_{\beta \gamma \epsilon }{}^{\nu } F_{\delta \varepsilon \mu 
\nu } FF R_{\alpha }{}^{\epsilon \varepsilon \mu } R^{\alpha 
\beta \gamma \delta }+\nn\\&&  m_{185}^{} FF^2 R_{\alpha \gamma \beta \delta } 
R^{\alpha \beta \gamma \delta }\labell{T3}
\eeqa

There are 24 couplings with structure of three Riemann curvatures and two $F$, \ie
\beqa
{\cal L}_6^{F^2R^3}&=&m_{211}^{} F_{\delta \nu 
}{}^{\sigma \lambda } F_{\epsilon \mu \sigma \lambda } 
R_{\alpha }{}^{\epsilon }{}_{\gamma }{}^{\varepsilon } R^{\alpha 
\beta \gamma \delta } R_{\beta }{}^{\mu }{}_{\varepsilon 
}{}^{\nu } +  m_{212}^{} F_{\delta \mu }{}^{\sigma \lambda } 
F_{\epsilon \nu \sigma \lambda } R_{\alpha }{}^{\epsilon 
}{}_{\gamma }{}^{\varepsilon } R^{\alpha \beta \gamma \delta } 
R_{\beta }{}^{\mu }{}_{\varepsilon }{}^{\nu } + \nn\\&& m_{213}^{} 
F_{\delta \epsilon }{}^{\sigma \lambda } F_{\mu \nu \sigma 
\lambda } R_{\alpha }{}^{\epsilon }{}_{\gamma }{}^{\varepsilon } 
R^{\alpha \beta \gamma \delta } R_{\beta }{}^{\mu 
}{}_{\varepsilon }{}^{\nu } +  m_{215}^{} 
F_{\varepsilon }{}^{\nu \sigma \lambda } F_{\mu \nu \sigma 
\lambda } R_{\alpha \beta }{}^{\epsilon \varepsilon } R^{\alpha 
\beta \gamma \delta } R_{\gamma \epsilon \delta }{}^{\mu } + 
\nn\\&& m_{216}^{} F_{\delta \nu }{}^{\sigma \lambda } F_{\varepsilon 
\mu \sigma \lambda } R_{\alpha \beta }{}^{\epsilon \varepsilon 
} R^{\alpha \beta \gamma \delta } R_{\gamma }{}^{\mu 
}{}_{\epsilon }{}^{\nu } +  m_{217}^{} F_{\delta \mu }{}^{\sigma 
\lambda } F_{\varepsilon \nu \sigma \lambda } R_{\alpha \beta 
}{}^{\epsilon \varepsilon } R^{\alpha \beta \gamma \delta } R_{
\gamma }{}^{\mu }{}_{\epsilon }{}^{\nu } + \nn\\&& m_{218}^{} F_{\delta 
\varepsilon }{}^{\sigma \lambda } F_{\mu \nu \sigma \lambda } 
R_{\alpha \beta }{}^{\epsilon \varepsilon } R^{\alpha \beta 
\gamma \delta } R_{\gamma }{}^{\mu }{}_{\epsilon }{}^{\nu } + 
m_{219}^{} F_{\beta \mu \nu }{}^{\lambda } F_{\delta \epsilon 
\sigma \lambda } R_{\alpha }{}^{\epsilon \varepsilon \mu } 
R^{\alpha \beta \gamma \delta } R_{\gamma }{}^{\nu 
}{}_{\varepsilon }{}^{\sigma } + \nn\\&& m_{220}^{} F_{\beta \delta \mu 
}{}^{\lambda } F_{\epsilon \nu \sigma \lambda } R_{\alpha }{}^{
\epsilon \varepsilon \mu } R^{\alpha \beta \gamma \delta } R_{
\gamma }{}^{\nu }{}_{\varepsilon }{}^{\sigma } +  m_{221}^{} 
F_{\beta \delta \epsilon }{}^{\lambda } F_{\mu \nu \sigma 
\lambda } R_{\alpha }{}^{\epsilon \varepsilon \mu } R^{\alpha 
\beta \gamma \delta } R_{\gamma }{}^{\nu }{}_{\varepsilon }{}^{
\sigma } + \nn\\&& m_{222}^{} F_{\beta \epsilon \varepsilon \nu } 
F_{\delta \mu \sigma \lambda } R_{\alpha }{}^{\epsilon 
\varepsilon \mu } R^{\alpha \beta \gamma \delta } R_{\gamma 
}{}^{\nu \sigma \lambda } + m_{223}^{} F_{\beta \delta \epsilon 
\nu } F_{\varepsilon \mu \sigma \lambda } R_{\alpha 
}{}^{\epsilon \varepsilon \mu } R^{\alpha \beta \gamma \delta 
} R_{\gamma }{}^{\nu \sigma \lambda } + \nn\\&& m_{224}^{} 
F_{\varepsilon }{}^{\nu \sigma \lambda } F_{\mu \nu \sigma 
\lambda } R_{\alpha \gamma \beta }{}^{\epsilon } R^{\alpha 
\beta \gamma \delta } R_{\delta }{}^{\varepsilon }{}_{\epsilon 
}{}^{\mu } +  m_{225}^{} F_{\epsilon \varepsilon }{}^{\sigma 
\lambda } F_{\mu \nu \sigma \lambda } R_{\alpha \gamma \beta 
}{}^{\epsilon } R^{\alpha \beta \gamma \delta } R_{\delta }{}^{
\varepsilon \mu \nu } + \nn\\&& m_{231}^{} F_{\beta \mu }{}^{\sigma 
\lambda } F_{\delta \nu \sigma \lambda } R_{\alpha \gamma 
}{}^{\epsilon \varepsilon } R^{\alpha \beta \gamma \delta } R_{
\epsilon }{}^{\mu }{}_{\varepsilon }{}^{\nu } +  m_{232}^{} 
F_{\beta \delta }{}^{\sigma \lambda } F_{\mu \nu \sigma 
\lambda } R_{\alpha \gamma }{}^{\epsilon \varepsilon } 
R^{\alpha \beta \gamma \delta } R_{\epsilon }{}^{\mu 
}{}_{\varepsilon }{}^{\nu } + \nn\\&& m_{233}^{} F_{\beta \delta \mu 
}{}^{\lambda } F_{\varepsilon \nu \sigma \lambda } R_{\alpha 
\gamma }{}^{\epsilon \varepsilon } R^{\alpha \beta \gamma 
\delta } R_{\epsilon }{}^{\mu \nu \sigma } + m_{270}^{} 
F_{\epsilon \varepsilon }{}^{\sigma \lambda } F_{\mu \nu 
\sigma \lambda } R_{\alpha \gamma \beta \delta } R^{\alpha 
\beta \gamma \delta } R^{\epsilon \varepsilon \mu \nu } + 
\nn\\&& m_{272}^{} F_{\delta \varepsilon \mu }{}^{\lambda } F_{\epsilon 
\nu \sigma \lambda } R_{\alpha \gamma \beta }{}^{\epsilon } 
R^{\alpha \beta \gamma \delta } R^{\varepsilon \mu \nu \sigma 
} + m_{273}^{} F_{\beta \varepsilon \mu \nu } F_{\delta 
\epsilon \sigma \lambda } R_{\alpha }{}^{\epsilon }{}_{\gamma 
}{}^{\varepsilon } R^{\alpha \beta \gamma \delta } R^{\mu \nu 
\sigma \lambda } + \nn\\&& m_{274}^{} F_{\beta \epsilon \mu \nu } 
F_{\delta \varepsilon \sigma \lambda } R_{\alpha }{}^{\epsilon 
}{}_{\gamma }{}^{\varepsilon } R^{\alpha \beta \gamma \delta } 
R^{\mu \nu \sigma \lambda } + m_{275}^{} F_{\beta \delta \mu 
\nu } F_{\epsilon \varepsilon \sigma \lambda } R_{\alpha 
}{}^{\epsilon }{}_{\gamma }{}^{\varepsilon } R^{\alpha \beta 
\gamma \delta } R^{\mu \nu \sigma \lambda }+\nn\\&& m_{210}^{} FF R_{\alpha }{}^{\epsilon }{}_{\gamma 
}{}^{\varepsilon } R^{\alpha \beta \gamma \delta } R_{\beta 
\varepsilon \delta \epsilon } +  m_{214}^{} FF R_{\alpha \beta }{}^{
\epsilon \varepsilon } R^{\alpha \beta \gamma \delta } 
R_{\gamma \epsilon \delta \varepsilon }\labell{T4}
\eeqa

There are 7 couplings with structure of four Riemann curvatures, \ie
\beqa
{\cal L}_6^{R^4}&=& m_{226}^{} R_{\alpha \beta }{}^{\epsilon \varepsilon } 
R^{\alpha \beta \gamma \delta } R_{\gamma }{}^{\mu 
}{}_{\epsilon }{}^{\nu } R_{\delta \mu \varepsilon \nu } + 
 m_{227}^{} R_{\alpha }{}^{\epsilon }{}_{\gamma }{}^{\varepsilon } 
R^{\alpha \beta \gamma \delta } R_{\beta }{}^{\mu 
}{}_{\epsilon }{}^{\nu } R_{\delta \nu \varepsilon \mu } + 
\nn\\&& m_{228}^{} R_{\alpha \beta }{}^{\epsilon \varepsilon } R^{\alpha 
\beta \gamma \delta } R_{\gamma }{}^{\mu }{}_{\epsilon 
}{}^{\nu } R_{\delta \nu \varepsilon \mu } + m_{229}^{} 
R_{\alpha \beta }{}^{\epsilon \varepsilon } R^{\alpha \beta 
\gamma \delta } R_{\gamma }{}^{\mu }{}_{\delta }{}^{\nu } 
R_{\epsilon \mu \varepsilon \nu } + \nn\\&& m_{230}^{} R_{\alpha \gamma 
\beta }{}^{\epsilon } R^{\alpha \beta \gamma \delta } 
R_{\delta }{}^{\varepsilon \mu \nu } R_{\epsilon \mu 
\varepsilon \nu } +  m_{234}^{} R_{\alpha }{}^{\epsilon 
}{}_{\gamma }{}^{\varepsilon } R^{\alpha \beta \gamma \delta } 
R_{\beta }{}^{\mu }{}_{\delta }{}^{\nu } R_{\epsilon \nu 
\varepsilon \mu } + \nn\\&& m_{271}^{} R_{\alpha \gamma \beta \delta } 
R^{\alpha \beta \gamma \delta } R_{\epsilon \mu \varepsilon 
\nu } R^{\epsilon \varepsilon \mu \nu }\labell{T5}
\eeqa

There are 24 couplings with structure of two Riemann curvatures and  two $\nabla F$, \ie
\beqa
\cL_6^{R^2(\prt F)^2}&=& m_{282}^{} R_{\alpha }{}^{\epsilon }{}_{\gamma }{}^{\varepsilon 
} R^{\alpha \beta \gamma \delta } \nabla_{\epsilon }F_{\beta 
}{}^{\mu \nu \sigma } \nabla_{\varepsilon }F_{\delta \mu \nu 
\sigma } +  m_{283}^{} R_{\alpha \beta }{}^{\epsilon \varepsilon 
} R^{\alpha \beta \gamma \delta } \nabla_{\epsilon }F_{\gamma 
}{}^{\mu \nu \sigma } \nabla_{\varepsilon }F_{\delta \mu \nu 
\sigma } + \nn\\&& m_{284}^{} R_{\alpha }{}^{\epsilon }{}_{\gamma 
}{}^{\varepsilon } R^{\alpha \beta \gamma \delta } 
\nabla_{\delta }F_{\beta }{}^{\mu \nu \sigma } 
\nabla_{\varepsilon }F_{\epsilon \mu \nu \sigma } +  m_{285}^{} 
R_{\alpha }{}^{\epsilon \varepsilon \mu } R^{\alpha \beta 
\gamma \delta } \nabla_{\gamma }F_{\beta \varepsilon }{}^{\nu 
\sigma } \nabla_{\mu }F_{\delta \epsilon \nu \sigma } + 
\nn\\&& m_{286}^{} R_{\alpha }{}^{\epsilon \varepsilon \mu } R^{\alpha 
\beta \gamma \delta } \nabla_{\gamma }F_{\beta \epsilon 
}{}^{\nu \sigma } \nabla_{\mu }F_{\delta \varepsilon \nu 
\sigma } + m_{287}^{} R_{\alpha }{}^{\epsilon \varepsilon \mu } 
R^{\alpha \beta \gamma \delta } \nabla_{\epsilon }F_{\beta 
\gamma }{}^{\nu \sigma } \nabla_{\mu }F_{\delta \varepsilon 
\nu \sigma } + \nn\\&& m_{292}^{} R_{\alpha }{}^{\epsilon \varepsilon 
\mu } R^{\alpha \beta \gamma \delta } \nabla_{\delta 
}F_{\beta \gamma }{}^{\nu \sigma } \nabla_{\mu }F_{\epsilon 
\varepsilon \nu \sigma } + m_{294}^{} R^{\alpha \beta \gamma 
\delta } R^{\epsilon \varepsilon \mu \nu } \nabla_{\mu 
}F_{\alpha \beta \epsilon }{}^{\sigma } \nabla_{\nu }F_{\gamma 
\delta \varepsilon \sigma } + \nn\\&& m_{295}^{} R^{\alpha \beta \gamma 
\delta } R^{\epsilon \varepsilon \mu \nu } \nabla_{\varepsilon 
}F_{\alpha \beta \epsilon }{}^{\sigma } \nabla_{\nu }F_{\gamma 
\delta \mu \sigma } + m_{296}^{} R^{\alpha \beta \gamma \delta 
} R^{\epsilon \varepsilon \mu \nu } \nabla_{\gamma }F_{\alpha 
\beta \epsilon }{}^{\sigma } \nabla_{\nu }F_{\delta 
\varepsilon \mu \sigma } + \nn\\&& m_{316}^{} R^{\alpha \beta \gamma 
\delta } R^{\epsilon \varepsilon \mu \nu } \nabla_{\sigma }F_{
\beta \delta \varepsilon \nu } \nabla^{\sigma }F_{\alpha 
\gamma \epsilon \mu } +  m_{317}^{} R_{\alpha }{}^{\epsilon 
\varepsilon \mu } R^{\alpha \beta \gamma \delta } \nabla_{\mu 
}F_{\delta \epsilon \nu \sigma } \nabla^{\sigma }F_{\beta 
\gamma \varepsilon }{}^{\nu } + \nn\\&& m_{318}^{} R_{\alpha 
}{}^{\epsilon \varepsilon \mu } R^{\alpha \beta \gamma \delta 
} \nabla_{\epsilon }F_{\gamma \delta \nu \sigma } 
\nabla^{\sigma }F_{\beta \varepsilon \mu }{}^{\nu } + 
 m_{319}^{} R_{\alpha }{}^{\epsilon \varepsilon \mu } R^{\alpha 
\beta \gamma \delta } \nabla_{\nu }F_{\gamma \delta \epsilon 
\sigma } \nabla^{\sigma }F_{\beta \varepsilon \mu }{}^{\nu } + 
\nn\\&& m_{320}^{} R_{\alpha }{}^{\epsilon \varepsilon \mu } R^{\alpha 
\beta \gamma \delta } \nabla_{\sigma }F_{\gamma \delta 
\epsilon \nu } \nabla^{\sigma }F_{\beta \varepsilon \mu 
}{}^{\nu } +  m_{321}^{} R_{\alpha }{}^{\epsilon }{}_{\gamma }{}^{
\varepsilon } R^{\alpha \beta \gamma \delta } \nabla_{\nu }F_{
\delta \epsilon \mu \sigma } \nabla^{\sigma }F_{\beta 
\varepsilon }{}^{\mu \nu } + \nn\\&& m_{322}^{} R_{\alpha }{}^{\epsilon 
}{}_{\gamma }{}^{\varepsilon } R^{\alpha \beta \gamma \delta } 
\nabla_{\sigma }F_{\delta \epsilon \mu \nu } \nabla^{\sigma 
}F_{\beta \varepsilon }{}^{\mu \nu } + m_{323}^{} R_{\alpha 
\beta }{}^{\epsilon \varepsilon } R^{\alpha \beta \gamma 
\delta } \nabla_{\nu }F_{\epsilon \varepsilon \mu \sigma } 
\nabla^{\sigma }F_{\gamma \delta }{}^{\mu \nu } + \nn\\&& m_{324}^{} 
R_{\alpha \beta }{}^{\epsilon \varepsilon } R^{\alpha \beta 
\gamma \delta } \nabla_{\sigma }F_{\epsilon \varepsilon \mu 
\nu } \nabla^{\sigma }F_{\gamma \delta }{}^{\mu \nu } + 
 m_{325}^{} R_{\alpha \beta }{}^{\epsilon \varepsilon } R^{\alpha 
\beta \gamma \delta } \nabla_{\nu }F_{\delta \varepsilon \mu 
\sigma } \nabla^{\sigma }F_{\gamma \epsilon }{}^{\mu \nu } + 
\nn\\&& m_{326}^{} R_{\alpha \beta }{}^{\epsilon \varepsilon } R^{\alpha 
\beta \gamma \delta } \nabla_{\sigma }F_{\delta \varepsilon 
\mu \nu } \nabla^{\sigma }F_{\gamma \epsilon }{}^{\mu \nu } + 
 m_{327}^{} R_{\alpha \gamma \beta }{}^{\epsilon } R^{\alpha 
\beta \gamma \delta } \nabla_{\nu }F_{\epsilon \varepsilon 
\mu \sigma } \nabla^{\sigma }F_{\delta }{}^{\varepsilon \mu 
\nu } + \nn\\&& m_{328}^{} R_{\alpha \gamma \beta }{}^{\epsilon } 
R^{\alpha \beta \gamma \delta } \nabla_{\sigma }F_{\epsilon 
\varepsilon \mu \nu } \nabla^{\sigma }F_{\delta 
}{}^{\varepsilon \mu \nu } +  m_{329}^{} R_{\alpha \gamma \beta 
\delta } R^{\alpha \beta \gamma \delta } \nabla_{\sigma 
}F_{\epsilon \varepsilon \mu \nu } \nabla^{\sigma }F^{\epsilon 
\varepsilon \mu \nu }\labell{T8}
\eeqa

There are 15 couplings with structure of four $\nabla F$, \ie
\beqa
\cL_6^{(\prt F)^4}&=& m_{375}^{} \nabla_{\epsilon }F_{\alpha \beta }{}^{\varepsilon 
\mu } \nabla^{\epsilon }F^{\alpha \beta \gamma \delta } 
\nabla_{\lambda }F_{\varepsilon \mu \nu \sigma } 
\nabla^{\lambda }F_{\gamma \delta }{}^{\nu \sigma } + 
 m_{378}^{} \nabla_{\epsilon }F_{\alpha \beta }{}^{\varepsilon 
\mu } \nabla^{\epsilon }F^{\alpha \beta \gamma \delta } 
\nabla_{\lambda }F_{\delta \mu \nu \sigma } \nabla^{\lambda 
}F_{\gamma \varepsilon }{}^{\nu \sigma } + \nn\\&& m_{384}^{} 
\nabla_{\epsilon }F_{\alpha \beta \gamma }{}^{\varepsilon } 
\nabla^{\epsilon }F^{\alpha \beta \gamma \delta } 
\nabla_{\lambda }F_{\varepsilon \mu \nu \sigma } 
\nabla^{\lambda }F_{\delta }{}^{\mu \nu \sigma } +  m_{385}^{} 
\nabla^{\epsilon }F^{\alpha \beta \gamma \delta } 
\nabla^{\varepsilon }F_{\alpha \beta \gamma \delta } 
\nabla_{\lambda }F_{\varepsilon \mu \nu \sigma } 
\nabla^{\lambda }F_{\epsilon }{}^{\mu \nu \sigma } + \nn\\&& m_{386}^{} 
\nabla_{\delta }F_{\mu \nu \sigma \lambda } \nabla^{\epsilon 
}F^{\alpha \beta \gamma \delta } \nabla^{\varepsilon 
}F_{\alpha \beta \gamma \epsilon } \nabla^{\lambda 
}F_{\varepsilon }{}^{\mu \nu \sigma } + m_{387}^{} 
\nabla_{\epsilon }F_{\alpha \beta \gamma \delta } 
\nabla^{\epsilon }F^{\alpha \beta \gamma \delta } 
\nabla_{\lambda }F_{\varepsilon \mu \nu \sigma } 
\nabla^{\lambda }F^{\varepsilon \mu \nu \sigma } + \nn\\&& m_{379}^{} 
\nabla^{\epsilon }F^{\alpha \beta \gamma \delta } 
\nabla_{\lambda }F_{\varepsilon \mu \nu \sigma } 
\nabla^{\lambda }F_{\delta \epsilon }{}^{\nu \sigma } \nabla^{
\mu }F_{\alpha \beta \gamma }{}^{\varepsilon } + m_{380}^{} 
\nabla^{\epsilon }F^{\alpha \beta \gamma \delta } 
\nabla_{\lambda }F_{\epsilon \mu \nu \sigma } \nabla^{\lambda 
}F_{\delta \varepsilon }{}^{\nu \sigma } \nabla^{\mu 
}F_{\alpha \beta \gamma }{}^{\varepsilon } + \nn\\&& m_{381}^{} \nabla^{
\epsilon }F^{\alpha \beta \gamma \delta } \nabla_{\varepsilon 
}F_{\epsilon \nu \sigma \lambda } \nabla^{\lambda }F_{\delta 
\mu }{}^{\nu \sigma } \nabla^{\mu }F_{\alpha \beta \gamma 
}{}^{\varepsilon } + m_{383}^{} \nabla^{\epsilon }F^{\alpha 
\beta \gamma \delta } \nabla_{\lambda }F_{\epsilon \varepsilon 
\nu \sigma } \nabla^{\lambda }F_{\delta \mu }{}^{\nu \sigma } 
\nabla^{\mu }F_{\alpha \beta \gamma }{}^{\varepsilon } + 
\nn\\&& m_{293}^{} \nabla_{\epsilon }F_{\delta }{}^{\nu \sigma \lambda 
} \nabla^{\epsilon }F^{\alpha \beta \gamma \delta } 
\nabla_{\mu }F_{\varepsilon \nu \sigma \lambda } \nabla^{\mu 
}F_{\alpha \beta \gamma }{}^{\varepsilon } + m_{374}^{} \nabla^{
\epsilon }F^{\alpha \beta \gamma \delta } \nabla_{\lambda }F_{
\epsilon \varepsilon \mu \sigma } \nabla^{\lambda }F_{\gamma 
\delta \nu }{}^{\sigma } \nabla^{\nu }F_{\alpha \beta 
}{}^{\varepsilon \mu } + \nn\\&& m_{377}^{} \nabla^{\epsilon }F^{\alpha 
\beta \gamma \delta } \nabla^{\lambda }F_{\gamma \epsilon \nu 
}{}^{\sigma } \nabla_{\mu }F_{\delta \varepsilon \sigma 
\lambda } \nabla^{\nu }F_{\alpha \beta }{}^{\varepsilon \mu } 
+  m_{376}^{} \nabla^{\epsilon }F^{\alpha \beta \gamma \delta } 
\nabla^{\lambda }F_{\gamma \epsilon \varepsilon }{}^{\sigma } 
\nabla^{\nu }F_{\alpha \beta }{}^{\varepsilon \mu } 
\nabla_{\sigma }F_{\delta \mu \nu \lambda } + \nn\\&& m_{382}^{} 
\nabla^{\epsilon }F^{\alpha \beta \gamma \delta } 
\nabla^{\lambda }F_{\delta \mu }{}^{\nu \sigma } \nabla^{\mu 
}F_{\alpha \beta \gamma }{}^{\varepsilon } \nabla_{\sigma 
}F_{\epsilon \varepsilon \nu \lambda }\labell{T9}
\eeqa

We have also found there are 134 couplings with structure $F^8$, 530 couplings with structure $F^4(\nabla F)^2$ and  217 couplings with structure $RF^2(\nabla F)^2$ that appear in the Appendix.

 Even though the total number of minimal gauge invariant  couplings at order $\ell_p^6$ are fixed, \ie 1062,  the number of couplings in each structure are not fixed. In different schemes, one may find  different structures and different number of couplings in each structure. The above structures and the number of terms in each structure are fixed in the scheme that we have chosen. Note, however, that 104 couplings in the structure  $\cL_6^{F^8}$  are invariant under field redefinition, Bianchi identities and total derivative terms. They are scheme independent. All other couplings dependent on the scheme that one uses for the  couplings. The values of the 1062 parameters may be  fixed  by various techniques in  M-theory. 

They may be fixed by reducing the couplings on a circle to produce the type IIA couplings at one-loop. Then one may find the 1062 parameters by calculating various S-matrix elements in the resulting  type IIA effective field theory and comparing them with the corresponding S-matrix elements in the type IIA superstring theory which has no arbitrary parameters. In this method one has to calculate in the  string theory various S-matrix elements which produces 1062 independent contact terms. In the next section we briefly discuss the dimensional reduction of the couplings on a circle to fix some of the parameters.

\section{Reduction on a circle}\label{sec.3}
The  dimensional reduction of  the 11-dimensional couplings on a circle can be done by using the following   Kaluza-Kelin (KK) reduction of the metric:
\beqa
g_{\mu\nu}=e^{-2\Phi/3}\left(\matrix{G_{ab}+e^{2\Phi}C_aC_b& e^{2\Phi}C_a&\cr e^{2\Phi}C_b&e^{2\Phi}&}\right)\,\,;\,\,g^{\mu\nu}=e^{2\Phi/3}\left(\matrix{G^{ab} &  -C^a&\cr -C^b&e^{-2\Phi}+C_aC^a&}\right)\labell{reduc}\eeqa
where $G^{ab}$ is the inverse of the 10-dimensional   metric which raises the index of the  the R-R vector $C_a$, and the following  reductions for the three-form:
\beqa
A_{abc}=C_{abc}\,\,;\,\,
A_{aby}=B_{ab}\labell{CC}
\eeqa
where $C^{(3)}$ is the R-R three-form and $B$ is the antisymmetric $B$-field of the type IIA superstring theory. Using these reduction one can calculate the reduction of different 11-dimensional couplings to the 10 dimensions, \eg the reduction of the overall factor $\sqrt{-g}$ and the scalar curvature in $S_0$ are
\beqa
\sqrt{-g}&=&e^{-8\Phi/3}\sqrt{-G}\nn\\
R&=&e^{2\Phi/3}\left(R-\frac{16}{3}\nabla_a\Phi\nabla^a\Phi+\frac{14}{3}\nabla_a\nabla^a\Phi-\frac{1}{2.2!}e^{2\Phi}F_{ab}F^{ab}\right)
\eeqa
where $F_{ac}$ if field strength of the R-R one-form.
Up to a total derivative term they produce the standard reduction, \ie
\beqa
\sqrt{-g}R&=&e^{-2\Phi}\sqrt{-G}\left(R+4\nabla_a\Phi\nabla^a\Phi-\frac{1}{2.2!}e^{2\Phi}F_{ab}F^{ab}\right)
\eeqa 
The reduction of the  coupling in the action $S_0$ involving the field strength of the three-form is
\beqa
-\frac{1}{2.4!}\sqrt{-g}F_{\mu\nu\alpha\beta}F^{\mu\nu\alpha\beta}&=&e^{-2\Phi}\sqrt{-G}\left(-
\frac{1}{2.3!}H_{abc}H^{abc}-\frac{1}{2.4!}e^{2\Phi}\bar{F}_{abcd}\bar{F}^{abcd}\right)
\eeqa
where the R-R field strength $\bar{F}^{(4)}$ is $\bar{F}_{abcd}=F_{abcd}+H_{[abc}C_{d]}$. 
Note that the dilaton factor indicates that the reduction of $S_0$ correspond to the sphere-level effective action of type IIA. There are stringy corrections to the sphere-level effective action of type IIA which are related to the non-zero modes of the KK mass spectrum \cite{Green:1997as}. 

Using the relation between type IIA coupling $g_s$, the string length $\ell_s$ and the 11-dimensional Plank length $\ell_p$, \ie $\ell_p=g_s^{1/3}\ell_s$, and the fact that the dilaton factor in the $n_h$-loop effective action of type IIA superstring theory is given by $e^{-(2-2n_h)\Phi}$, one finds the relation $6n_h=n$  between the derivative couplings in the M-theory  at the level $\ell_p^n$, and the $n_h$-loop couplings in the type IIA theory. Then the  allowed couplings in the $\ell_p$-expansion are at $n=0,6,12,18,24,\cdots$. They are correspond to the loop-level couplings in type IIA theory with $n_h=0,1,2,3,4,\cdots$, respectively.  In other words, the couplings at each loop-level has no higher-loop corrections. However, there are stringy corrections at each loop-level which are related to the non-zero modes of the KK mass spectrum.  

The reduction of the couplings in  $S_6$  which have no three-form is
\beqa
\frac{\ell_p^6}{2\kappa_{11}^2}\int d^{11}x\sqrt{-g}\cL_6^{(R^4)}&=&\frac{2\pi R_{11}\ell_p^6}{2\kappa_{11}^2}\int d^{10}x\sqrt{-G}\Big[ m_{226}^{} R_{\alpha \beta }{}^{\epsilon \varepsilon } 
R^{\alpha \beta \gamma \delta } R_{\gamma }{}^{\mu 
}{}_{\epsilon }{}^{\nu } R_{\delta \mu \varepsilon \nu }   
\labell{RA}\\&&+ m_{228}^{} R_{\alpha \beta }{}^{\epsilon \varepsilon } R^{\alpha 
\beta \gamma \delta } R_{\gamma }{}^{\mu }{}_{\epsilon 
}{}^{\nu } R_{\delta \nu \varepsilon \mu } + m_{229}^{} 
R_{\alpha \beta }{}^{\epsilon \varepsilon } R^{\alpha \beta 
\gamma \delta } R_{\gamma }{}^{\mu }{}_{\delta }{}^{\nu } 
R_{\epsilon \mu \varepsilon \nu }  \nn\\&&+ m_{230}^{} R_{\alpha \gamma 
\beta }{}^{\epsilon } R^{\alpha \beta \gamma \delta } 
R_{\delta }{}^{\varepsilon \mu \nu } R_{\epsilon \mu 
\varepsilon \nu } +  m_{234}^{} R_{\alpha }{}^{\epsilon 
}{}_{\gamma }{}^{\varepsilon } R^{\alpha \beta \gamma \delta } 
R_{\beta }{}^{\mu }{}_{\delta }{}^{\nu } R_{\epsilon \nu 
\varepsilon \mu }  \nn\\&&+ m_{271}^{} R_{\alpha \gamma \beta \delta } 
R^{\alpha \beta \gamma \delta } R_{\epsilon \mu \varepsilon 
\nu } R^{\epsilon \varepsilon \mu \nu }+ 
 m_{227}^{} R_{\alpha }{}^{\epsilon }{}_{\gamma }{}^{\varepsilon } 
R^{\alpha \beta \gamma \delta } R_{\beta }{}^{\mu 
}{}_{\epsilon }{}^{\nu } R_{\delta \nu \varepsilon \mu }+\cdots\Big]\nn
\eeqa
where $R_{11}=\ell_sg_s$ is the radious of the circle and dots represent the R-R one-form and the dilaton couplings in the  effective action of the type IIA theory. Note that as expected there is no overall dilaton factor which indicates that the above couplings correspond to the tours-level effective action of type IIA. On the other hand, the one-loop gravity couplings in type IIA theory are given in a scheme which includes the  Ricci curvature and the Ricci scalar, as  \cite{Sakai:1986bi,Antoniadis:1997eg,Kiritsis:1997em}
\beqa
\bS_3(G)&=&\frac{\ell_s^6g_2^2}{2\kappa^2}\frac{a}{3.2^7}\int d^{10}x\,\sqrt{-G}(t_8t_8-\frac{1}{4}\epsilon_8\epsilon_8)R^4\labell{S31}
\eeqa
where $a$ is  a constant number, $\kappa_{11}^2=2\pi\ell_sg_s\kappa^2$  and the tensors $\epsilon_8\epsilon_8$ and $t_8$ are defined as
\beqa
\epsilon_8{}^{\mu_1\cdots\mu_8}\epsilon_8{}^{\nu_1\cdots\nu_8}&=&\frac{1}{2}\epsilon_{10}{}^{\mu_1\cdots\mu_8\alpha\beta}\epsilon_{10}{}^{\nu_1\cdots\nu_8}{}_{\alpha\beta}\\
t_8^{\mu_1\cdots\mu_8}M^1_{\mu1\mu2}M^2_{\mu3\mu4}M^3_{\mu5\mu6}M^4_{\mu7\mu8}&=&8\Tr(M^1M^2M^3M^4)+8\Tr(M^1M^3M^4M^2)\nn\\&&+8\Tr(M^1M^3M^2M^4)-2\Tr(M^1M^2)\Tr(M^3M^4)\nn\\&&-2\Tr(M^1M^3)\Tr(M^2M^4)-2\Tr(M^1M^4)\Tr(M^2M^3)\nn
\eeqa
where $M^1,\cdots, M^4$ are four arbitrary antisymmetric matrices. The Ricci curvature and the Ricci scalar in above couplings can be removed by a field redefinition. The  Riemann curvature  couplings can then be compared with the couplings in \reef{RA}. One finds the following parameters for the couplings in \reef{T5}:
\beqa
m_{227}=0\,;\,m_{226}=m_{228}=-m_{229}=\frac{1}{4}m_{230}=-m_{234}=-4m_{271}=- a
\eeqa
The gravity couplings are then fixed as
\beqa
{\cal L}_6^{R^4}&\!\!\!\!\!=\!\!\!\!\!& a\Bigg[- R_{\alpha \beta }{}^{\epsilon \varepsilon } 
R^{\alpha \beta \gamma \delta } R_{\gamma }{}^{\mu 
}{}_{\epsilon }{}^{\nu } R_{\delta \mu \varepsilon \nu }-R_{\alpha \beta }{}^{\epsilon \varepsilon } R^{\alpha 
\beta \gamma \delta } R_{\gamma }{}^{\mu }{}_{\epsilon 
}{}^{\nu } R_{\delta \nu \varepsilon \mu } +  
R_{\alpha \beta }{}^{\epsilon \varepsilon } R^{\alpha \beta 
\gamma \delta } R_{\gamma }{}^{\mu }{}_{\delta }{}^{\nu } 
R_{\epsilon \mu \varepsilon \nu }  \nn\\&& -4 R_{\alpha \gamma 
\beta }{}^{\epsilon } R^{\alpha \beta \gamma \delta } 
R_{\delta }{}^{\varepsilon \mu \nu } R_{\epsilon \mu 
\varepsilon \nu } +   R_{\alpha }{}^{\epsilon 
}{}_{\gamma }{}^{\varepsilon } R^{\alpha \beta \gamma \delta } 
R_{\beta }{}^{\mu }{}_{\delta }{}^{\nu } R_{\epsilon \nu 
\varepsilon \mu } +\frac{1}{4} R_{\alpha \gamma \beta \delta } 
R^{\alpha \beta \gamma \delta } R_{\epsilon \mu \varepsilon 
\nu } R^{\epsilon \varepsilon \mu \nu }\Bigg]
\eeqa

The complete one-loop effective action of type IIA for other NS-NS or R-R fields are not known. Hence, the other parameters in the M-theory effective action can not be fixed completely in this way.
However, the  couplings involving  four NS-NS fields are known to be given by \reef{S31} in which the Riemann curvature is replaced by the following expression \cite{Gross:1986mw}:
\beqa
R_{\mu\nu\alpha\beta}&=& \cR_{\mu\nu\alpha\beta}+\frac{1}{2}\prt_{\beta}H_{\mu\nu\alpha}-\frac{1}{2}\prt_{\alpha}H_{\mu\nu\beta}
\eeqa 
where $\cR_{\mu\nu\alpha\beta}$ is the linearised Riemann curvature.  The last term in \reef{S31} has no effect in four-point functions.  One can compare the   four-point functions resulting from the first term in \reef{T53} with the corresponding four-point functions in the dimensional reduction of the couplings in \reef{T53}, \reef{T8} and \reef{T9}. This S-matrix constraint  fixes the parameter in \reef{T53} to be zero, \ie  
\beqa
m_{315}= 0
\eeqa
 and  fixes the following relations between the couplings in \reef{T8}:
\beqa
&& m_{283}= -a/6 + m_{282}/2,\, m_{284}= a/3 - m_{282},\, m_{286}= -2a + 2 m_{285},\, 
 m_{287}= a - 2 m_{285},\,\nn\\&&
  m_{292}= -a + m_{285},\, m_{296}= 2 m_{295},\, 
 m_{317}= 4a - 4 m_{285} - 4 m_{294} - 4 m_{295},\, m_{318}= -a/2,\, \nn\\&&
 m_{320}= -a/2 - m_{285}/2 + m_{294} - m_{295},\, 
 m_{321}= 3 m_{282} - 2 m_{285} - 2 m_{295},\,\nn\\&&
  m_{322}= -a/2 - 3 m_{282}/2 + m_{285},\,
  m_{324}= a/8 + 3 m_{282}/8 - m_{285}/4,\,\nn\\&&
   m_{325}= m_{285} - m_{294} - 2 m_{323},\, 
 m_{326}= -3 m_{282}/4,\, m_{327}= a - 6 m_{282} + m_{285},\, \nn\\&&
 m_{328}= -2a/3 + 2 m_{282},\, m_{329}= a/48 - m_{282}/16
\eeqa
and the following relations between the couplings in \reef{T9}:
\beqa
&&m_{375} = a/32 - 9 m_{293}/4 - m_{374}/2,\, m_{379} = -a/48 + 2 m_{374}/3 - 
  2 m_{378}/3,\,\nn\\&& m_{381} = a/24 - m_{377}/6 + 4 m_{378}/3 + 2 m_{380},\, m_{382} = 
 a/12 - 2 m_{376}/3 + 8 m_{378}/3 + 4 m_{380},\,\nn\\&& m_{383} = -a/48 + 2 m_{374}/
  3 + 2 m_{378},\, m_{384} = -a/48 + m_{293} + 2 m_{374}/9,\,\nn\\&& m_{385} = 
 a/288 + m_{374}/18 - m_{376}/36 + 4 m_{378}/9 + m_{380}/3,\,\nn\\&& m_{386} = 
 5a/144 + m_{293} - 2 m_{374}/9 - m_{376}/9 + 10 m_{378}/9 + 4 m_{380}/
  3,\,\nn\\&& m_{387} = a/576 - m_{293}/16 - m_{374}/48
\eeqa
It is extremely difficult to fix all 1062 parameters by the S-matrix method. One may   use symmetries of the effective action to fix them all.

The sphere-level gravity couplings in  type II theory at order $\alpha'^3$ is known to be 
\beqa
\int d^{10}x\,\sqrt{-G}e^{-2\Phi}(t_8t_8+\frac{1}{4}\epsilon_8\epsilon_8)R^4\labell{S32}
\eeqa
In this case the reduction of the classical theory on a circle has a $O(1,1)$-symmetry \cite{Sen:1991zi,Hohm:2014sxa}. This  symmetry may be  used  to find all couplings in type II effective action. In fact the $Z_2$-subgroup of this  symmetry has been used in \cite{Garousi:2020gio} to fix  all 872 parameters of the NS-NS couplings. There is no such bosonic symmetry in the  one-loop effective action.  The sypersymmetry of the effective actions, however, exsists in the classical and loop levels.   It has been shown in \cite{Green:1997di,Hyakutake:2007sm} that  the $R^4$ couplings and the Chern-Simons couplings $C\wedge R\wedge R\wedge R\wedge R$ transform into each other under the supersymmetry transformations.   It would be interesting to impose the supersymmetry constraint to fix all 1062 parameters in $\cL_6$ and also the parameters in the Chern-Simons sector $\cL_6^{CS}$.

%{\bf Acknowledgments}:   This work is supported by Ferdowsi University of Mashhad under grant  1/51021(1398/09/05).
 
\newpage
\appendix
{\bf \Large Appendix}\label{appen}

 \vskip 0.5 cm
 
In this appendix we write the independent couplings with the structures  $F^8$, $F^4(\nabla F)^2$, and $RF^2(\nabla F)^2$. There are 134 couplings with structure $F^8$, \ie
\beqa
{\cal L}_6^{F^8}&=& m_{1}^{} F_{\alpha }{}^{\epsilon \varepsilon \mu } F^{\alpha 
\beta \gamma \delta } F_{\beta }{}^{\nu \sigma \lambda } 
F_{\gamma }{}^{\kappa \tau \omega } F_{\delta }{}^{\varphi \xi 
\zeta } F_{\epsilon \nu \kappa \varphi } F_{\varepsilon \sigma 
\tau \xi } F_{\mu \lambda \omega \zeta } + \nn\\&& m_{28}^{} F_{\alpha 
\beta }{}^{\epsilon \varepsilon } F^{\alpha \beta \gamma 
\delta } F_{\gamma \epsilon }{}^{\mu \nu } F_{\delta 
}{}^{\sigma \lambda \kappa } F_{\varepsilon }{}^{\tau \omega 
\varphi } F_{\kappa \omega \varphi \zeta } F_{\mu \sigma \tau 
}{}^{\xi } F_{\nu \lambda \xi }{}^{\zeta } + \nn\\&& m_{39}^{} 
F_{\alpha \beta }{}^{\epsilon \varepsilon } F^{\alpha \beta 
\gamma \delta } F_{\gamma \epsilon }{}^{\mu \nu } F_{\delta 
}{}^{\sigma \lambda \kappa } F_{\varepsilon }{}^{\tau \omega 
\varphi } F_{\kappa \varphi \xi \zeta } F_{\mu \sigma \tau 
}{}^{\xi } F_{\nu \lambda \omega }{}^{\zeta } + \nn\\&& m_{40}^{} 
F_{\alpha \beta }{}^{\epsilon \varepsilon } F^{\alpha \beta 
\gamma \delta } F_{\gamma \delta }{}^{\mu \nu } F_{\epsilon 
}{}^{\sigma \lambda \kappa } F_{\varepsilon }{}^{\tau \omega 
\varphi } F_{\kappa \varphi \xi \zeta } F_{\mu \sigma \tau 
}{}^{\xi } F_{\nu \lambda \omega }{}^{\zeta } + \nn\\&& m_{29}^{} 
F_{\alpha \beta }{}^{\epsilon \varepsilon } F^{\alpha \beta 
\gamma \delta } F_{\gamma \epsilon }{}^{\mu \nu } F_{\delta 
}{}^{\sigma \lambda \kappa } F_{\varepsilon }{}^{\tau \omega 
\varphi } F_{\kappa \omega \varphi \zeta } F_{\mu \sigma 
\lambda }{}^{\xi } F_{\nu \tau \xi }{}^{\zeta } + \nn\\&& m_{31}^{} F_{
\alpha \beta }{}^{\epsilon \varepsilon } F^{\alpha \beta 
\gamma \delta } F_{\gamma \epsilon }{}^{\mu \nu } F_{\delta 
}{}^{\sigma \lambda \kappa } F_{\varepsilon \sigma }{}^{\tau 
\omega } F_{\kappa \omega \xi \zeta } F_{\mu \lambda 
}{}^{\varphi \xi } F_{\nu \tau \varphi }{}^{\zeta } + \nn\\&& m_{32}^{} 
F_{\alpha \beta }{}^{\epsilon \varepsilon } F^{\alpha \beta 
\gamma \delta } F_{\gamma \delta }{}^{\mu \nu } F_{\epsilon 
}{}^{\sigma \lambda \kappa } F_{\varepsilon \sigma }{}^{\tau 
\omega } F_{\kappa \omega \xi \zeta } F_{\mu \lambda 
}{}^{\varphi \xi } F_{\nu \tau \varphi }{}^{\zeta } + \nn\\&& m_{41}^{} 
F_{\alpha \beta }{}^{\epsilon \varepsilon } F^{\alpha \beta 
\gamma \delta } F_{\gamma \delta }{}^{\mu \nu } F_{\epsilon 
}{}^{\sigma \lambda \kappa } F_{\varepsilon }{}^{\tau \omega 
\varphi } F_{\kappa \varphi \xi \zeta } F_{\mu \sigma \lambda 
}{}^{\xi } F_{\nu \tau \omega }{}^{\zeta } + \nn\\&& m_{18}^{} 
F_{\alpha \beta }{}^{\epsilon \varepsilon } F^{\alpha \beta 
\gamma \delta } F_{\gamma \epsilon }{}^{\mu \nu } F_{\delta 
}{}^{\sigma \lambda \kappa } F_{\varepsilon \sigma }{}^{\tau 
\omega } F_{\kappa \tau \omega \zeta } F_{\mu \lambda 
}{}^{\varphi \xi } F_{\nu \varphi \xi }{}^{\zeta } + \nn\\&& m_{42}^{} 
F_{\alpha \beta }{}^{\epsilon \varepsilon } F^{\alpha \beta 
\gamma \delta } F_{\gamma \epsilon }{}^{\mu \nu } F_{\delta 
}{}^{\sigma \lambda \kappa } F_{\varepsilon }{}^{\tau \omega 
\varphi } F_{\kappa \varphi \xi \zeta } F_{\mu \sigma \lambda 
\tau } F_{\nu \omega }{}^{\xi \zeta } + \nn\\&& m_{43}^{} F_{\alpha 
\beta }{}^{\epsilon \varepsilon } F^{\alpha \beta \gamma 
\delta } F_{\gamma \delta }{}^{\mu \nu } F_{\epsilon 
}{}^{\sigma \lambda \kappa } F_{\varepsilon }{}^{\tau \omega 
\varphi } F_{\kappa \varphi \xi \zeta } F_{\mu \sigma \lambda 
\tau } F_{\nu \omega }{}^{\xi \zeta } + \nn\\&& m_{23}^{} F_{\alpha 
\beta }{}^{\epsilon \varepsilon } F^{\alpha \beta \gamma 
\delta } F_{\gamma \delta }{}^{\mu \nu } F_{\epsilon 
}{}^{\sigma \lambda \kappa } F_{\varepsilon \sigma \lambda 
}{}^{\tau } F_{\kappa \tau \xi \zeta } F_{\mu }{}^{\omega 
\varphi \xi } F_{\nu \omega \varphi }{}^{\zeta } + \nn\\&& m_{24}^{} 
F_{\alpha \beta }{}^{\epsilon \varepsilon } F^{\alpha \beta 
\gamma \delta } F_{\gamma \epsilon }{}^{\mu \nu } F_{\delta 
\mu }{}^{\sigma \lambda } F_{\varepsilon \sigma }{}^{\kappa 
\tau } F_{\kappa \tau \xi \zeta } F_{\lambda \omega \varphi 
}{}^{\zeta } F_{\nu }{}^{\omega \varphi \xi } + \nn\\&& m_{25}^{} 
F_{\alpha \beta }{}^{\epsilon \varepsilon } F^{\alpha \beta 
\gamma \delta } F_{\gamma \delta }{}^{\mu \nu } F_{\epsilon 
\mu }{}^{\sigma \lambda } F_{\varepsilon \sigma }{}^{\kappa 
\tau } F_{\kappa \tau \xi \zeta } F_{\lambda \omega \varphi 
}{}^{\zeta } F_{\nu }{}^{\omega \varphi \xi } + \nn\\&& m_{26}^{} 
F_{\alpha \beta }{}^{\epsilon \varepsilon } F^{\alpha \beta 
\gamma \delta } F_{\gamma \epsilon }{}^{\mu \nu } F_{\delta 
\varepsilon }{}^{\sigma \lambda } F_{\kappa \tau \xi \zeta } 
F_{\lambda \omega \varphi }{}^{\zeta } F_{\mu \sigma 
}{}^{\kappa \tau } F_{\nu }{}^{\omega \varphi \xi } + \nn\\&& m_{27}^{} 
F_{\alpha \beta }{}^{\epsilon \varepsilon } F^{\alpha \beta 
\gamma \delta } F_{\gamma \delta }{}^{\mu \nu } F_{\epsilon 
\varepsilon }{}^{\sigma \lambda } F_{\kappa \tau \xi \zeta } 
F_{\lambda \omega \varphi }{}^{\zeta } F_{\mu \sigma 
}{}^{\kappa \tau } F_{\nu }{}^{\omega \varphi \xi } + \nn\\&& m_{5}^{} 
F_{\alpha \beta }{}^{\epsilon \varepsilon } F^{\alpha \beta 
\gamma \delta } F_{\gamma \epsilon }{}^{\mu \nu } F_{\delta 
\mu }{}^{\sigma \lambda } F_{\varepsilon }{}^{\kappa \tau 
\omega } F_{\lambda \omega \xi \zeta } F_{\nu }{}^{\varphi 
\xi \zeta } F_{\sigma \kappa \tau \varphi } + \nn\\&& m_{6}^{} 
F_{\alpha \beta }{}^{\epsilon \varepsilon } F^{\alpha \beta 
\gamma \delta } F_{\gamma \delta }{}^{\mu \nu } F_{\epsilon 
\mu }{}^{\sigma \lambda } F_{\varepsilon }{}^{\kappa \tau 
\omega } F_{\lambda \omega \xi \zeta } F_{\nu }{}^{\varphi 
\xi \zeta } F_{\sigma \kappa \tau \varphi } + \nn\\&& m_{7}^{} 
F_{\alpha \beta }{}^{\epsilon \varepsilon } F^{\alpha \beta 
\gamma \delta } F_{\gamma \epsilon }{}^{\mu \nu } F_{\delta 
\varepsilon }{}^{\sigma \lambda } F_{\lambda \omega \xi \zeta 
} F_{\mu }{}^{\kappa \tau \omega } F_{\nu }{}^{\varphi \xi 
\zeta } F_{\sigma \kappa \tau \varphi } + \nn\\&& m_{8}^{} F_{\alpha 
\beta }{}^{\epsilon \varepsilon } F^{\alpha \beta \gamma 
\delta } F_{\gamma \delta }{}^{\mu \nu } F_{\epsilon 
\varepsilon }{}^{\sigma \lambda } F_{\lambda \omega \xi \zeta 
} F_{\mu }{}^{\kappa \tau \omega } F_{\nu }{}^{\varphi \xi 
\zeta } F_{\sigma \kappa \tau \varphi } + \nn\\&& m_{2}^{} F_{\alpha 
}{}^{\epsilon \varepsilon \mu } F^{\alpha \beta \gamma \delta 
} F_{\beta \epsilon }{}^{\nu \sigma } F_{\gamma \varepsilon 
}{}^{\lambda \kappa } F_{\delta }{}^{\tau \omega \varphi } 
F_{\mu \tau }{}^{\xi \zeta } F_{\nu \lambda \omega \xi } 
F_{\sigma \kappa \varphi \zeta } + \nn\\&& m_{30}^{} F_{\alpha \beta 
}{}^{\epsilon \varepsilon } F^{\alpha \beta \gamma \delta } F_{
\gamma \epsilon }{}^{\mu \nu } F_{\delta }{}^{\sigma \lambda 
\kappa } F_{\varepsilon }{}^{\tau \omega \varphi } F_{\kappa 
\omega \varphi \zeta } F_{\mu \nu }{}^{\xi \zeta } F_{\sigma 
\lambda \tau \xi } + \nn\\&& m_{33}^{} F_{\alpha \beta }{}^{\epsilon 
\varepsilon } F^{\alpha \beta \gamma \delta } F_{\gamma 
}{}^{\mu \nu \sigma } F_{\delta \mu }{}^{\lambda \kappa } 
F_{\epsilon \nu }{}^{\tau \omega } F_{\varepsilon }{}^{\varphi 
\xi \zeta } F_{\kappa \omega \xi \zeta } F_{\sigma \lambda 
\tau \varphi } + \nn\\&& m_{34}^{} F_{\alpha \beta }{}^{\epsilon 
\varepsilon } F^{\alpha \beta \gamma \delta } F_{\gamma 
}{}^{\mu \nu \sigma } F_{\delta \mu }{}^{\lambda \kappa } 
F_{\epsilon \nu }{}^{\tau \omega } F_{\varepsilon \tau 
}{}^{\varphi \xi } F_{\kappa \omega \xi \zeta } F_{\sigma 
\lambda \varphi }{}^{\zeta } + \nn\\&& m_{35}^{} F_{\alpha }{}^{\epsilon 
\varepsilon \mu } F^{\alpha \beta \gamma \delta } F_{\beta 
\epsilon }{}^{\nu \sigma } F_{\gamma \varepsilon }{}^{\lambda 
\kappa } F_{\delta \nu }{}^{\tau \omega } F_{\kappa \omega 
\xi \zeta } F_{\mu \tau }{}^{\varphi \xi } F_{\sigma \lambda 
\varphi }{}^{\zeta } + \nn\\&& m_{19}^{} F_{\alpha \beta }{}^{\epsilon 
\varepsilon } F^{\alpha \beta \gamma \delta } F_{\gamma 
}{}^{\mu \nu \sigma } F_{\delta }{}^{\lambda \kappa \tau } F_{
\epsilon \mu \nu }{}^{\omega } F_{\varepsilon }{}^{\varphi \xi 
\zeta } F_{\kappa \tau \omega \zeta } F_{\sigma \lambda 
\varphi \xi } + \nn\\&& m_{20}^{} F_{\alpha \beta }{}^{\epsilon 
\varepsilon } F^{\alpha \beta \gamma \delta } F_{\gamma 
}{}^{\mu \nu \sigma } F_{\delta \mu }{}^{\lambda \kappa } 
F_{\epsilon \nu }{}^{\tau \omega } F_{\varepsilon }{}^{\varphi 
\xi \zeta } F_{\kappa \tau \omega \zeta } F_{\sigma \lambda 
\varphi \xi } + \nn\\&& m_{44}^{} F_{\alpha \beta }{}^{\epsilon 
\varepsilon } F^{\alpha \beta \gamma \delta } F_{\gamma 
}{}^{\mu \nu \sigma } F_{\delta \mu }{}^{\lambda \kappa } 
F_{\epsilon \nu }{}^{\tau \omega } F_{\varepsilon \tau 
}{}^{\varphi \xi } F_{\kappa \varphi \xi \zeta } F_{\sigma 
\lambda \omega }{}^{\zeta } + \nn\\&& m_{4}^{} F_{\alpha \beta 
}{}^{\epsilon \varepsilon } F^{\alpha \beta \gamma \delta } F_{
\gamma }{}^{\mu \nu \sigma } F_{\delta }{}^{\lambda \kappa 
\tau } F_{\epsilon \mu \lambda }{}^{\omega } F_{\varepsilon 
}{}^{\varphi \xi \zeta } F_{\nu \kappa \omega \varphi } 
F_{\sigma \tau \xi \zeta } + \nn\\&& m_{3}^{} F_{\alpha \beta 
}{}^{\epsilon \varepsilon } F^{\alpha \beta \gamma \delta } F_{
\gamma }{}^{\mu \nu \sigma } F_{\delta }{}^{\lambda \kappa 
\tau } F_{\epsilon \mu }{}^{\omega \varphi } F_{\varepsilon 
\lambda }{}^{\xi \zeta } F_{\nu \kappa \omega \xi } F_{\sigma 
\tau \varphi \zeta } + \nn\\&& m_{36}^{} F_{\alpha \beta }{}^{\epsilon 
\varepsilon } F^{\alpha \beta \gamma \delta } F_{\gamma 
}{}^{\mu \nu \sigma } F_{\delta \mu }{}^{\lambda \kappa } 
F_{\epsilon \nu }{}^{\tau \omega } F_{\varepsilon \lambda }{}^{
\varphi \xi } F_{\kappa \omega \xi \zeta } F_{\sigma \tau 
\varphi }{}^{\zeta } + \nn\\&& m_{9}^{} F_{\alpha }{}^{\epsilon 
\varepsilon \mu } F^{\alpha \beta \gamma \delta } F_{\beta 
\epsilon }{}^{\nu \sigma } F_{\gamma \varepsilon \nu 
}{}^{\lambda } F_{\delta }{}^{\kappa \tau \omega } F_{\lambda 
\omega \xi \zeta } F_{\mu \kappa }{}^{\varphi \xi } F_{\sigma 
\tau \varphi }{}^{\zeta } + \nn\\&& m_{37}^{} F_{\alpha }{}^{\epsilon 
\varepsilon \mu } F^{\alpha \beta \gamma \delta } F_{\beta 
\epsilon }{}^{\nu \sigma } F_{\gamma \varepsilon }{}^{\lambda 
\kappa } F_{\delta \nu }{}^{\tau \omega } F_{\kappa \omega 
\xi \zeta } F_{\mu \lambda }{}^{\varphi \xi } F_{\sigma \tau 
\varphi }{}^{\zeta } + \nn\\&& m_{10}^{} F_{\alpha \beta }{}^{\epsilon 
\varepsilon } F^{\alpha \beta \gamma \delta } F_{\gamma 
\epsilon }{}^{\mu \nu } F_{\delta \mu }{}^{\sigma \lambda } 
F_{\varepsilon }{}^{\kappa \tau \omega } F_{\lambda \varphi 
\xi \zeta } F_{\nu \kappa }{}^{\varphi \xi } F_{\sigma \tau 
\omega }{}^{\zeta } + \nn\\&& m_{11}^{} F_{\alpha \beta }{}^{\epsilon 
\varepsilon } F^{\alpha \beta \gamma \delta } F_{\gamma \delta 
}{}^{\mu \nu } F_{\epsilon \mu }{}^{\sigma \lambda } 
F_{\varepsilon }{}^{\kappa \tau \omega } F_{\lambda \varphi 
\xi \zeta } F_{\nu \kappa }{}^{\varphi \xi } F_{\sigma \tau 
\omega }{}^{\zeta } + \nn\\&& m_{12}^{} F_{\alpha \beta }{}^{\epsilon 
\varepsilon } F^{\alpha \beta \gamma \delta } F_{\gamma 
\epsilon }{}^{\mu \nu } F_{\delta \varepsilon }{}^{\sigma 
\lambda } F_{\lambda \varphi \xi \zeta } F_{\mu }{}^{\kappa 
\tau \omega } F_{\nu \kappa }{}^{\varphi \xi } F_{\sigma \tau 
\omega }{}^{\zeta } + \nn\\&& m_{13}^{} F_{\alpha \beta }{}^{\epsilon 
\varepsilon } F^{\alpha \beta \gamma \delta } F_{\gamma \delta 
}{}^{\mu \nu } F_{\epsilon \varepsilon }{}^{\sigma \lambda } 
F_{\lambda \varphi \xi \zeta } F_{\mu }{}^{\kappa \tau \omega 
} F_{\nu \kappa }{}^{\varphi \xi } F_{\sigma \tau \omega }{}^{
\zeta } + \nn\\&& m_{21}^{} F_{\alpha \beta }{}^{\epsilon \varepsilon } 
F^{\alpha \beta \gamma \delta } F_{\gamma }{}^{\mu \nu \sigma 
} F_{\delta }{}^{\lambda \kappa \tau } F_{\epsilon \mu \nu 
}{}^{\omega } F_{\varepsilon \lambda }{}^{\varphi \xi } 
F_{\kappa \tau \omega \zeta } F_{\sigma \varphi \xi 
}{}^{\zeta } + \nn\\&& m_{22}^{} F_{\alpha \beta }{}^{\epsilon 
\varepsilon } F^{\alpha \beta \gamma \delta } F_{\gamma 
}{}^{\mu \nu \sigma } F_{\delta \mu }{}^{\lambda \kappa } 
F_{\epsilon \nu }{}^{\tau \omega } F_{\varepsilon \lambda }{}^{
\varphi \xi } F_{\kappa \tau \omega \zeta } F_{\sigma \varphi 
\xi }{}^{\zeta } + \nn\\&& m_{38}^{} F_{\alpha \beta }{}^{\epsilon 
\varepsilon } F^{\alpha \beta \gamma \delta } F_{\gamma 
}{}^{\mu \nu \sigma } F_{\delta \mu }{}^{\lambda \kappa } 
F_{\epsilon \nu }{}^{\tau \omega } F_{\varepsilon \lambda \tau 
}{}^{\varphi } F_{\kappa \omega \xi \zeta } F_{\sigma \varphi 
}{}^{\xi \zeta } + \nn\\&& m_{45}^{} F_{\alpha \beta }{}^{\epsilon 
\varepsilon } F^{\alpha \beta \gamma \delta } F_{\gamma 
}{}^{\mu \nu \sigma } F_{\delta \mu }{}^{\lambda \kappa } 
F_{\epsilon \nu }{}^{\tau \omega } F_{\varepsilon \lambda \tau 
}{}^{\varphi } F_{\kappa \varphi \xi \zeta } F_{\sigma \omega 
}{}^{\xi \zeta } + \nn\\&& m_{14}^{} F_{\alpha \beta }{}^{\epsilon 
\varepsilon } F^{\alpha \beta \gamma \delta } F_{\gamma 
\epsilon }{}^{\mu \nu } F_{\delta \mu }{}^{\sigma \lambda } 
F_{\varepsilon }{}^{\kappa \tau \omega } F_{\lambda \varphi 
\xi \zeta } F_{\nu \kappa \tau }{}^{\varphi } F_{\sigma 
\omega }{}^{\xi \zeta } + \nn\\&& m_{15}^{} F_{\alpha \beta 
}{}^{\epsilon \varepsilon } F^{\alpha \beta \gamma \delta } F_{
\gamma \delta }{}^{\mu \nu } F_{\epsilon \mu }{}^{\sigma 
\lambda } F_{\varepsilon }{}^{\kappa \tau \omega } F_{\lambda 
\varphi \xi \zeta } F_{\nu \kappa \tau }{}^{\varphi } 
F_{\sigma \omega }{}^{\xi \zeta } + \nn\\&& m_{16}^{} F_{\alpha \beta 
}{}^{\epsilon \varepsilon } F^{\alpha \beta \gamma \delta } F_{
\gamma \epsilon }{}^{\mu \nu } F_{\delta \varepsilon 
}{}^{\sigma \lambda } F_{\lambda \varphi \xi \zeta } F_{\mu 
}{}^{\kappa \tau \omega } F_{\nu \kappa \tau }{}^{\varphi } 
F_{\sigma \omega }{}^{\xi \zeta } + \nn\\&& m_{17}^{} F_{\alpha \beta 
}{}^{\epsilon \varepsilon } F^{\alpha \beta \gamma \delta } F_{
\gamma \delta }{}^{\mu \nu } F_{\epsilon \varepsilon 
}{}^{\sigma \lambda } F_{\lambda \varphi \xi \zeta } F_{\mu 
}{}^{\kappa \tau \omega } F_{\nu \kappa \tau }{}^{\varphi } 
F_{\sigma \omega }{}^{\xi \zeta } + \nn\\&& m_{71}^{} F_{\alpha \beta 
}{}^{\epsilon \varepsilon } F^{\alpha \beta \gamma \delta } F_{
\gamma \epsilon }{}^{\mu \nu } F_{\delta }{}^{\sigma \lambda 
\kappa } F_{\varepsilon \sigma \lambda }{}^{\tau } F_{\kappa 
\omega }{}^{\xi \zeta } F_{\mu \nu }{}^{\omega \varphi } 
F_{\tau \varphi \xi \zeta } + \nn\\&& m_{61}^{} F_{\alpha \beta 
}{}^{\epsilon \varepsilon } F^{\alpha \beta \gamma \delta } F_{
\gamma \epsilon }{}^{\mu \nu } F_{\delta \mu }{}^{\sigma 
\lambda } F_{\varepsilon \sigma }{}^{\kappa \tau } F_{\lambda 
\omega }{}^{\xi \zeta } F_{\nu \kappa }{}^{\omega \varphi } 
F_{\tau \varphi \xi \zeta } + \nn\\&& m_{62}^{} F_{\alpha \beta 
}{}^{\epsilon \varepsilon } F^{\alpha \beta \gamma \delta } F_{
\gamma \delta }{}^{\mu \nu } F_{\epsilon \mu }{}^{\sigma 
\lambda } F_{\varepsilon \sigma }{}^{\kappa \tau } F_{\lambda 
\omega }{}^{\xi \zeta } F_{\nu \kappa }{}^{\omega \varphi } 
F_{\tau \varphi \xi \zeta } + \nn\\&& m_{63}^{} F_{\alpha \beta 
}{}^{\epsilon \varepsilon } F^{\alpha \beta \gamma \delta } F_{
\gamma \epsilon }{}^{\mu \nu } F_{\delta \varepsilon 
}{}^{\sigma \lambda } F_{\lambda \omega }{}^{\xi \zeta } 
F_{\mu \sigma }{}^{\kappa \tau } F_{\nu \kappa }{}^{\omega 
\varphi } F_{\tau \varphi \xi \zeta } + \nn\\&& m_{64}^{} F_{\alpha 
\beta }{}^{\epsilon \varepsilon } F^{\alpha \beta \gamma 
\delta } F_{\gamma \delta }{}^{\mu \nu } F_{\epsilon 
\varepsilon }{}^{\sigma \lambda } F_{\lambda \omega }{}^{\xi 
\zeta } F_{\mu \sigma }{}^{\kappa \tau } F_{\nu \kappa 
}{}^{\omega \varphi } F_{\tau \varphi \xi \zeta } + \nn\\&& m_{65}^{} 
F_{\alpha \beta \gamma }{}^{\epsilon } F^{\alpha \beta \gamma 
\delta } F_{\delta }{}^{\varepsilon \mu \nu } F_{\epsilon 
\varepsilon }{}^{\sigma \lambda } F_{\lambda \omega }{}^{\xi 
\zeta } F_{\mu \sigma }{}^{\kappa \tau } F_{\nu \kappa 
}{}^{\omega \varphi } F_{\tau \varphi \xi \zeta } + \nn\\&& m_{72}^{} 
F_{\alpha \beta }{}^{\epsilon \varepsilon } F^{\alpha \beta 
\gamma \delta } F_{\gamma \delta }{}^{\mu \nu } F_{\epsilon 
\mu }{}^{\sigma \lambda } F_{\varepsilon \sigma }{}^{\kappa 
\tau } F_{\kappa \omega }{}^{\xi \zeta } F_{\nu \lambda 
}{}^{\omega \varphi } F_{\tau \varphi \xi \zeta } + \nn\\&& m_{73}^{} 
F_{\alpha \beta }{}^{\epsilon \varepsilon } F^{\alpha \beta 
\gamma \delta } F_{\gamma \epsilon }{}^{\mu \nu } F_{\delta 
\varepsilon }{}^{\sigma \lambda } F_{\kappa \omega }{}^{\xi 
\zeta } F_{\mu \sigma }{}^{\kappa \tau } F_{\nu \lambda 
}{}^{\omega \varphi } F_{\tau \varphi \xi \zeta } + \nn\\&& m_{74}^{} 
F_{\alpha \beta }{}^{\epsilon \varepsilon } F^{\alpha \beta 
\gamma \delta } F_{\gamma \delta }{}^{\mu \nu } F_{\epsilon 
\varepsilon }{}^{\sigma \lambda } F_{\kappa \omega }{}^{\xi 
\zeta } F_{\mu \sigma }{}^{\kappa \tau } F_{\nu \lambda 
}{}^{\omega \varphi } F_{\tau \varphi \xi \zeta } + \nn\\&& m_{75}^{} 
F_{\alpha \beta \gamma }{}^{\epsilon } F^{\alpha \beta \gamma 
\delta } F_{\delta }{}^{\varepsilon \mu \nu } F_{\epsilon 
\varepsilon }{}^{\sigma \lambda } F_{\kappa \omega }{}^{\xi 
\zeta } F_{\mu \sigma }{}^{\kappa \tau } F_{\nu \lambda 
}{}^{\omega \varphi } F_{\tau \varphi \xi \zeta } + \nn\\&& m_{54}^{} 
F_{\alpha \beta }{}^{\epsilon \varepsilon } F^{\alpha \beta 
\gamma \delta } F_{\gamma \delta }{}^{\mu \nu } F_{\epsilon 
}{}^{\sigma \lambda \kappa } F_{\varepsilon \sigma \lambda 
}{}^{\tau } F_{\mu \kappa }{}^{\omega \varphi } F_{\nu \omega 
}{}^{\xi \zeta } F_{\tau \varphi \xi \zeta } + \nn\\&& m_{55}^{} 
F_{\alpha \beta }{}^{\epsilon \varepsilon } F^{\alpha \beta 
\gamma \delta } F_{\gamma \epsilon }{}^{\mu \nu } F_{\delta 
\mu }{}^{\sigma \lambda } F_{\varepsilon \sigma }{}^{\kappa 
\tau } F_{\lambda \kappa \omega }{}^{\zeta } F_{\nu 
}{}^{\omega \varphi \xi } F_{\tau \varphi \xi \zeta } + 
\nn\\&& m_{56}^{} F_{\alpha \beta }{}^{\epsilon \varepsilon } F^{\alpha 
\beta \gamma \delta } F_{\gamma \delta }{}^{\mu \nu } 
F_{\epsilon \mu }{}^{\sigma \lambda } F_{\varepsilon \sigma 
}{}^{\kappa \tau } F_{\lambda \kappa \omega }{}^{\zeta } 
F_{\nu }{}^{\omega \varphi \xi } F_{\tau \varphi \xi \zeta } 
+ \nn\\&& m_{66}^{} F_{\alpha \beta }{}^{\epsilon \varepsilon } 
F^{\alpha \beta \gamma \delta } F_{\gamma \epsilon }{}^{\mu 
\nu } F_{\delta \mu }{}^{\sigma \lambda } F_{\varepsilon \nu 
}{}^{\kappa \tau } F_{\lambda \omega }{}^{\xi \zeta } 
F_{\sigma \kappa }{}^{\omega \varphi } F_{\tau \varphi \xi 
\zeta } + \nn\\&& m_{67}^{} F_{\alpha \beta }{}^{\epsilon \varepsilon } 
F^{\alpha \beta \gamma \delta } F_{\gamma \delta }{}^{\mu \nu 
} F_{\epsilon \mu }{}^{\sigma \lambda } F_{\varepsilon \nu 
}{}^{\kappa \tau } F_{\lambda \omega }{}^{\xi \zeta } 
F_{\sigma \kappa }{}^{\omega \varphi } F_{\tau \varphi \xi 
\zeta } + \nn\\&& m_{68}^{} F_{\alpha \beta }{}^{\epsilon \varepsilon } 
F^{\alpha \beta \gamma \delta } F_{\gamma \epsilon }{}^{\mu 
\nu } F_{\delta \varepsilon }{}^{\sigma \lambda } F_{\lambda 
\omega }{}^{\xi \zeta } F_{\mu \nu }{}^{\kappa \tau } 
F_{\sigma \kappa }{}^{\omega \varphi } F_{\tau \varphi \xi 
\zeta } + \nn\\&& m_{69}^{} F_{\alpha \beta }{}^{\epsilon \varepsilon } 
F^{\alpha \beta \gamma \delta } F_{\gamma \delta }{}^{\mu \nu 
} F_{\epsilon \varepsilon }{}^{\sigma \lambda } F_{\lambda 
\omega }{}^{\xi \zeta } F_{\mu \nu }{}^{\kappa \tau } 
F_{\sigma \kappa }{}^{\omega \varphi } F_{\tau \varphi \xi 
\zeta } + \nn\\&& m_{70}^{} F_{\alpha \beta \gamma }{}^{\epsilon } 
F^{\alpha \beta \gamma \delta } F_{\delta }{}^{\varepsilon \mu 
\nu } F_{\epsilon \varepsilon }{}^{\sigma \lambda } F_{\lambda 
\omega }{}^{\xi \zeta } F_{\mu \nu }{}^{\kappa \tau } 
F_{\sigma \kappa }{}^{\omega \varphi } F_{\tau \varphi \xi 
\zeta } + \nn\\&& m_{76}^{} F_{\alpha \beta }{}^{\epsilon \varepsilon } 
F^{\alpha \beta \gamma \delta } F_{\gamma \delta }{}^{\mu \nu 
} F_{\epsilon \mu }{}^{\sigma \lambda } F_{\varepsilon \nu 
}{}^{\kappa \tau } F_{\kappa \omega }{}^{\xi \zeta } F_{\sigma 
\lambda }{}^{\omega \varphi } F_{\tau \varphi \xi \zeta } + 
\nn\\&& m_{77}^{} F_{\alpha \beta }{}^{\epsilon \varepsilon } F^{\alpha 
\beta \gamma \delta } F_{\gamma \epsilon }{}^{\mu \nu } 
F_{\delta \varepsilon }{}^{\sigma \lambda } F_{\kappa \omega 
}{}^{\xi \zeta } F_{\mu \nu }{}^{\kappa \tau } F_{\sigma 
\lambda }{}^{\omega \varphi } F_{\tau \varphi \xi \zeta } + 
\nn\\&& m_{78}^{} F_{\alpha \beta }{}^{\epsilon \varepsilon } F^{\alpha 
\beta \gamma \delta } F_{\gamma \delta }{}^{\mu \nu } 
F_{\epsilon \varepsilon }{}^{\sigma \lambda } F_{\kappa \omega 
}{}^{\xi \zeta } F_{\mu \nu }{}^{\kappa \tau } F_{\sigma 
\lambda }{}^{\omega \varphi } F_{\tau \varphi \xi \zeta } + 
\nn\\&& m_{79}^{} F_{\alpha \beta \gamma }{}^{\epsilon } F^{\alpha 
\beta \gamma \delta } F_{\delta }{}^{\varepsilon \mu \nu } F_{
\epsilon \varepsilon }{}^{\sigma \lambda } F_{\kappa \omega 
}{}^{\xi \zeta } F_{\mu \nu }{}^{\kappa \tau } F_{\sigma 
\lambda }{}^{\omega \varphi } F_{\tau \varphi \xi \zeta } + 
\nn\\&& m_{80}^{} F_{\alpha \beta }{}^{\epsilon \varepsilon } F^{\alpha 
\beta \gamma \delta } F_{\gamma \epsilon }{}^{\mu \nu } 
F_{\delta \varepsilon \mu }{}^{\sigma } F_{\kappa \omega 
}{}^{\xi \zeta } F_{\nu }{}^{\lambda \kappa \tau } F_{\sigma 
\lambda }{}^{\omega \varphi } F_{\tau \varphi \xi \zeta } + 
\nn\\&& m_{81}^{} F_{\alpha \beta }{}^{\epsilon \varepsilon } F^{\alpha 
\beta \gamma \delta } F_{\gamma \delta }{}^{\mu \nu } 
F_{\epsilon \varepsilon \mu }{}^{\sigma } F_{\kappa \omega 
}{}^{\xi \zeta } F_{\nu }{}^{\lambda \kappa \tau } F_{\sigma 
\lambda }{}^{\omega \varphi } F_{\tau \varphi \xi \zeta } + 
\nn\\&& m_{82}^{} F_{\alpha \beta \gamma }{}^{\epsilon } F^{\alpha 
\beta \gamma \delta } F_{\delta }{}^{\varepsilon \mu \nu } F_{
\epsilon \varepsilon \mu }{}^{\sigma } F_{\kappa \omega 
}{}^{\xi \zeta } F_{\nu }{}^{\lambda \kappa \tau } F_{\sigma 
\lambda }{}^{\omega \varphi } F_{\tau \varphi \xi \zeta } + 
\nn\\&& m_{57}^{} F_{\alpha \beta }{}^{\epsilon \varepsilon } F^{\alpha 
\beta \gamma \delta } F_{\gamma \epsilon }{}^{\mu \nu } 
F_{\delta \mu \nu }{}^{\sigma } F_{\varepsilon }{}^{\lambda 
\kappa \tau } F_{\lambda \kappa \omega }{}^{\zeta } F_{\sigma 
}{}^{\omega \varphi \xi } F_{\tau \varphi \xi \zeta } + 
\nn\\&& m_{58}^{} F_{\alpha \beta }{}^{\epsilon \varepsilon } F^{\alpha 
\beta \gamma \delta } F_{\gamma \epsilon }{}^{\mu \nu } 
F_{\delta \varepsilon \mu }{}^{\sigma } F_{\lambda \kappa 
\omega }{}^{\zeta } F_{\nu }{}^{\lambda \kappa \tau } 
F_{\sigma }{}^{\omega \varphi \xi } F_{\tau \varphi \xi \zeta 
} + \nn\\&& m_{59}^{} F_{\alpha \beta }{}^{\epsilon \varepsilon } 
F^{\alpha \beta \gamma \delta } F_{\gamma \delta }{}^{\mu \nu 
} F_{\epsilon \varepsilon \mu }{}^{\sigma } F_{\lambda \kappa 
\omega }{}^{\zeta } F_{\nu }{}^{\lambda \kappa \tau } 
F_{\sigma }{}^{\omega \varphi \xi } F_{\tau \varphi \xi \zeta 
} + \nn\\&& m_{60}^{} F_{\alpha \beta \gamma }{}^{\epsilon } F^{\alpha 
\beta \gamma \delta } F_{\delta }{}^{\varepsilon \mu \nu } F_{
\epsilon \varepsilon \mu }{}^{\sigma } F_{\lambda \kappa 
\omega }{}^{\zeta } F_{\nu }{}^{\lambda \kappa \tau } 
F_{\sigma }{}^{\omega \varphi \xi } F_{\tau \varphi \xi \zeta 
} + \nn\\&& m_{83}^{} F_{\alpha \beta }{}^{\epsilon \varepsilon } 
F^{\alpha \beta \gamma \delta } F_{\gamma \epsilon }{}^{\mu 
\nu } F_{\delta \varepsilon \mu \nu } F_{\kappa \omega 
}{}^{\xi \zeta } F_{\sigma \lambda }{}^{\omega \varphi } 
F^{\sigma \lambda \kappa \tau } F_{\tau \varphi \xi \zeta } + 
\nn\\&& m_{84}^{} F_{\alpha \beta }{}^{\epsilon \varepsilon } F^{\alpha 
\beta \gamma \delta } F_{\gamma \delta }{}^{\mu \nu } 
F_{\epsilon \varepsilon \mu \nu } F_{\kappa \omega }{}^{\xi 
\zeta } F_{\sigma \lambda }{}^{\omega \varphi } F^{\sigma 
\lambda \kappa \tau } F_{\tau \varphi \xi \zeta } + \nn\\&& m_{85}^{} 
F_{\alpha \beta \gamma }{}^{\epsilon } F^{\alpha \beta \gamma 
\delta } F_{\delta }{}^{\varepsilon \mu \nu } F_{\epsilon 
\varepsilon \mu \nu } F_{\kappa \omega }{}^{\xi \zeta } 
F_{\sigma \lambda }{}^{\omega \varphi } F^{\sigma \lambda 
\kappa \tau } F_{\tau \varphi \xi \zeta } + \nn\\&& m_{52}^{} 
F_{\alpha \beta }{}^{\epsilon \varepsilon } F^{\alpha \beta 
\gamma \delta } F_{\gamma \epsilon }{}^{\mu \nu } F_{\delta 
}{}^{\sigma \lambda \kappa } F_{\varepsilon \sigma }{}^{\tau 
\omega } F_{\lambda \kappa \varphi }{}^{\zeta } F_{\mu \nu 
}{}^{\varphi \xi } F_{\tau \omega \xi \zeta } + \nn\\&& m_{53}^{} 
F_{\alpha \beta }{}^{\epsilon \varepsilon } F^{\alpha \beta 
\gamma \delta } F_{\gamma \epsilon }{}^{\mu \nu } F_{\delta 
\mu }{}^{\sigma \lambda } F_{\varepsilon }{}^{\kappa \tau 
\omega } F_{\lambda \kappa \varphi }{}^{\zeta } F_{\nu \sigma 
}{}^{\varphi \xi } F_{\tau \omega \xi \zeta } + \nn\\&& m_{46}^{} 
F_{\alpha \beta }{}^{\epsilon \varepsilon } F^{\alpha \beta 
\gamma \delta } F_{\gamma \delta }{}^{\mu \nu } F_{\epsilon 
}{}^{\sigma \lambda \kappa } F_{\varepsilon \sigma }{}^{\tau 
\omega } F_{\mu \lambda \kappa }{}^{\varphi } F_{\nu \varphi 
}{}^{\xi \zeta } F_{\tau \omega \xi \zeta } + \nn\\&& m_{50}^{} 
F_{\alpha \beta }{}^{\epsilon \varepsilon } F^{\alpha \beta 
\gamma \delta } F_{\gamma }{}^{\mu \nu \sigma } F_{\delta 
}{}^{\lambda \kappa \tau } F_{\epsilon \mu \nu }{}^{\omega } 
F_{\varepsilon \lambda }{}^{\varphi \xi } F_{\sigma \kappa 
\varphi }{}^{\zeta } F_{\tau \omega \xi \zeta } + \nn\\&& m_{49}^{} F_{
\alpha \beta }{}^{\epsilon \varepsilon } F^{\alpha \beta 
\gamma \delta } F_{\gamma }{}^{\mu \nu \sigma } F_{\delta 
}{}^{\lambda \kappa \tau } F_{\epsilon \mu \lambda }{}^{\omega 
} F_{\varepsilon \nu }{}^{\varphi \xi } F_{\sigma \kappa 
\varphi }{}^{\zeta } F_{\tau \omega \xi \zeta } + \nn\\&& m_{47}^{} F_{
\alpha \beta }{}^{\epsilon \varepsilon } F^{\alpha \beta 
\gamma \delta } F_{\gamma \delta }{}^{\mu \nu } F_{\epsilon 
\mu }{}^{\sigma \lambda } F_{\varepsilon }{}^{\kappa \tau 
\omega } F_{\nu }{}^{\varphi \xi \zeta } F_{\sigma \lambda 
\kappa \varphi } F_{\tau \omega \xi \zeta } + \nn\\&& m_{48}^{} 
F_{\alpha \beta }{}^{\epsilon \varepsilon } F^{\alpha \beta 
\gamma \delta } F_{\gamma \epsilon }{}^{\mu \nu } F_{\delta 
\mu }{}^{\sigma \lambda } F_{\varepsilon }{}^{\kappa \tau 
\omega } F_{\nu \kappa }{}^{\varphi \xi } F_{\sigma \lambda 
\varphi }{}^{\zeta } F_{\tau \omega \xi \zeta } + \nn\\&& m_{51}^{} F_{
\alpha \beta }{}^{\epsilon \varepsilon } F^{\alpha \beta 
\gamma \delta } F_{\gamma }{}^{\mu \nu \sigma } F_{\delta 
}{}^{\lambda \kappa \tau } F_{\epsilon \mu \nu }{}^{\omega } 
F_{\varepsilon \lambda \kappa }{}^{\varphi } F_{\sigma \varphi 
}{}^{\xi \zeta } F_{\tau \omega \xi \zeta } + \nn\\&& m_{101}^{} 
F_{\alpha \beta }{}^{\epsilon \varepsilon } F^{\alpha \beta 
\gamma \delta } F_{\gamma }{}^{\mu \nu \sigma } F_{\delta \mu 
}{}^{\lambda \kappa } F_{\epsilon \nu }{}^{\tau \omega } 
F_{\varepsilon \sigma }{}^{\varphi \xi } F_{\lambda \kappa 
\tau }{}^{\zeta } F_{\omega \varphi \xi \zeta } + \nn\\&& m_{102}^{} 
F_{\alpha \beta }{}^{\epsilon \varepsilon } F^{\alpha \beta 
\gamma \delta } F_{\gamma \epsilon }{}^{\mu \nu } F_{\delta 
}{}^{\sigma \lambda \kappa } F_{\varepsilon }{}^{\tau \omega 
\varphi } F_{\lambda \kappa \tau }{}^{\zeta } F_{\mu \nu 
\sigma }{}^{\xi } F_{\omega \varphi \xi \zeta } + \nn\\&& m_{88}^{} F_{
\alpha \beta }{}^{\epsilon \varepsilon } F^{\alpha \beta 
\gamma \delta } F_{\gamma \epsilon }{}^{\mu \nu } F_{\delta 
}{}^{\sigma \lambda \kappa } F_{\varepsilon \sigma }{}^{\tau 
\omega } F_{\mu \lambda \tau }{}^{\varphi } F_{\nu \kappa 
}{}^{\xi \zeta } F_{\omega \varphi \xi \zeta } + \nn\\&& m_{89}^{} 
F_{\alpha \beta }{}^{\epsilon \varepsilon } F^{\alpha \beta 
\gamma \delta } F_{\gamma \delta }{}^{\mu \nu } F_{\epsilon 
}{}^{\sigma \lambda \kappa } F_{\varepsilon \sigma }{}^{\tau 
\omega } F_{\mu \lambda \tau }{}^{\varphi } F_{\nu \kappa 
}{}^{\xi \zeta } F_{\omega \varphi \xi \zeta } + \nn\\&& m_{87}^{} 
F_{\alpha \beta }{}^{\epsilon \varepsilon } F^{\alpha \beta 
\gamma \delta } F_{\gamma \delta }{}^{\mu \nu } F_{\epsilon 
}{}^{\sigma \lambda \kappa } F_{\varepsilon }{}^{\tau \omega 
\varphi } F_{\mu \sigma \lambda \tau } F_{\nu \kappa }{}^{\xi 
\zeta } F_{\omega \varphi \xi \zeta } + \nn\\&& m_{86}^{} F_{\alpha 
\beta }{}^{\epsilon \varepsilon } F^{\alpha \beta \gamma 
\delta } F_{\gamma \epsilon }{}^{\mu \nu } F_{\delta 
}{}^{\sigma \lambda \kappa } F_{\varepsilon }{}^{\tau \omega 
\varphi } F_{\mu \sigma \lambda }{}^{\xi } F_{\nu \kappa \tau 
}{}^{\zeta } F_{\omega \varphi \xi \zeta } + \nn\\&& m_{104}^{} 
F_{\alpha \beta }{}^{\epsilon \varepsilon } F^{\alpha \beta 
\gamma \delta } F_{\gamma \delta }{}^{\mu \nu } F_{\epsilon 
\mu }{}^{\sigma \lambda } F_{\varepsilon \sigma }{}^{\kappa 
\tau } F_{\lambda \tau }{}^{\xi \zeta } F_{\nu \kappa 
}{}^{\omega \varphi } F_{\omega \varphi \xi \zeta } + 
\nn\\&& m_{105}^{} F_{\alpha \beta }{}^{\epsilon \varepsilon } F^{\alpha 
\beta \gamma \delta } F_{\gamma \delta }{}^{\mu \nu } 
F_{\epsilon \varepsilon }{}^{\sigma \lambda } F_{\lambda \tau 
}{}^{\xi \zeta } F_{\mu \sigma }{}^{\kappa \tau } F_{\nu 
\kappa }{}^{\omega \varphi } F_{\omega \varphi \xi \zeta } + 
\nn\\&& m_{106}^{} F_{\alpha \beta \gamma }{}^{\epsilon } F^{\alpha 
\beta \gamma \delta } F_{\delta }{}^{\varepsilon \mu \nu } F_{
\epsilon \varepsilon }{}^{\sigma \lambda } F_{\lambda \tau 
}{}^{\xi \zeta } F_{\mu \sigma }{}^{\kappa \tau } F_{\nu 
\kappa }{}^{\omega \varphi } F_{\omega \varphi \xi \zeta } + 
\nn\\&& m_{110}^{} F_{\alpha \beta \gamma }{}^{\epsilon } F^{\alpha 
\beta \gamma \delta } F_{\delta }{}^{\varepsilon \mu \nu } F_{
\epsilon }{}^{\sigma \lambda \kappa } F_{\varepsilon \mu 
\sigma }{}^{\tau } F_{\kappa \tau }{}^{\xi \zeta } F_{\nu 
\lambda }{}^{\omega \varphi } F_{\omega \varphi \xi \zeta } + 
\nn\\&& m_{111}^{} F_{\alpha \beta }{}^{\epsilon \varepsilon } F^{\alpha 
\beta \gamma \delta } F_{\gamma \delta }{}^{\mu \nu } 
F_{\epsilon \varepsilon }{}^{\sigma \lambda } F_{\kappa \tau 
}{}^{\xi \zeta } F_{\mu \sigma }{}^{\kappa \tau } F_{\nu 
\lambda }{}^{\omega \varphi } F_{\omega \varphi \xi \zeta } + 
\nn\\&& m_{112}^{} F_{\alpha \beta \gamma }{}^{\epsilon } F^{\alpha 
\beta \gamma \delta } F_{\delta }{}^{\varepsilon \mu \nu } F_{
\epsilon \varepsilon }{}^{\sigma \lambda } F_{\kappa \tau }{}^{
\xi \zeta } F_{\mu \sigma }{}^{\kappa \tau } F_{\nu \lambda 
}{}^{\omega \varphi } F_{\omega \varphi \xi \zeta } + 
\nn\\&& m_{103}^{} F_{\alpha \beta }{}^{\epsilon \varepsilon } F^{\alpha 
\beta \gamma \delta } F_{\gamma \epsilon }{}^{\mu \nu } 
F_{\delta \mu }{}^{\sigma \lambda } F_{\varepsilon }{}^{\kappa 
\tau \omega } F_{\lambda \kappa \tau }{}^{\zeta } F_{\nu 
\sigma }{}^{\varphi \xi } F_{\omega \varphi \xi \zeta } + 
\nn\\&& m_{90}^{} F_{\alpha \beta }{}^{\epsilon \varepsilon } F^{\alpha 
\beta \gamma \delta } F_{\gamma \epsilon }{}^{\mu \nu } 
F_{\delta }{}^{\sigma \lambda \kappa } F_{\varepsilon \sigma 
}{}^{\tau \omega } F_{\mu \lambda \kappa }{}^{\varphi } F_{\nu 
\tau }{}^{\xi \zeta } F_{\omega \varphi \xi \zeta } + 
\nn\\&& m_{91}^{} F_{\alpha \beta }{}^{\epsilon \varepsilon } F^{\alpha 
\beta \gamma \delta } F_{\gamma \delta }{}^{\mu \nu } 
F_{\epsilon }{}^{\sigma \lambda \kappa } F_{\varepsilon \sigma 
}{}^{\tau \omega } F_{\mu \lambda \kappa }{}^{\varphi } F_{\nu 
\tau }{}^{\xi \zeta } F_{\omega \varphi \xi \zeta } + 
\nn\\&& m_{98}^{} F_{\alpha \beta }{}^{\epsilon \varepsilon } F^{\alpha 
\beta \gamma \delta } F_{\gamma }{}^{\mu \nu \sigma } 
F_{\delta \mu }{}^{\lambda \kappa } F_{\epsilon \nu }{}^{\tau 
\omega } F_{\varepsilon \lambda \tau }{}^{\varphi } F_{\sigma 
\kappa }{}^{\xi \zeta } F_{\omega \varphi \xi \zeta } + 
\nn\\&& m_{96}^{} F_{\alpha \beta }{}^{\epsilon \varepsilon } F^{\alpha 
\beta \gamma \delta } F_{\gamma }{}^{\mu \nu \sigma } 
F_{\delta }{}^{\lambda \kappa \tau } F_{\epsilon \mu \nu }{}^{
\omega } F_{\varepsilon \lambda }{}^{\varphi \xi } F_{\sigma 
\kappa \tau }{}^{\zeta } F_{\omega \varphi \xi \zeta } + 
\nn\\&& m_{97}^{} F_{\alpha \beta }{}^{\epsilon \varepsilon } F^{\alpha 
\beta \gamma \delta } F_{\gamma }{}^{\mu \nu \sigma } 
F_{\delta \mu }{}^{\lambda \kappa } F_{\epsilon \nu }{}^{\tau 
\omega } F_{\varepsilon \lambda }{}^{\varphi \xi } F_{\sigma 
\kappa \tau }{}^{\zeta } F_{\omega \varphi \xi \zeta } + 
\nn\\&& m_{95}^{} F_{\alpha \beta }{}^{\epsilon \varepsilon } F^{\alpha 
\beta \gamma \delta } F_{\gamma }{}^{\mu \nu \sigma } 
F_{\delta }{}^{\lambda \kappa \tau } F_{\epsilon \mu \lambda 
}{}^{\omega } F_{\varepsilon \nu }{}^{\varphi \xi } F_{\sigma 
\kappa \tau }{}^{\zeta } F_{\omega \varphi \xi \zeta } + 
\nn\\&& m_{107}^{} F_{\alpha \beta }{}^{\epsilon \varepsilon } F^{\alpha 
\beta \gamma \delta } F_{\gamma \delta }{}^{\mu \nu } 
F_{\epsilon \mu }{}^{\sigma \lambda } F_{\varepsilon \nu 
}{}^{\kappa \tau } F_{\lambda \tau }{}^{\xi \zeta } F_{\sigma 
\kappa }{}^{\omega \varphi } F_{\omega \varphi \xi \zeta } + 
\nn\\&& m_{108}^{} F_{\alpha \beta }{}^{\epsilon \varepsilon } F^{\alpha 
\beta \gamma \delta } F_{\gamma \delta }{}^{\mu \nu } 
F_{\epsilon \varepsilon }{}^{\sigma \lambda } F_{\lambda \tau 
}{}^{\xi \zeta } F_{\mu \nu }{}^{\kappa \tau } F_{\sigma 
\kappa }{}^{\omega \varphi } F_{\omega \varphi \xi \zeta } + 
\nn\\&& m_{109}^{} F_{\alpha \beta \gamma }{}^{\epsilon } F^{\alpha 
\beta \gamma \delta } F_{\delta }{}^{\varepsilon \mu \nu } F_{
\epsilon \varepsilon }{}^{\sigma \lambda } F_{\lambda \tau 
}{}^{\xi \zeta } F_{\mu \nu }{}^{\kappa \tau } F_{\sigma 
\kappa }{}^{\omega \varphi } F_{\omega \varphi \xi \zeta } + 
\nn\\&& m_{94}^{} F_{\alpha \beta }{}^{\epsilon \varepsilon } F^{\alpha 
\beta \gamma \delta } F_{\gamma \delta }{}^{\mu \nu } 
F_{\epsilon \mu }{}^{\sigma \lambda } F_{\varepsilon 
}{}^{\kappa \tau \omega } F_{\nu \kappa \tau }{}^{\varphi } 
F_{\sigma \lambda }{}^{\xi \zeta } F_{\omega \varphi \xi 
\zeta } + \nn\\&& m_{92}^{} F_{\alpha \beta }{}^{\epsilon \varepsilon } 
F^{\alpha \beta \gamma \delta } F_{\gamma \epsilon }{}^{\mu 
\nu } F_{\delta \mu }{}^{\sigma \lambda } F_{\varepsilon 
}{}^{\kappa \tau \omega } F_{\nu \kappa }{}^{\varphi \xi } F_{
\sigma \lambda \tau }{}^{\zeta } F_{\omega \varphi \xi \zeta 
} + \nn\\&& m_{93}^{} F_{\alpha \beta }{}^{\epsilon \varepsilon } 
F^{\alpha \beta \gamma \delta } F_{\gamma \delta }{}^{\mu \nu 
} F_{\epsilon \mu }{}^{\sigma \lambda } F_{\varepsilon 
}{}^{\kappa \tau \omega } F_{\nu \kappa }{}^{\varphi \xi } F_{
\sigma \lambda \tau }{}^{\zeta } F_{\omega \varphi \xi \zeta 
} + \nn\\&& m_{113}^{} F_{\alpha \beta }{}^{\epsilon \varepsilon } 
F^{\alpha \beta \gamma \delta } F_{\gamma \delta }{}^{\mu \nu 
} F_{\epsilon \varepsilon }{}^{\sigma \lambda } F_{\kappa \tau 
}{}^{\xi \zeta } F_{\mu \nu }{}^{\kappa \tau } F_{\sigma 
\lambda }{}^{\omega \varphi } F_{\omega \varphi \xi \zeta } + 
\nn\\&& m_{114}^{} F_{\alpha \beta \gamma }{}^{\epsilon } F^{\alpha 
\beta \gamma \delta } F_{\delta }{}^{\varepsilon \mu \nu } F_{
\epsilon \varepsilon }{}^{\sigma \lambda } F_{\kappa \tau }{}^{
\xi \zeta } F_{\mu \nu }{}^{\kappa \tau } F_{\sigma \lambda 
}{}^{\omega \varphi } F_{\omega \varphi \xi \zeta } + 
\nn\\&& m_{115}^{} F_{\alpha \beta }{}^{\epsilon \varepsilon } F^{\alpha 
\beta \gamma \delta } F_{\gamma \delta }{}^{\mu \nu } 
F_{\epsilon \varepsilon \mu }{}^{\sigma } F_{\kappa \tau 
}{}^{\xi \zeta } F_{\nu }{}^{\lambda \kappa \tau } F_{\sigma 
\lambda }{}^{\omega \varphi } F_{\omega \varphi \xi \zeta } + 
\nn\\&& m_{116}^{} F_{\alpha \beta \gamma }{}^{\epsilon } F^{\alpha 
\beta \gamma \delta } F_{\delta }{}^{\varepsilon \mu \nu } F_{
\epsilon \varepsilon \mu }{}^{\sigma } F_{\kappa \tau }{}^{\xi 
\zeta } F_{\nu }{}^{\lambda \kappa \tau } F_{\sigma \lambda 
}{}^{\omega \varphi } F_{\omega \varphi \xi \zeta } + 
\nn\\&& m_{100}^{} F_{\alpha \beta }{}^{\epsilon \varepsilon } F^{\alpha 
\beta \gamma \delta } F_{\gamma }{}^{\mu \nu \sigma } 
F_{\delta }{}^{\lambda \kappa \tau } F_{\epsilon \mu \nu }{}^{
\omega } F_{\varepsilon \lambda \kappa }{}^{\varphi } F_{\sigma 
\tau }{}^{\xi \zeta } F_{\omega \varphi \xi \zeta } + 
\nn\\&& m_{99}^{} F_{\alpha \beta }{}^{\epsilon \varepsilon } F^{\alpha 
\beta \gamma \delta } F_{\gamma }{}^{\mu \nu \sigma } 
F_{\delta }{}^{\lambda \kappa \tau } F_{\epsilon \mu \lambda 
}{}^{\omega } F_{\varepsilon \nu \kappa }{}^{\varphi } 
F_{\sigma \tau }{}^{\xi \zeta } F_{\omega \varphi \xi \zeta } 
+ \nn\\&& m_{117}^{} F_{\alpha \beta }{}^{\epsilon \varepsilon } 
F^{\alpha \beta \gamma \delta } F_{\gamma \delta }{}^{\mu \nu 
} F_{\epsilon \varepsilon \mu \nu } F_{\kappa \tau }{}^{\xi 
\zeta } F_{\sigma \lambda }{}^{\omega \varphi } F^{\sigma 
\lambda \kappa \tau } F_{\omega \varphi \xi \zeta } + 
\nn\\&& m_{118}^{} F_{\alpha \beta \gamma }{}^{\epsilon } F^{\alpha 
\beta \gamma \delta } F_{\delta }{}^{\varepsilon \mu \nu } F_{
\epsilon \varepsilon \mu \nu } F_{\kappa \tau }{}^{\xi \zeta 
} F_{\sigma \lambda }{}^{\omega \varphi } F^{\sigma \lambda 
\kappa \tau } F_{\omega \varphi \xi \zeta } + \nn\\&& m_{120}^{} 
F_{\alpha \beta }{}^{\epsilon \varepsilon } F^{\alpha \beta 
\gamma \delta } F_{\gamma }{}^{\mu \nu \sigma } F_{\delta \mu 
}{}^{\lambda \kappa } F_{\epsilon \nu \sigma }{}^{\tau } 
F_{\varepsilon \lambda \kappa \tau } FF + \nn\\&& m_{119}^{} F_{\alpha 
\beta }{}^{\epsilon \varepsilon } F^{\alpha \beta \gamma 
\delta } F_{\gamma }{}^{\mu \nu \sigma } F_{\delta 
}{}^{\lambda \kappa \tau } F_{\epsilon \mu \nu \lambda } 
F_{\varepsilon \sigma \kappa \tau } FF + \nn\\&& m_{121}^{} F_{\alpha 
\beta }{}^{\epsilon \varepsilon } F^{\alpha \beta \gamma 
\delta } F_{\gamma \epsilon }{}^{\mu \nu } F_{\delta 
}{}^{\sigma \lambda \kappa } F_{\varepsilon \sigma \lambda 
}{}^{\tau } F_{\mu \nu \kappa \tau } FF + \nn\\&& m_{122}^{} F_{\alpha 
}{}^{\epsilon \varepsilon \mu } F^{\alpha \beta \gamma \delta 
} F_{\beta \epsilon }{}^{\nu \sigma } F_{\gamma \varepsilon 
}{}^{\lambda \kappa } F_{\delta \nu \lambda }{}^{\tau } F_{\mu 
\sigma \kappa \tau } FF + \nn\\&& m_{123}^{} F_{\alpha \beta 
}{}^{\epsilon \varepsilon } F^{\alpha \beta \gamma \delta } F_{
\gamma \epsilon }{}^{\mu \nu } F_{\delta \mu }{}^{\sigma 
\lambda } F_{\varepsilon \sigma }{}^{\kappa \tau } F_{\nu 
\lambda \kappa \tau } FF + \nn\\&& m_{124}^{} F_{\alpha \beta 
}{}^{\epsilon \varepsilon } F^{\alpha \beta \gamma \delta } F_{
\gamma \delta }{}^{\mu \nu } F_{\epsilon \mu }{}^{\sigma 
\lambda } F_{\varepsilon \sigma }{}^{\kappa \tau } F_{\nu 
\lambda \kappa \tau } FF + \nn\\&& m_{125}^{} F_{\alpha \beta 
}{}^{\epsilon \varepsilon } F^{\alpha \beta \gamma \delta } F_{
\gamma \epsilon }{}^{\mu \nu } F_{\delta \varepsilon 
}{}^{\sigma \lambda } F_{\mu \sigma }{}^{\kappa \tau } F_{\nu 
\lambda \kappa \tau } FF + \nn\\&& m_{126}^{} F_{\alpha \beta 
}{}^{\epsilon \varepsilon } F^{\alpha \beta \gamma \delta } F_{
\gamma \delta }{}^{\mu \nu } F_{\epsilon \varepsilon 
}{}^{\sigma \lambda } F_{\mu \sigma }{}^{\kappa \tau } F_{\nu 
\lambda \kappa \tau } FF + \nn\\&& m_{127}^{} F_{\alpha \beta 
}{}^{\epsilon \varepsilon } F^{\alpha \beta \gamma \delta } F_{
\gamma \delta }{}^{\mu \nu } F_{\epsilon \mu }{}^{\sigma 
\lambda } F_{\varepsilon \nu }{}^{\kappa \tau } F_{\sigma 
\lambda \kappa \tau } FF + \nn\\&& m_{128}^{} F_{\alpha \beta 
}{}^{\epsilon \varepsilon } F^{\alpha \beta \gamma \delta } F_{
\gamma \delta }{}^{\mu \nu } F_{\epsilon \varepsilon 
}{}^{\sigma \lambda } F_{\mu \nu }{}^{\kappa \tau } F_{\sigma 
\lambda \kappa \tau } FF + \nn\\&& m_{129}^{} F_{\alpha \beta \gamma 
}{}^{\epsilon } F^{\alpha \beta \gamma \delta } F_{\delta }{}^{
\varepsilon \mu \nu } F_{\epsilon \varepsilon }{}^{\sigma 
\lambda } F_{\mu \nu }{}^{\kappa \tau } F_{\sigma \lambda 
\kappa \tau } FF +\nn\\&& m_{133}^{} F_{\alpha \beta }{}^{\epsilon \varepsilon } 
F^{\alpha \beta \gamma \delta } F_{\gamma \delta \epsilon 
}{}^{\kappa } F_{\varepsilon }{}^{\lambda \mu \nu } F_{\kappa 
\lambda }{}^{\sigma \tau } F_{\mu \nu \sigma \tau } FF 
+ \nn\\&& m_{134}^{} F_{\alpha \beta }{}^{\epsilon \varepsilon } 
F^{\alpha \beta \gamma \delta } F_{\gamma \delta \epsilon 
\varepsilon } F_{\kappa \lambda }{}^{\sigma \tau } F^{\kappa 
\lambda \mu \nu } F_{\mu \nu \sigma \tau } FF+ \nn\\&& m_{130}^{} F_{\alpha \beta }{}^{\epsilon 
\varepsilon } F^{\alpha \beta \gamma \delta } F_{\gamma 
\epsilon }{}^{\mu \nu } F_{\delta \varepsilon \mu \nu } FF^2 + 
\nn\\&& m_{131}^{} F_{\alpha \beta }{}^{\epsilon \varepsilon } F^{\alpha 
\beta \gamma \delta } F_{\gamma \delta }{}^{\mu \nu } 
F_{\epsilon \varepsilon \mu \nu } FF^2 +  m_{132}^{} FF^4\labell{T1}
\eeqa

There are 530 couplings with structure of four $F$ and two $\nabla F$, \ie
\beqa
{\cal L}_6^{F^4(\prt F)^2}&=& m_{433}^{} F_{\alpha \beta }{}^{\epsilon \varepsilon } 
F^{\alpha \beta \gamma \delta } F_{\gamma }{}^{\mu \nu \sigma 
} F_{\mu \nu }{}^{\lambda \kappa } \nabla_{\epsilon }F_{\delta 
\sigma }{}^{\tau \omega } \nabla_{\kappa }F_{\varepsilon 
\lambda \tau \omega } + \nn\\&& m_{464}^{} F_{\alpha \beta \gamma }{}^{
\epsilon } F^{\alpha \beta \gamma \delta } F_{\varepsilon \mu 
}{}^{\lambda \kappa } F^{\varepsilon \mu \nu \sigma } \nabla_{
\epsilon }F_{\delta \nu }{}^{\tau \omega } \nabla_{\kappa }F_{
\sigma \lambda \tau \omega } + \nn\\&& m_{465}^{} F_{\alpha \beta }{}^{
\epsilon \varepsilon } F^{\alpha \beta \gamma \delta } 
F_{\gamma }{}^{\mu \nu \sigma } F_{\mu \nu }{}^{\lambda 
\kappa } \nabla_{\varepsilon }F_{\delta \epsilon }{}^{\tau 
\omega } \nabla_{\kappa }F_{\sigma \lambda \tau \omega } + 
\nn\\&& m_{466}^{} F_{\alpha \beta }{}^{\epsilon \varepsilon } F^{\alpha 
\beta \gamma \delta } F_{\gamma }{}^{\mu \nu \sigma } 
F_{\epsilon \mu }{}^{\lambda \kappa } \nabla_{\varepsilon 
}F_{\delta \nu }{}^{\tau \omega } \nabla_{\kappa }F_{\sigma 
\lambda \tau \omega } + \nn\\&& m_{431}^{} F_{\alpha \beta 
}{}^{\epsilon \varepsilon } F^{\alpha \beta \gamma \delta } F_{
\gamma }{}^{\mu \nu \sigma } F_{\mu \nu }{}^{\lambda \kappa } 
\nabla_{\kappa }F_{\varepsilon \sigma \tau \omega } 
\nabla_{\lambda }F_{\delta \epsilon }{}^{\tau \omega } + 
\nn\\&& m_{394}^{} F_{\alpha \beta }{}^{\epsilon \varepsilon } F^{\alpha 
\beta \gamma \delta } F_{\mu \nu }{}^{\kappa \tau } F^{\mu 
\nu \sigma \lambda } \nabla_{\kappa }F_{\gamma \epsilon 
\sigma }{}^{\omega } \nabla_{\lambda }F_{\delta \varepsilon 
\tau \omega } + \nn\\&& m_{396}^{} F_{\alpha }{}^{\epsilon \varepsilon 
\mu } F^{\alpha \beta \gamma \delta } F_{\beta }{}^{\nu 
\sigma \lambda } F_{\epsilon }{}^{\kappa \tau \omega } 
\nabla_{\kappa }F_{\gamma \varepsilon \nu \sigma } 
\nabla_{\lambda }F_{\delta \mu \tau \omega } + \nn\\&& m_{397}^{} 
F_{\alpha }{}^{\epsilon \varepsilon \mu } F^{\alpha \beta 
\gamma \delta } F_{\beta }{}^{\nu \sigma \lambda } F_{\epsilon 
\nu }{}^{\kappa \tau } \nabla_{\kappa }F_{\gamma \varepsilon 
\sigma }{}^{\omega } \nabla_{\lambda }F_{\delta \mu \tau 
\omega } + \nn\\&& m_{421}^{} F_{\alpha \beta \gamma }{}^{\epsilon } F^{
\alpha \beta \gamma \delta } F_{\varepsilon \mu }{}^{\lambda 
\kappa } F^{\varepsilon \mu \nu \sigma } \nabla_{\kappa 
}F_{\epsilon \sigma \tau \omega } \nabla_{\lambda }F_{\delta 
\nu }{}^{\tau \omega } + \nn\\&& m_{432}^{} F_{\alpha \beta 
}{}^{\epsilon \varepsilon } F^{\alpha \beta \gamma \delta } F_{
\gamma }{}^{\mu \nu \sigma } F_{\epsilon \mu }{}^{\lambda 
\kappa } \nabla_{\kappa }F_{\varepsilon \sigma \tau \omega } 
\nabla_{\lambda }F_{\delta \nu }{}^{\tau \omega } + \nn\\&& m_{453}^{} 
F_{\alpha }{}^{\epsilon \varepsilon \mu } F^{\alpha \beta 
\gamma \delta } F_{\beta \epsilon }{}^{\nu \sigma } F_{\gamma 
\varepsilon }{}^{\lambda \kappa } \nabla_{\kappa }F_{\mu 
\sigma \tau \omega } \nabla_{\lambda }F_{\delta \nu }{}^{\tau 
\omega } + \nn\\&& m_{450}^{} F_{\alpha \beta \gamma }{}^{\epsilon } F^{
\alpha \beta \gamma \delta } F_{\delta }{}^{\varepsilon \mu 
\nu } F_{\varepsilon }{}^{\sigma \lambda \kappa } 
\nabla_{\kappa }F_{\mu \nu \tau \omega } \nabla_{\lambda }F_{
\epsilon \sigma }{}^{\tau \omega } + \nn\\&& m_{451}^{} F_{\alpha \beta 
}{}^{\epsilon \varepsilon } F^{\alpha \beta \gamma \delta } F_{
\gamma \epsilon }{}^{\mu \nu } F_{\delta }{}^{\sigma \lambda 
\kappa } \nabla_{\kappa }F_{\mu \nu \tau \omega } 
\nabla_{\lambda }F_{\varepsilon \sigma }{}^{\tau \omega } + 
\nn\\&& m_{452}^{} F_{\alpha \beta }{}^{\epsilon \varepsilon } F^{\alpha 
\beta \gamma \delta } F_{\gamma \delta }{}^{\mu \nu } 
F_{\epsilon }{}^{\sigma \lambda \kappa } \nabla_{\kappa 
}F_{\mu \nu \tau \omega } \nabla_{\lambda }F_{\varepsilon 
\sigma }{}^{\tau \omega } + \nn\\&& m_{363}^{} F_{\alpha }{}^{\epsilon 
\varepsilon \mu } F^{\alpha \beta \gamma \delta } F_{\beta 
}{}^{\nu \sigma \lambda } F_{\epsilon }{}^{\kappa \tau \omega 
} \nabla_{\varepsilon }F_{\gamma \delta \nu \sigma } 
\nabla_{\lambda }F_{\mu \kappa \tau \omega } + \nn\\&& m_{371}^{} 
F_{\alpha \beta \gamma }{}^{\epsilon } F^{\alpha \beta \gamma 
\delta } F_{\varepsilon \mu \nu }{}^{\lambda } F^{\varepsilon 
\mu \nu \sigma } \nabla_{\epsilon }F_{\delta }{}^{\kappa \tau 
\omega } \nabla_{\lambda }F_{\sigma \kappa \tau \omega } + 
\nn\\&& m_{372}^{} F_{\alpha \beta }{}^{\epsilon \varepsilon } F^{\alpha 
\beta \gamma \delta } F_{\gamma }{}^{\mu \nu \sigma } 
F_{\epsilon \mu \nu }{}^{\lambda } \nabla_{\varepsilon 
}F_{\delta }{}^{\kappa \tau \omega } \nabla_{\lambda 
}F_{\sigma \kappa \tau \omega } + \nn\\&& m_{373}^{} F_{\alpha 
}{}^{\epsilon \varepsilon \mu } F^{\alpha \beta \gamma \delta 
} F_{\beta \epsilon }{}^{\nu \sigma } F_{\gamma \varepsilon 
\nu }{}^{\lambda } \nabla_{\lambda }F_{\sigma \kappa \tau 
\omega } \nabla_{\mu }F_{\delta }{}^{\kappa \tau \omega } + 
\nn\\&& m_{395}^{} F_{\alpha }{}^{\epsilon \varepsilon \mu } F^{\alpha 
\beta \gamma \delta } F_{\beta }{}^{\nu \sigma \lambda } 
F_{\epsilon }{}^{\kappa \tau \omega } \nabla_{\kappa 
}F_{\gamma \varepsilon \nu \sigma } \nabla_{\mu }F_{\delta 
\lambda \tau \omega } + \nn\\&& m_{467}^{} F_{\alpha }{}^{\epsilon 
\varepsilon \mu } F^{\alpha \beta \gamma \delta } F_{\beta 
\epsilon }{}^{\nu \sigma } F_{\gamma \varepsilon }{}^{\lambda 
\kappa } \nabla_{\kappa }F_{\sigma \lambda \tau \omega } 
\nabla_{\mu }F_{\delta \nu }{}^{\tau \omega } + \nn\\&& m_{456}^{} 
F_{\alpha \beta \gamma }{}^{\epsilon } F^{\alpha \beta \gamma 
\delta } F_{\delta }{}^{\varepsilon \mu \nu } F_{\varepsilon 
}{}^{\sigma \lambda \kappa } \nabla_{\kappa }F_{\nu \lambda 
\tau \omega } \nabla_{\mu }F_{\epsilon \sigma }{}^{\tau 
\omega } + \nn\\&& m_{457}^{} F_{\alpha \beta }{}^{\epsilon \varepsilon 
} F^{\alpha \beta \gamma \delta } F_{\gamma \epsilon }{}^{\mu 
\nu } F_{\delta }{}^{\sigma \lambda \kappa } \nabla_{\kappa 
}F_{\nu \lambda \tau \omega } \nabla_{\mu }F_{\varepsilon 
\sigma }{}^{\tau \omega } + \nn\\&& m_{458}^{} F_{\alpha \beta 
}{}^{\epsilon \varepsilon } F^{\alpha \beta \gamma \delta } F_{
\gamma \delta }{}^{\mu \nu } F_{\epsilon }{}^{\sigma \lambda 
\kappa } \nabla_{\kappa }F_{\nu \lambda \tau \omega } 
\nabla_{\mu }F_{\varepsilon \sigma }{}^{\tau \omega } + 
\nn\\&& m_{468}^{} F_{\alpha \beta }{}^{\epsilon \varepsilon } F^{\alpha 
\beta \gamma \delta } F_{\gamma }{}^{\mu \nu \sigma } 
F_{\epsilon \mu }{}^{\lambda \kappa } \nabla_{\kappa 
}F_{\sigma \lambda \tau \omega } \nabla_{\nu }F_{\delta 
\varepsilon }{}^{\tau \omega } + \nn\\&& m_{430}^{} F_{\alpha \beta 
}{}^{\epsilon \varepsilon } F^{\alpha \beta \gamma \delta } F_{
\gamma }{}^{\mu \nu \sigma } F_{\epsilon \mu }{}^{\lambda 
\kappa } \nabla_{\kappa }F_{\varepsilon \sigma \tau \omega } 
\nabla_{\nu }F_{\delta \lambda }{}^{\tau \omega } + \nn\\&& m_{434}^{} 
F_{\alpha \beta }{}^{\epsilon \varepsilon } F^{\alpha \beta 
\gamma \delta } F_{\gamma \epsilon }{}^{\mu \nu } F_{\mu }{}^{
\sigma \lambda \kappa } \nabla_{\kappa }F_{\varepsilon \lambda 
\tau \omega } \nabla_{\nu }F_{\delta \sigma }{}^{\tau \omega 
} + \nn\\&& m_{469}^{} F_{\alpha \beta }{}^{\epsilon \varepsilon } 
F^{\alpha \beta \gamma \delta } F_{\gamma }{}^{\mu \nu \sigma 
} F_{\delta \mu }{}^{\lambda \kappa } \nabla_{\kappa 
}F_{\sigma \lambda \tau \omega } \nabla_{\nu }F_{\epsilon 
\varepsilon }{}^{\tau \omega } + \nn\\&& m_{359}^{} F_{\alpha \beta 
}{}^{\epsilon \varepsilon } F^{\alpha \beta \gamma \delta } F_{
\mu \nu }{}^{\kappa \tau } F^{\mu \nu \sigma \lambda } 
\nabla_{\lambda }F_{\varepsilon \kappa \tau \omega } 
\nabla_{\sigma }F_{\gamma \delta \epsilon }{}^{\omega } + 
\nn\\&& m_{364}^{} F_{\alpha }{}^{\epsilon \varepsilon \mu } F^{\alpha 
\beta \gamma \delta } F_{\beta }{}^{\nu \sigma \lambda } 
F_{\gamma }{}^{\kappa \tau \omega } \nabla_{\lambda }F_{\mu 
\kappa \tau \omega } \nabla_{\sigma }F_{\delta \epsilon 
\varepsilon \nu } + \nn\\&& m_{435}^{} F_{\alpha \beta }{}^{\epsilon 
\varepsilon } F^{\alpha \beta \gamma \delta } F_{\gamma 
}{}^{\mu \nu \sigma } F_{\mu \nu }{}^{\lambda \kappa } 
\nabla_{\kappa }F_{\varepsilon \lambda \tau \omega } 
\nabla_{\sigma }F_{\delta \epsilon }{}^{\tau \omega } + 
\nn\\&& m_{459}^{} F_{\alpha \beta }{}^{\epsilon \varepsilon } F^{\alpha 
\beta \gamma \delta } F_{\gamma \epsilon }{}^{\mu \nu } 
F_{\mu }{}^{\sigma \lambda \kappa } \nabla_{\kappa }F_{\nu 
\lambda \tau \omega } \nabla_{\sigma }F_{\delta \varepsilon 
}{}^{\tau \omega } + \nn\\&& m_{354}^{} F_{\alpha \beta \gamma 
}{}^{\epsilon } F^{\alpha \beta \gamma \delta } F_{\varepsilon 
\mu \nu }{}^{\lambda } F^{\varepsilon \mu \nu \sigma } 
\nabla_{\lambda }F_{\epsilon \kappa \tau \omega } 
\nabla_{\sigma }F_{\delta }{}^{\kappa \tau \omega } + 
\nn\\&& m_{360}^{} F_{\alpha \beta }{}^{\epsilon \varepsilon } F^{\alpha 
\beta \gamma \delta } F_{\gamma }{}^{\mu \nu \sigma } 
F_{\epsilon \mu \nu }{}^{\lambda } \nabla_{\lambda 
}F_{\varepsilon \kappa \tau \omega } \nabla_{\sigma }F_{\delta 
}{}^{\kappa \tau \omega } + \nn\\&& m_{422}^{} F_{\alpha \beta \gamma 
}{}^{\epsilon } F^{\alpha \beta \gamma \delta } F_{\varepsilon 
\mu }{}^{\lambda \kappa } F^{\varepsilon \mu \nu \sigma } 
\nabla_{\kappa }F_{\epsilon \lambda \tau \omega } 
\nabla_{\sigma }F_{\delta \nu }{}^{\tau \omega } + \nn\\&& m_{436}^{} 
F_{\alpha \beta }{}^{\epsilon \varepsilon } F^{\alpha \beta 
\gamma \delta } F_{\gamma }{}^{\mu \nu \sigma } F_{\epsilon 
\mu }{}^{\lambda \kappa } \nabla_{\kappa }F_{\varepsilon 
\lambda \tau \omega } \nabla_{\sigma }F_{\delta \nu }{}^{\tau 
\omega } + \nn\\&& m_{437}^{} F_{\alpha \beta }{}^{\epsilon \varepsilon 
} F^{\alpha \beta \gamma \delta } F_{\gamma \epsilon }{}^{\mu 
\nu } F_{\mu }{}^{\sigma \lambda \kappa } \nabla_{\kappa 
}F_{\varepsilon \lambda \tau \omega } \nabla_{\sigma 
}F_{\delta \nu }{}^{\tau \omega } + \nn\\&& m_{454}^{} F_{\alpha 
}{}^{\epsilon \varepsilon \mu } F^{\alpha \beta \gamma \delta 
} F_{\beta \epsilon }{}^{\nu \sigma } F_{\gamma \varepsilon 
}{}^{\lambda \kappa } \nabla_{\kappa }F_{\mu \lambda \tau 
\omega } \nabla_{\sigma }F_{\delta \nu }{}^{\tau \omega } + 
\nn\\&& m_{418}^{} F_{\alpha \beta }{}^{\epsilon \varepsilon } F^{\alpha 
\beta \gamma \delta } F_{\gamma }{}^{\mu \nu \sigma } F_{\mu 
}{}^{\lambda \kappa \tau } \nabla_{\kappa }F_{\delta \nu 
\lambda }{}^{\omega } \nabla_{\sigma }F_{\epsilon \varepsilon 
\tau \omega } + \nn\\&& m_{333}^{} F_{\alpha \beta \gamma }{}^{\epsilon 
} F^{\alpha \beta \gamma \delta } F^{\varepsilon \mu \nu 
\sigma } F^{\lambda \kappa \tau \omega } \nabla_{\lambda 
}F_{\delta \varepsilon \mu \nu } \nabla_{\sigma }F_{\epsilon 
\kappa \tau \omega } + \nn\\&& m_{365}^{} F_{\alpha \beta \gamma 
}{}^{\epsilon } F^{\alpha \beta \gamma \delta } F_{\delta }{}^{
\varepsilon \mu \nu } F_{\varepsilon \mu }{}^{\sigma \lambda } 
\nabla_{\lambda }F_{\nu \kappa \tau \omega } \nabla_{\sigma 
}F_{\epsilon }{}^{\kappa \tau \omega } + \nn\\&& m_{460}^{} F_{\alpha 
\beta \gamma }{}^{\epsilon } F^{\alpha \beta \gamma \delta } 
F_{\delta }{}^{\varepsilon \mu \nu } F_{\varepsilon }{}^{\sigma 
\lambda \kappa } \nabla_{\kappa }F_{\nu \lambda \tau \omega } 
\nabla_{\sigma }F_{\epsilon \mu }{}^{\tau \omega } + \nn\\&& m_{438}^{} 
F_{\alpha \beta }{}^{\epsilon \varepsilon } F^{\alpha \beta 
\gamma \delta } F_{\gamma }{}^{\mu \nu \sigma } F_{\delta \mu 
}{}^{\lambda \kappa } \nabla_{\kappa }F_{\varepsilon \lambda 
\tau \omega } \nabla_{\sigma }F_{\epsilon \nu }{}^{\tau 
\omega } + \nn\\&& m_{351}^{} F_{\alpha \beta }{}^{\epsilon \varepsilon 
} F^{\alpha \beta \gamma \delta } F_{\gamma }{}^{\mu \nu 
\sigma } F_{\epsilon \mu \nu }{}^{\lambda } \nabla_{\lambda 
}F_{\delta }{}^{\kappa \tau \omega } \nabla_{\sigma 
}F_{\varepsilon \kappa \tau \omega } + \nn\\&& m_{349}^{} F_{\alpha 
\beta }{}^{\epsilon \varepsilon } F^{\alpha \beta \gamma 
\delta } F_{\gamma }{}^{\mu \nu \sigma } F_{\epsilon \mu }{}^{
\lambda \kappa } \nabla_{\lambda }F_{\delta \nu }{}^{\tau 
\omega } \nabla_{\sigma }F_{\varepsilon \kappa \tau \omega } + 
\nn\\&& m_{353}^{} F_{\alpha \beta }{}^{\epsilon \varepsilon } F^{\alpha 
\beta \gamma \delta } F_{\gamma }{}^{\mu \nu \sigma } 
F_{\delta \mu }{}^{\lambda \kappa } \nabla_{\lambda 
}F_{\epsilon \nu }{}^{\tau \omega } \nabla_{\sigma 
}F_{\varepsilon \kappa \tau \omega } + \nn\\&& m_{311}^{} F_{\alpha 
\beta }{}^{\epsilon \varepsilon } F^{\alpha \beta \gamma 
\delta } F_{\gamma }{}^{\mu \nu \sigma } F_{\mu }{}^{\lambda 
\kappa \tau } \nabla_{\nu }F_{\delta \epsilon \lambda 
}{}^{\omega } \nabla_{\sigma }F_{\varepsilon \kappa \tau 
\omega } + \nn\\&& m_{366}^{} F_{\alpha \beta }{}^{\epsilon \varepsilon 
} F^{\alpha \beta \gamma \delta } F_{\gamma \epsilon }{}^{\mu 
\nu } F_{\delta \mu }{}^{\sigma \lambda } \nabla_{\lambda }F_{
\nu \kappa \tau \omega } \nabla_{\sigma }F_{\varepsilon 
}{}^{\kappa \tau \omega } + \nn\\&& m_{461}^{} F_{\alpha \beta 
}{}^{\epsilon \varepsilon } F^{\alpha \beta \gamma \delta } F_{
\gamma \epsilon }{}^{\mu \nu } F_{\delta }{}^{\sigma \lambda 
\kappa } \nabla_{\kappa }F_{\nu \lambda \tau \omega } 
\nabla_{\sigma }F_{\varepsilon \mu }{}^{\tau \omega } + 
\nn\\&& m_{462}^{} F_{\alpha \beta \gamma }{}^{\epsilon } F^{\alpha 
\beta \gamma \delta } F_{\delta }{}^{\varepsilon \mu \nu } F_{
\epsilon }{}^{\sigma \lambda \kappa } \nabla_{\kappa }F_{\nu 
\lambda \tau \omega } \nabla_{\sigma }F_{\varepsilon \mu }{}^{
\tau \omega } + \nn\\&& m_{401}^{} F_{\alpha \beta }{}^{\epsilon 
\varepsilon } F^{\alpha \beta \gamma \delta } F_{\gamma 
}{}^{\mu \nu \sigma } F_{\mu }{}^{\lambda \kappa \tau } 
\nabla_{\kappa }F_{\delta \epsilon \lambda }{}^{\omega } 
\nabla_{\sigma }F_{\varepsilon \nu \tau \omega } + \nn\\&& m_{400}^{} 
F_{\alpha \beta }{}^{\epsilon \varepsilon } F^{\alpha \beta 
\gamma \delta } F_{\gamma }{}^{\mu \nu \sigma } F^{\lambda 
\kappa \tau \omega } \nabla_{\kappa }F_{\delta \epsilon \mu 
\lambda } \nabla_{\sigma }F_{\varepsilon \nu \tau \omega } + 
\nn\\&& m_{410}^{} F_{\alpha \beta }{}^{\epsilon \varepsilon } F^{\alpha 
\beta \gamma \delta } F_{\gamma }{}^{\mu \nu \sigma } 
F_{\epsilon }{}^{\lambda \kappa \tau } \nabla_{\kappa 
}F_{\delta \mu \lambda }{}^{\omega } \nabla_{\sigma 
}F_{\varepsilon \nu \tau \omega } + \nn\\&& m_{455}^{} F_{\alpha \beta 
}{}^{\epsilon \varepsilon } F^{\alpha \beta \gamma \delta } F_{
\gamma \delta \epsilon }{}^{\mu } F^{\nu \sigma \lambda 
\kappa } \nabla_{\kappa }F_{\mu \lambda \tau \omega } 
\nabla_{\sigma }F_{\varepsilon \nu }{}^{\tau \omega } + 
\nn\\&& m_{350}^{} F_{\alpha }{}^{\epsilon \varepsilon \mu } F^{\alpha 
\beta \gamma \delta } F_{\beta \epsilon }{}^{\nu \sigma } 
F_{\gamma \varepsilon }{}^{\lambda \kappa } \nabla_{\lambda 
}F_{\delta \nu }{}^{\tau \omega } \nabla_{\sigma }F_{\mu 
\kappa \tau \omega } + \nn\\&& m_{367}^{} F_{\alpha \beta }{}^{\epsilon 
\varepsilon } F^{\alpha \beta \gamma \delta } F_{\gamma 
\epsilon }{}^{\mu \nu } F_{\delta \varepsilon }{}^{\sigma 
\lambda } \nabla_{\lambda }F_{\nu \kappa \tau \omega } 
\nabla_{\sigma }F_{\mu }{}^{\kappa \tau \omega } + \nn\\&& m_{368}^{} 
F_{\alpha \beta }{}^{\epsilon \varepsilon } F^{\alpha \beta 
\gamma \delta } F_{\gamma \delta }{}^{\mu \nu } F_{\epsilon 
\varepsilon }{}^{\sigma \lambda } \nabla_{\lambda }F_{\nu 
\kappa \tau \omega } \nabla_{\sigma }F_{\mu }{}^{\kappa \tau 
\omega } + \nn\\&& m_{369}^{} F_{\alpha \beta \gamma }{}^{\epsilon } F^{
\alpha \beta \gamma \delta } F_{\delta }{}^{\varepsilon \mu 
\nu } F_{\epsilon \varepsilon }{}^{\sigma \lambda } 
\nabla_{\lambda }F_{\nu \kappa \tau \omega } \nabla_{\sigma 
}F_{\mu }{}^{\kappa \tau \omega } + \nn\\&& m_{314}^{} F_{\alpha 
}{}^{\epsilon \varepsilon \mu } F^{\alpha \beta \gamma \delta 
} F_{\beta \epsilon }{}^{\nu \sigma } F^{\lambda \kappa \tau 
\omega } \nabla_{\varepsilon }F_{\gamma \delta \lambda \kappa 
} \nabla_{\sigma }F_{\mu \nu \tau \omega } + \nn\\&& m_{405}^{} 
F_{\alpha }{}^{\epsilon \varepsilon \mu } F^{\alpha \beta 
\gamma \delta } F_{\beta \epsilon }{}^{\nu \sigma } F_{\gamma 
}{}^{\lambda \kappa \tau } \nabla_{\kappa }F_{\delta 
\varepsilon \lambda }{}^{\omega } \nabla_{\sigma }F_{\mu \nu 
\tau \omega } + \nn\\&& m_{594}^{} F_{\alpha }{}^{\epsilon \varepsilon 
\mu } F^{\alpha \beta \gamma \delta } F_{\beta }{}^{\nu 
\sigma \lambda } F_{\epsilon }{}^{\kappa \tau \omega } 
\nabla_{\lambda }F_{\mu \nu \sigma \omega } \nabla_{\tau }F_{
\gamma \delta \varepsilon \kappa } + \nn\\&& m_{595}^{} F_{\alpha 
}{}^{\epsilon \varepsilon \mu } F^{\alpha \beta \gamma \delta 
} F_{\beta }{}^{\nu \sigma \lambda } F_{\epsilon }{}^{\kappa 
\tau \omega } \nabla_{\lambda }F_{\varepsilon \mu \sigma 
\omega } \nabla_{\tau }F_{\gamma \delta \nu \kappa } + 
\nn\\&& m_{597}^{} F_{\alpha \beta }{}^{\epsilon \varepsilon } F^{\alpha 
\beta \gamma \delta } F_{\mu \nu }{}^{\kappa \tau } F^{\mu 
\nu \sigma \lambda } \nabla_{\lambda }F_{\gamma \epsilon 
\sigma }{}^{\omega } \nabla_{\tau }F_{\delta \varepsilon 
\kappa \omega } + \nn\\&& m_{596}^{} F_{\alpha \beta }{}^{\epsilon 
\varepsilon } F^{\alpha \beta \gamma \delta } F_{\mu \nu }{}^{
\kappa \tau } F^{\mu \nu \sigma \lambda } \nabla_{\kappa }F_{
\gamma \epsilon \sigma }{}^{\omega } \nabla_{\tau }F_{\delta 
\varepsilon \lambda \omega } + \nn\\&& m_{599}^{} F_{\alpha 
}{}^{\epsilon \varepsilon \mu } F^{\alpha \beta \gamma \delta 
} F_{\beta }{}^{\nu \sigma \lambda } F_{\epsilon \nu 
}{}^{\kappa \tau } \nabla_{\mu }F_{\gamma \varepsilon \sigma 
}{}^{\omega } \nabla_{\tau }F_{\delta \lambda \kappa \omega } 
+ \nn\\&& m_{600}^{} F_{\alpha \beta }{}^{\epsilon \varepsilon } 
F^{\alpha \beta \gamma \delta } F_{\gamma }{}^{\mu \nu \sigma 
} F_{\mu }{}^{\lambda \kappa \tau } \nabla_{\sigma 
}F_{\epsilon \varepsilon \nu \omega } \nabla_{\tau }F_{\delta 
\lambda \kappa }{}^{\omega } + \nn\\&& m_{601}^{} F_{\alpha 
}{}^{\epsilon \varepsilon \mu } F^{\alpha \beta \gamma \delta 
} F_{\beta \epsilon }{}^{\nu \sigma } F_{\gamma }{}^{\lambda 
\kappa \tau } \nabla_{\sigma }F_{\varepsilon \mu \nu \omega } 
\nabla_{\tau }F_{\delta \lambda \kappa }{}^{\omega } + 
\nn\\&& m_{602}^{} F_{\alpha \beta }{}^{\epsilon \varepsilon } F^{\alpha 
\beta \gamma \delta } F_{\gamma }{}^{\mu \nu \sigma } 
F_{\epsilon }{}^{\lambda \kappa \tau } \nabla_{\sigma 
}F_{\varepsilon \mu \nu \omega } \nabla_{\tau }F_{\delta 
\lambda \kappa }{}^{\omega } + \nn\\&& m_{598}^{} F_{\alpha 
}{}^{\epsilon \varepsilon \mu } F^{\alpha \beta \gamma \delta 
} F_{\beta }{}^{\nu \sigma \lambda } F_{\epsilon \nu 
}{}^{\kappa \tau } \nabla_{\kappa }F_{\gamma \varepsilon 
\sigma }{}^{\omega } \nabla_{\tau }F_{\delta \mu \lambda 
\omega } + \nn\\&& m_{607}^{} F_{\alpha \beta }{}^{\epsilon \varepsilon 
} F^{\alpha \beta \gamma \delta } F_{\mu \nu }{}^{\kappa \tau 
} F^{\mu \nu \sigma \lambda } \nabla_{\lambda }F_{\gamma 
\delta \sigma }{}^{\omega } \nabla_{\tau }F_{\epsilon 
\varepsilon \kappa \omega } + \nn\\&& m_{608}^{} F_{\alpha \beta 
}{}^{\epsilon \varepsilon } F^{\alpha \beta \gamma \delta } F_{
\gamma }{}^{\mu \nu \sigma } F_{\mu }{}^{\lambda \kappa \tau 
} \nabla_{\lambda }F_{\delta \nu \sigma }{}^{\omega } \nabla_{
\tau }F_{\epsilon \varepsilon \kappa \omega } + \nn\\&& m_{606}^{} 
F_{\alpha \beta }{}^{\epsilon \varepsilon } F^{\alpha \beta 
\gamma \delta } F_{\gamma }{}^{\mu \nu \sigma } F_{\mu 
}{}^{\lambda \kappa \tau } \nabla_{\sigma }F_{\delta \nu 
\lambda }{}^{\omega } \nabla_{\tau }F_{\epsilon \varepsilon 
\kappa \omega } + \nn\\&& m_{605}^{} F_{\alpha \beta }{}^{\epsilon 
\varepsilon } F^{\alpha \beta \gamma \delta } F_{\mu \nu }{}^{
\kappa \tau } F^{\mu \nu \sigma \lambda } \nabla_{\kappa }F_{
\gamma \delta \sigma }{}^{\omega } \nabla_{\tau }F_{\epsilon 
\varepsilon \lambda \omega } + \nn\\&& m_{604}^{} F_{\alpha \beta 
}{}^{\epsilon \varepsilon } F^{\alpha \beta \gamma \delta } F_{
\gamma }{}^{\mu \nu \sigma } F_{\mu }{}^{\lambda \kappa \tau 
} \nabla_{\kappa }F_{\delta \nu \lambda }{}^{\omega } \nabla_{
\tau }F_{\epsilon \varepsilon \sigma \omega } + \nn\\&& m_{603}^{} 
F_{\alpha \beta }{}^{\epsilon \varepsilon } F^{\alpha \beta 
\gamma \delta } F_{\gamma }{}^{\mu \nu \sigma } F_{\mu 
}{}^{\lambda \kappa \tau } \nabla_{\nu }F_{\delta \lambda 
\kappa }{}^{\omega } \nabla_{\tau }F_{\epsilon \varepsilon 
\sigma \omega } + \nn\\&& m_{611}^{} F_{\alpha \beta \gamma 
}{}^{\epsilon } F^{\alpha \beta \gamma \delta } F_{\varepsilon 
}{}^{\lambda \kappa \tau } F^{\varepsilon \mu \nu \sigma } 
\nabla_{\sigma }F_{\delta \mu \nu }{}^{\omega } \nabla_{\tau 
}F_{\epsilon \lambda \kappa \omega } + \nn\\&& m_{610}^{} F_{\alpha 
\beta \gamma }{}^{\epsilon } F^{\alpha \beta \gamma \delta } 
F_{\varepsilon }{}^{\lambda \kappa \tau } F^{\varepsilon \mu 
\nu \sigma } \nabla_{\lambda }F_{\delta \mu \nu }{}^{\omega } 
\nabla_{\tau }F_{\epsilon \sigma \kappa \omega } + \nn\\&& m_{609}^{} 
F_{\alpha \beta \gamma }{}^{\epsilon } F^{\alpha \beta \gamma 
\delta } F_{\varepsilon }{}^{\lambda \kappa \tau } 
F^{\varepsilon \mu \nu \sigma } \nabla_{\nu }F_{\delta \mu 
\lambda }{}^{\omega } \nabla_{\tau }F_{\epsilon \sigma \kappa 
\omega } + \nn\\&& m_{625}^{} F_{\alpha \beta }{}^{\epsilon \varepsilon 
} F^{\alpha \beta \gamma \delta } F_{\mu \nu }{}^{\kappa \tau 
} F^{\mu \nu \sigma \lambda } \nabla_{\epsilon }F_{\gamma 
\delta \sigma }{}^{\omega } \nabla_{\tau }F_{\varepsilon 
\lambda \kappa \omega } + \nn\\&& m_{626}^{} F_{\alpha \beta 
}{}^{\epsilon \varepsilon } F^{\alpha \beta \gamma \delta } F_{
\gamma }{}^{\mu \nu \sigma } F_{\mu }{}^{\lambda \kappa \tau 
} \nabla_{\epsilon }F_{\delta \nu \sigma }{}^{\omega } 
\nabla_{\tau }F_{\varepsilon \lambda \kappa \omega } + 
\nn\\&& m_{627}^{} F_{\alpha \beta }{}^{\epsilon \varepsilon } F^{\alpha 
\beta \gamma \delta } F_{\gamma \epsilon }{}^{\mu \nu } 
F^{\sigma \lambda \kappa \tau } \nabla_{\nu }F_{\delta \mu 
\sigma }{}^{\omega } \nabla_{\tau }F_{\varepsilon \lambda 
\kappa \omega } + \nn\\&& m_{628}^{} F_{\alpha \beta }{}^{\epsilon 
\varepsilon } F^{\alpha \beta \gamma \delta } F_{\mu \nu }{}^{
\kappa \tau } F^{\mu \nu \sigma \lambda } \nabla_{\sigma }F_{
\gamma \delta \epsilon }{}^{\omega } \nabla_{\tau 
}F_{\varepsilon \lambda \kappa \omega } + \nn\\&& m_{629}^{} F_{\alpha 
\beta }{}^{\epsilon \varepsilon } F^{\alpha \beta \gamma 
\delta } F_{\gamma }{}^{\mu \nu \sigma } F_{\mu }{}^{\lambda 
\kappa \tau } \nabla_{\sigma }F_{\delta \epsilon \nu 
}{}^{\omega } \nabla_{\tau }F_{\varepsilon \lambda \kappa 
\omega } + \nn\\&& m_{630}^{} F_{\alpha \beta }{}^{\epsilon \varepsilon 
} F^{\alpha \beta \gamma \delta } F_{\gamma }{}^{\mu \nu 
\sigma } F_{\epsilon }{}^{\lambda \kappa \tau } \nabla_{\sigma 
}F_{\delta \mu \nu }{}^{\omega } \nabla_{\tau }F_{\varepsilon 
\lambda \kappa \omega } + \nn\\&& m_{631}^{} F_{\alpha \beta 
}{}^{\epsilon \varepsilon } F^{\alpha \beta \gamma \delta } F_{
\gamma \epsilon }{}^{\mu \nu } F^{\sigma \lambda \kappa \tau 
} \nabla_{\sigma }F_{\delta \mu \nu }{}^{\omega } 
\nabla_{\tau }F_{\varepsilon \lambda \kappa \omega } + 
\nn\\&& m_{632}^{} F_{\alpha \beta }{}^{\epsilon \varepsilon } F^{\alpha 
\beta \gamma \delta } F_{\gamma }{}^{\mu \nu \sigma } 
F_{\delta }{}^{\lambda \kappa \tau } \nabla_{\sigma 
}F_{\epsilon \mu \nu }{}^{\omega } \nabla_{\tau 
}F_{\varepsilon \lambda \kappa \omega } + \nn\\&& m_{612}^{} F_{\alpha 
\beta \gamma }{}^{\epsilon } F^{\alpha \beta \gamma \delta } 
F_{\delta }{}^{\varepsilon \mu \nu } F^{\sigma \lambda \kappa 
\tau } \nabla_{\kappa }F_{\epsilon \sigma \lambda }{}^{\omega 
} \nabla_{\tau }F_{\varepsilon \mu \nu \omega } + \nn\\&& m_{616}^{} 
F_{\alpha \beta }{}^{\epsilon \varepsilon } F^{\alpha \beta 
\gamma \delta } F_{\gamma \epsilon }{}^{\mu \nu } F^{\sigma 
\lambda \kappa \tau } \nabla_{\lambda }F_{\delta \mu \sigma 
}{}^{\omega } \nabla_{\tau }F_{\varepsilon \nu \kappa \omega } 
+ \nn\\&& m_{617}^{} F_{\alpha \beta }{}^{\epsilon \varepsilon } 
F^{\alpha \beta \gamma \delta } F_{\gamma \delta }{}^{\mu \nu 
} F^{\sigma \lambda \kappa \tau } \nabla_{\lambda }F_{\epsilon 
\mu \sigma }{}^{\omega } \nabla_{\tau }F_{\varepsilon \nu 
\kappa \omega } + \nn\\&& m_{613}^{} F_{\alpha \beta }{}^{\epsilon 
\varepsilon } F^{\alpha \beta \gamma \delta } F_{\gamma 
}{}^{\mu \nu \sigma } F_{\mu }{}^{\lambda \kappa \tau } 
\nabla_{\epsilon }F_{\delta \lambda \kappa }{}^{\omega } 
\nabla_{\tau }F_{\varepsilon \nu \sigma \omega } + \nn\\&& m_{615}^{} 
F_{\alpha \beta }{}^{\epsilon \varepsilon } F^{\alpha \beta 
\gamma \delta } F_{\gamma }{}^{\mu \nu \sigma } F_{\epsilon 
}{}^{\lambda \kappa \tau } \nabla_{\kappa }F_{\delta \mu 
\lambda }{}^{\omega } \nabla_{\tau }F_{\varepsilon \nu \sigma 
\omega } + \nn\\&& m_{614}^{} F_{\alpha \beta }{}^{\epsilon \varepsilon 
} F^{\alpha \beta \gamma \delta } F_{\gamma }{}^{\mu \nu 
\sigma } F_{\epsilon }{}^{\lambda \kappa \tau } \nabla_{\mu 
}F_{\delta \lambda \kappa }{}^{\omega } \nabla_{\tau 
}F_{\varepsilon \nu \sigma \omega } + \nn\\&& m_{618}^{} F_{\alpha 
\beta }{}^{\epsilon \varepsilon } F^{\alpha \beta \gamma 
\delta } F_{\gamma }{}^{\mu \nu \sigma } F_{\mu }{}^{\lambda 
\kappa \tau } \nabla_{\epsilon }F_{\delta \nu \lambda 
}{}^{\omega } \nabla_{\tau }F_{\varepsilon \sigma \kappa 
\omega } + \nn\\&& m_{622}^{} F_{\alpha \beta }{}^{\epsilon \varepsilon 
} F^{\alpha \beta \gamma \delta } F_{\gamma }{}^{\mu \nu 
\sigma } F_{\mu }{}^{\lambda \kappa \tau } \nabla_{\lambda 
}F_{\delta \epsilon \nu }{}^{\omega } \nabla_{\tau 
}F_{\varepsilon \sigma \kappa \omega } + \nn\\&& m_{623}^{} F_{\alpha 
\beta }{}^{\epsilon \varepsilon } F^{\alpha \beta \gamma 
\delta } F_{\gamma }{}^{\mu \nu \sigma } F_{\epsilon 
}{}^{\lambda \kappa \tau } \nabla_{\lambda }F_{\delta \mu \nu 
}{}^{\omega } \nabla_{\tau }F_{\varepsilon \sigma \kappa 
\omega } + \nn\\&& m_{624}^{} F_{\alpha \beta }{}^{\epsilon \varepsilon 
} F^{\alpha \beta \gamma \delta } F_{\gamma }{}^{\mu \nu 
\sigma } F_{\delta }{}^{\lambda \kappa \tau } \nabla_{\lambda 
}F_{\epsilon \mu \nu }{}^{\omega } \nabla_{\tau 
}F_{\varepsilon \sigma \kappa \omega } + \nn\\&& m_{619}^{} F_{\alpha 
\beta }{}^{\epsilon \varepsilon } F^{\alpha \beta \gamma 
\delta } F_{\gamma }{}^{\mu \nu \sigma } F_{\mu }{}^{\lambda 
\kappa \tau } \nabla_{\nu }F_{\delta \epsilon \lambda 
}{}^{\omega } \nabla_{\tau }F_{\varepsilon \sigma \kappa 
\omega } + \nn\\&& m_{620}^{} F_{\alpha \beta }{}^{\epsilon \varepsilon 
} F^{\alpha \beta \gamma \delta } F_{\gamma }{}^{\mu \nu 
\sigma } F_{\epsilon }{}^{\lambda \kappa \tau } \nabla_{\nu 
}F_{\delta \mu \lambda }{}^{\omega } \nabla_{\tau 
}F_{\varepsilon \sigma \kappa \omega } + \nn\\&& m_{621}^{} F_{\alpha 
\beta }{}^{\epsilon \varepsilon } F^{\alpha \beta \gamma 
\delta } F_{\gamma }{}^{\mu \nu \sigma } F_{\delta 
}{}^{\lambda \kappa \tau } \nabla_{\nu }F_{\epsilon \mu 
\lambda }{}^{\omega } \nabla_{\tau }F_{\varepsilon \sigma 
\kappa \omega } + \nn\\&& m_{643}^{} F_{\alpha }{}^{\epsilon \varepsilon 
\mu } F^{\alpha \beta \gamma \delta } F_{\beta }{}^{\nu 
\sigma \lambda } F_{\epsilon \nu }{}^{\kappa \tau } 
\nabla_{\sigma }F_{\gamma \delta \varepsilon }{}^{\omega } 
\nabla_{\tau }F_{\mu \lambda \kappa \omega } + \nn\\&& m_{635}^{} 
F_{\alpha \beta \gamma }{}^{\epsilon } F^{\alpha \beta \gamma 
\delta } F_{\delta }{}^{\varepsilon \mu \nu } F^{\sigma 
\lambda \kappa \tau } \nabla_{\varepsilon }F_{\epsilon \sigma 
\lambda }{}^{\omega } \nabla_{\tau }F_{\mu \nu \kappa \omega 
} + \nn\\&& m_{636}^{} F_{\alpha \beta }{}^{\epsilon \varepsilon } 
F^{\alpha \beta \gamma \delta } F_{\gamma \epsilon }{}^{\mu 
\nu } F^{\sigma \lambda \kappa \tau } \nabla_{\lambda 
}F_{\delta \varepsilon \sigma }{}^{\omega } \nabla_{\tau 
}F_{\mu \nu \kappa \omega } + \nn\\&& m_{637}^{} F_{\alpha \beta 
}{}^{\epsilon \varepsilon } F^{\alpha \beta \gamma \delta } F_{
\gamma \delta }{}^{\mu \nu } F^{\sigma \lambda \kappa \tau } 
\nabla_{\lambda }F_{\epsilon \varepsilon \sigma }{}^{\omega } 
\nabla_{\tau }F_{\mu \nu \kappa \omega } + \nn\\&& m_{638}^{} 
F_{\alpha \beta \gamma }{}^{\epsilon } F^{\alpha \beta \gamma 
\delta } F_{\delta }{}^{\varepsilon \mu \nu } F^{\sigma 
\lambda \kappa \tau } \nabla_{\lambda }F_{\epsilon \varepsilon 
\sigma }{}^{\omega } \nabla_{\tau }F_{\mu \nu \kappa \omega } 
+ \nn\\&& m_{633}^{} F_{\alpha }{}^{\epsilon \varepsilon \mu } F^{\alpha 
\beta \gamma \delta } F_{\beta \epsilon }{}^{\nu \sigma } 
F_{\gamma }{}^{\lambda \kappa \tau } \nabla_{\varepsilon 
}F_{\delta \lambda \kappa }{}^{\omega } \nabla_{\tau }F_{\mu 
\nu \sigma \omega } + \nn\\&& m_{634}^{} F_{\alpha }{}^{\epsilon 
\varepsilon \mu } F^{\alpha \beta \gamma \delta } F_{\beta 
\epsilon }{}^{\nu \sigma } F_{\gamma }{}^{\lambda \kappa \tau 
} \nabla_{\kappa }F_{\delta \varepsilon \lambda }{}^{\omega } 
\nabla_{\tau }F_{\mu \nu \sigma \omega } + \nn\\&& m_{642}^{} 
F_{\alpha }{}^{\epsilon \varepsilon \mu } F^{\alpha \beta 
\gamma \delta } F_{\beta \epsilon }{}^{\nu \sigma } F_{\gamma 
}{}^{\lambda \kappa \tau } \nabla_{\nu }F_{\delta \varepsilon 
\lambda }{}^{\omega } \nabla_{\tau }F_{\mu \sigma \kappa 
\omega } + \nn\\&& m_{640}^{} F_{\alpha }{}^{\epsilon \varepsilon \mu } 
F^{\alpha \beta \gamma \delta } F_{\beta }{}^{\nu \sigma 
\lambda } F_{\epsilon \nu }{}^{\kappa \tau } 
\nabla_{\varepsilon }F_{\gamma \delta \kappa }{}^{\omega } 
\nabla_{\tau }F_{\mu \sigma \lambda \omega } + \nn\\&& m_{641}^{} 
F_{\alpha }{}^{\epsilon \varepsilon \mu } F^{\alpha \beta 
\gamma \delta } F_{\beta }{}^{\nu \sigma \lambda } F_{\epsilon 
\nu }{}^{\kappa \tau } \nabla_{\kappa }F_{\gamma \delta 
\varepsilon }{}^{\omega } \nabla_{\tau }F_{\mu \sigma \lambda 
\omega } + \nn\\&& m_{652}^{} F_{\alpha \beta }{}^{\epsilon \varepsilon 
} F^{\alpha \beta \gamma \delta } F_{\gamma \epsilon }{}^{\mu 
\nu } F^{\sigma \lambda \kappa \tau } \nabla_{\mu }F_{\delta 
\varepsilon \sigma }{}^{\omega } \nabla_{\tau }F_{\nu \lambda 
\kappa \omega } + \nn\\&& m_{653}^{} F_{\alpha \beta }{}^{\epsilon 
\varepsilon } F^{\alpha \beta \gamma \delta } F_{\gamma \delta 
}{}^{\mu \nu } F^{\sigma \lambda \kappa \tau } \nabla_{\mu 
}F_{\epsilon \varepsilon \sigma }{}^{\omega } \nabla_{\tau }F_{
\nu \lambda \kappa \omega } + \nn\\&& m_{654}^{} F_{\alpha \beta 
\gamma }{}^{\epsilon } F^{\alpha \beta \gamma \delta } 
F_{\delta }{}^{\varepsilon \mu \nu } F^{\sigma \lambda \kappa 
\tau } \nabla_{\mu }F_{\epsilon \varepsilon \sigma }{}^{\omega 
} \nabla_{\tau }F_{\nu \lambda \kappa \omega } + \nn\\&& m_{655}^{} F_{
\alpha \beta }{}^{\epsilon \varepsilon } F^{\alpha \beta 
\gamma \delta } F_{\gamma \epsilon }{}^{\mu \nu } F^{\sigma 
\lambda \kappa \tau } \nabla_{\sigma }F_{\delta \varepsilon 
\mu }{}^{\omega } \nabla_{\tau }F_{\nu \lambda \kappa \omega 
} + \nn\\&& m_{656}^{} F_{\alpha \beta }{}^{\epsilon \varepsilon } 
F^{\alpha \beta \gamma \delta } F_{\gamma \delta }{}^{\mu \nu 
} F^{\sigma \lambda \kappa \tau } \nabla_{\sigma }F_{\epsilon 
\varepsilon \mu }{}^{\omega } \nabla_{\tau }F_{\nu \lambda 
\kappa \omega } + \nn\\&& m_{657}^{} F_{\alpha \beta \gamma 
}{}^{\epsilon } F^{\alpha \beta \gamma \delta } F_{\delta }{}^{
\varepsilon \mu \nu } F^{\sigma \lambda \kappa \tau } 
\nabla_{\sigma }F_{\epsilon \varepsilon \mu }{}^{\omega } 
\nabla_{\tau }F_{\nu \lambda \kappa \omega } + \nn\\&& m_{644}^{} 
F_{\alpha \beta \gamma }{}^{\epsilon } F^{\alpha \beta \gamma 
\delta } F_{\varepsilon }{}^{\lambda \kappa \tau } 
F^{\varepsilon \mu \nu \sigma } \nabla_{\epsilon }F_{\delta 
\mu \lambda }{}^{\omega } \nabla_{\tau }F_{\nu \sigma \kappa 
\omega } + \nn\\&& m_{645}^{} F_{\alpha \beta }{}^{\epsilon \varepsilon 
} F^{\alpha \beta \gamma \delta } F_{\gamma }{}^{\mu \nu 
\sigma } F_{\mu }{}^{\lambda \kappa \tau } \nabla_{\varepsilon 
}F_{\delta \epsilon \lambda }{}^{\omega } \nabla_{\tau }F_{\nu 
\sigma \kappa \omega } + \nn\\&& m_{646}^{} F_{\alpha \beta 
}{}^{\epsilon \varepsilon } F^{\alpha \beta \gamma \delta } F_{
\gamma }{}^{\mu \nu \sigma } F_{\epsilon }{}^{\lambda \kappa 
\tau } \nabla_{\varepsilon }F_{\delta \mu \lambda }{}^{\omega 
} \nabla_{\tau }F_{\nu \sigma \kappa \omega } + \nn\\&& m_{648}^{} 
F_{\alpha \beta }{}^{\epsilon \varepsilon } F^{\alpha \beta 
\gamma \delta } F_{\gamma }{}^{\mu \nu \sigma } F_{\mu 
}{}^{\lambda \kappa \tau } \nabla_{\lambda }F_{\delta \epsilon 
\varepsilon }{}^{\omega } \nabla_{\tau }F_{\nu \sigma \kappa 
\omega } + \nn\\&& m_{649}^{} F_{\alpha }{}^{\epsilon \varepsilon \mu } 
F^{\alpha \beta \gamma \delta } F_{\beta \epsilon }{}^{\nu 
\sigma } F_{\gamma }{}^{\lambda \kappa \tau } \nabla_{\lambda 
}F_{\delta \varepsilon \mu }{}^{\omega } \nabla_{\tau }F_{\nu 
\sigma \kappa \omega } + \nn\\&& m_{650}^{} F_{\alpha \beta 
}{}^{\epsilon \varepsilon } F^{\alpha \beta \gamma \delta } F_{
\gamma }{}^{\mu \nu \sigma } F_{\epsilon }{}^{\lambda \kappa 
\tau } \nabla_{\lambda }F_{\delta \varepsilon \mu }{}^{\omega 
} \nabla_{\tau }F_{\nu \sigma \kappa \omega } + \nn\\&& m_{651}^{} 
F_{\alpha \beta }{}^{\epsilon \varepsilon } F^{\alpha \beta 
\gamma \delta } F_{\gamma }{}^{\mu \nu \sigma } F_{\delta 
}{}^{\lambda \kappa \tau } \nabla_{\lambda }F_{\epsilon 
\varepsilon \mu }{}^{\omega } \nabla_{\tau }F_{\nu \sigma 
\kappa \omega } + \nn\\&& m_{647}^{} F_{\alpha }{}^{\epsilon \varepsilon 
\mu } F^{\alpha \beta \gamma \delta } F_{\beta \epsilon 
}{}^{\nu \sigma } F_{\gamma }{}^{\lambda \kappa \tau } 
\nabla_{\mu }F_{\delta \varepsilon \lambda }{}^{\omega } 
\nabla_{\tau }F_{\nu \sigma \kappa \omega } + \nn\\&& m_{659}^{} 
F_{\alpha \beta \gamma }{}^{\epsilon } F^{\alpha \beta \gamma 
\delta } F_{\varepsilon }{}^{\lambda \kappa \tau } 
F^{\varepsilon \mu \nu \sigma } \nabla_{\epsilon }F_{\delta 
\mu \nu }{}^{\omega } \nabla_{\tau }F_{\sigma \lambda \kappa 
\omega } + \nn\\&& m_{660}^{} F_{\alpha \beta }{}^{\epsilon \varepsilon 
} F^{\alpha \beta \gamma \delta } F_{\gamma }{}^{\mu \nu 
\sigma } F_{\mu }{}^{\lambda \kappa \tau } \nabla_{\varepsilon 
}F_{\delta \epsilon \nu }{}^{\omega } \nabla_{\tau }F_{\sigma 
\lambda \kappa \omega } + \nn\\&& m_{661}^{} F_{\alpha \beta 
}{}^{\epsilon \varepsilon } F^{\alpha \beta \gamma \delta } F_{
\gamma }{}^{\mu \nu \sigma } F_{\epsilon }{}^{\lambda \kappa 
\tau } \nabla_{\varepsilon }F_{\delta \mu \nu }{}^{\omega } 
\nabla_{\tau }F_{\sigma \lambda \kappa \omega } + \nn\\&& m_{662}^{} 
F_{\alpha }{}^{\epsilon \varepsilon \mu } F^{\alpha \beta 
\gamma \delta } F_{\beta }{}^{\nu \sigma \lambda } F_{\epsilon 
\nu }{}^{\kappa \tau } \nabla_{\mu }F_{\gamma \delta 
\varepsilon }{}^{\omega } \nabla_{\tau }F_{\sigma \lambda 
\kappa \omega } + \nn\\&& m_{663}^{} F_{\alpha }{}^{\epsilon \varepsilon 
\mu } F^{\alpha \beta \gamma \delta } F_{\beta \epsilon 
}{}^{\nu \sigma } F_{\gamma }{}^{\lambda \kappa \tau } 
\nabla_{\mu }F_{\delta \varepsilon \nu }{}^{\omega } 
\nabla_{\tau }F_{\sigma \lambda \kappa \omega } + \nn\\&& m_{664}^{} 
F_{\alpha \beta }{}^{\epsilon \varepsilon } F^{\alpha \beta 
\gamma \delta } F_{\gamma }{}^{\mu \nu \sigma } F_{\mu 
}{}^{\lambda \kappa \tau } \nabla_{\nu }F_{\delta \epsilon 
\varepsilon }{}^{\omega } \nabla_{\tau }F_{\sigma \lambda 
\kappa \omega } + \nn\\&& m_{665}^{} F_{\alpha }{}^{\epsilon \varepsilon 
\mu } F^{\alpha \beta \gamma \delta } F_{\beta \epsilon 
}{}^{\nu \sigma } F_{\gamma }{}^{\lambda \kappa \tau } 
\nabla_{\nu }F_{\delta \varepsilon \mu }{}^{\omega } 
\nabla_{\tau }F_{\sigma \lambda \kappa \omega } + \nn\\&& m_{666}^{} 
F_{\alpha \beta }{}^{\epsilon \varepsilon } F^{\alpha \beta 
\gamma \delta } F_{\gamma }{}^{\mu \nu \sigma } F_{\epsilon 
}{}^{\lambda \kappa \tau } \nabla_{\nu }F_{\delta \varepsilon 
\mu }{}^{\omega } \nabla_{\tau }F_{\sigma \lambda \kappa 
\omega } + \nn\\&& m_{667}^{} F_{\alpha \beta }{}^{\epsilon \varepsilon 
} F^{\alpha \beta \gamma \delta } F_{\gamma }{}^{\mu \nu 
\sigma } F_{\delta }{}^{\lambda \kappa \tau } \nabla_{\nu }F_{
\epsilon \varepsilon \mu }{}^{\omega } \nabla_{\tau }F_{\sigma 
\lambda \kappa \omega } + \nn\\&& m_{668}^{} F_{\alpha }{}^{\epsilon 
\varepsilon \mu } F^{\alpha \beta \gamma \delta } F_{\beta 
}{}^{\nu \sigma \lambda } F_{\epsilon }{}^{\kappa \tau \omega 
} \nabla_{\lambda }F_{\varepsilon \mu \nu \sigma } 
\nabla_{\omega }F_{\gamma \delta \kappa \tau } + \nn\\&& m_{671}^{} F_{
\alpha \beta }{}^{\epsilon \varepsilon } F^{\alpha \beta 
\gamma \delta } F_{\mu }{}^{\kappa \tau \omega } F^{\mu \nu 
\sigma \lambda } \nabla_{\lambda }F_{\gamma \epsilon \nu 
\sigma } \nabla_{\omega }F_{\delta \varepsilon \kappa \tau } + 
\nn\\&& m_{670}^{} F_{\alpha \beta }{}^{\epsilon \varepsilon } F^{\alpha 
\beta \gamma \delta } F_{\mu }{}^{\kappa \tau \omega } F^{\mu 
\nu \sigma \lambda } \nabla_{\kappa }F_{\gamma \epsilon \nu 
\sigma } \nabla_{\omega }F_{\delta \varepsilon \lambda \tau } 
+ \nn\\&& m_{669}^{} F_{\alpha \beta }{}^{\epsilon \varepsilon } 
F^{\alpha \beta \gamma \delta } F_{\mu }{}^{\kappa \tau 
\omega } F^{\mu \nu \sigma \lambda } \nabla_{\sigma 
}F_{\gamma \epsilon \nu \kappa } \nabla_{\omega }F_{\delta 
\varepsilon \lambda \tau } + \nn\\&& m_{675}^{} F_{\alpha }{}^{\epsilon 
\varepsilon \mu } F^{\alpha \beta \gamma \delta } F_{\beta 
}{}^{\nu \sigma \lambda } F_{\epsilon }{}^{\kappa \tau \omega 
} \nabla_{\mu }F_{\gamma \varepsilon \nu \sigma } 
\nabla_{\omega }F_{\delta \lambda \kappa \tau } + \nn\\&& m_{676}^{} 
F_{\alpha \beta }{}^{\epsilon \varepsilon } F^{\alpha \beta 
\gamma \delta } F_{\gamma }{}^{\mu \nu \sigma } F^{\lambda 
\kappa \tau \omega } \nabla_{\sigma }F_{\epsilon \varepsilon 
\mu \nu } \nabla_{\omega }F_{\delta \lambda \kappa \tau } + 
\nn\\&& m_{674}^{} F_{\alpha }{}^{\epsilon \varepsilon \mu } F^{\alpha 
\beta \gamma \delta } F_{\beta }{}^{\nu \sigma \lambda } 
F_{\epsilon }{}^{\kappa \tau \omega } \nabla_{\lambda 
}F_{\gamma \varepsilon \nu \sigma } \nabla_{\omega }F_{\delta 
\mu \kappa \tau } + \nn\\&& m_{673}^{} F_{\alpha }{}^{\epsilon 
\varepsilon \mu } F^{\alpha \beta \gamma \delta } F_{\beta 
}{}^{\nu \sigma \lambda } F_{\epsilon }{}^{\kappa \tau \omega 
} \nabla_{\kappa }F_{\gamma \varepsilon \nu \sigma } 
\nabla_{\omega }F_{\delta \mu \lambda \tau } + \nn\\&& m_{672}^{} 
F_{\alpha }{}^{\epsilon \varepsilon \mu } F^{\alpha \beta 
\gamma \delta } F_{\beta }{}^{\nu \sigma \lambda } F_{\epsilon 
}{}^{\kappa \tau \omega } \nabla_{\sigma }F_{\gamma 
\varepsilon \nu \kappa } \nabla_{\omega }F_{\delta \mu 
\lambda \tau } + \nn\\&& m_{685}^{} F_{\alpha \beta }{}^{\epsilon 
\varepsilon } F^{\alpha \beta \gamma \delta } F_{\mu 
}{}^{\kappa \tau \omega } F^{\mu \nu \sigma \lambda } 
\nabla_{\lambda }F_{\gamma \delta \nu \sigma } \nabla_{\omega 
}F_{\epsilon \varepsilon \kappa \tau } + \nn\\&& m_{686}^{} F_{\alpha 
\beta }{}^{\epsilon \varepsilon } F^{\alpha \beta \gamma 
\delta } F_{\gamma }{}^{\mu \nu \sigma } F^{\lambda \kappa 
\tau \omega } \nabla_{\lambda }F_{\delta \mu \nu \sigma } 
\nabla_{\omega }F_{\epsilon \varepsilon \kappa \tau } + 
\nn\\&& m_{687}^{} F_{\alpha \beta }{}^{\epsilon \varepsilon } F^{\alpha 
\beta \gamma \delta } F_{\gamma }{}^{\mu \nu \sigma } F_{\mu 
}{}^{\lambda \kappa \tau } \nabla_{\lambda }F_{\delta \nu 
\sigma }{}^{\omega } \nabla_{\omega }F_{\epsilon \varepsilon 
\kappa \tau } + \nn\\&& m_{684}^{} F_{\alpha \beta }{}^{\epsilon 
\varepsilon } F^{\alpha \beta \gamma \delta } F_{\gamma 
}{}^{\mu \nu \sigma } F^{\lambda \kappa \tau \omega } 
\nabla_{\sigma }F_{\delta \mu \nu \lambda } \nabla_{\omega 
}F_{\epsilon \varepsilon \kappa \tau } + \nn\\&& m_{683}^{} F_{\alpha 
\beta }{}^{\epsilon \varepsilon } F^{\alpha \beta \gamma 
\delta } F_{\mu }{}^{\kappa \tau \omega } F^{\mu \nu \sigma 
\lambda } \nabla_{\kappa }F_{\gamma \delta \nu \sigma } 
\nabla_{\omega }F_{\epsilon \varepsilon \lambda \tau } + 
\nn\\&& m_{682}^{} F_{\alpha \beta }{}^{\epsilon \varepsilon } F^{\alpha 
\beta \gamma \delta } F_{\mu }{}^{\kappa \tau \omega } F^{\mu 
\nu \sigma \lambda } \nabla_{\sigma }F_{\gamma \delta \nu 
\kappa } \nabla_{\omega }F_{\epsilon \varepsilon \lambda \tau 
} + \nn\\&& m_{677}^{} F_{\alpha \beta }{}^{\epsilon \varepsilon } 
F^{\alpha \beta \gamma \delta } F_{\gamma }{}^{\mu \nu \sigma 
} F^{\lambda \kappa \tau \omega } \nabla_{\mu }F_{\delta 
\lambda \kappa \tau } \nabla_{\omega }F_{\epsilon \varepsilon 
\nu \sigma } + \nn\\&& m_{678}^{} F_{\alpha \beta }{}^{\epsilon 
\varepsilon } F^{\alpha \beta \gamma \delta } F_{\gamma 
}{}^{\mu \nu \sigma } F^{\lambda \kappa \tau \omega } 
\nabla_{\tau }F_{\delta \mu \lambda \kappa } \nabla_{\omega 
}F_{\epsilon \varepsilon \nu \sigma } + \nn\\&& m_{680}^{} F_{\alpha 
\beta }{}^{\epsilon \varepsilon } F^{\alpha \beta \gamma 
\delta } F_{\gamma }{}^{\mu \nu \sigma } F^{\lambda \kappa 
\tau \omega } \nabla_{\kappa }F_{\delta \mu \nu \lambda } 
\nabla_{\omega }F_{\epsilon \varepsilon \sigma \tau } + 
\nn\\&& m_{681}^{} F_{\alpha \beta }{}^{\epsilon \varepsilon } F^{\alpha 
\beta \gamma \delta } F_{\gamma }{}^{\mu \nu \sigma } F_{\mu 
}{}^{\lambda \kappa \tau } \nabla_{\kappa }F_{\delta \nu 
\lambda }{}^{\omega } \nabla_{\omega }F_{\epsilon \varepsilon 
\sigma \tau } + \nn\\&& m_{679}^{} F_{\alpha \beta }{}^{\epsilon 
\varepsilon } F^{\alpha \beta \gamma \delta } F_{\gamma 
}{}^{\mu \nu \sigma } F^{\lambda \kappa \tau \omega } 
\nabla_{\nu }F_{\delta \mu \lambda \kappa } \nabla_{\omega 
}F_{\epsilon \varepsilon \sigma \tau } + \nn\\&& m_{691}^{} F_{\alpha 
\beta \gamma }{}^{\epsilon } F^{\alpha \beta \gamma \delta } 
F^{\varepsilon \mu \nu \sigma } F^{\lambda \kappa \tau \omega 
} \nabla_{\sigma }F_{\delta \varepsilon \mu \nu } 
\nabla_{\omega }F_{\epsilon \lambda \kappa \tau } + \nn\\&& m_{688}^{} 
F_{\alpha \beta \gamma }{}^{\epsilon } F^{\alpha \beta \gamma 
\delta } F^{\varepsilon \mu \nu \sigma } F^{\lambda \kappa 
\tau \omega } \nabla_{\kappa }F_{\delta \varepsilon \mu 
\lambda } \nabla_{\omega }F_{\epsilon \nu \sigma \tau } + 
\nn\\&& m_{690}^{} F_{\alpha \beta \gamma }{}^{\epsilon } F^{\alpha 
\beta \gamma \delta } F^{\varepsilon \mu \nu \sigma } 
F^{\lambda \kappa \tau \omega } \nabla_{\lambda }F_{\delta 
\varepsilon \mu \nu } \nabla_{\omega }F_{\epsilon \sigma 
\kappa \tau } + \nn\\&& m_{689}^{} F_{\alpha \beta \gamma }{}^{\epsilon 
} F^{\alpha \beta \gamma \delta } F^{\varepsilon \mu \nu 
\sigma } F^{\lambda \kappa \tau \omega } \nabla_{\nu 
}F_{\delta \varepsilon \mu \lambda } \nabla_{\omega 
}F_{\epsilon \sigma \kappa \tau } + \nn\\&& m_{711}^{} F_{\alpha \beta 
}{}^{\epsilon \varepsilon } F^{\alpha \beta \gamma \delta } F_{
\mu }{}^{\kappa \tau \omega } F^{\mu \nu \sigma \lambda } 
\nabla_{\epsilon }F_{\gamma \delta \nu \sigma } 
\nabla_{\omega }F_{\varepsilon \lambda \kappa \tau } + 
\nn\\&& m_{712}^{} F_{\alpha \beta }{}^{\epsilon \varepsilon } F^{\alpha 
\beta \gamma \delta } F_{\gamma }{}^{\mu \nu \sigma } 
F^{\lambda \kappa \tau \omega } \nabla_{\epsilon }F_{\delta 
\mu \nu \sigma } \nabla_{\omega }F_{\varepsilon \lambda 
\kappa \tau } + \nn\\&& m_{713}^{} F_{\alpha \beta }{}^{\epsilon 
\varepsilon } F^{\alpha \beta \gamma \delta } F_{\gamma 
}{}^{\mu \nu \sigma } F_{\mu }{}^{\lambda \kappa \tau } 
\nabla_{\epsilon }F_{\delta \nu \sigma }{}^{\omega } 
\nabla_{\omega }F_{\varepsilon \lambda \kappa \tau } + 
\nn\\&& m_{714}^{} F_{\alpha \beta }{}^{\epsilon \varepsilon } F^{\alpha 
\beta \gamma \delta } F_{\gamma }{}^{\mu \nu \sigma } F_{\mu 
\nu }{}^{\lambda \kappa } \nabla_{\epsilon }F_{\delta \sigma 
}{}^{\tau \omega } \nabla_{\omega }F_{\varepsilon \lambda 
\kappa \tau } + \nn\\&& m_{715}^{} F_{\alpha \beta }{}^{\epsilon 
\varepsilon } F^{\alpha \beta \gamma \delta } F_{\mu 
}{}^{\kappa \tau \omega } F^{\mu \nu \sigma \lambda } 
\nabla_{\sigma }F_{\gamma \delta \epsilon \nu } 
\nabla_{\omega }F_{\varepsilon \lambda \kappa \tau } + 
\nn\\&& m_{716}^{} F_{\alpha \beta }{}^{\epsilon \varepsilon } F^{\alpha 
\beta \gamma \delta } F_{\gamma }{}^{\mu \nu \sigma } 
F^{\lambda \kappa \tau \omega } \nabla_{\sigma }F_{\delta 
\epsilon \mu \nu } \nabla_{\omega }F_{\varepsilon \lambda 
\kappa \tau } + \nn\\&& m_{717}^{} F_{\alpha \beta }{}^{\epsilon 
\varepsilon } F^{\alpha \beta \gamma \delta } F_{\gamma 
}{}^{\mu \nu \sigma } F_{\mu }{}^{\lambda \kappa \tau } 
\nabla_{\sigma }F_{\delta \epsilon \nu }{}^{\omega } 
\nabla_{\omega }F_{\varepsilon \lambda \kappa \tau } + 
\nn\\&& m_{718}^{} F_{\alpha \beta }{}^{\epsilon \varepsilon } F^{\alpha 
\beta \gamma \delta } F_{\gamma }{}^{\mu \nu \sigma } F_{\mu 
\nu }{}^{\lambda \kappa } \nabla_{\sigma }F_{\delta \epsilon 
}{}^{\tau \omega } \nabla_{\omega }F_{\varepsilon \lambda 
\kappa \tau } + \nn\\&& m_{719}^{} F_{\alpha \beta }{}^{\epsilon 
\varepsilon } F^{\alpha \beta \gamma \delta } F_{\gamma 
\epsilon }{}^{\mu \nu } F^{\sigma \lambda \kappa \tau } 
\nabla_{\sigma }F_{\delta \mu \nu }{}^{\omega } 
\nabla_{\omega }F_{\varepsilon \lambda \kappa \tau } + 
\nn\\&& m_{698}^{} F_{\alpha }{}^{\epsilon \varepsilon \mu } F^{\alpha 
\beta \gamma \delta } F_{\beta }{}^{\nu \sigma \lambda } 
F_{\epsilon }{}^{\kappa \tau \omega } \nabla_{\lambda 
}F_{\gamma \delta \nu \sigma } \nabla_{\omega }F_{\varepsilon 
\mu \kappa \tau } + \nn\\&& m_{697}^{} F_{\alpha }{}^{\epsilon 
\varepsilon \mu } F^{\alpha \beta \gamma \delta } F_{\beta 
}{}^{\nu \sigma \lambda } F_{\epsilon }{}^{\kappa \tau \omega 
} \nabla_{\kappa }F_{\gamma \delta \nu \sigma } 
\nabla_{\omega }F_{\varepsilon \mu \lambda \tau } + \nn\\&& m_{696}^{} 
F_{\alpha }{}^{\epsilon \varepsilon \mu } F^{\alpha \beta 
\gamma \delta } F_{\beta }{}^{\nu \sigma \lambda } F_{\epsilon 
}{}^{\kappa \tau \omega } \nabla_{\sigma }F_{\gamma \delta 
\nu \kappa } \nabla_{\omega }F_{\varepsilon \mu \lambda \tau 
} + \nn\\&& m_{692}^{} F_{\alpha \beta }{}^{\epsilon \varepsilon } 
F^{\alpha \beta \gamma \delta } F_{\gamma }{}^{\mu \nu \sigma 
} F^{\lambda \kappa \tau \omega } \nabla_{\epsilon }F_{\delta 
\lambda \kappa \tau } \nabla_{\omega }F_{\varepsilon \mu \nu 
\sigma } + \nn\\&& m_{693}^{} F_{\alpha \beta }{}^{\epsilon \varepsilon 
} F^{\alpha \beta \gamma \delta } F_{\gamma }{}^{\mu \nu 
\sigma } F^{\lambda \kappa \tau \omega } \nabla_{\tau 
}F_{\delta \epsilon \lambda \kappa } \nabla_{\omega 
}F_{\varepsilon \mu \nu \sigma } + \nn\\&& m_{694}^{} F_{\alpha 
}{}^{\epsilon \varepsilon \mu } F^{\alpha \beta \gamma \delta 
} F_{\beta }{}^{\nu \sigma \lambda } F_{\epsilon }{}^{\kappa 
\tau \omega } \nabla_{\nu }F_{\gamma \delta \kappa \tau } 
\nabla_{\omega }F_{\varepsilon \mu \sigma \lambda } + 
\nn\\&& m_{695}^{} F_{\alpha }{}^{\epsilon \varepsilon \mu } F^{\alpha 
\beta \gamma \delta } F_{\beta }{}^{\nu \sigma \lambda } 
F_{\epsilon }{}^{\kappa \tau \omega } \nabla_{\tau }F_{\gamma 
\delta \nu \kappa } \nabla_{\omega }F_{\varepsilon \mu \sigma 
\lambda } + \nn\\&& m_{699}^{} F_{\alpha \beta }{}^{\epsilon \varepsilon 
} F^{\alpha \beta \gamma \delta } F_{\gamma }{}^{\mu \nu 
\sigma } F^{\lambda \kappa \tau \omega } \nabla_{\epsilon }F_{
\delta \mu \lambda \kappa } \nabla_{\omega }F_{\varepsilon 
\nu \sigma \tau } + \nn\\&& m_{702}^{} F_{\alpha \beta }{}^{\epsilon 
\varepsilon } F^{\alpha \beta \gamma \delta } F_{\gamma 
}{}^{\mu \nu \sigma } F_{\mu }{}^{\lambda \kappa \tau } 
\nabla_{\kappa }F_{\delta \epsilon \lambda }{}^{\omega } 
\nabla_{\omega }F_{\varepsilon \nu \sigma \tau } + \nn\\&& m_{701}^{} 
F_{\alpha \beta }{}^{\epsilon \varepsilon } F^{\alpha \beta 
\gamma \delta } F_{\gamma }{}^{\mu \nu \sigma } F^{\lambda 
\kappa \tau \omega } \nabla_{\kappa }F_{\delta \epsilon \mu 
\lambda } \nabla_{\omega }F_{\varepsilon \nu \sigma \tau } + 
\nn\\&& m_{700}^{} F_{\alpha \beta }{}^{\epsilon \varepsilon } F^{\alpha 
\beta \gamma \delta } F_{\gamma }{}^{\mu \nu \sigma } 
F^{\lambda \kappa \tau \omega } \nabla_{\mu }F_{\delta 
\epsilon \lambda \kappa } \nabla_{\omega }F_{\varepsilon \nu 
\sigma \tau } + \nn\\&& m_{706}^{} F_{\alpha \beta }{}^{\epsilon 
\varepsilon } F^{\alpha \beta \gamma \delta } F_{\gamma 
}{}^{\mu \nu \sigma } F^{\lambda \kappa \tau \omega } 
\nabla_{\epsilon }F_{\delta \mu \nu \lambda } \nabla_{\omega 
}F_{\varepsilon \sigma \kappa \tau } + \nn\\&& m_{709}^{} F_{\alpha 
\beta }{}^{\epsilon \varepsilon } F^{\alpha \beta \gamma 
\delta } F_{\gamma }{}^{\mu \nu \sigma } F^{\lambda \kappa 
\tau \omega } \nabla_{\lambda }F_{\delta \epsilon \mu \nu } 
\nabla_{\omega }F_{\varepsilon \sigma \kappa \tau } + 
\nn\\&& m_{710}^{} F_{\alpha \beta }{}^{\epsilon \varepsilon } F^{\alpha 
\beta \gamma \delta } F_{\gamma }{}^{\mu \nu \sigma } F_{\mu 
}{}^{\lambda \kappa \tau } \nabla_{\lambda }F_{\delta \epsilon 
\nu }{}^{\omega } \nabla_{\omega }F_{\varepsilon \sigma \kappa 
\tau } + \nn\\&& m_{708}^{} F_{\alpha \beta }{}^{\epsilon \varepsilon } 
F^{\alpha \beta \gamma \delta } F_{\gamma }{}^{\mu \nu \sigma 
} F_{\mu }{}^{\lambda \kappa \tau } \nabla_{\nu }F_{\delta 
\epsilon \lambda }{}^{\omega } \nabla_{\omega }F_{\varepsilon 
\sigma \kappa \tau } + \nn\\&& m_{707}^{} F_{\alpha \beta }{}^{\epsilon 
\varepsilon } F^{\alpha \beta \gamma \delta } F_{\gamma 
}{}^{\mu \nu \sigma } F^{\lambda \kappa \tau \omega } 
\nabla_{\nu }F_{\delta \epsilon \mu \lambda } \nabla_{\omega 
}F_{\varepsilon \sigma \kappa \tau } + \nn\\&& m_{703}^{} F_{\alpha 
\beta }{}^{\epsilon \varepsilon } F^{\alpha \beta \gamma 
\delta } F_{\mu }{}^{\kappa \tau \omega } F^{\mu \nu \sigma 
\lambda } \nabla_{\delta }F_{\gamma \epsilon \nu \kappa } 
\nabla_{\omega }F_{\varepsilon \sigma \lambda \tau } + 
\nn\\&& m_{704}^{} F_{\alpha \beta }{}^{\epsilon \varepsilon } F^{\alpha 
\beta \gamma \delta } F_{\mu }{}^{\kappa \tau \omega } F^{\mu 
\nu \sigma \lambda } \nabla_{\epsilon }F_{\gamma \delta \nu 
\kappa } \nabla_{\omega }F_{\varepsilon \sigma \lambda \tau } 
+ \nn\\&& m_{705}^{} F_{\alpha \beta }{}^{\epsilon \varepsilon } 
F^{\alpha \beta \gamma \delta } F_{\mu }{}^{\kappa \tau 
\omega } F^{\mu \nu \sigma \lambda } \nabla_{\kappa 
}F_{\gamma \delta \epsilon \nu } \nabla_{\omega 
}F_{\varepsilon \sigma \lambda \tau } + \nn\\&& m_{736}^{} F_{\alpha 
}{}^{\epsilon \varepsilon \mu } F^{\alpha \beta \gamma \delta 
} F_{\beta }{}^{\nu \sigma \lambda } F_{\epsilon }{}^{\kappa 
\tau \omega } \nabla_{\varepsilon }F_{\gamma \delta \nu 
\sigma } \nabla_{\omega }F_{\mu \lambda \kappa \tau } + 
\nn\\&& m_{737}^{} F_{\alpha }{}^{\epsilon \varepsilon \mu } F^{\alpha 
\beta \gamma \delta } F_{\beta }{}^{\nu \sigma \lambda } 
F_{\epsilon }{}^{\kappa \tau \omega } \nabla_{\sigma 
}F_{\gamma \delta \varepsilon \nu } \nabla_{\omega }F_{\mu 
\lambda \kappa \tau } + \nn\\&& m_{738}^{} F_{\alpha }{}^{\epsilon 
\varepsilon \mu } F^{\alpha \beta \gamma \delta } F_{\beta 
}{}^{\nu \sigma \lambda } F_{\gamma }{}^{\kappa \tau \omega } 
\nabla_{\sigma }F_{\delta \epsilon \varepsilon \nu } 
\nabla_{\omega }F_{\mu \lambda \kappa \tau } + \nn\\&& m_{726}^{} 
F_{\alpha \beta }{}^{\epsilon \varepsilon } F^{\alpha \beta 
\gamma \delta } F_{\gamma \epsilon }{}^{\mu \nu } F^{\sigma 
\lambda \kappa \tau } \nabla_{\lambda }F_{\delta \varepsilon 
\sigma }{}^{\omega } \nabla_{\omega }F_{\mu \nu \kappa \tau } 
+ \nn\\&& m_{727}^{} F_{\alpha \beta \gamma }{}^{\epsilon } F^{\alpha 
\beta \gamma \delta } F_{\delta }{}^{\varepsilon \mu \nu } F^{
\sigma \lambda \kappa \tau } \nabla_{\lambda }F_{\epsilon 
\varepsilon \sigma }{}^{\omega } \nabla_{\omega }F_{\mu \nu 
\kappa \tau } + \nn\\&& m_{720}^{} F_{\alpha }{}^{\epsilon \varepsilon 
\mu } F^{\alpha \beta \gamma \delta } F_{\beta }{}^{\nu 
\sigma \lambda } F_{\epsilon }{}^{\kappa \tau \omega } 
\nabla_{\varepsilon }F_{\gamma \delta \kappa \tau } 
\nabla_{\omega }F_{\mu \nu \sigma \lambda } + \nn\\&& m_{721}^{} 
F_{\alpha }{}^{\epsilon \varepsilon \mu } F^{\alpha \beta 
\gamma \delta } F_{\beta }{}^{\nu \sigma \lambda } F_{\epsilon 
}{}^{\kappa \tau \omega } \nabla_{\tau }F_{\gamma \delta 
\varepsilon \kappa } \nabla_{\omega }F_{\mu \nu \sigma 
\lambda } + \nn\\&& m_{722}^{} F_{\alpha \beta }{}^{\epsilon \varepsilon 
} F^{\alpha \beta \gamma \delta } F_{\gamma }{}^{\mu \nu 
\sigma } F^{\lambda \kappa \tau \omega } \nabla_{\varepsilon 
}F_{\delta \epsilon \lambda \kappa } \nabla_{\omega }F_{\mu 
\nu \sigma \tau } + \nn\\&& m_{723}^{} F_{\alpha }{}^{\epsilon 
\varepsilon \mu } F^{\alpha \beta \gamma \delta } F_{\beta 
\epsilon }{}^{\nu \sigma } F^{\lambda \kappa \tau \omega } 
\nabla_{\kappa }F_{\gamma \delta \varepsilon \lambda } 
\nabla_{\omega }F_{\mu \nu \sigma \tau } + \nn\\&& m_{724}^{} 
F_{\alpha \beta }{}^{\epsilon \varepsilon } F^{\alpha \beta 
\gamma \delta } F_{\gamma }{}^{\mu \nu \sigma } F^{\lambda 
\kappa \tau \omega } \nabla_{\kappa }F_{\delta \epsilon 
\varepsilon \lambda } \nabla_{\omega }F_{\mu \nu \sigma \tau 
} + \nn\\&& m_{725}^{} F_{\alpha }{}^{\epsilon \varepsilon \mu } 
F^{\alpha \beta \gamma \delta } F_{\beta \epsilon }{}^{\nu 
\sigma } F_{\gamma }{}^{\lambda \kappa \tau } \nabla_{\kappa 
}F_{\delta \varepsilon \lambda }{}^{\omega } \nabla_{\omega 
}F_{\mu \nu \sigma \tau } + \nn\\&& m_{734}^{} F_{\alpha }{}^{\epsilon 
\varepsilon \mu } F^{\alpha \beta \gamma \delta } F_{\beta 
\epsilon }{}^{\nu \sigma } F^{\lambda \kappa \tau \omega } 
\nabla_{\nu }F_{\gamma \delta \varepsilon \lambda } 
\nabla_{\omega }F_{\mu \sigma \kappa \tau } + \nn\\&& m_{735}^{} 
F_{\alpha }{}^{\epsilon \varepsilon \mu } F^{\alpha \beta 
\gamma \delta } F_{\beta \epsilon }{}^{\nu \sigma } F_{\gamma 
}{}^{\lambda \kappa \tau } \nabla_{\nu }F_{\delta \varepsilon 
\lambda }{}^{\omega } \nabla_{\omega }F_{\mu \sigma \kappa 
\tau } + \nn\\&& m_{728}^{} F_{\alpha }{}^{\epsilon \varepsilon \mu } 
F^{\alpha \beta \gamma \delta } F_{\beta }{}^{\nu \sigma 
\lambda } F_{\gamma }{}^{\kappa \tau \omega } \nabla_{\delta 
}F_{\epsilon \varepsilon \nu \kappa } \nabla_{\omega }F_{\mu 
\sigma \lambda \tau } + \nn\\&& m_{729}^{} F_{\alpha }{}^{\epsilon 
\varepsilon \mu } F^{\alpha \beta \gamma \delta } F_{\beta 
}{}^{\nu \sigma \lambda } F_{\epsilon }{}^{\kappa \tau \omega 
} \nabla_{\varepsilon }F_{\gamma \delta \nu \kappa } 
\nabla_{\omega }F_{\mu \sigma \lambda \tau } + \nn\\&& m_{730}^{} 
F_{\alpha }{}^{\epsilon \varepsilon \mu } F^{\alpha \beta 
\gamma \delta } F_{\beta }{}^{\nu \sigma \lambda } F_{\gamma 
}{}^{\kappa \tau \omega } \nabla_{\varepsilon }F_{\delta 
\epsilon \nu \kappa } \nabla_{\omega }F_{\mu \sigma \lambda 
\tau } + \nn\\&& m_{732}^{} F_{\alpha }{}^{\epsilon \varepsilon \mu } 
F^{\alpha \beta \gamma \delta } F_{\beta }{}^{\nu \sigma 
\lambda } F_{\epsilon }{}^{\kappa \tau \omega } \nabla_{\kappa 
}F_{\gamma \delta \varepsilon \nu } \nabla_{\omega }F_{\mu 
\sigma \lambda \tau } + \nn\\&& m_{733}^{} F_{\alpha }{}^{\epsilon 
\varepsilon \mu } F^{\alpha \beta \gamma \delta } F_{\beta 
}{}^{\nu \sigma \lambda } F_{\gamma }{}^{\kappa \tau \omega } 
\nabla_{\kappa }F_{\delta \epsilon \varepsilon \nu } 
\nabla_{\omega }F_{\mu \sigma \lambda \tau } + \nn\\&& m_{731}^{} 
F_{\alpha }{}^{\epsilon \varepsilon \mu } F^{\alpha \beta 
\gamma \delta } F_{\beta }{}^{\nu \sigma \lambda } F_{\epsilon 
}{}^{\kappa \tau \omega } \nabla_{\nu }F_{\gamma \delta 
\varepsilon \kappa } \nabla_{\omega }F_{\mu \sigma \lambda 
\tau } + \nn\\&& m_{749}^{} F_{\alpha \beta }{}^{\epsilon \varepsilon } 
F^{\alpha \beta \gamma \delta } F_{\gamma \epsilon }{}^{\mu 
\nu } F^{\sigma \lambda \kappa \tau } \nabla_{\mu }F_{\delta 
\varepsilon \sigma }{}^{\omega } \nabla_{\omega }F_{\nu 
\lambda \kappa \tau } + \nn\\&& m_{750}^{} F_{\alpha \beta 
}{}^{\epsilon \varepsilon } F^{\alpha \beta \gamma \delta } F_{
\gamma \delta }{}^{\mu \nu } F^{\sigma \lambda \kappa \tau } 
\nabla_{\mu }F_{\epsilon \varepsilon \sigma }{}^{\omega } 
\nabla_{\omega }F_{\nu \lambda \kappa \tau } + \nn\\&& m_{751}^{} 
F_{\alpha \beta \gamma }{}^{\epsilon } F^{\alpha \beta \gamma 
\delta } F_{\delta }{}^{\varepsilon \mu \nu } F^{\sigma 
\lambda \kappa \tau } \nabla_{\mu }F_{\epsilon \varepsilon 
\sigma }{}^{\omega } \nabla_{\omega }F_{\nu \lambda \kappa 
\tau } + \nn\\&& m_{752}^{} F_{\alpha \beta }{}^{\epsilon \varepsilon } 
F^{\alpha \beta \gamma \delta } F_{\gamma \epsilon }{}^{\mu 
\nu } F^{\sigma \lambda \kappa \tau } \nabla_{\sigma 
}F_{\delta \varepsilon \mu }{}^{\omega } \nabla_{\omega 
}F_{\nu \lambda \kappa \tau } + \nn\\&& m_{753}^{} F_{\alpha \beta 
}{}^{\epsilon \varepsilon } F^{\alpha \beta \gamma \delta } F_{
\gamma \epsilon }{}^{\mu \nu } F_{\mu }{}^{\sigma \lambda 
\kappa } \nabla_{\sigma }F_{\delta \varepsilon }{}^{\tau 
\omega } \nabla_{\omega }F_{\nu \lambda \kappa \tau } + 
\nn\\&& m_{754}^{} F_{\alpha \beta }{}^{\epsilon \varepsilon } F^{\alpha 
\beta \gamma \delta } F_{\gamma \delta }{}^{\mu \nu } 
F^{\sigma \lambda \kappa \tau } \nabla_{\sigma }F_{\epsilon 
\varepsilon \mu }{}^{\omega } \nabla_{\omega }F_{\nu \lambda 
\kappa \tau } + \nn\\&& m_{755}^{} F_{\alpha \beta \gamma }{}^{\epsilon 
} F^{\alpha \beta \gamma \delta } F_{\delta }{}^{\varepsilon 
\mu \nu } F^{\sigma \lambda \kappa \tau } \nabla_{\sigma }F_{
\epsilon \varepsilon \mu }{}^{\omega } \nabla_{\omega }F_{\nu 
\lambda \kappa \tau } + \nn\\&& m_{756}^{} F_{\alpha \beta \gamma }{}^{
\epsilon } F^{\alpha \beta \gamma \delta } F_{\delta 
}{}^{\varepsilon \mu \nu } F_{\varepsilon }{}^{\sigma \lambda 
\kappa } \nabla_{\sigma }F_{\epsilon \mu }{}^{\tau \omega } 
\nabla_{\omega }F_{\nu \lambda \kappa \tau } + \nn\\&& m_{757}^{} 
F_{\alpha \beta }{}^{\epsilon \varepsilon } F^{\alpha \beta 
\gamma \delta } F_{\gamma \epsilon }{}^{\mu \nu } F_{\delta 
}{}^{\sigma \lambda \kappa } \nabla_{\sigma }F_{\varepsilon 
\mu }{}^{\tau \omega } \nabla_{\omega }F_{\nu \lambda \kappa 
\tau } + \nn\\&& m_{758}^{} F_{\alpha \beta }{}^{\epsilon \varepsilon } 
F^{\alpha \beta \gamma \delta } F_{\gamma \delta }{}^{\mu \nu 
} F_{\epsilon }{}^{\sigma \lambda \kappa } \nabla_{\sigma 
}F_{\varepsilon \mu }{}^{\tau \omega } \nabla_{\omega }F_{\nu 
\lambda \kappa \tau } + \nn\\&& m_{739}^{} F_{\alpha \beta \gamma }{}^{
\epsilon } F^{\alpha \beta \gamma \delta } F^{\varepsilon \mu 
\nu \sigma } F^{\lambda \kappa \tau \omega } \nabla_{\epsilon 
}F_{\delta \varepsilon \mu \lambda } \nabla_{\omega }F_{\nu 
\sigma \kappa \tau } + \nn\\&& m_{741}^{} F_{\alpha \beta }{}^{\epsilon 
\varepsilon } F^{\alpha \beta \gamma \delta } F_{\gamma 
}{}^{\mu \nu \sigma } F_{\mu }{}^{\lambda \kappa \tau } 
\nabla_{\varepsilon }F_{\delta \epsilon \lambda }{}^{\omega } 
\nabla_{\omega }F_{\nu \sigma \kappa \tau } + \nn\\&& m_{740}^{} 
F_{\alpha \beta }{}^{\epsilon \varepsilon } F^{\alpha \beta 
\gamma \delta } F_{\gamma }{}^{\mu \nu \sigma } F^{\lambda 
\kappa \tau \omega } \nabla_{\varepsilon }F_{\delta \epsilon 
\mu \lambda } \nabla_{\omega }F_{\nu \sigma \kappa \tau } + 
\nn\\&& m_{744}^{} F_{\alpha \beta }{}^{\epsilon \varepsilon } F^{\alpha 
\beta \gamma \delta } F_{\gamma }{}^{\mu \nu \sigma } 
F^{\lambda \kappa \tau \omega } \nabla_{\lambda }F_{\delta 
\epsilon \varepsilon \mu } \nabla_{\omega }F_{\nu \sigma 
\kappa \tau } + \nn\\&& m_{745}^{} F_{\alpha \beta }{}^{\epsilon 
\varepsilon } F^{\alpha \beta \gamma \delta } F_{\gamma 
}{}^{\mu \nu \sigma } F_{\mu }{}^{\lambda \kappa \tau } 
\nabla_{\lambda }F_{\delta \epsilon \varepsilon }{}^{\omega } 
\nabla_{\omega }F_{\nu \sigma \kappa \tau } + \nn\\&& m_{746}^{} 
F_{\alpha }{}^{\epsilon \varepsilon \mu } F^{\alpha \beta 
\gamma \delta } F_{\beta \epsilon }{}^{\nu \sigma } F_{\gamma 
}{}^{\lambda \kappa \tau } \nabla_{\lambda }F_{\delta 
\varepsilon \mu }{}^{\omega } \nabla_{\omega }F_{\nu \sigma 
\kappa \tau } + \nn\\&& m_{747}^{} F_{\alpha \beta }{}^{\epsilon 
\varepsilon } F^{\alpha \beta \gamma \delta } F_{\gamma 
}{}^{\mu \nu \sigma } F_{\epsilon }{}^{\lambda \kappa \tau } 
\nabla_{\lambda }F_{\delta \varepsilon \mu }{}^{\omega } 
\nabla_{\omega }F_{\nu \sigma \kappa \tau } + \nn\\&& m_{748}^{} 
F_{\alpha \beta }{}^{\epsilon \varepsilon } F^{\alpha \beta 
\gamma \delta } F_{\gamma }{}^{\mu \nu \sigma } F_{\delta 
}{}^{\lambda \kappa \tau } \nabla_{\lambda }F_{\epsilon 
\varepsilon \mu }{}^{\omega } \nabla_{\omega }F_{\nu \sigma 
\kappa \tau } + \nn\\&& m_{742}^{} F_{\alpha }{}^{\epsilon \varepsilon 
\mu } F^{\alpha \beta \gamma \delta } F_{\beta \epsilon 
}{}^{\nu \sigma } F^{\lambda \kappa \tau \omega } \nabla_{\mu 
}F_{\gamma \delta \varepsilon \lambda } \nabla_{\omega }F_{\nu 
\sigma \kappa \tau } + \nn\\&& m_{743}^{} F_{\alpha \beta }{}^{\epsilon 
\varepsilon } F^{\alpha \beta \gamma \delta } F_{\gamma 
}{}^{\mu \nu \sigma } F^{\lambda \kappa \tau \omega } 
\nabla_{\mu }F_{\delta \epsilon \varepsilon \lambda } \nabla_{
\omega }F_{\nu \sigma \kappa \tau } + \nn\\&& m_{759}^{} F_{\alpha 
\beta \gamma }{}^{\epsilon } F^{\alpha \beta \gamma \delta } 
F^{\varepsilon \mu \nu \sigma } F^{\lambda \kappa \tau \omega 
} \nabla_{\epsilon }F_{\delta \varepsilon \mu \nu } 
\nabla_{\omega }F_{\sigma \lambda \kappa \tau } + \nn\\&& m_{760}^{} 
F_{\alpha \beta \gamma }{}^{\epsilon } F^{\alpha \beta \gamma 
\delta } F_{\varepsilon }{}^{\lambda \kappa \tau } 
F^{\varepsilon \mu \nu \sigma } \nabla_{\epsilon }F_{\delta 
\mu \nu }{}^{\omega } \nabla_{\omega }F_{\sigma \lambda 
\kappa \tau } + \nn\\&& m_{761}^{} F_{\alpha \beta \gamma }{}^{\epsilon 
} F^{\alpha \beta \gamma \delta } F_{\varepsilon \mu 
}{}^{\lambda \kappa } F^{\varepsilon \mu \nu \sigma } \nabla_{
\epsilon }F_{\delta \nu }{}^{\tau \omega } \nabla_{\omega }F_{
\sigma \lambda \kappa \tau } + \nn\\&& m_{762}^{} F_{\alpha \beta }{}^{
\epsilon \varepsilon } F^{\alpha \beta \gamma \delta } 
F_{\gamma }{}^{\mu \nu \sigma } F^{\lambda \kappa \tau \omega 
} \nabla_{\varepsilon }F_{\delta \epsilon \mu \nu } 
\nabla_{\omega }F_{\sigma \lambda \kappa \tau } + \nn\\&& m_{763}^{} 
F_{\alpha \beta }{}^{\epsilon \varepsilon } F^{\alpha \beta 
\gamma \delta } F_{\gamma }{}^{\mu \nu \sigma } F_{\mu 
}{}^{\lambda \kappa \tau } \nabla_{\varepsilon }F_{\delta 
\epsilon \nu }{}^{\omega } \nabla_{\omega }F_{\sigma \lambda 
\kappa \tau } + \nn\\&& m_{764}^{} F_{\alpha \beta }{}^{\epsilon 
\varepsilon } F^{\alpha \beta \gamma \delta } F_{\gamma 
}{}^{\mu \nu \sigma } F_{\epsilon }{}^{\lambda \kappa \tau } 
\nabla_{\varepsilon }F_{\delta \mu \nu }{}^{\omega } 
\nabla_{\omega }F_{\sigma \lambda \kappa \tau } + \nn\\&& m_{765}^{} 
F_{\alpha \beta }{}^{\epsilon \varepsilon } F^{\alpha \beta 
\gamma \delta } F_{\gamma }{}^{\mu \nu \sigma } F_{\epsilon 
\mu }{}^{\lambda \kappa } \nabla_{\varepsilon }F_{\delta \nu 
}{}^{\tau \omega } \nabla_{\omega }F_{\sigma \lambda \kappa 
\tau } + \nn\\&& m_{766}^{} F_{\alpha }{}^{\epsilon \varepsilon \mu } 
F^{\alpha \beta \gamma \delta } F_{\beta }{}^{\nu \sigma 
\lambda } F_{\epsilon }{}^{\kappa \tau \omega } \nabla_{\mu 
}F_{\gamma \delta \varepsilon \nu } \nabla_{\omega }F_{\sigma 
\lambda \kappa \tau } + \nn\\&& m_{767}^{} F_{\alpha }{}^{\epsilon 
\varepsilon \mu } F^{\alpha \beta \gamma \delta } F_{\beta 
\epsilon }{}^{\nu \sigma } F^{\lambda \kappa \tau \omega } 
\nabla_{\mu }F_{\gamma \delta \varepsilon \nu } 
\nabla_{\omega }F_{\sigma \lambda \kappa \tau } + \nn\\&& m_{768}^{} 
F_{\alpha }{}^{\epsilon \varepsilon \mu } F^{\alpha \beta 
\gamma \delta } F_{\beta }{}^{\nu \sigma \lambda } F_{\gamma 
}{}^{\kappa \tau \omega } \nabla_{\mu }F_{\delta \epsilon 
\varepsilon \nu } \nabla_{\omega }F_{\sigma \lambda \kappa 
\tau } + \nn\\&& m_{769}^{} F_{\alpha }{}^{\epsilon \varepsilon \mu } 
F^{\alpha \beta \gamma \delta } F_{\beta \epsilon }{}^{\nu 
\sigma } F_{\gamma }{}^{\lambda \kappa \tau } \nabla_{\mu }F_{
\delta \varepsilon \nu }{}^{\omega } \nabla_{\omega }F_{\sigma 
\lambda \kappa \tau } + \nn\\&& m_{770}^{} F_{\alpha }{}^{\epsilon 
\varepsilon \mu } F^{\alpha \beta \gamma \delta } F_{\beta 
\epsilon }{}^{\nu \sigma } F_{\gamma \varepsilon }{}^{\lambda 
\kappa } \nabla_{\mu }F_{\delta \nu }{}^{\tau \omega } 
\nabla_{\omega }F_{\sigma \lambda \kappa \tau } + \nn\\&& m_{771}^{} 
F_{\alpha \beta }{}^{\epsilon \varepsilon } F^{\alpha \beta 
\gamma \delta } F_{\gamma \delta \epsilon }{}^{\mu } F^{\nu 
\sigma \lambda \kappa } \nabla_{\mu }F_{\varepsilon \nu 
}{}^{\tau \omega } \nabla_{\omega }F_{\sigma \lambda \kappa 
\tau } + \nn\\&& m_{772}^{} F_{\alpha \beta }{}^{\epsilon \varepsilon } 
F^{\alpha \beta \gamma \delta } F_{\mu }{}^{\kappa \tau 
\omega } F^{\mu \nu \sigma \lambda } \nabla_{\nu }F_{\gamma 
\delta \epsilon \varepsilon } \nabla_{\omega }F_{\sigma 
\lambda \kappa \tau } + \nn\\&& m_{773}^{} F_{\alpha }{}^{\epsilon 
\varepsilon \mu } F^{\alpha \beta \gamma \delta } F_{\beta 
}{}^{\nu \sigma \lambda } F_{\epsilon }{}^{\kappa \tau \omega 
} \nabla_{\nu }F_{\gamma \delta \varepsilon \mu } 
\nabla_{\omega }F_{\sigma \lambda \kappa \tau } + \nn\\&& m_{774}^{} 
F_{\alpha }{}^{\epsilon \varepsilon \mu } F^{\alpha \beta 
\gamma \delta } F_{\beta }{}^{\nu \sigma \lambda } F_{\gamma 
}{}^{\kappa \tau \omega } \nabla_{\nu }F_{\delta \epsilon 
\varepsilon \mu } \nabla_{\omega }F_{\sigma \lambda \kappa 
\tau } + \nn\\&& m_{775}^{} F_{\alpha \beta }{}^{\epsilon \varepsilon } 
F^{\alpha \beta \gamma \delta } F_{\gamma }{}^{\mu \nu \sigma 
} F^{\lambda \kappa \tau \omega } \nabla_{\nu }F_{\delta 
\epsilon \varepsilon \mu } \nabla_{\omega }F_{\sigma \lambda 
\kappa \tau } + \nn\\&& m_{776}^{} F_{\alpha \beta }{}^{\epsilon 
\varepsilon } F^{\alpha \beta \gamma \delta } F_{\gamma 
}{}^{\mu \nu \sigma } F_{\mu }{}^{\lambda \kappa \tau } 
\nabla_{\nu }F_{\delta \epsilon \varepsilon }{}^{\omega } 
\nabla_{\omega }F_{\sigma \lambda \kappa \tau } + \nn\\&& m_{777}^{} 
F_{\alpha }{}^{\epsilon \varepsilon \mu } F^{\alpha \beta 
\gamma \delta } F_{\beta \epsilon }{}^{\nu \sigma } F_{\gamma 
}{}^{\lambda \kappa \tau } \nabla_{\nu }F_{\delta \varepsilon 
\mu }{}^{\omega } \nabla_{\omega }F_{\sigma \lambda \kappa 
\tau } + \nn\\&& m_{778}^{} F_{\alpha \beta }{}^{\epsilon \varepsilon } 
F^{\alpha \beta \gamma \delta } F_{\gamma }{}^{\mu \nu \sigma 
} F_{\epsilon }{}^{\lambda \kappa \tau } \nabla_{\nu 
}F_{\delta \varepsilon \mu }{}^{\omega } \nabla_{\omega 
}F_{\sigma \lambda \kappa \tau } + \nn\\&& m_{779}^{} F_{\alpha \beta 
}{}^{\epsilon \varepsilon } F^{\alpha \beta \gamma \delta } F_{
\gamma \epsilon }{}^{\mu \nu } F^{\sigma \lambda \kappa \tau 
} \nabla_{\nu }F_{\delta \varepsilon \mu }{}^{\omega } 
\nabla_{\omega }F_{\sigma \lambda \kappa \tau } + \nn\\&& m_{780}^{} 
F_{\alpha \beta }{}^{\epsilon \varepsilon } F^{\alpha \beta 
\gamma \delta } F_{\gamma }{}^{\mu \nu \sigma } F_{\epsilon 
\mu }{}^{\lambda \kappa } \nabla_{\nu }F_{\delta \varepsilon 
}{}^{\tau \omega } \nabla_{\omega }F_{\sigma \lambda \kappa 
\tau } + \nn\\&& m_{781}^{} F_{\alpha \beta }{}^{\epsilon \varepsilon } 
F^{\alpha \beta \gamma \delta } F_{\gamma \epsilon }{}^{\mu 
\nu } F_{\mu }{}^{\sigma \lambda \kappa } \nabla_{\nu 
}F_{\delta \varepsilon }{}^{\tau \omega } \nabla_{\omega 
}F_{\sigma \lambda \kappa \tau } + \nn\\&& m_{782}^{} F_{\alpha \beta 
}{}^{\epsilon \varepsilon } F^{\alpha \beta \gamma \delta } F_{
\gamma }{}^{\mu \nu \sigma } F_{\delta }{}^{\lambda \kappa 
\tau } \nabla_{\nu }F_{\epsilon \varepsilon \mu }{}^{\omega } 
\nabla_{\omega }F_{\sigma \lambda \kappa \tau } + \nn\\&& m_{783}^{} 
F_{\alpha \beta \gamma }{}^{\epsilon } F^{\alpha \beta \gamma 
\delta } F_{\delta }{}^{\varepsilon \mu \nu } F^{\sigma 
\lambda \kappa \tau } \nabla_{\nu }F_{\epsilon \varepsilon 
\mu }{}^{\omega } \nabla_{\omega }F_{\sigma \lambda \kappa 
\tau } + \nn\\&& m_{784}^{} F_{\alpha \beta }{}^{\epsilon \varepsilon } 
F^{\alpha \beta \gamma \delta } F_{\gamma }{}^{\mu \nu \sigma 
} F_{\delta \mu }{}^{\lambda \kappa } \nabla_{\nu }F_{\epsilon 
\varepsilon }{}^{\tau \omega } \nabla_{\omega }F_{\sigma 
\lambda \kappa \tau } + \nn\\&& m_{786}^{} F_{\alpha \beta \gamma }{}^{
\epsilon } F^{\alpha \beta \gamma \delta } F_{\delta 
}{}^{\varepsilon \mu \nu } F_{\varepsilon \mu }{}^{\sigma 
\lambda } \nabla_{\nu }F_{\epsilon }{}^{\kappa \tau \omega } 
\nabla_{\omega }F_{\sigma \lambda \kappa \tau } + \nn\\&& m_{785}^{} 
F_{\alpha \beta \gamma }{}^{\epsilon } F^{\alpha \beta \gamma 
\delta } F_{\delta }{}^{\varepsilon \mu \nu } F_{\varepsilon 
}{}^{\sigma \lambda \kappa } \nabla_{\nu }F_{\epsilon \mu 
}{}^{\tau \omega } \nabla_{\omega }F_{\sigma \lambda \kappa 
\tau } + \nn\\&& m_{789}^{} F_{\alpha \beta }{}^{\epsilon \varepsilon } 
F^{\alpha \beta \gamma \delta } F_{\gamma \epsilon }{}^{\mu 
\nu } F_{\delta \mu }{}^{\sigma \lambda } \nabla_{\nu 
}F_{\varepsilon }{}^{\kappa \tau \omega } \nabla_{\omega 
}F_{\sigma \lambda \kappa \tau } + \nn\\&& m_{787}^{} F_{\alpha \beta 
}{}^{\epsilon \varepsilon } F^{\alpha \beta \gamma \delta } F_{
\gamma \epsilon }{}^{\mu \nu } F_{\delta }{}^{\sigma \lambda 
\kappa } \nabla_{\nu }F_{\varepsilon \mu }{}^{\tau \omega } 
\nabla_{\omega }F_{\sigma \lambda \kappa \tau } + \nn\\&& m_{788}^{} 
F_{\alpha \beta }{}^{\epsilon \varepsilon } F^{\alpha \beta 
\gamma \delta } F_{\gamma \delta }{}^{\mu \nu } F_{\epsilon 
}{}^{\sigma \lambda \kappa } \nabla_{\nu }F_{\varepsilon \mu 
}{}^{\tau \omega } \nabla_{\omega }F_{\sigma \lambda \kappa 
\tau } + \nn\\&& m_{790}^{} F_{\alpha \beta }{}^{\epsilon \varepsilon } 
F^{\alpha \beta \gamma \delta } F_{\mu \nu }{}^{\kappa \tau } 
F^{\mu \nu \sigma \lambda } \nabla_{\omega }F_{\sigma \lambda 
\kappa \tau } \nabla^{\omega }F_{\gamma \delta \epsilon 
\varepsilon } + \nn\\&& m_{791}^{} F_{\alpha \beta }{}^{\epsilon 
\varepsilon } F^{\alpha \beta \gamma \delta } F_{\mu \nu }{}^{
\kappa \tau } F^{\mu \nu \sigma \lambda } \nabla_{\lambda 
}F_{\varepsilon \kappa \tau \omega } \nabla^{\omega }F_{\gamma 
\delta \epsilon \sigma } + \nn\\&& m_{792}^{} F_{\alpha \beta 
}{}^{\epsilon \varepsilon } F^{\alpha \beta \gamma \delta } F_{
\mu \nu }{}^{\kappa \tau } F^{\mu \nu \sigma \lambda } 
\nabla_{\tau }F_{\varepsilon \lambda \kappa \omega } 
\nabla^{\omega }F_{\gamma \delta \epsilon \sigma } + \nn\\&& m_{793}^{} 
F_{\alpha \beta }{}^{\epsilon \varepsilon } F^{\alpha \beta 
\gamma \delta } F_{\mu \nu }{}^{\kappa \tau } F^{\mu \nu 
\sigma \lambda } \nabla_{\omega }F_{\varepsilon \lambda \kappa 
\tau } \nabla^{\omega }F_{\gamma \delta \epsilon \sigma } + 
\nn\\&& m_{797}^{} F_{\alpha }{}^{\epsilon \varepsilon \mu } F^{\alpha 
\beta \gamma \delta } F_{\beta }{}^{\nu \sigma \lambda } 
F_{\epsilon \nu }{}^{\kappa \tau } \nabla_{\lambda }F_{\mu 
\sigma \tau \omega } \nabla^{\omega }F_{\gamma \delta 
\varepsilon \kappa } + \nn\\&& m_{798}^{} F_{\alpha }{}^{\epsilon 
\varepsilon \mu } F^{\alpha \beta \gamma \delta } F_{\beta 
}{}^{\nu \sigma \lambda } F_{\epsilon \nu }{}^{\kappa \tau } 
\nabla_{\tau }F_{\mu \sigma \lambda \omega } \nabla^{\omega 
}F_{\gamma \delta \varepsilon \kappa } + \nn\\&& m_{799}^{} F_{\alpha 
}{}^{\epsilon \varepsilon \mu } F^{\alpha \beta \gamma \delta 
} F_{\beta }{}^{\nu \sigma \lambda } F_{\epsilon \nu 
}{}^{\kappa \tau } \nabla_{\omega }F_{\mu \sigma \lambda \tau 
} \nabla^{\omega }F_{\gamma \delta \varepsilon \kappa } + 
\nn\\&& m_{794}^{} F_{\alpha }{}^{\epsilon \varepsilon \mu } F^{\alpha 
\beta \gamma \delta } F_{\beta }{}^{\nu \sigma \lambda } 
F_{\epsilon \nu }{}^{\kappa \tau } \nabla_{\omega }F_{\sigma 
\lambda \kappa \tau } \nabla^{\omega }F_{\gamma \delta 
\varepsilon \mu } + \nn\\&& m_{795}^{} F_{\alpha }{}^{\epsilon 
\varepsilon \mu } F^{\alpha \beta \gamma \delta } F_{\beta 
}{}^{\nu \sigma \lambda } F_{\epsilon \nu }{}^{\kappa \tau } 
\nabla_{\tau }F_{\mu \lambda \kappa \omega } \nabla^{\omega 
}F_{\gamma \delta \varepsilon \sigma } + \nn\\&& m_{796}^{} F_{\alpha 
}{}^{\epsilon \varepsilon \mu } F^{\alpha \beta \gamma \delta 
} F_{\beta }{}^{\nu \sigma \lambda } F_{\epsilon \nu 
}{}^{\kappa \tau } \nabla_{\omega }F_{\mu \lambda \kappa \tau 
} \nabla^{\omega }F_{\gamma \delta \varepsilon \sigma } + 
\nn\\&& m_{803}^{} F_{\alpha }{}^{\epsilon \varepsilon \mu } F^{\alpha 
\beta \gamma \delta } F_{\beta }{}^{\nu \sigma \lambda } 
F_{\epsilon \nu }{}^{\kappa \tau } \nabla_{\omega 
}F_{\varepsilon \mu \sigma \lambda } \nabla^{\omega }F_{\gamma 
\delta \kappa \tau } + \nn\\&& m_{801}^{} F_{\alpha \beta }{}^{\epsilon 
\varepsilon } F^{\alpha \beta \gamma \delta } F_{\mu \nu }{}^{
\kappa \tau } F^{\mu \nu \sigma \lambda } \nabla_{\tau 
}F_{\epsilon \varepsilon \lambda \omega } \nabla^{\omega 
}F_{\gamma \delta \sigma \kappa } + \nn\\&& m_{802}^{} F_{\alpha \beta 
}{}^{\epsilon \varepsilon } F^{\alpha \beta \gamma \delta } F_{
\mu \nu }{}^{\kappa \tau } F^{\mu \nu \sigma \lambda } 
\nabla_{\omega }F_{\epsilon \varepsilon \lambda \tau } 
\nabla^{\omega }F_{\gamma \delta \sigma \kappa } + \nn\\&& m_{800}^{} 
F_{\alpha \beta }{}^{\epsilon \varepsilon } F^{\alpha \beta 
\gamma \delta } F_{\mu \nu }{}^{\kappa \tau } F^{\mu \nu 
\sigma \lambda } \nabla_{\omega }F_{\epsilon \varepsilon 
\kappa \tau } \nabla^{\omega }F_{\gamma \delta \sigma \lambda 
} + \nn\\&& m_{804}^{} F_{\alpha \beta }{}^{\epsilon \varepsilon } 
F^{\alpha \beta \gamma \delta } F_{\mu \nu }{}^{\kappa \tau } 
F^{\mu \nu \sigma \lambda } \nabla_{\omega }F_{\delta 
\varepsilon \lambda \tau } \nabla^{\omega }F_{\gamma \epsilon 
\sigma \kappa } + \nn\\&& m_{805}^{} F_{\alpha }{}^{\epsilon \varepsilon 
\mu } F^{\alpha \beta \gamma \delta } F_{\beta }{}^{\nu 
\sigma \lambda } F_{\epsilon \nu }{}^{\kappa \tau } 
\nabla_{\omega }F_{\delta \mu \lambda \tau } \nabla^{\omega 
}F_{\gamma \varepsilon \sigma \kappa } + \nn\\&& m_{808}^{} F_{\alpha 
\beta }{}^{\epsilon \varepsilon } F^{\alpha \beta \gamma 
\delta } F_{\gamma }{}^{\mu \nu \sigma } F_{\mu }{}^{\lambda 
\kappa \tau } \nabla_{\tau }F_{\nu \sigma \kappa \omega } 
\nabla^{\omega }F_{\delta \epsilon \varepsilon \lambda } + 
\nn\\&& m_{809}^{} F_{\alpha \beta }{}^{\epsilon \varepsilon } F^{\alpha 
\beta \gamma \delta } F_{\gamma }{}^{\mu \nu \sigma } F_{\mu 
}{}^{\lambda \kappa \tau } \nabla_{\omega }F_{\nu \sigma 
\kappa \tau } \nabla^{\omega }F_{\delta \epsilon \varepsilon 
\lambda } + \nn\\&& m_{806}^{} F_{\alpha \beta }{}^{\epsilon \varepsilon 
} F^{\alpha \beta \gamma \delta } F_{\gamma }{}^{\mu \nu 
\sigma } F_{\mu }{}^{\lambda \kappa \tau } \nabla_{\tau 
}F_{\sigma \lambda \kappa \omega } \nabla^{\omega }F_{\delta 
\epsilon \varepsilon \nu } + \nn\\&& m_{807}^{} F_{\alpha \beta 
}{}^{\epsilon \varepsilon } F^{\alpha \beta \gamma \delta } F_{
\gamma }{}^{\mu \nu \sigma } F_{\mu }{}^{\lambda \kappa \tau 
} \nabla_{\omega }F_{\sigma \lambda \kappa \tau } 
\nabla^{\omega }F_{\delta \epsilon \varepsilon \nu } + 
\nn\\&& m_{810}^{} F_{\alpha \beta }{}^{\epsilon \varepsilon } F^{\alpha 
\beta \gamma \delta } F_{\gamma }{}^{\mu \nu \sigma } F_{\mu 
\nu }{}^{\lambda \kappa } \nabla_{\kappa }F_{\sigma \lambda 
\tau \omega } \nabla^{\omega }F_{\delta \epsilon \varepsilon 
}{}^{\tau } + \nn\\&& m_{811}^{} F_{\alpha \beta }{}^{\epsilon 
\varepsilon } F^{\alpha \beta \gamma \delta } F_{\gamma 
}{}^{\mu \nu \sigma } F_{\mu \nu }{}^{\lambda \kappa } 
\nabla_{\tau }F_{\sigma \lambda \kappa \omega } 
\nabla^{\omega }F_{\delta \epsilon \varepsilon }{}^{\tau } + 
\nn\\&& m_{812}^{} F_{\alpha \beta }{}^{\epsilon \varepsilon } F^{\alpha 
\beta \gamma \delta } F_{\gamma }{}^{\mu \nu \sigma } F_{\mu 
\nu }{}^{\lambda \kappa } \nabla_{\omega }F_{\sigma \lambda 
\kappa \tau } \nabla^{\omega }F_{\delta \epsilon \varepsilon 
}{}^{\tau } + \nn\\&& m_{821}^{} F_{\alpha \beta }{}^{\epsilon 
\varepsilon } F^{\alpha \beta \gamma \delta } F_{\gamma 
}{}^{\mu \nu \sigma } F_{\mu }{}^{\lambda \kappa \tau } 
\nabla_{\sigma }F_{\varepsilon \nu \tau \omega } 
\nabla^{\omega }F_{\delta \epsilon \lambda \kappa } + 
\nn\\&& m_{822}^{} F_{\alpha \beta }{}^{\epsilon \varepsilon } F^{\alpha 
\beta \gamma \delta } F_{\gamma }{}^{\mu \nu \sigma } F_{\mu 
}{}^{\lambda \kappa \tau } \nabla_{\tau }F_{\varepsilon \nu 
\sigma \omega } \nabla^{\omega }F_{\delta \epsilon \lambda 
\kappa } + \nn\\&& m_{823}^{} F_{\alpha \beta }{}^{\epsilon \varepsilon 
} F^{\alpha \beta \gamma \delta } F_{\gamma }{}^{\mu \nu 
\sigma } F_{\mu }{}^{\lambda \kappa \tau } \nabla_{\omega }F_{
\varepsilon \nu \sigma \tau } \nabla^{\omega }F_{\delta 
\epsilon \lambda \kappa } + \nn\\&& m_{825}^{} F_{\alpha \beta 
}{}^{\epsilon \varepsilon } F^{\alpha \beta \gamma \delta } F_{
\gamma }{}^{\mu \nu \sigma } F_{\mu \nu }{}^{\lambda \kappa } 
\nabla_{\kappa }F_{\varepsilon \sigma \tau \omega } 
\nabla^{\omega }F_{\delta \epsilon \lambda }{}^{\tau } + 
\nn\\&& m_{824}^{} F_{\alpha \beta }{}^{\epsilon \varepsilon } F^{\alpha 
\beta \gamma \delta } F_{\gamma }{}^{\mu \nu \sigma } F_{\mu 
\nu }{}^{\lambda \kappa } \nabla_{\sigma }F_{\varepsilon 
\kappa \tau \omega } \nabla^{\omega }F_{\delta \epsilon 
\lambda }{}^{\tau } + \nn\\&& m_{826}^{} F_{\alpha \beta }{}^{\epsilon 
\varepsilon } F^{\alpha \beta \gamma \delta } F_{\gamma 
}{}^{\mu \nu \sigma } F_{\mu \nu }{}^{\lambda \kappa } 
\nabla_{\tau }F_{\varepsilon \sigma \kappa \omega } 
\nabla^{\omega }F_{\delta \epsilon \lambda }{}^{\tau } + 
\nn\\&& m_{827}^{} F_{\alpha \beta }{}^{\epsilon \varepsilon } F^{\alpha 
\beta \gamma \delta } F_{\gamma }{}^{\mu \nu \sigma } F_{\mu 
\nu }{}^{\lambda \kappa } \nabla_{\omega }F_{\varepsilon 
\sigma \kappa \tau } \nabla^{\omega }F_{\delta \epsilon 
\lambda }{}^{\tau } + \nn\\&& m_{815}^{} F_{\alpha \beta }{}^{\epsilon 
\varepsilon } F^{\alpha \beta \gamma \delta } F_{\gamma 
}{}^{\mu \nu \sigma } F_{\mu }{}^{\lambda \kappa \tau } 
\nabla_{\sigma }F_{\varepsilon \kappa \tau \omega } 
\nabla^{\omega }F_{\delta \epsilon \nu \lambda } + \nn\\&& m_{816}^{} 
F_{\alpha \beta }{}^{\epsilon \varepsilon } F^{\alpha \beta 
\gamma \delta } F_{\gamma }{}^{\mu \nu \sigma } F_{\mu 
}{}^{\lambda \kappa \tau } \nabla_{\tau }F_{\varepsilon \sigma 
\kappa \omega } \nabla^{\omega }F_{\delta \epsilon \nu 
\lambda } + \nn\\&& m_{817}^{} F_{\alpha \beta }{}^{\epsilon \varepsilon 
} F^{\alpha \beta \gamma \delta } F_{\gamma }{}^{\mu \nu 
\sigma } F_{\mu }{}^{\lambda \kappa \tau } \nabla_{\omega }F_{
\varepsilon \sigma \kappa \tau } \nabla^{\omega }F_{\delta 
\epsilon \nu \lambda } + \nn\\&& m_{813}^{} F_{\alpha \beta 
}{}^{\epsilon \varepsilon } F^{\alpha \beta \gamma \delta } F_{
\gamma }{}^{\mu \nu \sigma } F_{\mu }{}^{\lambda \kappa \tau 
} \nabla_{\tau }F_{\varepsilon \lambda \kappa \omega } 
\nabla^{\omega }F_{\delta \epsilon \nu \sigma } + \nn\\&& m_{814}^{} 
F_{\alpha \beta }{}^{\epsilon \varepsilon } F^{\alpha \beta 
\gamma \delta } F_{\gamma }{}^{\mu \nu \sigma } F_{\mu 
}{}^{\lambda \kappa \tau } \nabla_{\omega }F_{\varepsilon 
\lambda \kappa \tau } \nabla^{\omega }F_{\delta \epsilon \nu 
\sigma } + \nn\\&& m_{818}^{} F_{\alpha \beta }{}^{\epsilon \varepsilon 
} F^{\alpha \beta \gamma \delta } F_{\gamma }{}^{\mu \nu 
\sigma } F_{\mu \nu }{}^{\lambda \kappa } \nabla_{\kappa 
}F_{\varepsilon \lambda \tau \omega } \nabla^{\omega 
}F_{\delta \epsilon \sigma }{}^{\tau } + \nn\\&& m_{819}^{} F_{\alpha 
\beta }{}^{\epsilon \varepsilon } F^{\alpha \beta \gamma 
\delta } F_{\gamma }{}^{\mu \nu \sigma } F_{\mu \nu 
}{}^{\lambda \kappa } \nabla_{\tau }F_{\varepsilon \lambda 
\kappa \omega } \nabla^{\omega }F_{\delta \epsilon \sigma 
}{}^{\tau } + \nn\\&& m_{820}^{} F_{\alpha \beta }{}^{\epsilon 
\varepsilon } F^{\alpha \beta \gamma \delta } F_{\gamma 
}{}^{\mu \nu \sigma } F_{\mu \nu }{}^{\lambda \kappa } 
\nabla_{\omega }F_{\varepsilon \lambda \kappa \tau } 
\nabla^{\omega }F_{\delta \epsilon \sigma }{}^{\tau } + 
\nn\\&& m_{852}^{} F_{\alpha \beta }{}^{\epsilon \varepsilon } F^{\alpha 
\beta \gamma \delta } F_{\gamma }{}^{\mu \nu \sigma } 
F_{\epsilon \mu \nu }{}^{\lambda } \nabla_{\tau }F_{\sigma 
\lambda \kappa \omega } \nabla^{\omega }F_{\delta \varepsilon 
}{}^{\kappa \tau } + \nn\\&& m_{853}^{} F_{\alpha \beta }{}^{\epsilon 
\varepsilon } F^{\alpha \beta \gamma \delta } F_{\gamma 
\epsilon }{}^{\mu \nu } F_{\mu \nu }{}^{\sigma \lambda } 
\nabla_{\tau }F_{\sigma \lambda \kappa \omega } 
\nabla^{\omega }F_{\delta \varepsilon }{}^{\kappa \tau } + 
\nn\\&& m_{854}^{} F_{\alpha \beta }{}^{\epsilon \varepsilon } F^{\alpha 
\beta \gamma \delta } F_{\gamma }{}^{\mu \nu \sigma } 
F_{\epsilon \mu \nu }{}^{\lambda } \nabla_{\omega }F_{\sigma 
\lambda \kappa \tau } \nabla^{\omega }F_{\delta \varepsilon 
}{}^{\kappa \tau } + \nn\\&& m_{855}^{} F_{\alpha \beta }{}^{\epsilon 
\varepsilon } F^{\alpha \beta \gamma \delta } F_{\gamma 
\epsilon }{}^{\mu \nu } F_{\mu \nu }{}^{\sigma \lambda } 
\nabla_{\omega }F_{\sigma \lambda \kappa \tau } 
\nabla^{\omega }F_{\delta \varepsilon }{}^{\kappa \tau } + 
\nn\\&& m_{849}^{} F_{\alpha }{}^{\epsilon \varepsilon \mu } F^{\alpha 
\beta \gamma \delta } F_{\beta \epsilon }{}^{\nu \sigma } 
F_{\gamma }{}^{\lambda \kappa \tau } \nabla_{\sigma }F_{\mu 
\nu \tau \omega } \nabla^{\omega }F_{\delta \varepsilon 
\lambda \kappa } + \nn\\&& m_{850}^{} F_{\alpha }{}^{\epsilon 
\varepsilon \mu } F^{\alpha \beta \gamma \delta } F_{\beta 
\epsilon }{}^{\nu \sigma } F_{\gamma }{}^{\lambda \kappa \tau 
} \nabla_{\tau }F_{\mu \nu \sigma \omega } \nabla^{\omega 
}F_{\delta \varepsilon \lambda \kappa } + \nn\\&& m_{851}^{} F_{\alpha 
}{}^{\epsilon \varepsilon \mu } F^{\alpha \beta \gamma \delta 
} F_{\beta \epsilon }{}^{\nu \sigma } F_{\gamma }{}^{\lambda 
\kappa \tau } \nabla_{\omega }F_{\mu \nu \sigma \tau } 
\nabla^{\omega }F_{\delta \varepsilon \lambda \kappa } + 
\nn\\&& m_{835}^{} F_{\alpha }{}^{\epsilon \varepsilon \mu } F^{\alpha 
\beta \gamma \delta } F_{\beta \epsilon }{}^{\nu \sigma } 
F_{\gamma }{}^{\lambda \kappa \tau } \nabla_{\tau }F_{\nu 
\sigma \kappa \omega } \nabla^{\omega }F_{\delta \varepsilon 
\mu \lambda } + \nn\\&& m_{836}^{} F_{\alpha }{}^{\epsilon \varepsilon 
\mu } F^{\alpha \beta \gamma \delta } F_{\beta \epsilon 
}{}^{\nu \sigma } F_{\gamma }{}^{\lambda \kappa \tau } 
\nabla_{\omega }F_{\nu \sigma \kappa \tau } \nabla^{\omega 
}F_{\delta \varepsilon \mu \lambda } + \nn\\&& m_{837}^{} F_{\alpha 
\beta }{}^{\epsilon \varepsilon } F^{\alpha \beta \gamma 
\delta } F_{\gamma }{}^{\mu \nu \sigma } F_{\epsilon 
}{}^{\lambda \kappa \tau } \nabla_{\omega }F_{\nu \sigma 
\kappa \tau } \nabla^{\omega }F_{\delta \varepsilon \mu 
\lambda } + \nn\\&& m_{828}^{} F_{\alpha }{}^{\epsilon \varepsilon \mu } 
F^{\alpha \beta \gamma \delta } F_{\beta \epsilon }{}^{\nu 
\sigma } F_{\gamma }{}^{\lambda \kappa \tau } \nabla_{\tau 
}F_{\sigma \lambda \kappa \omega } \nabla^{\omega }F_{\delta 
\varepsilon \mu \nu } + \nn\\&& m_{829}^{} F_{\alpha \beta 
}{}^{\epsilon \varepsilon } F^{\alpha \beta \gamma \delta } F_{
\gamma }{}^{\mu \nu \sigma } F_{\epsilon }{}^{\lambda \kappa 
\tau } \nabla_{\tau }F_{\sigma \lambda \kappa \omega } 
\nabla^{\omega }F_{\delta \varepsilon \mu \nu } + \nn\\&& m_{830}^{} 
F_{\alpha }{}^{\epsilon \varepsilon \mu } F^{\alpha \beta 
\gamma \delta } F_{\beta \epsilon }{}^{\nu \sigma } F_{\gamma 
}{}^{\lambda \kappa \tau } \nabla_{\omega }F_{\sigma \lambda 
\kappa \tau } \nabla^{\omega }F_{\delta \varepsilon \mu \nu } 
+ \nn\\&& m_{831}^{} F_{\alpha \beta }{}^{\epsilon \varepsilon } 
F^{\alpha \beta \gamma \delta } F_{\gamma }{}^{\mu \nu \sigma 
} F_{\epsilon }{}^{\lambda \kappa \tau } \nabla_{\omega 
}F_{\sigma \lambda \kappa \tau } \nabla^{\omega }F_{\delta 
\varepsilon \mu \nu } + \nn\\&& m_{832}^{} F_{\alpha \beta 
}{}^{\epsilon \varepsilon } F^{\alpha \beta \gamma \delta } F_{
\gamma \epsilon }{}^{\mu \nu } F^{\sigma \lambda \kappa \tau 
} \nabla_{\omega }F_{\sigma \lambda \kappa \tau } 
\nabla^{\omega }F_{\delta \varepsilon \mu \nu } + \nn\\&& m_{833}^{} 
F_{\alpha \beta }{}^{\epsilon \varepsilon } F^{\alpha \beta 
\gamma \delta } F_{\gamma \epsilon }{}^{\mu \nu } F^{\sigma 
\lambda \kappa \tau } \nabla_{\tau }F_{\nu \lambda \kappa 
\omega } \nabla^{\omega }F_{\delta \varepsilon \mu \sigma } + 
\nn\\&& m_{834}^{} F_{\alpha \beta }{}^{\epsilon \varepsilon } F^{\alpha 
\beta \gamma \delta } F_{\gamma \epsilon }{}^{\mu \nu } 
F^{\sigma \lambda \kappa \tau } \nabla_{\omega }F_{\nu 
\lambda \kappa \tau } \nabla^{\omega }F_{\delta \varepsilon 
\mu \sigma } + \nn\\&& m_{838}^{} F_{\alpha }{}^{\epsilon \varepsilon 
\mu } F^{\alpha \beta \gamma \delta } F_{\beta \epsilon 
}{}^{\nu \sigma } F_{\gamma }{}^{\lambda \kappa \tau } 
\nabla_{\sigma }F_{\mu \kappa \tau \omega } \nabla^{\omega 
}F_{\delta \varepsilon \nu \lambda } + \nn\\&& m_{839}^{} F_{\alpha 
\beta }{}^{\epsilon \varepsilon } F^{\alpha \beta \gamma 
\delta } F_{\gamma }{}^{\mu \nu \sigma } F_{\epsilon \mu }{}^{
\lambda \kappa } \nabla_{\kappa }F_{\sigma \lambda \tau 
\omega } \nabla^{\omega }F_{\delta \varepsilon \nu }{}^{\tau } 
+ \nn\\&& m_{840}^{} F_{\alpha \beta }{}^{\epsilon \varepsilon } 
F^{\alpha \beta \gamma \delta } F_{\gamma }{}^{\mu \nu \sigma 
} F_{\epsilon \mu }{}^{\lambda \kappa } \nabla_{\tau 
}F_{\sigma \lambda \kappa \omega } \nabla^{\omega }F_{\delta 
\varepsilon \nu }{}^{\tau } + \nn\\&& m_{841}^{} F_{\alpha \beta 
}{}^{\epsilon \varepsilon } F^{\alpha \beta \gamma \delta } F_{
\gamma \epsilon }{}^{\mu \nu } F_{\mu }{}^{\sigma \lambda 
\kappa } \nabla_{\tau }F_{\sigma \lambda \kappa \omega } 
\nabla^{\omega }F_{\delta \varepsilon \nu }{}^{\tau } + 
\nn\\&& m_{842}^{} F_{\alpha \beta }{}^{\epsilon \varepsilon } F^{\alpha 
\beta \gamma \delta } F_{\gamma }{}^{\mu \nu \sigma } 
F_{\epsilon \mu }{}^{\lambda \kappa } \nabla_{\omega 
}F_{\sigma \lambda \kappa \tau } \nabla^{\omega }F_{\delta 
\varepsilon \nu }{}^{\tau } + \nn\\&& m_{843}^{} F_{\alpha \beta 
}{}^{\epsilon \varepsilon } F^{\alpha \beta \gamma \delta } F_{
\gamma \epsilon }{}^{\mu \nu } F_{\mu }{}^{\sigma \lambda 
\kappa } \nabla_{\omega }F_{\sigma \lambda \kappa \tau } 
\nabla^{\omega }F_{\delta \varepsilon \nu }{}^{\tau } + 
\nn\\&& m_{844}^{} F_{\alpha \beta }{}^{\epsilon \varepsilon } F^{\alpha 
\beta \gamma \delta } F_{\gamma \epsilon }{}^{\mu \nu } 
F^{\sigma \lambda \kappa \tau } \nabla_{\tau }F_{\mu \nu 
\kappa \omega } \nabla^{\omega }F_{\delta \varepsilon \sigma 
\lambda } + \nn\\&& m_{845}^{} F_{\alpha \beta }{}^{\epsilon \varepsilon 
} F^{\alpha \beta \gamma \delta } F_{\gamma \epsilon }{}^{\mu 
\nu } F^{\sigma \lambda \kappa \tau } \nabla_{\omega }F_{\mu 
\nu \kappa \tau } \nabla^{\omega }F_{\delta \varepsilon 
\sigma \lambda } + \nn\\&& m_{846}^{} F_{\alpha \beta }{}^{\epsilon 
\varepsilon } F^{\alpha \beta \gamma \delta } F_{\gamma 
\epsilon }{}^{\mu \nu } F_{\mu }{}^{\sigma \lambda \kappa } 
\nabla_{\kappa }F_{\nu \lambda \tau \omega } \nabla^{\omega 
}F_{\delta \varepsilon \sigma }{}^{\tau } + \nn\\&& m_{847}^{} F_{\alpha 
\beta }{}^{\epsilon \varepsilon } F^{\alpha \beta \gamma 
\delta } F_{\gamma \epsilon }{}^{\mu \nu } F_{\mu }{}^{\sigma 
\lambda \kappa } \nabla_{\tau }F_{\nu \lambda \kappa \omega } 
\nabla^{\omega }F_{\delta \varepsilon \sigma }{}^{\tau } + 
\nn\\&& m_{848}^{} F_{\alpha \beta }{}^{\epsilon \varepsilon } F^{\alpha 
\beta \gamma \delta } F_{\gamma \epsilon }{}^{\mu \nu } 
F_{\mu }{}^{\sigma \lambda \kappa } \nabla_{\omega }F_{\nu 
\lambda \kappa \tau } \nabla^{\omega }F_{\delta \varepsilon 
\sigma }{}^{\tau } + \nn\\&& m_{911}^{} F_{\alpha \beta }{}^{\epsilon 
\varepsilon } F^{\alpha \beta \gamma \delta } F_{\gamma 
}{}^{\mu \nu \sigma } F_{\mu }{}^{\lambda \kappa \tau } 
\nabla_{\sigma }F_{\epsilon \varepsilon \nu \omega } 
\nabla^{\omega }F_{\delta \lambda \kappa \tau } + \nn\\&& m_{912}^{} 
F_{\alpha }{}^{\epsilon \varepsilon \mu } F^{\alpha \beta 
\gamma \delta } F_{\beta \epsilon }{}^{\nu \sigma } F_{\gamma 
}{}^{\lambda \kappa \tau } \nabla_{\sigma }F_{\varepsilon \mu 
\nu \omega } \nabla^{\omega }F_{\delta \lambda \kappa \tau } 
+ \nn\\&& m_{913}^{} F_{\alpha \beta }{}^{\epsilon \varepsilon } 
F^{\alpha \beta \gamma \delta } F_{\gamma }{}^{\mu \nu \sigma 
} F_{\epsilon }{}^{\lambda \kappa \tau } \nabla_{\sigma 
}F_{\varepsilon \mu \nu \omega } \nabla^{\omega }F_{\delta 
\lambda \kappa \tau } + \nn\\&& m_{914}^{} F_{\alpha \beta 
}{}^{\epsilon \varepsilon } F^{\alpha \beta \gamma \delta } F_{
\gamma }{}^{\mu \nu \sigma } F_{\mu }{}^{\lambda \kappa \tau 
} \nabla_{\omega }F_{\epsilon \varepsilon \nu \sigma } 
\nabla^{\omega }F_{\delta \lambda \kappa \tau } + \nn\\&& m_{915}^{} 
F_{\alpha \beta }{}^{\epsilon \varepsilon } F^{\alpha \beta 
\gamma \delta } F_{\gamma }{}^{\mu \nu \sigma } F_{\epsilon 
}{}^{\lambda \kappa \tau } \nabla_{\omega }F_{\varepsilon \mu 
\nu \sigma } \nabla^{\omega }F_{\delta \lambda \kappa \tau } 
+ \nn\\&& m_{916}^{} F_{\alpha \beta }{}^{\epsilon \varepsilon } 
F^{\alpha \beta \gamma \delta } F_{\gamma }{}^{\mu \nu \sigma 
} F_{\mu \nu }{}^{\lambda \kappa } \nabla_{\sigma }F_{\epsilon 
\varepsilon \tau \omega } \nabla^{\omega }F_{\delta \lambda 
\kappa }{}^{\tau } + \nn\\&& m_{917}^{} F_{\alpha \beta }{}^{\epsilon 
\varepsilon } F^{\alpha \beta \gamma \delta } F_{\gamma 
}{}^{\mu \nu \sigma } F_{\epsilon \mu }{}^{\lambda \kappa } 
\nabla_{\sigma }F_{\varepsilon \nu \tau \omega } 
\nabla^{\omega }F_{\delta \lambda \kappa }{}^{\tau } + 
\nn\\&& m_{918}^{} F_{\alpha \beta }{}^{\epsilon \varepsilon } F^{\alpha 
\beta \gamma \delta } F_{\gamma }{}^{\mu \nu \sigma } F_{\mu 
\nu }{}^{\lambda \kappa } \nabla_{\tau }F_{\epsilon 
\varepsilon \sigma \omega } \nabla^{\omega }F_{\delta \lambda 
\kappa }{}^{\tau } + \nn\\&& m_{919}^{} F_{\alpha \beta }{}^{\epsilon 
\varepsilon } F^{\alpha \beta \gamma \delta } F_{\gamma 
}{}^{\mu \nu \sigma } F_{\epsilon \mu }{}^{\lambda \kappa } 
\nabla_{\tau }F_{\varepsilon \nu \sigma \omega } 
\nabla^{\omega }F_{\delta \lambda \kappa }{}^{\tau } + 
\nn\\&& m_{920}^{} F_{\alpha \beta }{}^{\epsilon \varepsilon } F^{\alpha 
\beta \gamma \delta } F_{\gamma }{}^{\mu \nu \sigma } F_{\mu 
\nu }{}^{\lambda \kappa } \nabla_{\omega }F_{\epsilon 
\varepsilon \sigma \tau } \nabla^{\omega }F_{\delta \lambda 
\kappa }{}^{\tau } + \nn\\&& m_{921}^{} F_{\alpha \beta }{}^{\epsilon 
\varepsilon } F^{\alpha \beta \gamma \delta } F_{\gamma 
}{}^{\mu \nu \sigma } F_{\epsilon \mu }{}^{\lambda \kappa } 
\nabla_{\omega }F_{\varepsilon \nu \sigma \tau } 
\nabla^{\omega }F_{\delta \lambda \kappa }{}^{\tau } + 
\nn\\&& m_{922}^{} F_{\alpha \beta }{}^{\epsilon \varepsilon } F^{\alpha 
\beta \gamma \delta } F_{\gamma }{}^{\mu \nu \sigma } 
F_{\epsilon \mu \nu }{}^{\lambda } \nabla_{\tau 
}F_{\varepsilon \sigma \kappa \omega } \nabla^{\omega 
}F_{\delta \lambda }{}^{\kappa \tau } + \nn\\&& m_{923}^{} F_{\alpha 
\beta }{}^{\epsilon \varepsilon } F^{\alpha \beta \gamma 
\delta } F_{\gamma }{}^{\mu \nu \sigma } F_{\epsilon \mu \nu 
}{}^{\lambda } \nabla_{\omega }F_{\varepsilon \sigma \kappa 
\tau } \nabla^{\omega }F_{\delta \lambda }{}^{\kappa \tau } + 
\nn\\&& m_{871}^{} F_{\alpha \beta }{}^{\epsilon \varepsilon } F^{\alpha 
\beta \gamma \delta } F_{\gamma }{}^{\mu \nu \sigma } 
F_{\epsilon }{}^{\lambda \kappa \tau } \nabla_{\sigma 
}F_{\varepsilon \nu \tau \omega } \nabla^{\omega }F_{\delta 
\mu \lambda \kappa } + \nn\\&& m_{872}^{} F_{\alpha \beta }{}^{\epsilon 
\varepsilon } F^{\alpha \beta \gamma \delta } F_{\gamma 
}{}^{\mu \nu \sigma } F_{\epsilon }{}^{\lambda \kappa \tau } 
\nabla_{\tau }F_{\varepsilon \nu \sigma \omega } 
\nabla^{\omega }F_{\delta \mu \lambda \kappa } + \nn\\&& m_{873}^{} F_{
\alpha \beta }{}^{\epsilon \varepsilon } F^{\alpha \beta 
\gamma \delta } F_{\gamma }{}^{\mu \nu \sigma } F_{\epsilon 
}{}^{\lambda \kappa \tau } \nabla_{\omega }F_{\varepsilon \nu 
\sigma \tau } \nabla^{\omega }F_{\delta \mu \lambda \kappa } 
+ \nn\\&& m_{862}^{} F_{\alpha \beta \gamma }{}^{\epsilon } F^{\alpha 
\beta \gamma \delta } F_{\varepsilon }{}^{\lambda \kappa \tau 
} F^{\varepsilon \mu \nu \sigma } \nabla_{\sigma }F_{\epsilon 
\kappa \tau \omega } \nabla^{\omega }F_{\delta \mu \nu 
\lambda } + \nn\\&& m_{863}^{} F_{\alpha \beta }{}^{\epsilon \varepsilon 
} F^{\alpha \beta \gamma \delta } F_{\gamma }{}^{\mu \nu 
\sigma } F_{\epsilon }{}^{\lambda \kappa \tau } \nabla_{\sigma 
}F_{\varepsilon \kappa \tau \omega } \nabla^{\omega }F_{\delta 
\mu \nu \lambda } + \nn\\&& m_{864}^{} F_{\alpha \beta \gamma 
}{}^{\epsilon } F^{\alpha \beta \gamma \delta } F_{\varepsilon 
}{}^{\lambda \kappa \tau } F^{\varepsilon \mu \nu \sigma } 
\nabla_{\tau }F_{\epsilon \sigma \kappa \omega } 
\nabla^{\omega }F_{\delta \mu \nu \lambda } + \nn\\&& m_{865}^{} 
F_{\alpha \beta }{}^{\epsilon \varepsilon } F^{\alpha \beta 
\gamma \delta } F_{\gamma }{}^{\mu \nu \sigma } F_{\epsilon 
}{}^{\lambda \kappa \tau } \nabla_{\tau }F_{\varepsilon \sigma 
\kappa \omega } \nabla^{\omega }F_{\delta \mu \nu \lambda } + 
\nn\\&& m_{866}^{} F_{\alpha \beta \gamma }{}^{\epsilon } F^{\alpha 
\beta \gamma \delta } F_{\varepsilon }{}^{\lambda \kappa \tau 
} F^{\varepsilon \mu \nu \sigma } \nabla_{\omega }F_{\epsilon 
\sigma \kappa \tau } \nabla^{\omega }F_{\delta \mu \nu 
\lambda } + \nn\\&& m_{867}^{} F_{\alpha \beta }{}^{\epsilon \varepsilon 
} F^{\alpha \beta \gamma \delta } F_{\gamma }{}^{\mu \nu 
\sigma } F_{\epsilon }{}^{\lambda \kappa \tau } \nabla_{\omega 
}F_{\varepsilon \sigma \kappa \tau } \nabla^{\omega }F_{\delta 
\mu \nu \lambda } + \nn\\&& m_{856}^{} F_{\alpha \beta \gamma 
}{}^{\epsilon } F^{\alpha \beta \gamma \delta } F_{\varepsilon 
}{}^{\lambda \kappa \tau } F^{\varepsilon \mu \nu \sigma } 
\nabla_{\tau }F_{\epsilon \lambda \kappa \omega } 
\nabla^{\omega }F_{\delta \mu \nu \sigma } + \nn\\&& m_{857}^{} 
F_{\alpha \beta }{}^{\epsilon \varepsilon } F^{\alpha \beta 
\gamma \delta } F_{\gamma }{}^{\mu \nu \sigma } F_{\epsilon 
}{}^{\lambda \kappa \tau } \nabla_{\tau }F_{\varepsilon 
\lambda \kappa \omega } \nabla^{\omega }F_{\delta \mu \nu 
\sigma } + \nn\\&& m_{858}^{} F_{\alpha \beta }{}^{\epsilon \varepsilon 
} F^{\alpha \beta \gamma \delta } F_{\gamma \epsilon }{}^{\mu 
\nu } F^{\sigma \lambda \kappa \tau } \nabla_{\tau 
}F_{\varepsilon \lambda \kappa \omega } \nabla^{\omega 
}F_{\delta \mu \nu \sigma } + \nn\\&& m_{859}^{} F_{\alpha \beta 
\gamma }{}^{\epsilon } F^{\alpha \beta \gamma \delta } 
F_{\varepsilon }{}^{\lambda \kappa \tau } F^{\varepsilon \mu 
\nu \sigma } \nabla_{\omega }F_{\epsilon \lambda \kappa \tau 
} \nabla^{\omega }F_{\delta \mu \nu \sigma } + \nn\\&& m_{860}^{} 
F_{\alpha \beta }{}^{\epsilon \varepsilon } F^{\alpha \beta 
\gamma \delta } F_{\gamma }{}^{\mu \nu \sigma } F_{\epsilon 
}{}^{\lambda \kappa \tau } \nabla_{\omega }F_{\varepsilon 
\lambda \kappa \tau } \nabla^{\omega }F_{\delta \mu \nu 
\sigma } + \nn\\&& m_{861}^{} F_{\alpha \beta }{}^{\epsilon \varepsilon 
} F^{\alpha \beta \gamma \delta } F_{\gamma \epsilon }{}^{\mu 
\nu } F^{\sigma \lambda \kappa \tau } \nabla_{\omega 
}F_{\varepsilon \lambda \kappa \tau } \nabla^{\omega 
}F_{\delta \mu \nu \sigma } + \nn\\&& m_{868}^{} F_{\alpha \beta 
}{}^{\epsilon \varepsilon } F^{\alpha \beta \gamma \delta } F_{
\gamma \epsilon }{}^{\mu \nu } F^{\sigma \lambda \kappa \tau 
} \nabla_{\nu }F_{\varepsilon \kappa \tau \omega } 
\nabla^{\omega }F_{\delta \mu \sigma \lambda } + \nn\\&& m_{869}^{} F_{
\alpha \beta }{}^{\epsilon \varepsilon } F^{\alpha \beta 
\gamma \delta } F_{\gamma \epsilon }{}^{\mu \nu } F^{\sigma 
\lambda \kappa \tau } \nabla_{\tau }F_{\varepsilon \nu \kappa 
\omega } \nabla^{\omega }F_{\delta \mu \sigma \lambda } + 
\nn\\&& m_{870}^{} F_{\alpha \beta }{}^{\epsilon \varepsilon } F^{\alpha 
\beta \gamma \delta } F_{\gamma \epsilon }{}^{\mu \nu } 
F^{\sigma \lambda \kappa \tau } \nabla_{\omega }F_{\varepsilon 
\nu \kappa \tau } \nabla^{\omega }F_{\delta \mu \sigma 
\lambda } + \nn\\&& m_{888}^{} F_{\alpha \beta }{}^{\epsilon \varepsilon 
} F^{\alpha \beta \gamma \delta } F_{\gamma }{}^{\mu \nu 
\sigma } F_{\mu }{}^{\lambda \kappa \tau } \nabla_{\sigma }F_{
\epsilon \varepsilon \tau \omega } \nabla^{\omega }F_{\delta 
\nu \lambda \kappa } + \nn\\&& m_{889}^{} F_{\alpha \beta }{}^{\epsilon 
\varepsilon } F^{\alpha \beta \gamma \delta } F_{\gamma 
}{}^{\mu \nu \sigma } F_{\mu }{}^{\lambda \kappa \tau } 
\nabla_{\tau }F_{\epsilon \varepsilon \sigma \omega } \nabla^{
\omega }F_{\delta \nu \lambda \kappa } + \nn\\&& m_{890}^{} F_{\alpha 
\beta }{}^{\epsilon \varepsilon } F^{\alpha \beta \gamma 
\delta } F_{\gamma }{}^{\mu \nu \sigma } F_{\mu }{}^{\lambda 
\kappa \tau } \nabla_{\omega }F_{\epsilon \varepsilon \sigma 
\tau } \nabla^{\omega }F_{\delta \nu \lambda \kappa } + 
\nn\\&& m_{892}^{} F_{\alpha \beta \gamma }{}^{\epsilon } F^{\alpha 
\beta \gamma \delta } F_{\varepsilon \mu }{}^{\lambda \kappa } 
F^{\varepsilon \mu \nu \sigma } \nabla_{\kappa }F_{\epsilon 
\sigma \tau \omega } \nabla^{\omega }F_{\delta \nu \lambda 
}{}^{\tau } + \nn\\&& m_{893}^{} F_{\alpha \beta }{}^{\epsilon 
\varepsilon } F^{\alpha \beta \gamma \delta } F_{\gamma 
}{}^{\mu \nu \sigma } F_{\epsilon \mu }{}^{\lambda \kappa } 
\nabla_{\kappa }F_{\varepsilon \sigma \tau \omega } 
\nabla^{\omega }F_{\delta \nu \lambda }{}^{\tau } + \nn\\&& m_{894}^{} 
F_{\alpha }{}^{\epsilon \varepsilon \mu } F^{\alpha \beta 
\gamma \delta } F_{\beta \epsilon }{}^{\nu \sigma } F_{\gamma 
\varepsilon }{}^{\lambda \kappa } \nabla_{\kappa }F_{\mu 
\sigma \tau \omega } \nabla^{\omega }F_{\delta \nu \lambda 
}{}^{\tau } + \nn\\&& m_{891}^{} F_{\alpha \beta }{}^{\epsilon 
\varepsilon } F^{\alpha \beta \gamma \delta } F_{\gamma 
}{}^{\mu \nu \sigma } F_{\epsilon \mu }{}^{\lambda \kappa } 
\nabla_{\sigma }F_{\varepsilon \kappa \tau \omega } 
\nabla^{\omega }F_{\delta \nu \lambda }{}^{\tau } + \nn\\&& m_{895}^{} 
F_{\alpha \beta \gamma }{}^{\epsilon } F^{\alpha \beta \gamma 
\delta } F_{\varepsilon \mu }{}^{\lambda \kappa } 
F^{\varepsilon \mu \nu \sigma } \nabla_{\tau }F_{\epsilon 
\sigma \kappa \omega } \nabla^{\omega }F_{\delta \nu \lambda 
}{}^{\tau } + \nn\\&& m_{896}^{} F_{\alpha \beta }{}^{\epsilon 
\varepsilon } F^{\alpha \beta \gamma \delta } F_{\gamma 
}{}^{\mu \nu \sigma } F_{\epsilon \mu }{}^{\lambda \kappa } 
\nabla_{\tau }F_{\varepsilon \sigma \kappa \omega } 
\nabla^{\omega }F_{\delta \nu \lambda }{}^{\tau } + \nn\\&& m_{897}^{} 
F_{\alpha }{}^{\epsilon \varepsilon \mu } F^{\alpha \beta 
\gamma \delta } F_{\beta \epsilon }{}^{\nu \sigma } F_{\gamma 
\varepsilon }{}^{\lambda \kappa } \nabla_{\tau }F_{\mu \sigma 
\kappa \omega } \nabla^{\omega }F_{\delta \nu \lambda 
}{}^{\tau } + \nn\\&& m_{898}^{} F_{\alpha \beta \gamma }{}^{\epsilon } 
F^{\alpha \beta \gamma \delta } F_{\varepsilon \mu 
}{}^{\lambda \kappa } F^{\varepsilon \mu \nu \sigma } \nabla_{
\omega }F_{\epsilon \sigma \kappa \tau } \nabla^{\omega 
}F_{\delta \nu \lambda }{}^{\tau } + \nn\\&& m_{899}^{} F_{\alpha \beta 
}{}^{\epsilon \varepsilon } F^{\alpha \beta \gamma \delta } F_{
\gamma }{}^{\mu \nu \sigma } F_{\epsilon \mu }{}^{\lambda 
\kappa } \nabla_{\omega }F_{\varepsilon \sigma \kappa \tau } 
\nabla^{\omega }F_{\delta \nu \lambda }{}^{\tau } + \nn\\&& m_{900}^{} 
F_{\alpha }{}^{\epsilon \varepsilon \mu } F^{\alpha \beta 
\gamma \delta } F_{\beta \epsilon }{}^{\nu \sigma } F_{\gamma 
\varepsilon }{}^{\lambda \kappa } \nabla_{\omega }F_{\mu 
\sigma \kappa \tau } \nabla^{\omega }F_{\delta \nu \lambda 
}{}^{\tau } + \nn\\&& m_{874}^{} F_{\alpha \beta }{}^{\epsilon 
\varepsilon } F^{\alpha \beta \gamma \delta } F_{\gamma 
}{}^{\mu \nu \sigma } F_{\mu }{}^{\lambda \kappa \tau } 
\nabla_{\tau }F_{\epsilon \varepsilon \kappa \omega } \nabla^{
\omega }F_{\delta \nu \sigma \lambda } + \nn\\&& m_{875}^{} F_{\alpha 
\beta }{}^{\epsilon \varepsilon } F^{\alpha \beta \gamma 
\delta } F_{\gamma }{}^{\mu \nu \sigma } F_{\mu }{}^{\lambda 
\kappa \tau } \nabla_{\omega }F_{\epsilon \varepsilon \kappa 
\tau } \nabla^{\omega }F_{\delta \nu \sigma \lambda } + 
\nn\\&& m_{876}^{} F_{\alpha \beta \gamma }{}^{\epsilon } F^{\alpha 
\beta \gamma \delta } F_{\varepsilon \mu }{}^{\lambda \kappa } 
F^{\varepsilon \mu \nu \sigma } \nabla_{\kappa }F_{\epsilon 
\lambda \tau \omega } \nabla^{\omega }F_{\delta \nu \sigma 
}{}^{\tau } + \nn\\&& m_{877}^{} F_{\alpha \beta }{}^{\epsilon 
\varepsilon } F^{\alpha \beta \gamma \delta } F_{\gamma 
}{}^{\mu \nu \sigma } F_{\epsilon \mu }{}^{\lambda \kappa } 
\nabla_{\kappa }F_{\varepsilon \lambda \tau \omega } 
\nabla^{\omega }F_{\delta \nu \sigma }{}^{\tau } + \nn\\&& m_{878}^{} 
F_{\alpha \beta }{}^{\epsilon \varepsilon } F^{\alpha \beta 
\gamma \delta } F_{\gamma \epsilon }{}^{\mu \nu } F_{\mu }{}^{
\sigma \lambda \kappa } \nabla_{\kappa }F_{\varepsilon \lambda 
\tau \omega } \nabla^{\omega }F_{\delta \nu \sigma }{}^{\tau 
} + \nn\\&& m_{879}^{} F_{\alpha }{}^{\epsilon \varepsilon \mu } 
F^{\alpha \beta \gamma \delta } F_{\beta \epsilon }{}^{\nu 
\sigma } F_{\gamma \varepsilon }{}^{\lambda \kappa } 
\nabla_{\kappa }F_{\mu \lambda \tau \omega } \nabla^{\omega 
}F_{\delta \nu \sigma }{}^{\tau } + \nn\\&& m_{880}^{} F_{\alpha \beta 
\gamma }{}^{\epsilon } F^{\alpha \beta \gamma \delta } 
F_{\varepsilon \mu }{}^{\lambda \kappa } F^{\varepsilon \mu 
\nu \sigma } \nabla_{\tau }F_{\epsilon \lambda \kappa \omega 
} \nabla^{\omega }F_{\delta \nu \sigma }{}^{\tau } + \nn\\&& m_{881}^{} 
F_{\alpha \beta }{}^{\epsilon \varepsilon } F^{\alpha \beta 
\gamma \delta } F_{\gamma }{}^{\mu \nu \sigma } F_{\epsilon 
\mu }{}^{\lambda \kappa } \nabla_{\tau }F_{\varepsilon \lambda 
\kappa \omega } \nabla^{\omega }F_{\delta \nu \sigma 
}{}^{\tau } + \nn\\&& m_{882}^{} F_{\alpha \beta }{}^{\epsilon 
\varepsilon } F^{\alpha \beta \gamma \delta } F_{\gamma 
\epsilon }{}^{\mu \nu } F_{\mu }{}^{\sigma \lambda \kappa } 
\nabla_{\tau }F_{\varepsilon \lambda \kappa \omega } 
\nabla^{\omega }F_{\delta \nu \sigma }{}^{\tau } + \nn\\&& m_{883}^{} 
F_{\alpha }{}^{\epsilon \varepsilon \mu } F^{\alpha \beta 
\gamma \delta } F_{\beta \epsilon }{}^{\nu \sigma } F_{\gamma 
\varepsilon }{}^{\lambda \kappa } \nabla_{\tau }F_{\mu \lambda 
\kappa \omega } \nabla^{\omega }F_{\delta \nu \sigma 
}{}^{\tau } + \nn\\&& m_{884}^{} F_{\alpha \beta \gamma }{}^{\epsilon } 
F^{\alpha \beta \gamma \delta } F_{\varepsilon \mu 
}{}^{\lambda \kappa } F^{\varepsilon \mu \nu \sigma } \nabla_{
\omega }F_{\epsilon \lambda \kappa \tau } \nabla^{\omega 
}F_{\delta \nu \sigma }{}^{\tau } + \nn\\&& m_{885}^{} F_{\alpha \beta 
}{}^{\epsilon \varepsilon } F^{\alpha \beta \gamma \delta } F_{
\gamma }{}^{\mu \nu \sigma } F_{\epsilon \mu }{}^{\lambda 
\kappa } \nabla_{\omega }F_{\varepsilon \lambda \kappa \tau } 
\nabla^{\omega }F_{\delta \nu \sigma }{}^{\tau } + \nn\\&& m_{886}^{} 
F_{\alpha \beta }{}^{\epsilon \varepsilon } F^{\alpha \beta 
\gamma \delta } F_{\gamma \epsilon }{}^{\mu \nu } F_{\mu }{}^{
\sigma \lambda \kappa } \nabla_{\omega }F_{\varepsilon \lambda 
\kappa \tau } \nabla^{\omega }F_{\delta \nu \sigma }{}^{\tau 
} + \nn\\&& m_{887}^{} F_{\alpha }{}^{\epsilon \varepsilon \mu } 
F^{\alpha \beta \gamma \delta } F_{\beta \epsilon }{}^{\nu 
\sigma } F_{\gamma \varepsilon }{}^{\lambda \kappa } 
\nabla_{\omega }F_{\mu \lambda \kappa \tau } \nabla^{\omega 
}F_{\delta \nu \sigma }{}^{\tau } + \nn\\&& m_{904}^{} F_{\alpha \beta 
}{}^{\epsilon \varepsilon } F^{\alpha \beta \gamma \delta } F_{
\gamma \epsilon }{}^{\mu \nu } F_{\mu \nu }{}^{\sigma \lambda 
} \nabla_{\lambda }F_{\varepsilon \kappa \tau \omega } 
\nabla^{\omega }F_{\delta \sigma }{}^{\kappa \tau } + 
\nn\\&& m_{905}^{} F_{\alpha \beta \gamma }{}^{\epsilon } F^{\alpha 
\beta \gamma \delta } F_{\varepsilon \mu \nu }{}^{\lambda } 
F^{\varepsilon \mu \nu \sigma } \nabla_{\tau }F_{\epsilon 
\lambda \kappa \omega } \nabla^{\omega }F_{\delta \sigma }{}^{
\kappa \tau } + \nn\\&& m_{906}^{} F_{\alpha \beta }{}^{\epsilon 
\varepsilon } F^{\alpha \beta \gamma \delta } F_{\gamma 
}{}^{\mu \nu \sigma } F_{\epsilon \mu \nu }{}^{\lambda } 
\nabla_{\tau }F_{\varepsilon \lambda \kappa \omega } 
\nabla^{\omega }F_{\delta \sigma }{}^{\kappa \tau } + 
\nn\\&& m_{907}^{} F_{\alpha \beta }{}^{\epsilon \varepsilon } F^{\alpha 
\beta \gamma \delta } F_{\gamma \epsilon }{}^{\mu \nu } 
F_{\mu \nu }{}^{\sigma \lambda } \nabla_{\tau }F_{\varepsilon 
\lambda \kappa \omega } \nabla^{\omega }F_{\delta \sigma }{}^{
\kappa \tau } + \nn\\&& m_{908}^{} F_{\alpha \beta \gamma }{}^{\epsilon 
} F^{\alpha \beta \gamma \delta } F_{\varepsilon \mu \nu }{}^{
\lambda } F^{\varepsilon \mu \nu \sigma } \nabla_{\omega 
}F_{\epsilon \lambda \kappa \tau } \nabla^{\omega }F_{\delta 
\sigma }{}^{\kappa \tau } + \nn\\&& m_{909}^{} F_{\alpha \beta 
}{}^{\epsilon \varepsilon } F^{\alpha \beta \gamma \delta } F_{
\gamma }{}^{\mu \nu \sigma } F_{\epsilon \mu \nu }{}^{\lambda 
} \nabla_{\omega }F_{\varepsilon \lambda \kappa \tau } 
\nabla^{\omega }F_{\delta \sigma }{}^{\kappa \tau } + 
\nn\\&& m_{910}^{} F_{\alpha \beta }{}^{\epsilon \varepsilon } F^{\alpha 
\beta \gamma \delta } F_{\gamma \epsilon }{}^{\mu \nu } 
F_{\mu \nu }{}^{\sigma \lambda } \nabla_{\omega 
}F_{\varepsilon \lambda \kappa \tau } \nabla^{\omega 
}F_{\delta \sigma }{}^{\kappa \tau } + \nn\\&& m_{901}^{} F_{\alpha 
\beta }{}^{\epsilon \varepsilon } F^{\alpha \beta \gamma 
\delta } F_{\gamma \epsilon }{}^{\mu \nu } F^{\sigma \lambda 
\kappa \tau } \nabla_{\varepsilon }F_{\mu \nu \tau \omega } 
\nabla^{\omega }F_{\delta \sigma \lambda \kappa } + \nn\\&& m_{902}^{} 
F_{\alpha \beta }{}^{\epsilon \varepsilon } F^{\alpha \beta 
\gamma \delta } F_{\gamma \epsilon }{}^{\mu \nu } F_{\mu }{}^{
\sigma \lambda \kappa } \nabla_{\varepsilon }F_{\nu \kappa 
\tau \omega } \nabla^{\omega }F_{\delta \sigma \lambda 
}{}^{\tau } + \nn\\&& m_{903}^{} F_{\alpha \beta }{}^{\epsilon 
\varepsilon } F^{\alpha \beta \gamma \delta } F_{\gamma 
\epsilon }{}^{\mu \nu } F_{\mu }{}^{\sigma \lambda \kappa } 
\nabla_{\nu }F_{\varepsilon \kappa \tau \omega } 
\nabla^{\omega }F_{\delta \sigma \lambda }{}^{\tau } + 
\nn\\&& m_{946}^{} F_{\alpha \beta }{}^{\epsilon \varepsilon } F^{\alpha 
\beta \gamma \delta } F_{\gamma }{}^{\mu \nu \sigma } F_{\mu 
}{}^{\lambda \kappa \tau } \nabla_{\delta }F_{\nu \sigma \tau 
\omega } \nabla^{\omega }F_{\epsilon \varepsilon \lambda 
\kappa } + \nn\\&& m_{935}^{} F_{\alpha \beta }{}^{\epsilon \varepsilon 
} F^{\alpha \beta \gamma \delta } F_{\gamma }{}^{\mu \nu 
\sigma } F_{\delta }{}^{\lambda \kappa \tau } \nabla_{\omega 
}F_{\nu \sigma \kappa \tau } \nabla^{\omega }F_{\epsilon 
\varepsilon \mu \lambda } + \nn\\&& m_{924}^{} F_{\alpha \beta 
}{}^{\epsilon \varepsilon } F^{\alpha \beta \gamma \delta } F_{
\gamma }{}^{\mu \nu \sigma } F_{\delta }{}^{\lambda \kappa 
\tau } \nabla_{\tau }F_{\sigma \lambda \kappa \omega } 
\nabla^{\omega }F_{\epsilon \varepsilon \mu \nu } + \nn\\&& m_{925}^{} 
F_{\alpha \beta }{}^{\epsilon \varepsilon } F^{\alpha \beta 
\gamma \delta } F_{\gamma }{}^{\mu \nu \sigma } F_{\delta 
}{}^{\lambda \kappa \tau } \nabla_{\omega }F_{\sigma \lambda 
\kappa \tau } \nabla^{\omega }F_{\epsilon \varepsilon \mu \nu 
} + \nn\\&& m_{926}^{} F_{\alpha \beta }{}^{\epsilon \varepsilon } 
F^{\alpha \beta \gamma \delta } F_{\gamma \delta }{}^{\mu \nu 
} F^{\sigma \lambda \kappa \tau } \nabla_{\omega }F_{\sigma 
\lambda \kappa \tau } \nabla^{\omega }F_{\epsilon \varepsilon 
\mu \nu } + \nn\\&& m_{927}^{} F_{\alpha \beta \gamma }{}^{\epsilon } 
F^{\alpha \beta \gamma \delta } F_{\delta }{}^{\varepsilon \mu 
\nu } F^{\sigma \lambda \kappa \tau } \nabla_{\omega 
}F_{\sigma \lambda \kappa \tau } \nabla^{\omega }F_{\epsilon 
\varepsilon \mu \nu } + \nn\\&& m_{929}^{} F_{\alpha \beta 
}{}^{\epsilon \varepsilon } F^{\alpha \beta \gamma \delta } F_{
\gamma \delta }{}^{\mu \nu } F^{\sigma \lambda \kappa \tau } 
\nabla_{\tau }F_{\nu \lambda \kappa \omega } \nabla^{\omega 
}F_{\epsilon \varepsilon \mu \sigma } + \nn\\&& m_{930}^{} F_{\alpha 
\beta \gamma }{}^{\epsilon } F^{\alpha \beta \gamma \delta } 
F_{\delta }{}^{\varepsilon \mu \nu } F^{\sigma \lambda \kappa 
\tau } \nabla_{\tau }F_{\nu \lambda \kappa \omega } 
\nabla^{\omega }F_{\epsilon \varepsilon \mu \sigma } + 
\nn\\&& m_{932}^{} F_{\alpha \beta }{}^{\epsilon \varepsilon } F^{\alpha 
\beta \gamma \delta } F_{\gamma \delta }{}^{\mu \nu } 
F^{\sigma \lambda \kappa \tau } \nabla_{\omega }F_{\nu 
\lambda \kappa \tau } \nabla^{\omega }F_{\epsilon \varepsilon 
\mu \sigma } + \nn\\&& m_{933}^{} F_{\alpha \beta \gamma }{}^{\epsilon 
} F^{\alpha \beta \gamma \delta } F_{\delta }{}^{\varepsilon 
\mu \nu } F^{\sigma \lambda \kappa \tau } \nabla_{\omega }F_{
\nu \lambda \kappa \tau } \nabla^{\omega }F_{\epsilon 
\varepsilon \mu \sigma } + \nn\\&& m_{937}^{} F_{\alpha \beta 
}{}^{\epsilon \varepsilon } F^{\alpha \beta \gamma \delta } F_{
\gamma }{}^{\mu \nu \sigma } F_{\mu }{}^{\lambda \kappa \tau 
} \nabla_{\delta }F_{\sigma \kappa \tau \omega } 
\nabla^{\omega }F_{\epsilon \varepsilon \nu \lambda } + 
\nn\\&& m_{936}^{} F_{\alpha \beta }{}^{\epsilon \varepsilon } F^{\alpha 
\beta \gamma \delta } F_{\gamma }{}^{\mu \nu \sigma } F_{\mu 
}{}^{\lambda \kappa \tau } \nabla_{\delta }F_{\lambda \kappa 
\tau \omega } \nabla^{\omega }F_{\epsilon \varepsilon \nu 
\sigma } + \nn\\&& m_{938}^{} F_{\alpha \beta }{}^{\epsilon \varepsilon 
} F^{\alpha \beta \gamma \delta } F_{\gamma }{}^{\mu \nu 
\sigma } F_{\delta \mu }{}^{\lambda \kappa } \nabla_{\kappa 
}F_{\sigma \lambda \tau \omega } \nabla^{\omega }F_{\epsilon 
\varepsilon \nu }{}^{\tau } + \nn\\&& m_{939}^{} F_{\alpha \beta 
}{}^{\epsilon \varepsilon } F^{\alpha \beta \gamma \delta } F_{
\gamma }{}^{\mu \nu \sigma } F_{\delta \mu }{}^{\lambda 
\kappa } \nabla_{\tau }F_{\sigma \lambda \kappa \omega } 
\nabla^{\omega }F_{\epsilon \varepsilon \nu }{}^{\tau } + 
\nn\\&& m_{940}^{} F_{\alpha \beta }{}^{\epsilon \varepsilon } F^{\alpha 
\beta \gamma \delta } F_{\gamma }{}^{\mu \nu \sigma } 
F_{\delta \mu }{}^{\lambda \kappa } \nabla_{\omega }F_{\sigma 
\lambda \kappa \tau } \nabla^{\omega }F_{\epsilon \varepsilon 
\nu }{}^{\tau } + \nn\\&& m_{941}^{} F_{\alpha \beta }{}^{\epsilon 
\varepsilon } F^{\alpha \beta \gamma \delta } F_{\gamma \delta 
}{}^{\mu \nu } F^{\sigma \lambda \kappa \tau } \nabla_{\tau 
}F_{\mu \nu \kappa \omega } \nabla^{\omega }F_{\epsilon 
\varepsilon \sigma \lambda } + \nn\\&& m_{942}^{} F_{\alpha \beta 
\gamma }{}^{\epsilon } F^{\alpha \beta \gamma \delta } 
F_{\delta }{}^{\varepsilon \mu \nu } F^{\sigma \lambda \kappa 
\tau } \nabla_{\tau }F_{\mu \nu \kappa \omega } 
\nabla^{\omega }F_{\epsilon \varepsilon \sigma \lambda } + 
\nn\\&& m_{943}^{} F_{\alpha \beta }{}^{\epsilon \varepsilon } F^{\alpha 
\beta \gamma \delta } F_{\gamma \delta }{}^{\mu \nu } 
F^{\sigma \lambda \kappa \tau } \nabla_{\omega }F_{\mu \nu 
\kappa \tau } \nabla^{\omega }F_{\epsilon \varepsilon \sigma 
\lambda } + \nn\\&& m_{944}^{} F_{\alpha \beta \gamma }{}^{\epsilon } 
F^{\alpha \beta \gamma \delta } F_{\delta }{}^{\varepsilon \mu 
\nu } F^{\sigma \lambda \kappa \tau } \nabla_{\omega }F_{\mu 
\nu \kappa \tau } \nabla^{\omega }F_{\epsilon \varepsilon 
\sigma \lambda } + \nn\\&& m_{975}^{} F_{\alpha \beta \gamma 
}{}^{\epsilon } F^{\alpha \beta \gamma \delta } F_{\delta }{}^{
\varepsilon \mu \nu } F_{\varepsilon \mu \nu }{}^{\sigma } 
\nabla_{\tau }F_{\sigma \lambda \kappa \omega } 
\nabla^{\omega }F_{\epsilon }{}^{\lambda \kappa \tau } + 
\nn\\&& m_{976}^{} F_{\alpha \beta \gamma }{}^{\epsilon } F^{\alpha 
\beta \gamma \delta } F_{\delta }{}^{\varepsilon \mu \nu } F_{
\varepsilon \mu \nu }{}^{\sigma } \nabla_{\omega }F_{\sigma 
\lambda \kappa \tau } \nabla^{\omega }F_{\epsilon }{}^{\lambda 
\kappa \tau } + \nn\\&& m_{949}^{} F_{\alpha \beta }{}^{\epsilon 
\varepsilon } F^{\alpha \beta \gamma \delta } F_{\gamma 
}{}^{\mu \nu \sigma } F_{\delta }{}^{\lambda \kappa \tau } 
\nabla_{\sigma }F_{\varepsilon \kappa \tau \omega } 
\nabla^{\omega }F_{\epsilon \mu \nu \lambda } + \nn\\&& m_{950}^{} 
F_{\alpha \beta }{}^{\epsilon \varepsilon } F^{\alpha \beta 
\gamma \delta } F_{\gamma }{}^{\mu \nu \sigma } F_{\delta 
}{}^{\lambda \kappa \tau } \nabla_{\tau }F_{\varepsilon \sigma 
\kappa \omega } \nabla^{\omega }F_{\epsilon \mu \nu \lambda } 
+ \nn\\&& m_{951}^{} F_{\alpha \beta }{}^{\epsilon \varepsilon } 
F^{\alpha \beta \gamma \delta } F_{\gamma }{}^{\mu \nu \sigma 
} F_{\delta }{}^{\lambda \kappa \tau } \nabla_{\omega 
}F_{\varepsilon \sigma \kappa \tau } \nabla^{\omega 
}F_{\epsilon \mu \nu \lambda } + \nn\\&& m_{947}^{} F_{\alpha \beta 
}{}^{\epsilon \varepsilon } F^{\alpha \beta \gamma \delta } F_{
\gamma }{}^{\mu \nu \sigma } F_{\delta }{}^{\lambda \kappa 
\tau } \nabla_{\tau }F_{\varepsilon \lambda \kappa \omega } 
\nabla^{\omega }F_{\epsilon \mu \nu \sigma } + \nn\\&& m_{948}^{} 
F_{\alpha \beta }{}^{\epsilon \varepsilon } F^{\alpha \beta 
\gamma \delta } F_{\gamma }{}^{\mu \nu \sigma } F_{\delta 
}{}^{\lambda \kappa \tau } \nabla_{\omega }F_{\varepsilon 
\lambda \kappa \tau } \nabla^{\omega }F_{\epsilon \mu \nu 
\sigma } + \nn\\&& m_{952}^{} F_{\alpha \beta \gamma }{}^{\epsilon } F^{
\alpha \beta \gamma \delta } F_{\delta }{}^{\varepsilon \mu 
\nu } F_{\varepsilon }{}^{\sigma \lambda \kappa } \nabla_{\tau 
}F_{\sigma \lambda \kappa \omega } \nabla^{\omega }F_{\epsilon 
\mu \nu }{}^{\tau } + \nn\\&& m_{953}^{} F_{\alpha \beta \gamma 
}{}^{\epsilon } F^{\alpha \beta \gamma \delta } F_{\delta }{}^{
\varepsilon \mu \nu } F_{\varepsilon }{}^{\sigma \lambda 
\kappa } \nabla_{\omega }F_{\sigma \lambda \kappa \tau } 
\nabla^{\omega }F_{\epsilon \mu \nu }{}^{\tau } + \nn\\&& m_{954}^{} 
F_{\alpha \beta }{}^{\epsilon \varepsilon } F^{\alpha \beta 
\gamma \delta } F_{\gamma \delta }{}^{\mu \nu } F^{\sigma 
\lambda \kappa \tau } \nabla_{\tau }F_{\varepsilon \nu \kappa 
\omega } \nabla^{\omega }F_{\epsilon \mu \sigma \lambda } + 
\nn\\&& m_{955}^{} F_{\alpha \beta }{}^{\epsilon \varepsilon } F^{\alpha 
\beta \gamma \delta } F_{\gamma \delta }{}^{\mu \nu } 
F^{\sigma \lambda \kappa \tau } \nabla_{\omega }F_{\varepsilon 
\nu \kappa \tau } \nabla^{\omega }F_{\epsilon \mu \sigma 
\lambda } + \nn\\&& m_{956}^{} F_{\alpha \beta \gamma }{}^{\epsilon } 
F^{\alpha \beta \gamma \delta } F_{\delta }{}^{\varepsilon \mu 
\nu } F_{\varepsilon }{}^{\sigma \lambda \kappa } 
\nabla_{\kappa }F_{\nu \lambda \tau \omega } \nabla^{\omega 
}F_{\epsilon \mu \sigma }{}^{\tau } + \nn\\&& m_{957}^{} F_{\alpha 
\beta \gamma }{}^{\epsilon } F^{\alpha \beta \gamma \delta } 
F_{\delta }{}^{\varepsilon \mu \nu } F_{\varepsilon }{}^{\sigma 
\lambda \kappa } \nabla_{\tau }F_{\nu \lambda \kappa \omega } 
\nabla^{\omega }F_{\epsilon \mu \sigma }{}^{\tau } + \nn\\&& m_{958}^{} 
F_{\alpha \beta \gamma }{}^{\epsilon } F^{\alpha \beta \gamma 
\delta } F_{\delta }{}^{\varepsilon \mu \nu } F_{\varepsilon 
}{}^{\sigma \lambda \kappa } \nabla_{\omega }F_{\nu \lambda 
\kappa \tau } \nabla^{\omega }F_{\epsilon \mu \sigma 
}{}^{\tau } + \nn\\&& m_{965}^{} F_{\alpha \beta \gamma }{}^{\epsilon } 
F^{\alpha \beta \gamma \delta } F_{\delta }{}^{\varepsilon \mu 
\nu } F_{\varepsilon \mu }{}^{\sigma \lambda } \nabla_{\tau 
}F_{\sigma \lambda \kappa \omega } \nabla^{\omega }F_{\epsilon 
\nu }{}^{\kappa \tau } + \nn\\&& m_{966}^{} F_{\alpha \beta \gamma 
}{}^{\epsilon } F^{\alpha \beta \gamma \delta } F_{\delta }{}^{
\varepsilon \mu \nu } F_{\varepsilon \mu }{}^{\sigma \lambda } 
\nabla_{\omega }F_{\sigma \lambda \kappa \tau } 
\nabla^{\omega }F_{\epsilon \nu }{}^{\kappa \tau } + \nn\\&& m_{963}^{} 
F_{\alpha \beta }{}^{\epsilon \varepsilon } F^{\alpha \beta 
\gamma \delta } F_{\gamma }{}^{\mu \nu \sigma } F_{\mu 
}{}^{\lambda \kappa \tau } \nabla_{\delta }F_{\varepsilon 
\sigma \tau \omega } \nabla^{\omega }F_{\epsilon \nu \lambda 
\kappa } + \nn\\&& m_{964}^{} F_{\alpha \beta }{}^{\epsilon \varepsilon 
} F^{\alpha \beta \gamma \delta } F_{\gamma }{}^{\mu \nu 
\sigma } F_{\delta \mu }{}^{\lambda \kappa } \nabla_{\kappa 
}F_{\varepsilon \sigma \tau \omega } \nabla^{\omega 
}F_{\epsilon \nu \lambda }{}^{\tau } + \nn\\&& m_{959}^{} F_{\alpha 
\beta }{}^{\epsilon \varepsilon } F^{\alpha \beta \gamma 
\delta } F_{\gamma }{}^{\mu \nu \sigma } F_{\mu }{}^{\lambda 
\kappa \tau } \nabla_{\delta }F_{\varepsilon \kappa \tau 
\omega } \nabla^{\omega }F_{\epsilon \nu \sigma \lambda } + 
\nn\\&& m_{960}^{} F_{\alpha \beta }{}^{\epsilon \varepsilon } F^{\alpha 
\beta \gamma \delta } F_{\gamma }{}^{\mu \nu \sigma } 
F_{\delta \mu }{}^{\lambda \kappa } \nabla_{\kappa 
}F_{\varepsilon \lambda \tau \omega } \nabla^{\omega 
}F_{\epsilon \nu \sigma }{}^{\tau } + \nn\\&& m_{961}^{} F_{\alpha 
\beta }{}^{\epsilon \varepsilon } F^{\alpha \beta \gamma 
\delta } F_{\gamma }{}^{\mu \nu \sigma } F_{\delta \mu 
}{}^{\lambda \kappa } \nabla_{\tau }F_{\varepsilon \lambda 
\kappa \omega } \nabla^{\omega }F_{\epsilon \nu \sigma 
}{}^{\tau } + \nn\\&& m_{962}^{} F_{\alpha \beta }{}^{\epsilon 
\varepsilon } F^{\alpha \beta \gamma \delta } F_{\gamma 
}{}^{\mu \nu \sigma } F_{\delta \mu }{}^{\lambda \kappa } 
\nabla_{\omega }F_{\varepsilon \lambda \kappa \tau } 
\nabla^{\omega }F_{\epsilon \nu \sigma }{}^{\tau } + \nn\\&& m_{972}^{} 
F_{\alpha \beta \gamma }{}^{\epsilon } F^{\alpha \beta \gamma 
\delta } F_{\delta }{}^{\varepsilon \mu \nu } F_{\varepsilon 
\mu }{}^{\sigma \lambda } \nabla_{\lambda }F_{\nu \kappa \tau 
\omega } \nabla^{\omega }F_{\epsilon \sigma }{}^{\kappa \tau } 
+ \nn\\&& m_{973}^{} F_{\alpha \beta \gamma }{}^{\epsilon } F^{\alpha 
\beta \gamma \delta } F_{\delta }{}^{\varepsilon \mu \nu } F_{
\varepsilon \mu }{}^{\sigma \lambda } \nabla_{\tau }F_{\nu 
\lambda \kappa \omega } \nabla^{\omega }F_{\epsilon \sigma 
}{}^{\kappa \tau } + \nn\\&& m_{974}^{} F_{\alpha \beta \gamma 
}{}^{\epsilon } F^{\alpha \beta \gamma \delta } F_{\delta }{}^{
\varepsilon \mu \nu } F_{\varepsilon \mu }{}^{\sigma \lambda } 
\nabla_{\omega }F_{\nu \lambda \kappa \tau } \nabla^{\omega 
}F_{\epsilon \sigma }{}^{\kappa \tau } + \nn\\&& m_{967}^{} F_{\alpha 
\beta \gamma }{}^{\epsilon } F^{\alpha \beta \gamma \delta } 
F_{\delta }{}^{\varepsilon \mu \nu } F^{\sigma \lambda \kappa 
\tau } \nabla_{\tau }F_{\varepsilon \mu \nu \omega } \nabla^{
\omega }F_{\epsilon \sigma \lambda \kappa } + \nn\\&& m_{968}^{} 
F_{\alpha \beta \gamma }{}^{\epsilon } F^{\alpha \beta \gamma 
\delta } F_{\delta }{}^{\varepsilon \mu \nu } F^{\sigma 
\lambda \kappa \tau } \nabla_{\omega }F_{\varepsilon \mu \nu 
\tau } \nabla^{\omega }F_{\epsilon \sigma \lambda \kappa } + 
\nn\\&& m_{969}^{} F_{\alpha \beta \gamma }{}^{\epsilon } F^{\alpha 
\beta \gamma \delta } F_{\delta }{}^{\varepsilon \mu \nu } F_{
\varepsilon }{}^{\sigma \lambda \kappa } \nabla_{\kappa 
}F_{\mu \nu \tau \omega } \nabla^{\omega }F_{\epsilon \sigma 
\lambda }{}^{\tau } + \nn\\&& m_{970}^{} F_{\alpha \beta \gamma 
}{}^{\epsilon } F^{\alpha \beta \gamma \delta } F_{\delta }{}^{
\varepsilon \mu \nu } F_{\varepsilon }{}^{\sigma \lambda 
\kappa } \nabla_{\tau }F_{\mu \nu \kappa \omega } 
\nabla^{\omega }F_{\epsilon \sigma \lambda }{}^{\tau } + 
\nn\\&& m_{971}^{} F_{\alpha \beta \gamma }{}^{\epsilon } F^{\alpha 
\beta \gamma \delta } F_{\delta }{}^{\varepsilon \mu \nu } F_{
\varepsilon }{}^{\sigma \lambda \kappa } \nabla_{\omega 
}F_{\mu \nu \kappa \tau } \nabla^{\omega }F_{\epsilon \sigma 
\lambda }{}^{\tau } + \nn\\&& m_{1020}^{} F_{\alpha \beta }{}^{\epsilon 
\varepsilon } F^{\alpha \beta \gamma \delta } F_{\gamma 
\epsilon }{}^{\mu \nu } F_{\delta \mu \nu }{}^{\sigma } 
\nabla_{\tau }F_{\sigma \lambda \kappa \omega } 
\nabla^{\omega }F_{\varepsilon }{}^{\lambda \kappa \tau } + 
\nn\\&& m_{1021}^{} F_{\alpha \beta }{}^{\epsilon \varepsilon } 
F^{\alpha \beta \gamma \delta } F_{\gamma \epsilon }{}^{\mu 
\nu } F_{\delta \mu \nu }{}^{\sigma } \nabla_{\omega 
}F_{\sigma \lambda \kappa \tau } \nabla^{\omega 
}F_{\varepsilon }{}^{\lambda \kappa \tau } + \nn\\&& m_{1000}^{} 
F_{\alpha }{}^{\epsilon \varepsilon \mu } F^{\alpha \beta 
\gamma \delta } F_{\beta \epsilon }{}^{\nu \sigma } F_{\gamma 
}{}^{\lambda \kappa \tau } \nabla_{\delta }F_{\nu \sigma \tau 
\omega } \nabla^{\omega }F_{\varepsilon \mu \lambda \kappa } + 
\nn\\&& m_{978}^{} F_{\alpha }{}^{\epsilon \varepsilon \mu } F^{\alpha 
\beta \gamma \delta } F_{\beta \epsilon }{}^{\nu \sigma } 
F_{\gamma }{}^{\lambda \kappa \tau } \nabla_{\delta }F_{\sigma 
\kappa \tau \omega } \nabla^{\omega }F_{\varepsilon \mu \nu 
\lambda } + \nn\\&& m_{977}^{} F_{\alpha \beta \gamma }{}^{\epsilon } 
F^{\alpha \beta \gamma \delta } F_{\delta }{}^{\varepsilon \mu 
\nu } F^{\sigma \lambda \kappa \tau } \nabla_{\epsilon 
}F_{\lambda \kappa \tau \omega } \nabla^{\omega 
}F_{\varepsilon \mu \nu \sigma } + \nn\\&& m_{979}^{} F_{\alpha \beta 
}{}^{\epsilon \varepsilon } F^{\alpha \beta \gamma \delta } F_{
\gamma \epsilon }{}^{\mu \nu } F_{\delta }{}^{\sigma \lambda 
\kappa } \nabla_{\tau }F_{\sigma \lambda \kappa \omega } 
\nabla^{\omega }F_{\varepsilon \mu \nu }{}^{\tau } + \nn\\&& m_{980}^{} 
F_{\alpha \beta }{}^{\epsilon \varepsilon } F^{\alpha \beta 
\gamma \delta } F_{\gamma \delta }{}^{\mu \nu } F_{\epsilon 
}{}^{\sigma \lambda \kappa } \nabla_{\tau }F_{\sigma \lambda 
\kappa \omega } \nabla^{\omega }F_{\varepsilon \mu \nu 
}{}^{\tau } + \nn\\&& m_{981}^{} F_{\alpha \beta \gamma }{}^{\epsilon } 
F^{\alpha \beta \gamma \delta } F_{\delta }{}^{\varepsilon \mu 
\nu } F_{\epsilon }{}^{\sigma \lambda \kappa } \nabla_{\tau 
}F_{\sigma \lambda \kappa \omega } \nabla^{\omega 
}F_{\varepsilon \mu \nu }{}^{\tau } + \nn\\&& m_{983}^{} F_{\alpha 
\beta }{}^{\epsilon \varepsilon } F^{\alpha \beta \gamma 
\delta } F_{\gamma \epsilon }{}^{\mu \nu } F_{\delta 
}{}^{\sigma \lambda \kappa } \nabla_{\omega }F_{\sigma \lambda 
\kappa \tau } \nabla^{\omega }F_{\varepsilon \mu \nu 
}{}^{\tau } + \nn\\&& m_{984}^{} F_{\alpha \beta }{}^{\epsilon 
\varepsilon } F^{\alpha \beta \gamma \delta } F_{\gamma \delta 
}{}^{\mu \nu } F_{\epsilon }{}^{\sigma \lambda \kappa } 
\nabla_{\omega }F_{\sigma \lambda \kappa \tau } 
\nabla^{\omega }F_{\varepsilon \mu \nu }{}^{\tau } + \nn\\&& m_{985}^{} 
F_{\alpha \beta \gamma }{}^{\epsilon } F^{\alpha \beta \gamma 
\delta } F_{\delta }{}^{\varepsilon \mu \nu } F_{\epsilon }{}^{
\sigma \lambda \kappa } \nabla_{\omega }F_{\sigma \lambda 
\kappa \tau } \nabla^{\omega }F_{\varepsilon \mu \nu 
}{}^{\tau } + \nn\\&& m_{987}^{} F_{\alpha \beta \gamma }{}^{\epsilon } 
F^{\alpha \beta \gamma \delta } F_{\delta }{}^{\varepsilon \mu 
\nu } F^{\sigma \lambda \kappa \tau } \nabla_{\epsilon 
}F_{\nu \kappa \tau \omega } \nabla^{\omega }F_{\varepsilon 
\mu \sigma \lambda } + \nn\\&& m_{988}^{} F_{\alpha \beta }{}^{\epsilon 
\varepsilon } F^{\alpha \beta \gamma \delta } F_{\gamma 
\epsilon }{}^{\mu \nu } F_{\delta }{}^{\sigma \lambda \kappa } 
\nabla_{\kappa }F_{\nu \lambda \tau \omega } \nabla^{\omega 
}F_{\varepsilon \mu \sigma }{}^{\tau } + \nn\\&& m_{989}^{} F_{\alpha 
\beta }{}^{\epsilon \varepsilon } F^{\alpha \beta \gamma 
\delta } F_{\gamma \delta }{}^{\mu \nu } F_{\epsilon 
}{}^{\sigma \lambda \kappa } \nabla_{\kappa }F_{\nu \lambda 
\tau \omega } \nabla^{\omega }F_{\varepsilon \mu \sigma 
}{}^{\tau } + \nn\\&& m_{990}^{} F_{\alpha \beta \gamma }{}^{\epsilon } 
F^{\alpha \beta \gamma \delta } F_{\delta }{}^{\varepsilon \mu 
\nu } F_{\epsilon }{}^{\sigma \lambda \kappa } \nabla_{\kappa 
}F_{\nu \lambda \tau \omega } \nabla^{\omega }F_{\varepsilon 
\mu \sigma }{}^{\tau } + \nn\\&& m_{992}^{} F_{\alpha \beta 
}{}^{\epsilon \varepsilon } F^{\alpha \beta \gamma \delta } F_{
\gamma \epsilon }{}^{\mu \nu } F_{\delta }{}^{\sigma \lambda 
\kappa } \nabla_{\tau }F_{\nu \lambda \kappa \omega } 
\nabla^{\omega }F_{\varepsilon \mu \sigma }{}^{\tau } + 
\nn\\&& m_{993}^{} F_{\alpha \beta }{}^{\epsilon \varepsilon } F^{\alpha 
\beta \gamma \delta } F_{\gamma \delta }{}^{\mu \nu } 
F_{\epsilon }{}^{\sigma \lambda \kappa } \nabla_{\tau }F_{\nu 
\lambda \kappa \omega } \nabla^{\omega }F_{\varepsilon \mu 
\sigma }{}^{\tau } + \nn\\&& m_{994}^{} F_{\alpha \beta \gamma 
}{}^{\epsilon } F^{\alpha \beta \gamma \delta } F_{\delta }{}^{
\varepsilon \mu \nu } F_{\epsilon }{}^{\sigma \lambda \kappa } 
\nabla_{\tau }F_{\nu \lambda \kappa \omega } \nabla^{\omega 
}F_{\varepsilon \mu \sigma }{}^{\tau } + \nn\\&& m_{996}^{} F_{\alpha 
\beta }{}^{\epsilon \varepsilon } F^{\alpha \beta \gamma 
\delta } F_{\gamma \epsilon }{}^{\mu \nu } F_{\delta 
}{}^{\sigma \lambda \kappa } \nabla_{\omega }F_{\nu \lambda 
\kappa \tau } \nabla^{\omega }F_{\varepsilon \mu \sigma 
}{}^{\tau } + \nn\\&& m_{997}^{} F_{\alpha \beta }{}^{\epsilon 
\varepsilon } F^{\alpha \beta \gamma \delta } F_{\gamma \delta 
}{}^{\mu \nu } F_{\epsilon }{}^{\sigma \lambda \kappa } 
\nabla_{\omega }F_{\nu \lambda \kappa \tau } \nabla^{\omega 
}F_{\varepsilon \mu \sigma }{}^{\tau } + \nn\\&& m_{998}^{} F_{\alpha 
\beta \gamma }{}^{\epsilon } F^{\alpha \beta \gamma \delta } 
F_{\delta }{}^{\varepsilon \mu \nu } F_{\epsilon }{}^{\sigma 
\lambda \kappa } \nabla_{\omega }F_{\nu \lambda \kappa \tau } 
\nabla^{\omega }F_{\varepsilon \mu \sigma }{}^{\tau } + 
\nn\\&& m_{1004}^{} F_{\alpha \beta }{}^{\epsilon \varepsilon } 
F^{\alpha \beta \gamma \delta } F_{\gamma \epsilon }{}^{\mu 
\nu } F_{\delta \mu }{}^{\sigma \lambda } \nabla_{\tau 
}F_{\sigma \lambda \kappa \omega } \nabla^{\omega 
}F_{\varepsilon \nu }{}^{\kappa \tau } + \nn\\&& m_{1005}^{} F_{\alpha 
\beta }{}^{\epsilon \varepsilon } F^{\alpha \beta \gamma 
\delta } F_{\gamma \delta }{}^{\mu \nu } F_{\epsilon \mu }{}^{
\sigma \lambda } \nabla_{\tau }F_{\sigma \lambda \kappa 
\omega } \nabla^{\omega }F_{\varepsilon \nu }{}^{\kappa \tau } 
+ \nn\\&& m_{1006}^{} F_{\alpha \beta }{}^{\epsilon \varepsilon } 
F^{\alpha \beta \gamma \delta } F_{\gamma \epsilon }{}^{\mu 
\nu } F_{\delta \mu }{}^{\sigma \lambda } \nabla_{\omega 
}F_{\sigma \lambda \kappa \tau } \nabla^{\omega 
}F_{\varepsilon \nu }{}^{\kappa \tau } + \nn\\&& m_{1007}^{} F_{\alpha 
\beta }{}^{\epsilon \varepsilon } F^{\alpha \beta \gamma 
\delta } F_{\gamma \delta }{}^{\mu \nu } F_{\epsilon \mu }{}^{
\sigma \lambda } \nabla_{\omega }F_{\sigma \lambda \kappa 
\tau } \nabla^{\omega }F_{\varepsilon \nu }{}^{\kappa \tau } + 
\nn\\&& m_{1001}^{} F_{\alpha \beta }{}^{\epsilon \varepsilon } 
F^{\alpha \beta \gamma \delta } F_{\gamma \delta \epsilon 
}{}^{\mu } F^{\nu \sigma \lambda \kappa } \nabla_{\kappa 
}F_{\mu \lambda \tau \omega } \nabla^{\omega }F_{\varepsilon 
\nu \sigma }{}^{\tau } + \nn\\&& m_{1002}^{} F_{\alpha \beta 
}{}^{\epsilon \varepsilon } F^{\alpha \beta \gamma \delta } F_{
\gamma \delta \epsilon }{}^{\mu } F^{\nu \sigma \lambda 
\kappa } \nabla_{\tau }F_{\mu \lambda \kappa \omega } 
\nabla^{\omega }F_{\varepsilon \nu \sigma }{}^{\tau } + 
\nn\\&& m_{1003}^{} F_{\alpha \beta }{}^{\epsilon \varepsilon } 
F^{\alpha \beta \gamma \delta } F_{\gamma \delta \epsilon 
}{}^{\mu } F^{\nu \sigma \lambda \kappa } \nabla_{\omega 
}F_{\mu \lambda \kappa \tau } \nabla^{\omega }F_{\varepsilon 
\nu \sigma }{}^{\tau } + \nn\\&& m_{1014}^{} F_{\alpha \beta 
}{}^{\epsilon \varepsilon } F^{\alpha \beta \gamma \delta } F_{
\gamma \epsilon }{}^{\mu \nu } F_{\delta \mu }{}^{\sigma 
\lambda } \nabla_{\lambda }F_{\nu \kappa \tau \omega } 
\nabla^{\omega }F_{\varepsilon \sigma }{}^{\kappa \tau } + 
\nn\\&& m_{1015}^{} F_{\alpha \beta }{}^{\epsilon \varepsilon } 
F^{\alpha \beta \gamma \delta } F_{\gamma \delta }{}^{\mu \nu 
} F_{\epsilon \mu }{}^{\sigma \lambda } \nabla_{\lambda 
}F_{\nu \kappa \tau \omega } \nabla^{\omega }F_{\varepsilon 
\sigma }{}^{\kappa \tau } + \nn\\&& m_{1016}^{} F_{\alpha \beta 
}{}^{\epsilon \varepsilon } F^{\alpha \beta \gamma \delta } F_{
\gamma \epsilon }{}^{\mu \nu } F_{\delta \mu }{}^{\sigma 
\lambda } \nabla_{\tau }F_{\nu \lambda \kappa \omega } 
\nabla^{\omega }F_{\varepsilon \sigma }{}^{\kappa \tau } + 
\nn\\&& m_{1017}^{} F_{\alpha \beta }{}^{\epsilon \varepsilon } 
F^{\alpha \beta \gamma \delta } F_{\gamma \delta }{}^{\mu \nu 
} F_{\epsilon \mu }{}^{\sigma \lambda } \nabla_{\tau }F_{\nu 
\lambda \kappa \omega } \nabla^{\omega }F_{\varepsilon \sigma 
}{}^{\kappa \tau } + \nn\\&& m_{1018}^{} F_{\alpha \beta }{}^{\epsilon 
\varepsilon } F^{\alpha \beta \gamma \delta } F_{\gamma 
\epsilon }{}^{\mu \nu } F_{\delta \mu }{}^{\sigma \lambda } 
\nabla_{\omega }F_{\nu \lambda \kappa \tau } \nabla^{\omega 
}F_{\varepsilon \sigma }{}^{\kappa \tau } + \nn\\&& m_{1019}^{} 
F_{\alpha \beta }{}^{\epsilon \varepsilon } F^{\alpha \beta 
\gamma \delta } F_{\gamma \delta }{}^{\mu \nu } F_{\epsilon 
\mu }{}^{\sigma \lambda } \nabla_{\omega }F_{\nu \lambda 
\kappa \tau } \nabla^{\omega }F_{\varepsilon \sigma 
}{}^{\kappa \tau } + \nn\\&& m_{1008}^{} F_{\alpha \beta }{}^{\epsilon 
\varepsilon } F^{\alpha \beta \gamma \delta } F_{\gamma 
\epsilon }{}^{\mu \nu } F_{\delta }{}^{\sigma \lambda \kappa } 
\nabla_{\kappa }F_{\mu \nu \tau \omega } \nabla^{\omega 
}F_{\varepsilon \sigma \lambda }{}^{\tau } + \nn\\&& m_{1009}^{} 
F_{\alpha \beta }{}^{\epsilon \varepsilon } F^{\alpha \beta 
\gamma \delta } F_{\gamma \delta }{}^{\mu \nu } F_{\epsilon 
}{}^{\sigma \lambda \kappa } \nabla_{\kappa }F_{\mu \nu \tau 
\omega } \nabla^{\omega }F_{\varepsilon \sigma \lambda 
}{}^{\tau } + \nn\\&& m_{1010}^{} F_{\alpha \beta }{}^{\epsilon 
\varepsilon } F^{\alpha \beta \gamma \delta } F_{\gamma 
\epsilon }{}^{\mu \nu } F_{\delta }{}^{\sigma \lambda \kappa } 
\nabla_{\tau }F_{\mu \nu \kappa \omega } \nabla^{\omega 
}F_{\varepsilon \sigma \lambda }{}^{\tau } + \nn\\&& m_{1011}^{} 
F_{\alpha \beta }{}^{\epsilon \varepsilon } F^{\alpha \beta 
\gamma \delta } F_{\gamma \delta }{}^{\mu \nu } F_{\epsilon 
}{}^{\sigma \lambda \kappa } \nabla_{\tau }F_{\mu \nu \kappa 
\omega } \nabla^{\omega }F_{\varepsilon \sigma \lambda 
}{}^{\tau } + \nn\\&& m_{1012}^{} F_{\alpha \beta }{}^{\epsilon 
\varepsilon } F^{\alpha \beta \gamma \delta } F_{\gamma 
\epsilon }{}^{\mu \nu } F_{\delta }{}^{\sigma \lambda \kappa } 
\nabla_{\omega }F_{\mu \nu \kappa \tau } \nabla^{\omega 
}F_{\varepsilon \sigma \lambda }{}^{\tau } + \nn\\&& m_{1013}^{} 
F_{\alpha \beta }{}^{\epsilon \varepsilon } F^{\alpha \beta 
\gamma \delta } F_{\gamma \delta }{}^{\mu \nu } F_{\epsilon 
}{}^{\sigma \lambda \kappa } \nabla_{\omega }F_{\mu \nu 
\kappa \tau } \nabla^{\omega }F_{\varepsilon \sigma \lambda 
}{}^{\tau } + \nn\\&& m_{1025}^{} F_{\alpha \beta }{}^{\epsilon 
\varepsilon } F^{\alpha \beta \gamma \delta } F_{\gamma 
\epsilon }{}^{\mu \nu } F_{\delta \varepsilon }{}^{\sigma 
\lambda } \nabla_{\tau }F_{\sigma \lambda \kappa \omega } 
\nabla^{\omega }F_{\mu \nu }{}^{\kappa \tau } + \nn\\&& m_{1026}^{} F_{
\alpha \beta }{}^{\epsilon \varepsilon } F^{\alpha \beta 
\gamma \delta } F_{\gamma \delta }{}^{\mu \nu } F_{\epsilon 
\varepsilon }{}^{\sigma \lambda } \nabla_{\tau }F_{\sigma 
\lambda \kappa \omega } \nabla^{\omega }F_{\mu \nu 
}{}^{\kappa \tau } + \nn\\&& m_{1027}^{} F_{\alpha \beta \gamma 
}{}^{\epsilon } F^{\alpha \beta \gamma \delta } F_{\delta }{}^{
\varepsilon \mu \nu } F_{\epsilon \varepsilon }{}^{\sigma 
\lambda } \nabla_{\tau }F_{\sigma \lambda \kappa \omega } 
\nabla^{\omega }F_{\mu \nu }{}^{\kappa \tau } + \nn\\&& m_{1028}^{} F_{
\alpha \beta }{}^{\epsilon \varepsilon } F^{\alpha \beta 
\gamma \delta } F_{\gamma \delta \epsilon }{}^{\mu } 
F_{\varepsilon }{}^{\nu \sigma \lambda } \nabla_{\tau 
}F_{\sigma \lambda \kappa \omega } \nabla^{\omega }F_{\mu \nu 
}{}^{\kappa \tau } + \nn\\&& m_{1029}^{} F_{\alpha \beta }{}^{\epsilon 
\varepsilon } F^{\alpha \beta \gamma \delta } F_{\gamma \delta 
\epsilon \varepsilon } F^{\mu \nu \sigma \lambda } 
\nabla_{\tau }F_{\sigma \lambda \kappa \omega } 
\nabla^{\omega }F_{\mu \nu }{}^{\kappa \tau } + \nn\\&& m_{1031}^{} F_{
\alpha \beta }{}^{\epsilon \varepsilon } F^{\alpha \beta 
\gamma \delta } F_{\gamma \epsilon }{}^{\mu \nu } F_{\delta 
\varepsilon }{}^{\sigma \lambda } \nabla_{\omega }F_{\sigma 
\lambda \kappa \tau } \nabla^{\omega }F_{\mu \nu }{}^{\kappa 
\tau } + \nn\\&& m_{1032}^{} F_{\alpha \beta }{}^{\epsilon \varepsilon } 
F^{\alpha \beta \gamma \delta } F_{\gamma \delta }{}^{\mu \nu 
} F_{\epsilon \varepsilon }{}^{\sigma \lambda } \nabla_{\omega 
}F_{\sigma \lambda \kappa \tau } \nabla^{\omega }F_{\mu \nu 
}{}^{\kappa \tau } + \nn\\&& m_{1033}^{} F_{\alpha \beta \gamma 
}{}^{\epsilon } F^{\alpha \beta \gamma \delta } F_{\delta }{}^{
\varepsilon \mu \nu } F_{\epsilon \varepsilon }{}^{\sigma 
\lambda } \nabla_{\omega }F_{\sigma \lambda \kappa \tau } 
\nabla^{\omega }F_{\mu \nu }{}^{\kappa \tau } + \nn\\&& m_{1034}^{} F_{
\alpha \beta }{}^{\epsilon \varepsilon } F^{\alpha \beta 
\gamma \delta } F_{\gamma \delta \epsilon }{}^{\mu } 
F_{\varepsilon }{}^{\nu \sigma \lambda } \nabla_{\omega 
}F_{\sigma \lambda \kappa \tau } \nabla^{\omega }F_{\mu \nu 
}{}^{\kappa \tau } + \nn\\&& m_{1035}^{} F_{\alpha \beta }{}^{\epsilon 
\varepsilon } F^{\alpha \beta \gamma \delta } F_{\gamma \delta 
\epsilon \varepsilon } F^{\mu \nu \sigma \lambda } 
\nabla_{\omega }F_{\sigma \lambda \kappa \tau } 
\nabla^{\omega }F_{\mu \nu }{}^{\kappa \tau } + \nn\\&& m_{1022}^{} F_{
\alpha \beta \gamma }{}^{\epsilon } F^{\alpha \beta \gamma 
\delta } F_{\delta }{}^{\varepsilon \mu \nu } F_{\varepsilon 
}{}^{\sigma \lambda \kappa } \nabla_{\epsilon }F_{\lambda 
\kappa \tau \omega } \nabla^{\omega }F_{\mu \nu \sigma 
}{}^{\tau } + \nn\\&& m_{1023}^{} F_{\alpha \beta }{}^{\epsilon 
\varepsilon } F^{\alpha \beta \gamma \delta } F_{\gamma 
\epsilon }{}^{\mu \nu } F_{\delta }{}^{\sigma \lambda \kappa } 
\nabla_{\varepsilon }F_{\lambda \kappa \tau \omega } 
\nabla^{\omega }F_{\mu \nu \sigma }{}^{\tau } + \nn\\&& m_{1024}^{} F_{
\alpha \beta }{}^{\epsilon \varepsilon } F^{\alpha \beta 
\gamma \delta } F_{\gamma \delta }{}^{\mu \nu } F_{\epsilon 
}{}^{\sigma \lambda \kappa } \nabla_{\varepsilon }F_{\lambda 
\kappa \tau \omega } \nabla^{\omega }F_{\mu \nu \sigma 
}{}^{\tau } + \nn\\&& m_{1040}^{} F_{\alpha \beta }{}^{\epsilon 
\varepsilon } F^{\alpha \beta \gamma \delta } F_{\gamma 
\epsilon }{}^{\mu \nu } F_{\delta \varepsilon }{}^{\sigma 
\lambda } \nabla_{\tau }F_{\nu \lambda \kappa \omega } 
\nabla^{\omega }F_{\mu \sigma }{}^{\kappa \tau } + \nn\\&& m_{1041}^{} 
F_{\alpha \beta }{}^{\epsilon \varepsilon } F^{\alpha \beta 
\gamma \delta } F_{\gamma \delta }{}^{\mu \nu } F_{\epsilon 
\varepsilon }{}^{\sigma \lambda } \nabla_{\tau }F_{\nu \lambda 
\kappa \omega } \nabla^{\omega }F_{\mu \sigma }{}^{\kappa 
\tau } + \nn\\&& m_{1042}^{} F_{\alpha \beta \gamma }{}^{\epsilon } 
F^{\alpha \beta \gamma \delta } F_{\delta }{}^{\varepsilon \mu 
\nu } F_{\epsilon \varepsilon }{}^{\sigma \lambda } 
\nabla_{\tau }F_{\nu \lambda \kappa \omega } \nabla^{\omega 
}F_{\mu \sigma }{}^{\kappa \tau } + \nn\\&& m_{1044}^{} F_{\alpha \beta 
}{}^{\epsilon \varepsilon } F^{\alpha \beta \gamma \delta } F_{
\gamma \epsilon }{}^{\mu \nu } F_{\delta \varepsilon 
}{}^{\sigma \lambda } \nabla_{\omega }F_{\nu \lambda \kappa 
\tau } \nabla^{\omega }F_{\mu \sigma }{}^{\kappa \tau } + 
\nn\\&& m_{1045}^{} F_{\alpha \beta }{}^{\epsilon \varepsilon } 
F^{\alpha \beta \gamma \delta } F_{\gamma \delta }{}^{\mu \nu 
} F_{\epsilon \varepsilon }{}^{\sigma \lambda } \nabla_{\omega 
}F_{\nu \lambda \kappa \tau } \nabla^{\omega }F_{\mu \sigma 
}{}^{\kappa \tau } + \nn\\&& m_{1046}^{} F_{\alpha \beta \gamma 
}{}^{\epsilon } F^{\alpha \beta \gamma \delta } F_{\delta }{}^{
\varepsilon \mu \nu } F_{\epsilon \varepsilon }{}^{\sigma 
\lambda } \nabla_{\omega }F_{\nu \lambda \kappa \tau } 
\nabla^{\omega }F_{\mu \sigma }{}^{\kappa \tau } + \nn\\&& m_{1037}^{} 
F_{\alpha \beta \gamma }{}^{\epsilon } F^{\alpha \beta \gamma 
\delta } F_{\delta }{}^{\varepsilon \mu \nu } F_{\varepsilon 
}{}^{\sigma \lambda \kappa } \nabla_{\epsilon }F_{\nu \kappa 
\tau \omega } \nabla^{\omega }F_{\mu \sigma \lambda }{}^{\tau 
} + \nn\\&& m_{1038}^{} F_{\alpha \beta }{}^{\epsilon \varepsilon } 
F^{\alpha \beta \gamma \delta } F_{\gamma \epsilon }{}^{\mu 
\nu } F_{\delta }{}^{\sigma \lambda \kappa } 
\nabla_{\varepsilon }F_{\nu \kappa \tau \omega } 
\nabla^{\omega }F_{\mu \sigma \lambda }{}^{\tau } + \nn\\&& m_{1039}^{} 
F_{\alpha \beta }{}^{\epsilon \varepsilon } F^{\alpha \beta 
\gamma \delta } F_{\gamma \delta }{}^{\mu \nu } F_{\epsilon 
}{}^{\sigma \lambda \kappa } \nabla_{\varepsilon }F_{\nu 
\kappa \tau \omega } \nabla^{\omega }F_{\mu \sigma \lambda 
}{}^{\tau } + \nn\\&& m_{1051}^{} F_{\alpha \beta }{}^{\epsilon 
\varepsilon } F^{\alpha \beta \gamma \delta } F_{\gamma 
\epsilon }{}^{\mu \nu } F_{\delta \varepsilon \mu }{}^{\sigma 
} \nabla_{\tau }F_{\sigma \lambda \kappa \omega } 
\nabla^{\omega }F_{\nu }{}^{\lambda \kappa \tau } + \nn\\&& m_{1052}^{} 
F_{\alpha \beta }{}^{\epsilon \varepsilon } F^{\alpha \beta 
\gamma \delta } F_{\gamma \delta }{}^{\mu \nu } F_{\epsilon 
\varepsilon \mu }{}^{\sigma } \nabla_{\tau }F_{\sigma \lambda 
\kappa \omega } \nabla^{\omega }F_{\nu }{}^{\lambda \kappa 
\tau } + \nn\\&& m_{1053}^{} F_{\alpha \beta \gamma }{}^{\epsilon } 
F^{\alpha \beta \gamma \delta } F_{\delta }{}^{\varepsilon \mu 
\nu } F_{\epsilon \varepsilon \mu }{}^{\sigma } \nabla_{\tau 
}F_{\sigma \lambda \kappa \omega } \nabla^{\omega }F_{\nu 
}{}^{\lambda \kappa \tau } + \nn\\&& m_{1055}^{} F_{\alpha \beta 
}{}^{\epsilon \varepsilon } F^{\alpha \beta \gamma \delta } F_{
\gamma \epsilon }{}^{\mu \nu } F_{\delta \varepsilon \mu }{}^{
\sigma } \nabla_{\omega }F_{\sigma \lambda \kappa \tau } 
\nabla^{\omega }F_{\nu }{}^{\lambda \kappa \tau } + \nn\\&& m_{1056}^{} 
F_{\alpha \beta }{}^{\epsilon \varepsilon } F^{\alpha \beta 
\gamma \delta } F_{\gamma \delta }{}^{\mu \nu } F_{\epsilon 
\varepsilon \mu }{}^{\sigma } \nabla_{\omega }F_{\sigma 
\lambda \kappa \tau } \nabla^{\omega }F_{\nu }{}^{\lambda 
\kappa \tau } + \nn\\&& m_{1057}^{} F_{\alpha \beta \gamma 
}{}^{\epsilon } F^{\alpha \beta \gamma \delta } F_{\delta }{}^{
\varepsilon \mu \nu } F_{\epsilon \varepsilon \mu }{}^{\sigma 
} \nabla_{\omega }F_{\sigma \lambda \kappa \tau } 
\nabla^{\omega }F_{\nu }{}^{\lambda \kappa \tau } + \nn\\&& m_{1048}^{} 
F_{\alpha \beta \gamma }{}^{\epsilon } F^{\alpha \beta \gamma 
\delta } F_{\delta }{}^{\varepsilon \mu \nu } F_{\varepsilon 
\mu }{}^{\sigma \lambda } \nabla_{\epsilon }F_{\lambda \kappa 
\tau \omega } \nabla^{\omega }F_{\nu \sigma }{}^{\kappa \tau 
} + \nn\\&& m_{1049}^{} F_{\alpha \beta }{}^{\epsilon \varepsilon } 
F^{\alpha \beta \gamma \delta } F_{\gamma \epsilon }{}^{\mu 
\nu } F_{\delta \mu }{}^{\sigma \lambda } \nabla_{\varepsilon 
}F_{\lambda \kappa \tau \omega } \nabla^{\omega }F_{\nu 
\sigma }{}^{\kappa \tau } + \nn\\&& m_{1050}^{} F_{\alpha \beta 
}{}^{\epsilon \varepsilon } F^{\alpha \beta \gamma \delta } F_{
\gamma \delta \epsilon }{}^{\mu } F_{\varepsilon }{}^{\nu 
\sigma \lambda } \nabla_{\mu }F_{\lambda \kappa \tau \omega } 
\nabla^{\omega }F_{\nu \sigma }{}^{\kappa \tau } + \nn\\&& m_{1059}^{} 
F_{\alpha \beta }{}^{\epsilon \varepsilon } F^{\alpha \beta 
\gamma \delta } F_{\gamma \epsilon }{}^{\mu \nu } F_{\delta 
\varepsilon \mu \nu } \nabla_{\omega }F_{\sigma \lambda 
\kappa \tau } \nabla^{\omega }F^{\sigma \lambda \kappa \tau } 
+ \nn\\&& m_{1060}^{} F_{\alpha \beta }{}^{\epsilon \varepsilon } 
F^{\alpha \beta \gamma \delta } F_{\gamma \delta }{}^{\mu \nu 
} F_{\epsilon \varepsilon \mu \nu } \nabla_{\omega }F_{\sigma 
\lambda \kappa \tau } \nabla^{\omega }F^{\sigma \lambda 
\kappa \tau } + \nn\\&& m_{1061}^{} F_{\alpha \beta \gamma 
}{}^{\epsilon } F^{\alpha \beta \gamma \delta } F_{\delta }{}^{
\varepsilon \mu \nu } F_{\epsilon \varepsilon \mu \nu } 
\nabla_{\omega }F_{\sigma \lambda \kappa \tau } 
\nabla^{\omega }F^{\sigma \lambda \kappa \tau }+\nn\\&& m_{463}^{} F_{\epsilon }{}^{\sigma \lambda \kappa } F^{\epsilon 
\varepsilon \mu \nu } FF \nabla_{\kappa }F_{\nu \lambda \tau 
\omega } \nabla_{\sigma }F_{\varepsilon \mu }{}^{\tau \omega } 
+ \nn\\&& m_{370}^{} F_{\epsilon \varepsilon }{}^{\sigma \lambda } 
F^{\epsilon \varepsilon \mu \nu } FF \nabla_{\lambda }F_{\nu 
\kappa \tau \omega } \nabla_{\sigma }F_{\mu }{}^{\kappa \tau 
\omega } + \nn\\&& m_{639}^{} F^{\epsilon \varepsilon \mu \nu } 
F^{\sigma \lambda \kappa \tau } FF \nabla_{\lambda 
}F_{\epsilon \varepsilon \sigma }{}^{\omega } \nabla_{\tau }F_{
\mu \nu \kappa \omega } + \nn\\&& m_{658}^{} F^{\epsilon \varepsilon 
\mu \nu } F^{\sigma \lambda \kappa \tau } FF \nabla_{\sigma 
}F_{\epsilon \varepsilon \mu }{}^{\omega } \nabla_{\tau 
}F_{\nu \lambda \kappa \omega } + \nn\\&& m_{928}^{} F^{\epsilon 
\varepsilon \mu \nu } F^{\sigma \lambda \kappa \tau } FF 
\nabla_{\omega }F_{\sigma \lambda \kappa \tau } 
\nabla^{\omega }F_{\epsilon \varepsilon \mu \nu } + \nn\\&& m_{931}^{} 
F^{\epsilon \varepsilon \mu \nu } F^{\sigma \lambda \kappa 
\tau } FF \nabla_{\tau }F_{\nu \lambda \kappa \omega } 
\nabla^{\omega }F_{\epsilon \varepsilon \mu \sigma } + 
\nn\\&& m_{934}^{} F^{\epsilon \varepsilon \mu \nu } F^{\sigma \lambda 
\kappa \tau } FF \nabla_{\omega }F_{\nu \lambda \kappa \tau } 
\nabla^{\omega }F_{\epsilon \varepsilon \mu \sigma } + 
\nn\\&& m_{945}^{} F^{\epsilon \varepsilon \mu \nu } F^{\sigma \lambda 
\kappa \tau } FF \nabla_{\omega }F_{\mu \nu \kappa \tau } 
\nabla^{\omega }F_{\epsilon \varepsilon \sigma \lambda } + 
\nn\\&& m_{982}^{} F_{\epsilon }{}^{\sigma \lambda \kappa } F^{\epsilon 
\varepsilon \mu \nu } FF \nabla_{\tau }F_{\sigma \lambda 
\kappa \omega } \nabla^{\omega }F_{\varepsilon \mu \nu 
}{}^{\tau } + \nn\\&& m_{986}^{} F_{\epsilon }{}^{\sigma \lambda \kappa 
} F^{\epsilon \varepsilon \mu \nu } FF \nabla_{\omega 
}F_{\sigma \lambda \kappa \tau } \nabla^{\omega 
}F_{\varepsilon \mu \nu }{}^{\tau } + \nn\\&& m_{991}^{} F_{\epsilon 
}{}^{\sigma \lambda \kappa } F^{\epsilon \varepsilon \mu \nu } 
FF \nabla_{\kappa }F_{\nu \lambda \tau \omega } 
\nabla^{\omega }F_{\varepsilon \mu \sigma }{}^{\tau } + 
\nn\\&& m_{995}^{} F_{\epsilon }{}^{\sigma \lambda \kappa } F^{\epsilon 
\varepsilon \mu \nu } FF \nabla_{\tau }F_{\nu \lambda \kappa 
\omega } \nabla^{\omega }F_{\varepsilon \mu \sigma }{}^{\tau } 
+ \nn\\&& m_{999}^{} F_{\epsilon }{}^{\sigma \lambda \kappa } 
F^{\epsilon \varepsilon \mu \nu } FF \nabla_{\omega }F_{\nu 
\lambda \kappa \tau } \nabla^{\omega }F_{\varepsilon \mu 
\sigma }{}^{\tau } + \nn\\&& m_{1030}^{} F_{\epsilon \varepsilon 
}{}^{\sigma \lambda } F^{\epsilon \varepsilon \mu \nu } FF 
\nabla_{\tau }F_{\sigma \lambda \kappa \omega } 
\nabla^{\omega }F_{\mu \nu }{}^{\kappa \tau } + \nn\\&& m_{1036}^{} F_{
\epsilon \varepsilon }{}^{\sigma \lambda } F^{\epsilon 
\varepsilon \mu \nu } FF \nabla_{\omega }F_{\sigma \lambda 
\kappa \tau } \nabla^{\omega }F_{\mu \nu }{}^{\kappa \tau } + 
\nn\\&& m_{1043}^{} F_{\epsilon \varepsilon }{}^{\sigma \lambda } 
F^{\epsilon \varepsilon \mu \nu } FF \nabla_{\tau }F_{\nu 
\lambda \kappa \omega } \nabla^{\omega }F_{\mu \sigma 
}{}^{\kappa \tau } + \nn\\&& m_{1047}^{} F_{\epsilon \varepsilon 
}{}^{\sigma \lambda } F^{\epsilon \varepsilon \mu \nu } FF 
\nabla_{\omega }F_{\nu \lambda \kappa \tau } \nabla^{\omega 
}F_{\mu \sigma }{}^{\kappa \tau } + \nn\\&& m_{1054}^{} F_{\epsilon 
\varepsilon \mu }{}^{\sigma } F^{\epsilon \varepsilon \mu \nu 
} FF \nabla_{\tau }F_{\sigma \lambda \kappa \omega } 
\nabla^{\omega }F_{\nu }{}^{\lambda \kappa \tau } + \nn\\&& m_{1058}^{} 
F_{\epsilon \varepsilon \mu }{}^{\sigma } F^{\epsilon 
\varepsilon \mu \nu } FF \nabla_{\omega }F_{\sigma \lambda 
\kappa \tau } \nabla^{\omega }F_{\nu }{}^{\lambda \kappa \tau 
}+\nn\\&&m_{1062}^{} FF^2 \nabla_{\omega }F_{\sigma \lambda \kappa \tau 
} \nabla^{\omega }F^{\sigma \lambda \kappa \tau }\labell{T6}
\eeqa

Finally, there are 217 couplings with structure of one Riemann curvature, two $F$ and  two $\nabla F$, \ie
\beqa
{\cal L}_6^{RF^2(\prt F)^2}&=&m_{276}^{} F_{\epsilon }{}^{\sigma \lambda \kappa } F^{\epsilon 
\varepsilon \mu \nu } R^{\alpha \beta \gamma \delta } 
\nabla_{\gamma }F_{\alpha \varepsilon \mu \sigma } 
\nabla_{\delta }F_{\beta \nu \lambda \kappa } + \nn\\&& m_{277}^{} 
F_{\epsilon \varepsilon }{}^{\sigma \lambda } F^{\epsilon 
\varepsilon \mu \nu } R^{\alpha \beta \gamma \delta } 
\nabla_{\gamma }F_{\alpha \mu \sigma }{}^{\kappa } 
\nabla_{\delta }F_{\beta \nu \lambda \kappa } + \nn\\&& m_{278}^{} 
F_{\epsilon }{}^{\sigma \lambda \kappa } F^{\epsilon 
\varepsilon \mu \nu } R^{\alpha \beta \gamma \delta } 
\nabla_{\gamma }F_{\alpha \varepsilon \mu \nu } 
\nabla_{\delta }F_{\beta \sigma \lambda \kappa } + \nn\\&& m_{279}^{} 
F_{\epsilon \varepsilon }{}^{\sigma \lambda } F^{\epsilon 
\varepsilon \mu \nu } R^{\alpha \beta \gamma \delta } 
\nabla_{\gamma }F_{\alpha \mu \nu }{}^{\kappa } 
\nabla_{\delta }F_{\beta \sigma \lambda \kappa } + \nn\\&& m_{280}^{} 
F_{\epsilon \varepsilon \mu }{}^{\sigma } F^{\epsilon 
\varepsilon \mu \nu } R^{\alpha \beta \gamma \delta } 
\nabla_{\gamma }F_{\alpha \nu }{}^{\lambda \kappa } 
\nabla_{\delta }F_{\beta \sigma \lambda \kappa }  + \nn\\&& m_{388}^{} F_{\alpha }{}^{\epsilon 
\varepsilon \mu } F^{\nu \sigma \lambda \kappa } R^{\alpha 
\beta \gamma \delta } \nabla_{\epsilon }F_{\beta \nu \sigma 
\lambda } \nabla_{\kappa }F_{\gamma \delta \varepsilon \mu } + 
\nn\\&& m_{389}^{} F_{\alpha }{}^{\epsilon \varepsilon \mu } F^{\nu 
\sigma \lambda \kappa } R^{\alpha \beta \gamma \delta } 
\nabla_{\varepsilon }F_{\beta \epsilon \nu \sigma } 
\nabla_{\kappa }F_{\gamma \delta \mu \lambda } + \nn\\&& m_{398}^{} F_{
\alpha }{}^{\epsilon \varepsilon \mu } F^{\nu \sigma \lambda 
\kappa } R^{\alpha \beta \gamma \delta } \nabla_{\gamma 
}F_{\beta \nu \sigma \lambda } \nabla_{\kappa }F_{\delta 
\epsilon \varepsilon \mu } + \nn\\&& m_{402}^{} F_{\alpha }{}^{\epsilon 
\varepsilon \mu } F^{\nu \sigma \lambda \kappa } R^{\alpha 
\beta \gamma \delta } \nabla_{\gamma }F_{\beta \epsilon \nu 
\sigma } \nabla_{\kappa }F_{\delta \varepsilon \mu \lambda } + 
\nn\\&& m_{403}^{} F_{\alpha }{}^{\epsilon \varepsilon \mu } F^{\nu 
\sigma \lambda \kappa } R^{\alpha \beta \gamma \delta } 
\nabla_{\epsilon }F_{\beta \gamma \nu \sigma } \nabla_{\kappa 
}F_{\delta \varepsilon \mu \lambda } + \nn\\&& m_{406}^{} F_{\epsilon 
}{}^{\sigma \lambda \kappa } F^{\epsilon \varepsilon \mu \nu } 
R^{\alpha \beta \gamma \delta } \nabla_{\gamma }F_{\alpha 
\beta \varepsilon \sigma } \nabla_{\kappa }F_{\delta \mu \nu 
\lambda } + \nn\\&& m_{407}^{} F_{\alpha }{}^{\epsilon \varepsilon \mu } 
F^{\nu \sigma \lambda \kappa } R^{\alpha \beta \gamma \delta 
} \nabla_{\gamma }F_{\beta \epsilon \varepsilon \nu } \nabla_{
\kappa }F_{\delta \mu \sigma \lambda } + \nn\\&& m_{408}^{} F_{\alpha 
}{}^{\epsilon \varepsilon \mu } F^{\nu \sigma \lambda \kappa } 
R^{\alpha \beta \gamma \delta } \nabla_{\varepsilon }F_{\beta 
\gamma \epsilon \nu } \nabla_{\kappa }F_{\delta \mu \sigma 
\lambda } + \nn\\&& m_{411}^{} F_{\epsilon }{}^{\sigma \lambda \kappa } 
F^{\epsilon \varepsilon \mu \nu } R^{\alpha \beta \gamma 
\delta } \nabla_{\gamma }F_{\alpha \beta \varepsilon \mu } 
\nabla_{\kappa }F_{\delta \nu \sigma \lambda } + \nn\\&& m_{412}^{} F_{
\epsilon \varepsilon }{}^{\sigma \lambda } F^{\epsilon 
\varepsilon \mu \nu } R^{\alpha \beta \gamma \delta } 
\nabla_{\gamma }F_{\alpha \beta \mu }{}^{\kappa } 
\nabla_{\kappa }F_{\delta \nu \sigma \lambda } + \nn\\&& m_{413}^{} F_{
\alpha }{}^{\epsilon \varepsilon \mu } F^{\nu \sigma \lambda 
\kappa } R^{\alpha \beta \gamma \delta } \nabla_{\gamma 
}F_{\beta \epsilon \varepsilon \mu } \nabla_{\kappa }F_{\delta 
\nu \sigma \lambda } + \nn\\&& m_{414}^{} F_{\alpha }{}^{\epsilon 
\varepsilon \mu } F_{\epsilon }{}^{\nu \sigma \lambda } 
R^{\alpha \beta \gamma \delta } \nabla_{\gamma }F_{\beta 
\varepsilon \mu }{}^{\kappa } \nabla_{\kappa }F_{\delta \nu 
\sigma \lambda } + \nn\\&& m_{415}^{} F_{\alpha }{}^{\epsilon 
\varepsilon \mu } F_{\epsilon \varepsilon }{}^{\nu \sigma } R^{
\alpha \beta \gamma \delta } \nabla_{\gamma }F_{\beta \mu 
}{}^{\lambda \kappa } \nabla_{\kappa }F_{\delta \nu \sigma 
\lambda } + \nn\\&& m_{419}^{} F_{\alpha }{}^{\epsilon \varepsilon \mu } 
F^{\nu \sigma \lambda \kappa } R^{\alpha \beta \gamma \delta 
} \nabla_{\delta }F_{\beta \gamma \nu \sigma } \nabla_{\kappa 
}F_{\epsilon \varepsilon \mu \lambda } + \nn\\&& m_{423}^{} F_{\alpha 
}{}^{\epsilon \varepsilon \mu } F^{\nu \sigma \lambda \kappa } 
R^{\alpha \beta \gamma \delta } \nabla_{\delta }F_{\beta 
\gamma \epsilon \nu } \nabla_{\kappa }F_{\varepsilon \mu 
\sigma \lambda } + \nn\\&& m_{424}^{} F_{\alpha }{}^{\epsilon 
\varepsilon \mu } F_{\epsilon }{}^{\nu \sigma \lambda } 
R^{\alpha \beta \gamma \delta } \nabla_{\delta }F_{\beta 
\gamma \nu }{}^{\kappa } \nabla_{\kappa }F_{\varepsilon \mu 
\sigma \lambda } + \nn\\&& m_{426}^{} F_{\alpha \beta }{}^{\epsilon 
\varepsilon } F^{\mu \nu \sigma \lambda } R^{\alpha \beta 
\gamma \delta } \nabla_{\epsilon }F_{\gamma \delta \mu 
}{}^{\kappa } \nabla_{\kappa }F_{\varepsilon \nu \sigma 
\lambda } + \nn\\&& m_{439}^{} F_{\alpha }{}^{\epsilon \varepsilon \mu } 
F^{\nu \sigma \lambda \kappa } R^{\alpha \beta \gamma \delta 
} \nabla_{\delta }F_{\beta \gamma \epsilon \varepsilon } 
\nabla_{\kappa }F_{\mu \nu \sigma \lambda } + \nn\\&& m_{440}^{} 
F_{\alpha }{}^{\epsilon \varepsilon \mu } F_{\epsilon }{}^{\nu 
\sigma \lambda } R^{\alpha \beta \gamma \delta } 
\nabla_{\delta }F_{\beta \gamma \varepsilon }{}^{\kappa } 
\nabla_{\kappa }F_{\mu \nu \sigma \lambda } + \nn\\&& m_{441}^{} 
F_{\alpha }{}^{\epsilon \varepsilon \mu } F_{\epsilon 
\varepsilon }{}^{\nu \sigma } R^{\alpha \beta \gamma \delta } 
\nabla_{\delta }F_{\beta \gamma }{}^{\lambda \kappa } \nabla_{
\kappa }F_{\mu \nu \sigma \lambda } + \nn\\&& m_{442}^{} F_{\alpha 
}{}^{\epsilon \varepsilon \mu } F_{\gamma }{}^{\nu \sigma 
\lambda } R^{\alpha \beta \gamma \delta } \nabla_{\delta 
}F_{\beta \epsilon \varepsilon }{}^{\kappa } \nabla_{\kappa 
}F_{\mu \nu \sigma \lambda } + \nn\\&& m_{443}^{} F_{\alpha 
}{}^{\epsilon \varepsilon \mu } F_{\gamma \epsilon }{}^{\nu 
\sigma } R^{\alpha \beta \gamma \delta } \nabla_{\delta 
}F_{\beta \varepsilon }{}^{\lambda \kappa } \nabla_{\kappa }F_{
\mu \nu \sigma \lambda } + \nn\\&& m_{444}^{} F_{\alpha }{}^{\epsilon 
\varepsilon \mu } F_{\beta }{}^{\nu \sigma \lambda } R^{\alpha 
\beta \gamma \delta } \nabla_{\varepsilon }F_{\gamma \delta 
\epsilon }{}^{\kappa } \nabla_{\kappa }F_{\mu \nu \sigma 
\lambda } + \nn\\&& m_{445}^{} F_{\alpha \beta }{}^{\epsilon \varepsilon 
} F^{\mu \nu \sigma \lambda } R^{\alpha \beta \gamma \delta } 
\nabla_{\varepsilon }F_{\gamma \delta \epsilon }{}^{\kappa } 
\nabla_{\kappa }F_{\mu \nu \sigma \lambda } + \nn\\&& m_{446}^{} 
F_{\alpha }{}^{\epsilon \varepsilon \mu } F_{\beta \epsilon 
}{}^{\nu \sigma } R^{\alpha \beta \gamma \delta } 
\nabla_{\varepsilon }F_{\gamma \delta }{}^{\lambda \kappa } 
\nabla_{\kappa }F_{\mu \nu \sigma \lambda } + \nn\\&& m_{447}^{} 
F_{\alpha \beta }{}^{\epsilon \varepsilon } F_{\epsilon 
}{}^{\mu \nu \sigma } R^{\alpha \beta \gamma \delta } 
\nabla_{\varepsilon }F_{\gamma \delta }{}^{\lambda \kappa } 
\nabla_{\kappa }F_{\mu \nu \sigma \lambda } + \nn\\&& m_{448}^{} 
F_{\alpha \beta }{}^{\epsilon \varepsilon } F_{\gamma }{}^{\mu 
\nu \sigma } R^{\alpha \beta \gamma \delta } 
\nabla_{\varepsilon }F_{\delta \epsilon }{}^{\lambda \kappa } 
\nabla_{\kappa }F_{\mu \nu \sigma \lambda } + \nn\\&& m_{449}^{} 
F_{\alpha \beta }{}^{\epsilon \varepsilon } F_{\gamma \epsilon 
}{}^{\mu \nu } R^{\alpha \beta \gamma \delta } 
\nabla_{\varepsilon }F_{\delta }{}^{\sigma \lambda \kappa } 
\nabla_{\kappa }F_{\mu \nu \sigma \lambda } + \nn\\&& m_{471}^{} 
F_{\epsilon \varepsilon }{}^{\sigma \lambda } F^{\epsilon 
\varepsilon \mu \nu } R^{\alpha \beta \gamma \delta } 
\nabla_{\kappa }F_{\beta \delta \sigma \lambda } 
\nabla^{\kappa }F_{\alpha \gamma \mu \nu } + \nn\\&& m_{473}^{} 
F_{\epsilon \varepsilon }{}^{\sigma \lambda } F^{\epsilon 
\varepsilon \mu \nu } R^{\alpha \beta \gamma \delta } 
\nabla_{\kappa }F_{\beta \delta \nu \lambda } \nabla^{\kappa 
}F_{\alpha \gamma \mu \sigma } + \nn\\&& m_{476}^{} F_{\epsilon 
\varepsilon \mu }{}^{\sigma } F^{\epsilon \varepsilon \mu \nu 
} R^{\alpha \beta \gamma \delta } \nabla_{\kappa }F_{\beta 
\delta \sigma \lambda } \nabla^{\kappa }F_{\alpha \gamma \nu 
}{}^{\lambda }  + 
\nn\\&& m_{484}^{} F_{\alpha }{}^{\epsilon \varepsilon \mu } F_{\gamma 
}{}^{\nu \sigma \lambda } R^{\alpha \beta \gamma \delta } 
\nabla_{\kappa }F_{\delta \varepsilon \mu \lambda } 
\nabla^{\kappa }F_{\beta \epsilon \nu \sigma } + \nn\\&& m_{486}^{} F_{
\alpha }{}^{\epsilon \varepsilon \mu } F_{\epsilon }{}^{\nu 
\sigma \lambda } R^{\alpha \beta \gamma \delta } 
\nabla_{\kappa }F_{\gamma \delta \sigma \lambda } 
\nabla^{\kappa }F_{\beta \varepsilon \mu \nu } + \nn\\&& m_{492}^{} F_{
\alpha }{}^{\epsilon \varepsilon \mu } F_{\gamma \epsilon }{}^{
\nu \sigma } R^{\alpha \beta \gamma \delta } \nabla_{\kappa 
}F_{\delta \mu \sigma \lambda } \nabla^{\kappa }F_{\beta 
\varepsilon \nu }{}^{\lambda } + \nn\\&& m_{489}^{} F_{\alpha 
}{}^{\epsilon \varepsilon \mu } F_{\epsilon }{}^{\nu \sigma 
\lambda } R^{\alpha \beta \gamma \delta } \nabla_{\kappa 
}F_{\gamma \delta \mu \lambda } \nabla^{\kappa }F_{\beta 
\varepsilon \nu \sigma } + \nn\\&& m_{495}^{} F_{\alpha }{}^{\epsilon 
\varepsilon \mu } F_{\epsilon \varepsilon }{}^{\nu \sigma } R^{
\alpha \beta \gamma \delta } \nabla_{\kappa }F_{\gamma \delta 
\sigma \lambda } \nabla^{\kappa }F_{\beta \mu \nu 
}{}^{\lambda } + \nn\\&& m_{498}^{} F_{\alpha }{}^{\epsilon \varepsilon 
\mu } F_{\epsilon }{}^{\nu \sigma \lambda } R^{\alpha \beta 
\gamma \delta } \nabla_{\kappa }F_{\gamma \delta \varepsilon 
\mu } \nabla^{\kappa }F_{\beta \nu \sigma \lambda } + 
\nn\\&& m_{499}^{} F_{\alpha }{}^{\epsilon \varepsilon \mu } F_{\gamma 
}{}^{\nu \sigma \lambda } R^{\alpha \beta \gamma \delta } 
\nabla_{\kappa }F_{\delta \epsilon \varepsilon \mu } 
\nabla^{\kappa }F_{\beta \nu \sigma \lambda } + \nn\\&& m_{504}^{} 
F_{\alpha }{}^{\epsilon \varepsilon \mu } F_{\epsilon 
\varepsilon }{}^{\nu \sigma } R^{\alpha \beta \gamma \delta } 
\nabla_{\kappa }F_{\gamma \delta \mu \lambda } \nabla^{\kappa 
}F_{\beta \nu \sigma }{}^{\lambda } + \nn\\&& m_{505}^{} F_{\alpha }{}^{
\epsilon \varepsilon \mu } F_{\gamma \epsilon }{}^{\nu \sigma 
} R^{\alpha \beta \gamma \delta } \nabla_{\kappa }F_{\delta 
\varepsilon \mu \lambda } \nabla^{\kappa }F_{\beta \nu \sigma 
}{}^{\lambda } + \nn\\&& m_{509}^{} F_{\alpha }{}^{\epsilon \varepsilon 
\mu } F_{\epsilon \varepsilon \mu }{}^{\nu } R^{\alpha \beta 
\gamma \delta } \nabla_{\kappa }F_{\gamma \delta \sigma 
\lambda } \nabla^{\kappa }F_{\beta \nu }{}^{\sigma \lambda } + 
\nn\\&& m_{510}^{} F_{\alpha }{}^{\epsilon \varepsilon \mu } F_{\gamma 
\epsilon \varepsilon }{}^{\nu } R^{\alpha \beta \gamma \delta 
} \nabla_{\kappa }F_{\delta \mu \sigma \lambda } 
\nabla^{\kappa }F_{\beta \nu }{}^{\sigma \lambda } + \nn\\&& m_{512}^{} 
F_{\alpha }{}^{\epsilon \varepsilon \mu } F_{\gamma \epsilon 
\varepsilon \mu } R^{\alpha \beta \gamma \delta } 
\nabla_{\kappa }F_{\delta \nu \sigma \lambda } \nabla^{\kappa 
}F_{\beta }{}^{\nu \sigma \lambda } + \nn\\&& m_{514}^{} F_{\alpha }{}^{
\epsilon \varepsilon \mu } F_{\beta }{}^{\nu \sigma \lambda } 
R^{\alpha \beta \gamma \delta } \nabla_{\kappa }F_{\mu \nu 
\sigma \lambda } \nabla^{\kappa }F_{\gamma \delta \epsilon 
\varepsilon } + \nn\\&& m_{515}^{} F_{\alpha \beta }{}^{\epsilon 
\varepsilon } F^{\mu \nu \sigma \lambda } R^{\alpha \beta 
\gamma \delta } \nabla_{\kappa }F_{\mu \nu \sigma \lambda } 
\nabla^{\kappa }F_{\gamma \delta \epsilon \varepsilon } + 
\nn\\&& m_{517}^{} F_{\alpha \beta }{}^{\epsilon \varepsilon } F^{\mu 
\nu \sigma \lambda } R^{\alpha \beta \gamma \delta } \nabla_{
\kappa }F_{\varepsilon \nu \sigma \lambda } \nabla^{\kappa 
}F_{\gamma \delta \epsilon \mu } + \nn\\&& m_{518}^{} F_{\alpha 
}{}^{\epsilon \varepsilon \mu } F_{\beta }{}^{\nu \sigma 
\lambda } R^{\alpha \beta \gamma \delta } \nabla_{\kappa 
}F_{\varepsilon \mu \sigma \lambda } \nabla^{\kappa }F_{\gamma 
\delta \epsilon \nu } + \nn\\&& m_{524}^{} F_{\alpha }{}^{\epsilon 
\varepsilon \mu } F_{\beta \epsilon }{}^{\nu \sigma } 
R^{\alpha \beta \gamma \delta } \nabla_{\kappa }F_{\mu \nu 
\sigma \lambda } \nabla^{\kappa }F_{\gamma \delta \varepsilon 
}{}^{\lambda } + \nn\\&& m_{525}^{} F_{\alpha \beta }{}^{\epsilon 
\varepsilon } F_{\epsilon }{}^{\mu \nu \sigma } R^{\alpha 
\beta \gamma \delta } \nabla_{\kappa }F_{\mu \nu \sigma 
\lambda } \nabla^{\kappa }F_{\gamma \delta \varepsilon 
}{}^{\lambda } + \nn\\&& m_{519}^{} F_{\alpha }{}^{\epsilon \varepsilon 
\mu } F_{\epsilon }{}^{\nu \sigma \lambda } R^{\alpha \beta 
\gamma \delta } \nabla_{\beta }F_{\nu \sigma \lambda \kappa } 
\nabla^{\kappa }F_{\gamma \delta \varepsilon \mu } + \nn\\&& m_{520}^{} 
F_{\alpha }{}^{\epsilon \varepsilon \mu } F_{\epsilon }{}^{\nu 
\sigma \lambda } R^{\alpha \beta \gamma \delta } 
\nabla_{\beta }F_{\mu \sigma \lambda \kappa } \nabla^{\kappa 
}F_{\gamma \delta \varepsilon \nu } + \nn\\&& m_{528}^{} F_{\alpha }{}^{
\epsilon \varepsilon \mu } F_{\epsilon \varepsilon }{}^{\nu 
\sigma } R^{\alpha \beta \gamma \delta } \nabla_{\beta 
}F_{\nu \sigma \lambda \kappa } \nabla^{\kappa }F_{\gamma 
\delta \mu }{}^{\lambda } + \nn\\&& m_{531}^{} F_{\alpha \beta 
}{}^{\epsilon \varepsilon } F_{\epsilon }{}^{\mu \nu \sigma } 
R^{\alpha \beta \gamma \delta } \nabla_{\kappa }F_{\varepsilon 
\nu \sigma \lambda } \nabla^{\kappa }F_{\gamma \delta \mu 
}{}^{\lambda } + \nn\\&& m_{527}^{} F_{\alpha \beta }{}^{\epsilon 
\varepsilon } F^{\mu \nu \sigma \lambda } R^{\alpha \beta 
\gamma \delta } \nabla_{\kappa }F_{\epsilon \varepsilon \sigma 
\lambda } \nabla^{\kappa }F_{\gamma \delta \mu \nu } + 
\nn\\&& m_{533}^{} F_{\alpha }{}^{\epsilon \varepsilon \mu } F_{\epsilon 
\varepsilon }{}^{\nu \sigma } R^{\alpha \beta \gamma \delta } 
\nabla_{\beta }F_{\mu \sigma \lambda \kappa } \nabla^{\kappa 
}F_{\gamma \delta \nu }{}^{\lambda } + \nn\\&& m_{532}^{} F_{\alpha 
}{}^{\epsilon \varepsilon \mu } F_{\epsilon }{}^{\nu \sigma 
\lambda } R^{\alpha \beta \gamma \delta } \nabla_{\beta 
}F_{\varepsilon \mu \lambda \kappa } \nabla^{\kappa }F_{\gamma 
\delta \nu \sigma } + \nn\\&& m_{534}^{} F_{\alpha }{}^{\epsilon 
\varepsilon \mu } F_{\epsilon \varepsilon \mu }{}^{\nu } 
R^{\alpha \beta \gamma \delta } \nabla_{\beta }F_{\nu \sigma 
\lambda \kappa } \nabla^{\kappa }F_{\gamma \delta }{}^{\sigma 
\lambda } + \nn\\&& m_{537}^{} F_{\alpha }{}^{\epsilon \varepsilon \mu } 
F_{\beta \epsilon \varepsilon }{}^{\nu } R^{\alpha \beta 
\gamma \delta } \nabla_{\kappa }F_{\mu \nu \sigma \lambda } 
\nabla^{\kappa }F_{\gamma \delta }{}^{\sigma \lambda } + 
\nn\\&& m_{538}^{} F_{\alpha \beta }{}^{\epsilon \varepsilon } 
F_{\epsilon \varepsilon }{}^{\mu \nu } R^{\alpha \beta \gamma 
\delta } \nabla_{\kappa }F_{\mu \nu \sigma \lambda } \nabla^{
\kappa }F_{\gamma \delta }{}^{\sigma \lambda } + \nn\\&& m_{541}^{} 
F_{\alpha }{}^{\epsilon \varepsilon \mu } F_{\beta }{}^{\nu 
\sigma \lambda } R^{\alpha \beta \gamma \delta } 
\nabla_{\kappa }F_{\delta \nu \sigma \lambda } \nabla^{\kappa 
}F_{\gamma \epsilon \varepsilon \mu } + \nn\\&& m_{542}^{} F_{\alpha 
\beta }{}^{\epsilon \varepsilon } F^{\mu \nu \sigma \lambda } 
R^{\alpha \beta \gamma \delta } \nabla_{\kappa }F_{\delta \nu 
\sigma \lambda } \nabla^{\kappa }F_{\gamma \epsilon 
\varepsilon \mu } + \nn\\&& m_{545}^{} F_{\alpha }{}^{\epsilon 
\varepsilon \mu } F_{\beta }{}^{\nu \sigma \lambda } R^{\alpha 
\beta \gamma \delta } \nabla_{\kappa }F_{\delta \mu \sigma 
\lambda } \nabla^{\kappa }F_{\gamma \epsilon \varepsilon \nu } 
+ \nn\\&& m_{546}^{} F_{\alpha \beta }{}^{\epsilon \varepsilon } F^{\mu 
\nu \sigma \lambda } R^{\alpha \beta \gamma \delta } \nabla_{
\varepsilon }F_{\delta \sigma \lambda \kappa } \nabla^{\kappa 
}F_{\gamma \epsilon \mu \nu } + \nn\\&& m_{548}^{} F_{\alpha \beta 
}{}^{\epsilon \varepsilon } F^{\mu \nu \sigma \lambda } 
R^{\alpha \beta \gamma \delta } \nabla_{\kappa }F_{\delta 
\varepsilon \sigma \lambda } \nabla^{\kappa }F_{\gamma 
\epsilon \mu \nu } + \nn\\&& m_{554}^{} F_{\alpha }{}^{\epsilon 
\varepsilon \mu } F_{\beta \epsilon }{}^{\nu \sigma } 
R^{\alpha \beta \gamma \delta } \nabla_{\kappa }F_{\delta \nu 
\sigma \lambda } \nabla^{\kappa }F_{\gamma \varepsilon \mu 
}{}^{\lambda } + \nn\\&& m_{555}^{} F_{\alpha \beta }{}^{\epsilon 
\varepsilon } F_{\epsilon }{}^{\mu \nu \sigma } R^{\alpha 
\beta \gamma \delta } \nabla_{\kappa }F_{\delta \nu \sigma 
\lambda } \nabla^{\kappa }F_{\gamma \varepsilon \mu 
}{}^{\lambda } + \nn\\&& m_{549}^{} F_{\alpha }{}^{\epsilon \varepsilon 
\mu } F_{\epsilon }{}^{\nu \sigma \lambda } R^{\alpha \beta 
\gamma \delta } \nabla_{\beta }F_{\delta \sigma \lambda 
\kappa } \nabla^{\kappa }F_{\gamma \varepsilon \mu \nu } + 
\nn\\&& m_{556}^{} F_{\alpha }{}^{\epsilon \varepsilon \mu } F_{\epsilon 
}{}^{\nu \sigma \lambda } R^{\alpha \beta \gamma \delta } 
\nabla_{\beta }F_{\delta \mu \lambda \kappa } \nabla^{\kappa 
}F_{\gamma \varepsilon \nu \sigma } + \nn\\&& m_{559}^{} F_{\alpha }{}^{
\epsilon \varepsilon \mu } F_{\epsilon \varepsilon }{}^{\nu 
\sigma } R^{\alpha \beta \gamma \delta } \nabla_{\beta 
}F_{\delta \sigma \lambda \kappa } \nabla^{\kappa }F_{\gamma 
\mu \nu }{}^{\lambda } + \nn\\&& m_{560}^{} F_{\alpha \beta 
}{}^{\epsilon \varepsilon } F_{\epsilon }{}^{\mu \nu \sigma } 
R^{\alpha \beta \gamma \delta } \nabla_{\delta }F_{\varepsilon 
\sigma \lambda \kappa } \nabla^{\kappa }F_{\gamma \mu \nu 
}{}^{\lambda } + \nn\\&& m_{561}^{} F_{\alpha \beta }{}^{\epsilon 
\varepsilon } F_{\epsilon }{}^{\mu \nu \sigma } R^{\alpha 
\beta \gamma \delta } \nabla_{\varepsilon }F_{\delta \sigma 
\lambda \kappa } \nabla^{\kappa }F_{\gamma \mu \nu 
}{}^{\lambda } + \nn\\&& m_{558}^{} F_{\alpha \beta }{}^{\epsilon 
\varepsilon } F^{\mu \nu \sigma \lambda } R^{\alpha \beta 
\gamma \delta } \nabla_{\delta }F_{\epsilon \varepsilon 
\lambda \kappa } \nabla^{\kappa }F_{\gamma \mu \nu \sigma } + 
\nn\\&& m_{566}^{} F_{\alpha }{}^{\epsilon \varepsilon \mu } F_{\beta 
\epsilon \varepsilon }{}^{\nu } R^{\alpha \beta \gamma \delta 
} \nabla_{\kappa }F_{\delta \nu \sigma \lambda } 
\nabla^{\kappa }F_{\gamma \mu }{}^{\sigma \lambda } + 
\nn\\&& m_{567}^{} F_{\alpha \beta }{}^{\epsilon \varepsilon } 
F_{\epsilon \varepsilon }{}^{\mu \nu } R^{\alpha \beta \gamma 
\delta } \nabla_{\kappa }F_{\delta \nu \sigma \lambda } 
\nabla^{\kappa }F_{\gamma \mu }{}^{\sigma \lambda } + 
\nn\\&& m_{568}^{} F_{\alpha }{}^{\epsilon \varepsilon \mu } F_{\epsilon 
\varepsilon \mu }{}^{\nu } R^{\alpha \beta \gamma \delta } 
\nabla_{\beta }F_{\delta \sigma \lambda \kappa } 
\nabla^{\kappa }F_{\gamma \nu }{}^{\sigma \lambda } + 
\nn\\&& m_{570}^{} F_{\alpha \beta }{}^{\epsilon \varepsilon } F_{\gamma 
}{}^{\mu \nu \sigma } R^{\alpha \beta \gamma \delta } 
\nabla_{\kappa }F_{\mu \nu \sigma \lambda } \nabla^{\kappa 
}F_{\delta \epsilon \varepsilon }{}^{\lambda } + \nn\\&& m_{573}^{} 
F_{\alpha \beta }{}^{\epsilon \varepsilon } F_{\gamma }{}^{\mu 
\nu \sigma } R^{\alpha \beta \gamma \delta } \nabla_{\kappa 
}F_{\varepsilon \nu \sigma \lambda } \nabla^{\kappa }F_{\delta 
\epsilon \mu }{}^{\lambda } + \nn\\&& m_{575}^{} F_{\alpha \beta 
}{}^{\epsilon \varepsilon } F_{\gamma \epsilon }{}^{\mu \nu } 
R^{\alpha \beta \gamma \delta } \nabla_{\kappa }F_{\mu \nu 
\sigma \lambda } \nabla^{\kappa }F_{\delta \varepsilon 
}{}^{\sigma \lambda } + \nn\\&& m_{578}^{} F_{\alpha \beta }{}^{\epsilon 
\varepsilon } F_{\gamma }{}^{\mu \nu \sigma } R^{\alpha \beta 
\gamma \delta } \nabla_{\kappa }F_{\epsilon \varepsilon \sigma 
\lambda } \nabla^{\kappa }F_{\delta \mu \nu }{}^{\lambda } + 
\nn\\&& m_{581}^{} F_{\alpha \beta }{}^{\epsilon \varepsilon } F_{\gamma 
\epsilon }{}^{\mu \nu } R^{\alpha \beta \gamma \delta } 
\nabla_{\kappa }F_{\varepsilon \nu \sigma \lambda } 
\nabla^{\kappa }F_{\delta \mu }{}^{\sigma \lambda } + 
\nn\\&& m_{582}^{} F_{\alpha \beta }{}^{\epsilon \varepsilon } F_{\gamma 
\epsilon \varepsilon }{}^{\mu } R^{\alpha \beta \gamma \delta 
} \nabla_{\kappa }F_{\mu \nu \sigma \lambda } \nabla^{\kappa 
}F_{\delta }{}^{\nu \sigma \lambda } + \nn\\&& m_{583}^{} F_{\alpha 
\beta }{}^{\epsilon \varepsilon } F_{\gamma }{}^{\mu \nu 
\sigma } R^{\alpha \beta \gamma \delta } \nabla_{\delta 
}F_{\nu \sigma \lambda \kappa } \nabla^{\kappa }F_{\epsilon 
\varepsilon \mu }{}^{\lambda } + \nn\\&& m_{585}^{} F_{\alpha \beta 
}{}^{\epsilon \varepsilon } F_{\gamma \delta }{}^{\mu \nu } R^{
\alpha \beta \gamma \delta } \nabla_{\kappa }F_{\mu \nu 
\sigma \lambda } \nabla^{\kappa }F_{\epsilon \varepsilon 
}{}^{\sigma \lambda } + \nn\\&& m_{586}^{} F_{\alpha \beta }{}^{\epsilon 
\varepsilon } F_{\gamma }{}^{\mu \nu \sigma } R^{\alpha \beta 
\gamma \delta } \nabla_{\delta }F_{\varepsilon \sigma \lambda 
\kappa } \nabla^{\kappa }F_{\epsilon \mu \nu }{}^{\lambda } + 
\nn\\&& m_{588}^{} F_{\alpha \beta }{}^{\epsilon \varepsilon } F_{\gamma 
\delta }{}^{\mu \nu } R^{\alpha \beta \gamma \delta } 
\nabla_{\kappa }F_{\varepsilon \nu \sigma \lambda } 
\nabla^{\kappa }F_{\epsilon \mu }{}^{\sigma \lambda } + 
\nn\\&& m_{589}^{} F_{\alpha \beta }{}^{\epsilon \varepsilon } F_{\gamma 
\epsilon }{}^{\mu \nu } R^{\alpha \beta \gamma \delta } 
\nabla_{\delta }F_{\nu \sigma \lambda \kappa } \nabla^{\kappa 
}F_{\varepsilon \mu }{}^{\sigma \lambda } + \nn\\&& m_{591}^{} F_{\alpha 
\beta }{}^{\epsilon \varepsilon } F_{\gamma \delta \epsilon 
}{}^{\mu } R^{\alpha \beta \gamma \delta } \nabla_{\kappa }F_{
\mu \nu \sigma \lambda } \nabla^{\kappa }F_{\varepsilon 
}{}^{\nu \sigma \lambda } + \nn\\&& m_{592}^{} F_{\alpha \beta 
}{}^{\epsilon \varepsilon } F_{\gamma \epsilon \varepsilon }{}^{
\mu } R^{\alpha \beta \gamma \delta } \nabla_{\delta }F_{\nu 
\sigma \lambda \kappa } \nabla^{\kappa }F_{\mu }{}^{\nu 
\sigma \lambda } + \nn\\&& m_{593}^{} F_{\alpha \beta }{}^{\epsilon 
\varepsilon } F_{\gamma \delta \epsilon \varepsilon } R^{\alpha 
\beta \gamma \delta } \nabla_{\kappa }F_{\mu \nu \sigma 
\lambda } \nabla^{\kappa }F^{\mu \nu \sigma \lambda } + 
\nn\\&& m_{399}^{} F_{\alpha }{}^{\epsilon \varepsilon \mu } F^{\nu 
\sigma \lambda \kappa } R^{\alpha \beta \gamma \delta } 
\nabla_{\kappa }F_{\delta \epsilon \varepsilon \mu } 
\nabla_{\lambda }F_{\beta \gamma \nu \sigma } + \nn\\&& m_{472}^{} 
F_{\epsilon \varepsilon }{}^{\sigma \lambda } F^{\epsilon 
\varepsilon \mu \nu } R^{\alpha \beta \gamma \delta } 
\nabla^{\kappa }F_{\alpha \gamma \mu \sigma } \nabla_{\lambda 
}F_{\beta \delta \nu \kappa } + \nn\\&& m_{470}^{} F_{\epsilon 
\varepsilon }{}^{\sigma \lambda } F^{\epsilon \varepsilon \mu 
\nu } R^{\alpha \beta \gamma \delta } \nabla^{\kappa 
}F_{\alpha \gamma \mu \nu } \nabla_{\lambda }F_{\beta \delta 
\sigma \kappa } + \nn\\&& m_{475}^{} F_{\epsilon \varepsilon \mu 
}{}^{\sigma } F^{\epsilon \varepsilon \mu \nu } R^{\alpha 
\beta \gamma \delta } \nabla^{\kappa }F_{\alpha \gamma \nu 
}{}^{\lambda } \nabla_{\lambda }F_{\beta \delta \sigma \kappa 
}  + \nn\\&& m_{330}^{} 
F_{\alpha }{}^{\epsilon \varepsilon \mu } F_{\epsilon }{}^{\nu 
\sigma \lambda } R^{\alpha \beta \gamma \delta } 
\nabla_{\varepsilon }F_{\beta \nu \sigma }{}^{\kappa } 
\nabla_{\lambda }F_{\gamma \delta \mu \kappa } + \nn\\&& m_{488}^{} F_{
\alpha }{}^{\epsilon \varepsilon \mu } F_{\epsilon }{}^{\nu 
\sigma \lambda } R^{\alpha \beta \gamma \delta } 
\nabla^{\kappa }F_{\beta \varepsilon \nu \sigma } 
\nabla_{\lambda }F_{\gamma \delta \mu \kappa } + \nn\\&& m_{502}^{} F_{
\alpha }{}^{\epsilon \varepsilon \mu } F_{\epsilon \varepsilon 
}{}^{\nu \sigma } R^{\alpha \beta \gamma \delta } 
\nabla^{\kappa }F_{\beta \nu \sigma }{}^{\lambda } 
\nabla_{\lambda }F_{\gamma \delta \mu \kappa } + \nn\\&& m_{485}^{} F_{
\alpha }{}^{\epsilon \varepsilon \mu } F_{\epsilon }{}^{\nu 
\sigma \lambda } R^{\alpha \beta \gamma \delta } 
\nabla^{\kappa }F_{\beta \varepsilon \mu \nu } 
\nabla_{\lambda }F_{\gamma \delta \sigma \kappa } + \nn\\&& m_{494}^{} 
F_{\alpha }{}^{\epsilon \varepsilon \mu } F_{\epsilon 
\varepsilon }{}^{\nu \sigma } R^{\alpha \beta \gamma \delta } 
\nabla^{\kappa }F_{\beta \mu \nu }{}^{\lambda } 
\nabla_{\lambda }F_{\gamma \delta \sigma \kappa } + \nn\\&& m_{507}^{} 
F_{\alpha }{}^{\epsilon \varepsilon \mu } F_{\epsilon 
\varepsilon \mu }{}^{\nu } R^{\alpha \beta \gamma \delta } 
\nabla^{\kappa }F_{\beta \nu }{}^{\sigma \lambda } 
\nabla_{\lambda }F_{\gamma \delta \sigma \kappa } + \nn\\&& m_{334}^{} 
F_{\alpha }{}^{\epsilon \varepsilon \mu } F_{\epsilon }{}^{\nu 
\sigma \lambda } R^{\alpha \beta \gamma \delta } 
\nabla_{\gamma }F_{\beta \nu \sigma }{}^{\kappa } 
\nabla_{\lambda }F_{\delta \varepsilon \mu \kappa } + 
\nn\\&& m_{335}^{} F_{\alpha }{}^{\epsilon \varepsilon \mu } F_{\gamma 
}{}^{\nu \sigma \lambda } R^{\alpha \beta \gamma \delta } 
\nabla_{\epsilon }F_{\beta \nu \sigma }{}^{\kappa } 
\nabla_{\lambda }F_{\delta \varepsilon \mu \kappa } + 
\nn\\&& m_{480}^{} F_{\alpha }{}^{\epsilon \varepsilon \mu } F_{\epsilon 
}{}^{\nu \sigma \lambda } R^{\alpha \beta \gamma \delta } 
\nabla^{\kappa }F_{\beta \gamma \nu \sigma } \nabla_{\lambda 
}F_{\delta \varepsilon \mu \kappa } + \nn\\&& m_{483}^{} F_{\alpha }{}^{
\epsilon \varepsilon \mu } F_{\gamma }{}^{\nu \sigma \lambda } 
R^{\alpha \beta \gamma \delta } \nabla^{\kappa }F_{\beta 
\epsilon \nu \sigma } \nabla_{\lambda }F_{\delta \varepsilon 
\mu \kappa } + \nn\\&& m_{503}^{} F_{\alpha }{}^{\epsilon \varepsilon 
\mu } F_{\gamma \epsilon }{}^{\nu \sigma } R^{\alpha \beta 
\gamma \delta } \nabla^{\kappa }F_{\beta \nu \sigma 
}{}^{\lambda } \nabla_{\lambda }F_{\delta \varepsilon \mu 
\kappa } + \nn\\&& m_{547}^{} F_{\alpha \beta }{}^{\epsilon \varepsilon 
} F^{\mu \nu \sigma \lambda } R^{\alpha \beta \gamma \delta } 
\nabla^{\kappa }F_{\gamma \epsilon \mu \nu } \nabla_{\lambda 
}F_{\delta \varepsilon \sigma \kappa } + \nn\\&& m_{337}^{} F_{\alpha 
}{}^{\epsilon \varepsilon \mu } F_{\epsilon }{}^{\nu \sigma 
\lambda } R^{\alpha \beta \gamma \delta } \nabla_{\gamma 
}F_{\beta \varepsilon \nu }{}^{\kappa } \nabla_{\lambda 
}F_{\delta \mu \sigma \kappa } + \nn\\&& m_{338}^{} F_{\alpha 
}{}^{\epsilon \varepsilon \mu } F_{\epsilon }{}^{\nu \sigma 
\lambda } R^{\alpha \beta \gamma \delta } \nabla_{\varepsilon 
}F_{\beta \gamma \nu }{}^{\kappa } \nabla_{\lambda }F_{\delta 
\mu \sigma \kappa } + \nn\\&& m_{339}^{} F_{\alpha }{}^{\epsilon 
\varepsilon \mu } F_{\gamma }{}^{\nu \sigma \lambda } 
R^{\alpha \beta \gamma \delta } \nabla_{\varepsilon }F_{\beta 
\epsilon \nu }{}^{\kappa } \nabla_{\lambda }F_{\delta \mu 
\sigma \kappa } + \nn\\&& m_{340}^{} F_{\alpha }{}^{\epsilon \varepsilon 
\mu } F_{\beta }{}^{\nu \sigma \lambda } R^{\alpha \beta 
\gamma \delta } \nabla_{\varepsilon }F_{\gamma \epsilon \nu 
}{}^{\kappa } \nabla_{\lambda }F_{\delta \mu \sigma \kappa } + 
\nn\\&& m_{479}^{} F_{\alpha }{}^{\epsilon \varepsilon \mu } F_{\epsilon 
}{}^{\nu \sigma \lambda } R^{\alpha \beta \gamma \delta } 
\nabla^{\kappa }F_{\beta \gamma \varepsilon \nu } 
\nabla_{\lambda }F_{\delta \mu \sigma \kappa } + \nn\\&& m_{491}^{} F_{
\alpha }{}^{\epsilon \varepsilon \mu } F_{\gamma \epsilon }{}^{
\nu \sigma } R^{\alpha \beta \gamma \delta } \nabla^{\kappa 
}F_{\beta \varepsilon \nu }{}^{\lambda } \nabla_{\lambda 
}F_{\delta \mu \sigma \kappa } + \nn\\&& m_{508}^{} F_{\alpha 
}{}^{\epsilon \varepsilon \mu } F_{\gamma \epsilon \varepsilon 
}{}^{\nu } R^{\alpha \beta \gamma \delta } \nabla^{\kappa }F_{
\beta \nu }{}^{\sigma \lambda } \nabla_{\lambda }F_{\delta 
\mu \sigma \kappa } + \nn\\&& m_{544}^{} F_{\alpha }{}^{\epsilon 
\varepsilon \mu } F_{\beta }{}^{\nu \sigma \lambda } R^{\alpha 
\beta \gamma \delta } \nabla^{\kappa }F_{\gamma \epsilon 
\varepsilon \nu } \nabla_{\lambda }F_{\delta \mu \sigma 
\kappa } + \nn\\&& m_{343}^{} F_{\epsilon \varepsilon }{}^{\sigma 
\lambda } F^{\epsilon \varepsilon \mu \nu } R^{\alpha \beta 
\gamma \delta } \nabla_{\gamma }F_{\alpha \beta \mu 
}{}^{\kappa } \nabla_{\lambda }F_{\delta \nu \sigma \kappa } + 
\nn\\&& m_{344}^{} F_{\alpha }{}^{\epsilon \varepsilon \mu } F_{\epsilon 
}{}^{\nu \sigma \lambda } R^{\alpha \beta \gamma \delta } 
\nabla_{\gamma }F_{\beta \varepsilon \mu }{}^{\kappa } 
\nabla_{\lambda }F_{\delta \nu \sigma \kappa } + \nn\\&& m_{511}^{} F_{
\alpha }{}^{\epsilon \varepsilon \mu } F_{\gamma \epsilon 
\varepsilon \mu } R^{\alpha \beta \gamma \delta } 
\nabla^{\kappa }F_{\beta }{}^{\nu \sigma \lambda } 
\nabla_{\lambda }F_{\delta \nu \sigma \kappa } + \nn\\&& m_{539}^{} F_{
\alpha }{}^{\epsilon \varepsilon \mu } F_{\beta }{}^{\nu 
\sigma \lambda } R^{\alpha \beta \gamma \delta } 
\nabla^{\kappa }F_{\gamma \epsilon \varepsilon \mu } 
\nabla_{\lambda }F_{\delta \nu \sigma \kappa } + \nn\\&& m_{540}^{} F_{
\alpha \beta }{}^{\epsilon \varepsilon } F^{\mu \nu \sigma 
\lambda } R^{\alpha \beta \gamma \delta } \nabla^{\kappa 
}F_{\gamma \epsilon \varepsilon \mu } \nabla_{\lambda 
}F_{\delta \nu \sigma \kappa } + \nn\\&& m_{552}^{} F_{\alpha 
}{}^{\epsilon \varepsilon \mu } F_{\beta \epsilon }{}^{\nu 
\sigma } R^{\alpha \beta \gamma \delta } \nabla^{\kappa 
}F_{\gamma \varepsilon \mu }{}^{\lambda } \nabla_{\lambda 
}F_{\delta \nu \sigma \kappa } + \nn\\&& m_{553}^{} F_{\alpha \beta 
}{}^{\epsilon \varepsilon } F_{\epsilon }{}^{\mu \nu \sigma } 
R^{\alpha \beta \gamma \delta } \nabla^{\kappa }F_{\gamma 
\varepsilon \mu }{}^{\lambda } \nabla_{\lambda }F_{\delta \nu 
\sigma \kappa } + \nn\\&& m_{564}^{} F_{\alpha }{}^{\epsilon \varepsilon 
\mu } F_{\beta \epsilon \varepsilon }{}^{\nu } R^{\alpha \beta 
\gamma \delta } \nabla^{\kappa }F_{\gamma \mu }{}^{\sigma 
\lambda } \nabla_{\lambda }F_{\delta \nu \sigma \kappa } + 
\nn\\&& m_{565}^{} F_{\alpha \beta }{}^{\epsilon \varepsilon } 
F_{\epsilon \varepsilon }{}^{\mu \nu } R^{\alpha \beta \gamma 
\delta } \nabla^{\kappa }F_{\gamma \mu }{}^{\sigma \lambda } 
\nabla_{\lambda }F_{\delta \nu \sigma \kappa } + \nn\\&& m_{526}^{} F_{
\alpha \beta }{}^{\epsilon \varepsilon } F^{\mu \nu \sigma 
\lambda } R^{\alpha \beta \gamma \delta } \nabla^{\kappa 
}F_{\gamma \delta \mu \nu } \nabla_{\lambda }F_{\epsilon 
\varepsilon \sigma \kappa } + \nn\\&& m_{577}^{} F_{\alpha \beta 
}{}^{\epsilon \varepsilon } F_{\gamma }{}^{\mu \nu \sigma } R^{
\alpha \beta \gamma \delta } \nabla^{\kappa }F_{\delta \mu 
\nu }{}^{\lambda } \nabla_{\lambda }F_{\epsilon \varepsilon 
\sigma \kappa } + \nn\\&& m_{355}^{} F_{\alpha }{}^{\epsilon \varepsilon 
\mu } F_{\epsilon }{}^{\nu \sigma \lambda } R^{\alpha \beta 
\gamma \delta } \nabla_{\delta }F_{\beta \gamma \nu 
}{}^{\kappa } \nabla_{\lambda }F_{\varepsilon \mu \sigma 
\kappa } + \nn\\&& m_{356}^{} F_{\alpha }{}^{\epsilon \varepsilon \mu } 
F_{\gamma }{}^{\nu \sigma \lambda } R^{\alpha \beta \gamma 
\delta } \nabla_{\delta }F_{\beta \epsilon \nu }{}^{\kappa } 
\nabla_{\lambda }F_{\varepsilon \mu \sigma \kappa } + 
\nn\\&& m_{516}^{} F_{\alpha \beta }{}^{\epsilon \varepsilon } F^{\mu 
\nu \sigma \lambda } R^{\alpha \beta \gamma \delta } \nabla^{
\kappa }F_{\gamma \delta \epsilon \mu } \nabla_{\lambda 
}F_{\varepsilon \nu \sigma \kappa } + \nn\\&& m_{530}^{} F_{\alpha 
\beta }{}^{\epsilon \varepsilon } F_{\epsilon }{}^{\mu \nu 
\sigma } R^{\alpha \beta \gamma \delta } \nabla^{\kappa 
}F_{\gamma \delta \mu }{}^{\lambda } \nabla_{\lambda 
}F_{\varepsilon \nu \sigma \kappa } + \nn\\&& m_{572}^{} F_{\alpha 
\beta }{}^{\epsilon \varepsilon } F_{\gamma }{}^{\mu \nu 
\sigma } R^{\alpha \beta \gamma \delta } \nabla^{\kappa 
}F_{\delta \epsilon \mu }{}^{\lambda } \nabla_{\lambda 
}F_{\varepsilon \nu \sigma \kappa } + \nn\\&& m_{580}^{} F_{\alpha 
\beta }{}^{\epsilon \varepsilon } F_{\gamma \epsilon }{}^{\mu 
\nu } R^{\alpha \beta \gamma \delta } \nabla^{\kappa 
}F_{\delta \mu }{}^{\sigma \lambda } \nabla_{\lambda 
}F_{\varepsilon \nu \sigma \kappa } + \nn\\&& m_{587}^{} F_{\alpha 
\beta }{}^{\epsilon \varepsilon } F_{\gamma \delta }{}^{\mu 
\nu } R^{\alpha \beta \gamma \delta } \nabla^{\kappa 
}F_{\epsilon \mu }{}^{\sigma \lambda } \nabla_{\lambda 
}F_{\varepsilon \nu \sigma \kappa } + \nn\\&& m_{361}^{} F_{\alpha }{}^{
\epsilon \varepsilon \mu } F_{\epsilon }{}^{\nu \sigma \lambda 
} R^{\alpha \beta \gamma \delta } \nabla_{\delta }F_{\beta 
\gamma \varepsilon }{}^{\kappa } \nabla_{\lambda }F_{\mu \nu 
\sigma \kappa } + \nn\\&& m_{362}^{} F_{\alpha }{}^{\epsilon \varepsilon 
\mu } F_{\gamma }{}^{\nu \sigma \lambda } R^{\alpha \beta 
\gamma \delta } \nabla_{\delta }F_{\beta \epsilon \varepsilon 
}{}^{\kappa } \nabla_{\lambda }F_{\mu \nu \sigma \kappa } + 
\nn\\&& m_{513}^{} F_{\alpha }{}^{\epsilon \varepsilon \mu } F_{\beta 
}{}^{\nu \sigma \lambda } R^{\alpha \beta \gamma \delta } 
\nabla^{\kappa }F_{\gamma \delta \epsilon \varepsilon } 
\nabla_{\lambda }F_{\mu \nu \sigma \kappa } + \nn\\&& m_{522}^{} 
F_{\alpha }{}^{\epsilon \varepsilon \mu } F_{\beta \epsilon 
}{}^{\nu \sigma } R^{\alpha \beta \gamma \delta } 
\nabla^{\kappa }F_{\gamma \delta \varepsilon }{}^{\lambda } 
\nabla_{\lambda }F_{\mu \nu \sigma \kappa } + \nn\\&& m_{523}^{} 
F_{\alpha \beta }{}^{\epsilon \varepsilon } F_{\epsilon 
}{}^{\mu \nu \sigma } R^{\alpha \beta \gamma \delta } 
\nabla^{\kappa }F_{\gamma \delta \varepsilon }{}^{\lambda } 
\nabla_{\lambda }F_{\mu \nu \sigma \kappa } + \nn\\&& m_{535}^{} 
F_{\alpha }{}^{\epsilon \varepsilon \mu } F_{\beta \epsilon 
\varepsilon }{}^{\nu } R^{\alpha \beta \gamma \delta } 
\nabla^{\kappa }F_{\gamma \delta }{}^{\sigma \lambda } 
\nabla_{\lambda }F_{\mu \nu \sigma \kappa } + \nn\\&& m_{536}^{} 
F_{\alpha \beta }{}^{\epsilon \varepsilon } F_{\epsilon 
\varepsilon }{}^{\mu \nu } R^{\alpha \beta \gamma \delta } 
\nabla^{\kappa }F_{\gamma \delta }{}^{\sigma \lambda } 
\nabla_{\lambda }F_{\mu \nu \sigma \kappa } + \nn\\&& m_{569}^{} 
F_{\alpha \beta }{}^{\epsilon \varepsilon } F_{\gamma }{}^{\mu 
\nu \sigma } R^{\alpha \beta \gamma \delta } \nabla^{\kappa 
}F_{\delta \epsilon \varepsilon }{}^{\lambda } \nabla_{\lambda 
}F_{\mu \nu \sigma \kappa } + \nn\\&& m_{574}^{} F_{\alpha \beta 
}{}^{\epsilon \varepsilon } F_{\gamma \epsilon }{}^{\mu \nu } 
R^{\alpha \beta \gamma \delta } \nabla^{\kappa }F_{\delta 
\varepsilon }{}^{\sigma \lambda } \nabla_{\lambda }F_{\mu \nu 
\sigma \kappa } + \nn\\&& m_{584}^{} F_{\alpha \beta }{}^{\epsilon 
\varepsilon } F_{\gamma \delta }{}^{\mu \nu } R^{\alpha \beta 
\gamma \delta } \nabla^{\kappa }F_{\epsilon \varepsilon 
}{}^{\sigma \lambda } \nabla_{\lambda }F_{\mu \nu \sigma 
\kappa } + \nn\\&& m_{590}^{} F_{\alpha \beta }{}^{\epsilon \varepsilon 
} F_{\gamma \delta \epsilon }{}^{\mu } R^{\alpha \beta \gamma 
\delta } \nabla^{\kappa }F_{\varepsilon }{}^{\nu \sigma 
\lambda } \nabla_{\lambda }F_{\mu \nu \sigma \kappa } + 
\nn\\&& m_{390}^{} F_{\epsilon }{}^{\sigma \lambda \kappa } F^{\epsilon 
\varepsilon \mu \nu } R^{\alpha \beta \gamma \delta } 
\nabla_{\kappa }F_{\gamma \delta \nu \lambda } \nabla_{\mu 
}F_{\alpha \beta \varepsilon \sigma } + \nn\\&& m_{416}^{} F_{\alpha 
}{}^{\epsilon \varepsilon \mu } F^{\nu \sigma \lambda \kappa } 
R^{\alpha \beta \gamma \delta } \nabla_{\kappa }F_{\delta \nu 
\sigma \lambda } \nabla_{\mu }F_{\beta \gamma \epsilon 
\varepsilon } + \nn\\&& m_{345}^{} F_{\alpha }{}^{\epsilon \varepsilon 
\mu } F_{\epsilon }{}^{\nu \sigma \lambda } R^{\alpha \beta 
\gamma \delta } \nabla_{\lambda }F_{\delta \nu \sigma \kappa 
} \nabla_{\mu }F_{\beta \gamma \varepsilon }{}^{\kappa } + 
\nn\\&& m_{346}^{} F_{\alpha }{}^{\epsilon \varepsilon \mu } F_{\gamma 
}{}^{\nu \sigma \lambda } R^{\alpha \beta \gamma \delta } 
\nabla_{\lambda }F_{\delta \nu \sigma \kappa } \nabla_{\mu 
}F_{\beta \epsilon \varepsilon }{}^{\kappa } + \nn\\&& m_{392}^{} 
F_{\alpha }{}^{\epsilon \varepsilon \mu } F^{\nu \sigma 
\lambda \kappa } R^{\alpha \beta \gamma \delta } 
\nabla_{\kappa }F_{\gamma \delta \sigma \lambda } \nabla_{\mu 
}F_{\beta \epsilon \varepsilon \nu } + \nn\\&& m_{331}^{} F_{\alpha 
}{}^{\epsilon \varepsilon \mu } F_{\epsilon }{}^{\nu \sigma 
\lambda } R^{\alpha \beta \gamma \delta } \nabla_{\lambda }F_{
\gamma \delta \sigma \kappa } \nabla_{\mu }F_{\beta 
\varepsilon \nu }{}^{\kappa } + \nn\\&& m_{427}^{} F_{\alpha \beta }{}^{
\epsilon \varepsilon } F^{\mu \nu \sigma \lambda } R^{\alpha 
\beta \gamma \delta } \nabla_{\kappa }F_{\varepsilon \nu 
\sigma \lambda } \nabla_{\mu }F_{\gamma \delta \epsilon 
}{}^{\kappa } + \nn\\&& m_{358}^{} F_{\alpha \beta }{}^{\epsilon 
\varepsilon } F^{\mu \nu \sigma \lambda } R^{\alpha \beta 
\gamma \delta } \nabla_{\lambda }F_{\varepsilon \nu \sigma 
\kappa } \nabla_{\mu }F_{\gamma \delta \epsilon }{}^{\kappa } 
+ \nn\\&& m_{496}^{} F_{\alpha }{}^{\epsilon \varepsilon \mu } 
F_{\epsilon }{}^{\nu \sigma \lambda } R^{\alpha \beta \gamma 
\delta } \nabla^{\kappa }F_{\beta \nu \sigma \lambda } 
\nabla_{\mu }F_{\gamma \delta \varepsilon \kappa } + \nn\\&& m_{487}^{} 
F_{\alpha }{}^{\epsilon \varepsilon \mu } F_{\epsilon }{}^{\nu 
\sigma \lambda } R^{\alpha \beta \gamma \delta } 
\nabla^{\kappa }F_{\beta \varepsilon \nu \sigma } \nabla_{\mu 
}F_{\gamma \delta \lambda \kappa } + \nn\\&& m_{500}^{} F_{\alpha 
}{}^{\epsilon \varepsilon \mu } F_{\epsilon \varepsilon 
}{}^{\nu \sigma } R^{\alpha \beta \gamma \delta } 
\nabla^{\kappa }F_{\beta \nu \sigma }{}^{\lambda } 
\nabla_{\mu }F_{\gamma \delta \lambda \kappa } + \nn\\&& m_{428}^{} F_{
\alpha \beta }{}^{\epsilon \varepsilon } F_{\epsilon }{}^{\mu 
\nu \sigma } R^{\alpha \beta \gamma \delta } \nabla_{\kappa 
}F_{\varepsilon \nu \sigma \lambda } \nabla_{\mu }F_{\gamma 
\delta }{}^{\lambda \kappa } + \nn\\&& m_{417}^{} F_{\alpha \beta 
}{}^{\epsilon \varepsilon } F^{\mu \nu \sigma \lambda } 
R^{\alpha \beta \gamma \delta } \nabla_{\kappa }F_{\delta \nu 
\sigma \lambda } \nabla_{\mu }F_{\gamma \epsilon \varepsilon 
}{}^{\kappa } + \nn\\&& m_{347}^{} F_{\alpha }{}^{\epsilon \varepsilon 
\mu } F_{\beta }{}^{\nu \sigma \lambda } R^{\alpha \beta 
\gamma \delta } \nabla_{\lambda }F_{\delta \nu \sigma \kappa 
} \nabla_{\mu }F_{\gamma \epsilon \varepsilon }{}^{\kappa } + 
\nn\\&& m_{348}^{} F_{\alpha \beta }{}^{\epsilon \varepsilon } F^{\mu 
\nu \sigma \lambda } R^{\alpha \beta \gamma \delta } \nabla_{
\lambda }F_{\delta \nu \sigma \kappa } \nabla_{\mu }F_{\gamma 
\epsilon \varepsilon }{}^{\kappa } + \nn\\&& m_{497}^{} F_{\alpha 
}{}^{\epsilon \varepsilon \mu } F_{\gamma }{}^{\nu \sigma 
\lambda } R^{\alpha \beta \gamma \delta } \nabla^{\kappa 
}F_{\beta \nu \sigma \lambda } \nabla_{\mu }F_{\delta 
\epsilon \varepsilon \kappa } + \nn\\&& m_{429}^{} F_{\alpha \beta }{}^{
\epsilon \varepsilon } F_{\gamma }{}^{\mu \nu \sigma } 
R^{\alpha \beta \gamma \delta } \nabla_{\kappa }F_{\varepsilon 
\nu \sigma \lambda } \nabla_{\mu }F_{\delta \epsilon 
}{}^{\lambda \kappa } + \nn\\&& m_{288}^{} F_{\alpha }{}^{\epsilon 
\varepsilon \mu } F^{\nu \sigma \lambda \kappa } R^{\alpha 
\beta \gamma \delta } \nabla_{\gamma }F_{\beta \epsilon \nu 
\sigma } \nabla_{\mu }F_{\delta \varepsilon \lambda \kappa } + 
\nn\\&& m_{482}^{} F_{\alpha }{}^{\epsilon \varepsilon \mu } F_{\gamma 
}{}^{\nu \sigma \lambda } R^{\alpha \beta \gamma \delta } 
\nabla^{\kappa }F_{\beta \epsilon \nu \sigma } \nabla_{\mu 
}F_{\delta \varepsilon \lambda \kappa } + \nn\\&& m_{501}^{} F_{\alpha 
}{}^{\epsilon \varepsilon \mu } F_{\gamma \epsilon }{}^{\nu 
\sigma } R^{\alpha \beta \gamma \delta } \nabla^{\kappa 
}F_{\beta \nu \sigma }{}^{\lambda } \nabla_{\mu }F_{\delta 
\varepsilon \lambda \kappa } + \nn\\&& m_{289}^{} F_{\alpha 
}{}^{\epsilon \varepsilon \mu } F^{\nu \sigma \lambda \kappa } 
R^{\alpha \beta \gamma \delta } \nabla_{\gamma }F_{\beta 
\epsilon \varepsilon \nu } \nabla_{\mu }F_{\delta \sigma 
\lambda \kappa } + \nn\\&& m_{290}^{} F_{\alpha }{}^{\epsilon 
\varepsilon \mu } F_{\epsilon }{}^{\nu \sigma \lambda } 
R^{\alpha \beta \gamma \delta } \nabla_{\gamma }F_{\beta 
\varepsilon \nu }{}^{\kappa } \nabla_{\mu }F_{\delta \sigma 
\lambda \kappa } + \nn\\&& m_{291}^{} F_{\alpha }{}^{\epsilon 
\varepsilon \mu } F_{\epsilon \varepsilon }{}^{\nu \sigma } R^{
\alpha \beta \gamma \delta } \nabla_{\gamma }F_{\beta \nu 
}{}^{\lambda \kappa } \nabla_{\mu }F_{\delta \sigma \lambda 
\kappa } + \nn\\&& m_{506}^{} F_{\alpha }{}^{\epsilon \varepsilon \mu } 
F_{\gamma \epsilon \varepsilon }{}^{\nu } R^{\alpha \beta 
\gamma \delta } \nabla^{\kappa }F_{\beta \nu }{}^{\sigma 
\lambda } \nabla_{\mu }F_{\delta \sigma \lambda \kappa } + 
\nn\\&& m_{543}^{} F_{\alpha }{}^{\epsilon \varepsilon \mu } F_{\beta 
}{}^{\nu \sigma \lambda } R^{\alpha \beta \gamma \delta } 
\nabla^{\kappa }F_{\gamma \epsilon \varepsilon \nu } 
\nabla_{\mu }F_{\delta \sigma \lambda \kappa } + \nn\\&& m_{393}^{} F_{
\epsilon }{}^{\sigma \lambda \kappa } F^{\epsilon \varepsilon 
\mu \nu } R^{\alpha \beta \gamma \delta } \nabla_{\kappa }F_{
\gamma \delta \sigma \lambda } \nabla_{\nu }F_{\alpha \beta 
\varepsilon \mu } + \nn\\&& m_{332}^{} F_{\epsilon \varepsilon 
}{}^{\sigma \lambda } F^{\epsilon \varepsilon \mu \nu } 
R^{\alpha \beta \gamma \delta } \nabla_{\lambda }F_{\gamma 
\delta \sigma \kappa } \nabla_{\nu }F_{\alpha \beta \mu }{}^{
\kappa } + \nn\\&& m_{409}^{} F_{\alpha }{}^{\epsilon \varepsilon \mu } 
F^{\nu \sigma \lambda \kappa } R^{\alpha \beta \gamma \delta 
} \nabla_{\kappa }F_{\delta \mu \sigma \lambda } \nabla_{\nu 
}F_{\beta \gamma \epsilon \varepsilon } + \nn\\&& m_{341}^{} F_{\alpha 
}{}^{\epsilon \varepsilon \mu } F_{\epsilon }{}^{\nu \sigma 
\lambda } R^{\alpha \beta \gamma \delta } \nabla_{\lambda }F_{
\delta \mu \sigma \kappa } \nabla_{\nu }F_{\beta \gamma 
\varepsilon }{}^{\kappa } + \nn\\&& m_{425}^{} F_{\alpha }{}^{\epsilon 
\varepsilon \mu } F_{\beta }{}^{\nu \sigma \lambda } R^{\alpha 
\beta \gamma \delta } \nabla_{\kappa }F_{\varepsilon \mu 
\sigma \lambda } \nabla_{\nu }F_{\gamma \delta \epsilon 
}{}^{\kappa } + \nn\\&& m_{357}^{} F_{\alpha }{}^{\epsilon \varepsilon 
\mu } F_{\beta }{}^{\nu \sigma \lambda } R^{\alpha \beta 
\gamma \delta } \nabla_{\lambda }F_{\varepsilon \mu \sigma 
\kappa } \nabla_{\nu }F_{\gamma \delta \epsilon }{}^{\kappa } 
+ \nn\\&& m_{420}^{} F_{\alpha \beta }{}^{\epsilon \varepsilon } F^{\mu 
\nu \sigma \lambda } R^{\alpha \beta \gamma \delta } \nabla_{
\kappa }F_{\epsilon \varepsilon \sigma \lambda } \nabla_{\nu 
}F_{\gamma \delta \mu }{}^{\kappa } + \nn\\&& m_{352}^{} F_{\alpha 
\beta }{}^{\epsilon \varepsilon } F^{\mu \nu \sigma \lambda } 
R^{\alpha \beta \gamma \delta } \nabla_{\lambda }F_{\epsilon 
\varepsilon \sigma \kappa } \nabla_{\nu }F_{\gamma \delta \mu 
}{}^{\kappa } + \nn\\&& m_{342}^{} F_{\alpha }{}^{\epsilon \varepsilon 
\mu } F_{\beta }{}^{\nu \sigma \lambda } R^{\alpha \beta 
\gamma \delta } \nabla_{\lambda }F_{\delta \mu \sigma \kappa 
} \nabla_{\nu }F_{\gamma \epsilon \varepsilon }{}^{\kappa } + 
\nn\\&& m_{336}^{} F_{\alpha \beta }{}^{\epsilon \varepsilon } F^{\mu 
\nu \sigma \lambda } R^{\alpha \beta \gamma \delta } \nabla_{
\lambda }F_{\delta \varepsilon \sigma \kappa } \nabla_{\nu 
}F_{\gamma \epsilon \mu }{}^{\kappa } + \nn\\&& m_{297}^{} F_{\epsilon 
\varepsilon }{}^{\sigma \lambda } F^{\epsilon \varepsilon \mu 
\nu } R^{\alpha \beta \gamma \delta } \nabla_{\gamma 
}F_{\alpha \beta \mu }{}^{\kappa } \nabla_{\nu }F_{\delta 
\sigma \lambda \kappa } + \nn\\&& m_{298}^{} F_{\alpha }{}^{\epsilon 
\varepsilon \mu } F_{\epsilon \varepsilon \mu }{}^{\nu } 
R^{\alpha \beta \gamma \delta } \nabla_{\gamma }F_{\beta }{}^{
\sigma \lambda \kappa } \nabla_{\nu }F_{\delta \sigma \lambda 
\kappa } + \nn\\&& m_{562}^{} F_{\alpha }{}^{\epsilon \varepsilon \mu } 
F_{\beta \epsilon \varepsilon }{}^{\nu } R^{\alpha \beta 
\gamma \delta } \nabla^{\kappa }F_{\gamma \mu }{}^{\sigma 
\lambda } \nabla_{\nu }F_{\delta \sigma \lambda \kappa } + 
\nn\\&& m_{563}^{} F_{\alpha \beta }{}^{\epsilon \varepsilon } 
F_{\epsilon \varepsilon }{}^{\mu \nu } R^{\alpha \beta \gamma 
\delta } \nabla^{\kappa }F_{\gamma \mu }{}^{\sigma \lambda } 
\nabla_{\nu }F_{\delta \sigma \lambda \kappa } + \nn\\&& m_{299}^{} F_{
\alpha }{}^{\epsilon \varepsilon \mu } F_{\beta \epsilon 
\varepsilon }{}^{\nu } R^{\alpha \beta \gamma \delta } 
\nabla_{\mu }F_{\gamma }{}^{\sigma \lambda \kappa } 
\nabla_{\nu }F_{\delta \sigma \lambda \kappa } + \nn\\&& m_{300}^{} F_{
\alpha \beta }{}^{\epsilon \varepsilon } F_{\epsilon 
\varepsilon }{}^{\mu \nu } R^{\alpha \beta \gamma \delta } 
\nabla_{\mu }F_{\gamma }{}^{\sigma \lambda \kappa } 
\nabla_{\nu }F_{\delta \sigma \lambda \kappa } + \nn\\&& m_{579}^{} F_{
\alpha \beta }{}^{\epsilon \varepsilon } F_{\gamma \epsilon 
}{}^{\mu \nu } R^{\alpha \beta \gamma \delta } \nabla^{\kappa 
}F_{\delta \mu }{}^{\sigma \lambda } \nabla_{\nu 
}F_{\varepsilon \sigma \lambda \kappa } + \nn\\&& m_{301}^{} F_{\alpha 
}{}^{\epsilon \varepsilon \mu } F_{\gamma \epsilon \varepsilon 
}{}^{\nu } R^{\alpha \beta \gamma \delta } \nabla_{\delta }F_{
\beta }{}^{\sigma \lambda \kappa } \nabla_{\nu }F_{\mu \sigma 
\lambda \kappa } + \nn\\&& m_{391}^{} F_{\epsilon }{}^{\sigma \lambda 
\kappa } F^{\epsilon \varepsilon \mu \nu } R^{\alpha \beta 
\gamma \delta } \nabla_{\kappa }F_{\gamma \delta \nu \lambda 
} \nabla_{\sigma }F_{\alpha \beta \varepsilon \mu } + 
\nn\\&& m_{404}^{} F_{\alpha }{}^{\epsilon \varepsilon \mu } F^{\nu 
\sigma \lambda \kappa } R^{\alpha \beta \gamma \delta } 
\nabla_{\kappa }F_{\delta \varepsilon \mu \lambda } 
\nabla_{\sigma }F_{\beta \gamma \epsilon \nu } + \nn\\&& m_{474}^{} F_{
\epsilon \varepsilon \mu }{}^{\sigma } F^{\epsilon \varepsilon 
\mu \nu } R^{\alpha \beta \gamma \delta } \nabla^{\kappa }F_{
\alpha \gamma \nu }{}^{\lambda } \nabla_{\sigma }F_{\beta 
\delta \lambda \kappa } + \nn\\&& m_{493}^{} F_{\alpha }{}^{\epsilon 
\varepsilon \mu } F_{\epsilon \varepsilon }{}^{\nu \sigma } R^{
\alpha \beta \gamma \delta } \nabla^{\kappa }F_{\beta \mu 
\nu }{}^{\lambda } \nabla_{\sigma }F_{\gamma \delta \lambda 
\kappa } + \nn\\&& m_{302}^{} F_{\alpha }{}^{\epsilon \varepsilon \mu } 
F_{\epsilon \varepsilon }{}^{\nu \sigma } R^{\alpha \beta 
\gamma \delta } \nabla_{\mu }F_{\beta \nu }{}^{\lambda \kappa 
} \nabla_{\sigma }F_{\gamma \delta \lambda \kappa } + 
\nn\\&& m_{303}^{} F_{\epsilon \varepsilon \mu }{}^{\sigma } F^{\epsilon 
\varepsilon \mu \nu } R^{\alpha \beta \gamma \delta } 
\nabla_{\nu }F_{\alpha \beta }{}^{\lambda \kappa } 
\nabla_{\sigma }F_{\gamma \delta \lambda \kappa } + \nn\\&& m_{304}^{} 
F_{\alpha }{}^{\epsilon \varepsilon \mu } F_{\epsilon 
\varepsilon }{}^{\nu \sigma } R^{\alpha \beta \gamma \delta } 
\nabla_{\gamma }F_{\beta \nu }{}^{\lambda \kappa } 
\nabla_{\sigma }F_{\delta \mu \lambda \kappa } + \nn\\&& m_{305}^{} F_{
\alpha }{}^{\epsilon \varepsilon \mu } F_{\gamma \epsilon }{}^{
\nu \sigma } R^{\alpha \beta \gamma \delta } 
\nabla_{\varepsilon }F_{\beta \nu }{}^{\lambda \kappa } 
\nabla_{\sigma }F_{\delta \mu \lambda \kappa } + \nn\\&& m_{481}^{} F_{
\alpha }{}^{\epsilon \varepsilon \mu } F_{\epsilon \varepsilon 
}{}^{\nu \sigma } R^{\alpha \beta \gamma \delta } 
\nabla^{\kappa }F_{\beta \gamma \nu }{}^{\lambda } 
\nabla_{\sigma }F_{\delta \mu \lambda \kappa } + \nn\\&& m_{490}^{} F_{
\alpha }{}^{\epsilon \varepsilon \mu } F_{\gamma \epsilon }{}^{
\nu \sigma } R^{\alpha \beta \gamma \delta } \nabla^{\kappa 
}F_{\beta \varepsilon \nu }{}^{\lambda } \nabla_{\sigma 
}F_{\delta \mu \lambda \kappa } + \nn\\&& m_{557}^{} F_{\alpha 
}{}^{\epsilon \varepsilon \mu } F_{\beta \epsilon }{}^{\nu 
\sigma } R^{\alpha \beta \gamma \delta } \nabla^{\kappa 
}F_{\gamma \varepsilon \nu }{}^{\lambda } \nabla_{\sigma 
}F_{\delta \mu \lambda \kappa } + \nn\\&& m_{306}^{} F_{\epsilon 
\varepsilon \mu }{}^{\sigma } F^{\epsilon \varepsilon \mu \nu 
} R^{\alpha \beta \gamma \delta } \nabla_{\gamma }F_{\alpha 
\beta }{}^{\lambda \kappa } \nabla_{\sigma }F_{\delta \nu 
\lambda \kappa } + \nn\\&& m_{307}^{} F_{\alpha }{}^{\epsilon 
\varepsilon \mu } F_{\epsilon \varepsilon }{}^{\nu \sigma } R^{
\alpha \beta \gamma \delta } \nabla_{\gamma }F_{\beta \mu 
}{}^{\lambda \kappa } \nabla_{\sigma }F_{\delta \nu \lambda 
\kappa } + \nn\\&& m_{550}^{} F_{\alpha }{}^{\epsilon \varepsilon \mu } 
F_{\beta \epsilon }{}^{\nu \sigma } R^{\alpha \beta \gamma 
\delta } \nabla^{\kappa }F_{\gamma \varepsilon \mu 
}{}^{\lambda } \nabla_{\sigma }F_{\delta \nu \lambda \kappa } 
+ \nn\\&& m_{551}^{} F_{\alpha \beta }{}^{\epsilon \varepsilon } 
F_{\epsilon }{}^{\mu \nu \sigma } R^{\alpha \beta \gamma 
\delta } \nabla^{\kappa }F_{\gamma \varepsilon \mu 
}{}^{\lambda } \nabla_{\sigma }F_{\delta \nu \lambda \kappa } 
+ \nn\\&& m_{308}^{} F_{\alpha }{}^{\epsilon \varepsilon \mu } 
F_{\epsilon \varepsilon }{}^{\nu \sigma } R^{\alpha \beta 
\gamma \delta } \nabla_{\mu }F_{\beta \gamma }{}^{\lambda 
\kappa } \nabla_{\sigma }F_{\delta \nu \lambda \kappa } + 
\nn\\&& m_{309}^{} F_{\alpha }{}^{\epsilon \varepsilon \mu } F_{\beta 
\epsilon }{}^{\nu \sigma } R^{\alpha \beta \gamma \delta } 
\nabla_{\mu }F_{\gamma \varepsilon }{}^{\lambda \kappa } 
\nabla_{\sigma }F_{\delta \nu \lambda \kappa } + \nn\\&& m_{310}^{} F_{
\alpha \beta }{}^{\epsilon \varepsilon } F_{\epsilon }{}^{\mu 
\nu \sigma } R^{\alpha \beta \gamma \delta } \nabla_{\mu }F_{
\gamma \varepsilon }{}^{\lambda \kappa } \nabla_{\sigma 
}F_{\delta \nu \lambda \kappa } + \nn\\&& m_{576}^{} F_{\alpha \beta 
}{}^{\epsilon \varepsilon } F_{\gamma }{}^{\mu \nu \sigma } R^{
\alpha \beta \gamma \delta } \nabla^{\kappa }F_{\delta \mu 
\nu }{}^{\lambda } \nabla_{\sigma }F_{\epsilon \varepsilon 
\lambda \kappa } + \nn\\&& m_{529}^{} F_{\alpha \beta }{}^{\epsilon 
\varepsilon } F_{\epsilon }{}^{\mu \nu \sigma } R^{\alpha 
\beta \gamma \delta } \nabla^{\kappa }F_{\gamma \delta \mu 
}{}^{\lambda } \nabla_{\sigma }F_{\varepsilon \nu \lambda 
\kappa } + \nn\\&& m_{571}^{} F_{\alpha \beta }{}^{\epsilon \varepsilon 
} F_{\gamma }{}^{\mu \nu \sigma } R^{\alpha \beta \gamma 
\delta } \nabla^{\kappa }F_{\delta \epsilon \mu }{}^{\lambda } 
\nabla_{\sigma }F_{\varepsilon \nu \lambda \kappa } + 
\nn\\&& m_{312}^{} F_{\alpha }{}^{\epsilon \varepsilon \mu } F_{\epsilon 
\varepsilon }{}^{\nu \sigma } R^{\alpha \beta \gamma \delta } 
\nabla_{\delta }F_{\beta \gamma }{}^{\lambda \kappa } \nabla_{
\sigma }F_{\mu \nu \lambda \kappa } + \nn\\&& m_{313}^{} F_{\alpha 
}{}^{\epsilon \varepsilon \mu } F_{\gamma \epsilon }{}^{\nu 
\sigma } R^{\alpha \beta \gamma \delta } \nabla_{\delta 
}F_{\beta \varepsilon }{}^{\lambda \kappa } \nabla_{\sigma }F_{
\mu \nu \lambda \kappa } + \nn\\&& m_{521}^{} F_{\alpha }{}^{\epsilon 
\varepsilon \mu } F_{\beta \epsilon }{}^{\nu \sigma } 
R^{\alpha \beta \gamma \delta } \nabla^{\kappa }F_{\gamma 
\delta \varepsilon }{}^{\lambda } \nabla_{\sigma }F_{\mu \nu 
\lambda \kappa }+ \nn\\&& m_{281}^{} 
FF R^{\alpha \beta \gamma \delta } \nabla_{\gamma }F_{\alpha 
}{}^{\sigma \lambda \kappa } \nabla_{\delta }F_{\beta \sigma 
\lambda \kappa }+ \nn\\&& m_{478}^{} FF R^{\alpha \beta \gamma \delta } 
\nabla_{\kappa }F_{\beta \delta \sigma \lambda } 
\nabla^{\kappa }F_{\alpha \gamma }{}^{\sigma \lambda }+ \nn\\&& m_{477}^{} FF R^{\alpha \beta \gamma \delta } 
\nabla^{\kappa }F_{\alpha \gamma }{}^{\sigma \lambda } 
\nabla_{\lambda }F_{\beta \delta \sigma \kappa }\labell{T7}
\eeqa
where $FF=F_{\mu\nu\alpha\beta}F^{\mu\nu\alpha\beta}$. Note that the number of all contractions of $F^8$ without imposing the field redefinition is 176. The are  104 couplings in \reef{T1} that their coefficients are unambiguous. The coefficients of the couplings in \reef{T1} which have $FF$ or $F_{\mu\alpha\beta\gamma}F_{\nu}{}^{\alpha\beta\gamma}$ are essential parameters.

\end{document}